\documentclass[11pt, a4paper, oneside]{Thesis} 

\graphicspath{{Pictures/}} 

\usepackage[square,sort,compress,numbers]{natbib}
\usepackage{cite}
\usepackage{amsmath,latexsym} 
\usepackage{float}
\usepackage[bottom]{footmisc}
\renewcommand{\thefootnote}{\arabic{footnote}}
\newtheorem*{theorem*}{\bf{Theorem}}
\usepackage{mathtools}
\usepackage{tasks}
\usepackage{acronym}
\usepackage{nomencl}
\usepackage{caption}

\title{\ttitle} 
\DeclareUnicodeCharacter{2212}{-}
\DeclareUnicodeCharacter{2212}{+}
\newcommand\blfootnote[1]{%
     \begingroup
     \renewcommand\thefootnote{}\footnote{#1}%
     \addtocounter{footnote}{-1}%
      \endgroup
    }
\begin{document}

\setstretch{1.7} 

\fancyhead{} 
\rhead{\thepage} 
\lhead{} 

%

\thesistitle{Late Time Acceleration with Observational Constraints in Modified Theories of Gravity}
\documenttype{Thesis}
\supervisor{Prof. Pradyumn Kumar Sahoo}
\supervisorposition{Professor}
\supervisorinstitute{BITS-Pilani, Hyderabad Campus}
\examiner{}
\degree{Ph.D. Research Scholar}
\coursecode{DOCTOR OF PHILOSOPHY}
\coursename{Thesis}
\authors{\textbf{SIMRAN ARORA}}
\IDNumber{2019PHXF0438H}
\addresses{}
\subject{}
\university{\texorpdfstring{\href{http://www.bits-pilani.ac.in/} 
                {Birla Institute of Technology and Science, Pilani}} 
                {Birla Institute of Technology and Science, Pilani}}
\UNIVERSITY{\texorpdfstring{\href{http://www.bits-pilani.ac.in/} 
                {BIRLA INSTITUTE OF TECHNOLOGY AND SCIENCE, PILANI}} 
                {BIRLA INSTITUTE OF TECHNOLOGY AND SCIENCE, PILANI}}



\department{\texorpdfstring{\href{http://www.bits-pilani.ac.in/pilani/Mathematics/Mathematics} 
                {Mathematics}} 
                {Mathematics}}
\DEPARTMENT{\texorpdfstring{\href{http://www.bits-pilani.ac.in/pilani/Mathematics/Mathematics} 
                {Mathematics}} 
                {Mathematics}}
\group{\texorpdfstring{\href{Research Group Web Site URL Here (include http://)}
                {Research Group Name}} 
                {Research Group Name}}
\GROUP{\texorpdfstring{\href{Research Group Web Site URL Here (include http://)}
                {RESEARCH GROUP NAME (IN BLOCK CAPITALS)}}
                {RESEARCH GROUP NAME (IN BLOCK CAPITALS)}}
\faculty{\texorpdfstring{\href{Faculty Web Site URL Here (include http://)}
                {Faculty Name}}
                {Faculty Name}}
\FACULTY{\texorpdfstring{\href{Faculty Web Site URL Here (include http://)}
                {FACULTY NAME (IN BLOCK CAPITALS)}}
                {FACULTY NAME (IN BLOCK CAPITALS)}}

\maketitle

\clearpage
\setstretch{1.3} 

\pagestyle{empty} 
\pagenumbering{gobble}


\addtocontents{toc}{\vspace{2em}} 

\frontmatter 
\Certificate

 \Declaration

\begin{acknowledgements}

Without the continuous guidance and support of numerous people, this journey would not be possible. This is to express my special thanks to them.

I would like to express my sincere thanks and gratitude to my supervisor, \textbf{Prof. Pradyumn Kumar Sahoo}, Professor, Department of Mathematics, BITS-Pilani, Hyderabad Campus, Hyderabad, for his unwavering support, guidance, and vast experience throughout my Ph.D. work. I am grateful to him for his endless patience and faith in me, which encouraged me to complete this thesis by developing a deeper understanding of the subject and an aptitude for scientific research. 

I sincerely thank my Doctoral Advisory Committee (DAC) members, \textbf{Prof. Bivudutta Mishra} and \textbf{Prof. K. Venkata Ratnam} for their guidance and valuable suggestions to improve my research work. 

I am privileged to extend my gratitude to the \textbf{Head} of the Mathematics Department, the \textbf{DRC} convener, \textbf{faculty} members, and my \textbf{colleagues} for their help, support, and encouragement in this amazing journey. I convey special thanks to my co-authors for their valuable discussions, suggestions, and collaborations. 

I acknowledge \textbf{BITS-Pilani, Hyderabad Campus}, for providing financial support and necessary facilities, and the Council of Scientific \& Industrial Research \textbf{(CSIR)}, India, for a sponsored research grant (No.03(1454)/19/EMR-II Dt.02/08/2019) that practically enabled the realization of my research. 

Finally, I thank my \textbf{parents} and \textbf{grandparents} for their unconditional love, support, and encouragement and for always being with me. I also thank \textbf{Mam Smrutirekha Sahoo} for her support during the COVID-19 lockdown. 

Last but not least, my heartiest thanks to my friends \textbf{Mridul Patel}, \textbf{Anmol Pruthi}, \textbf{Gaurav Gadbail}, \textbf{Gitika Juneja}, and \textbf{Sanjay Mandal} for their wholehearted support over the years.  

\vspace{1.4 cm}
Simran Arora,\\
ID: 2019PHXF0438H.

\end{acknowledgements}

\begin{abstract}
The late time acceleration of the Universe has challenged contemporary cosmology since its discovery. General Relativity explains this phenomenon by introducing the cosmological constant, named the standard cosmological model ($\Lambda$CDM). However, the cosmological constant solution has several drawbacks that have led cosmologists to explore and propose alternative models to explain the late time acceleration of the Universe. These alternatives span from models of a dynamical dark fluid, known as ``dark energy", to models of large-scale modifications of the gravitational interaction, known as ``modified gravity".

The first chapter briefly introduces background formulation, fundamental gravity theories, and cosmological observations. In chapters \ref{Chapter2}-\ref{Chapter5}, we investigate the dark sector of the Universe in modified gravity using Markov Chain Monte Carlo (MCMC) methods and large datasets derived from measurements of the background expansion of the Universe.

Chapter \ref{Chapter2} discusses the acceleration of the Universe by incorporating bulk viscosity in $f(R,\mathcal{T})$ gravity. Incorporating bulk viscosity into the $f(R,\mathcal{T})$ gravity model violated the strong energy condition describing the accelerated expansion. In chapters \ref{Chapter3} and \ref{Chapter4}, we examine the theoretical viability of $f(Q,\mathcal{T})$ gravity. We investigate $f(Q,\mathcal{T})$ gravity using the matter-dominated Universe and the effective equation of state. To achieve this, we constrain the two models with the Hubble dataset, Union 2.1 and Pantheon supernovae datasets, and the BAO dataset with the analyses of numerous cosmological parameters. The study indicates whether the $f(Q,\mathcal{T})$ gravity models are supported by the observational data in comparison to the $\Lambda$CDM scenario. The reconstructed models of dark energy exhibit accelerating behavior and deviate from the $\Lambda$CDM at certain redshifts.

In chapter \ref{Chapter5}, we analyze the exponential $f(Q)$ gravity to examine the formation of structures and the viable cosmology. The study aims to reproduce feasible results within $f(Q)$ gravity using MCMC constraints and N-body + SPH simulations. We deduce CDM+baryons over density/temperature/mean molecular weight fields, matter power spectrum, bispectrum, two-point correlation function, and halo mass function. Therefore, the outcomes for small and large simulation boxes are appropriately compared. Chapter \ref{Chapter6} finishes with concluding remarks and a discussion of the thesis with an eye toward the future.  
\end{abstract}

\Dedicatory{\bf \begin{LARGE}
Dedicated to
\end{LARGE} 
\\
\vspace{0.2cm}
\it Mom, Dad and Brother\\}



\lhead{\emph{Contents}} 
\tableofcontents 
\addtocontents{toc}{\vspace{1em}}
\lhead{\emph{List of Tables}}
\listoftables 
\addtocontents{toc}{\vspace{1em}}
\lhead{\emph{List of Figures}}
\listoffigures 
\addtocontents{toc}{\vspace{1em}}




\lhead{\emph{List of Symbols and Abbreviations}}
\listofsymbols{ll}{
$\mathbb{R}^{n}:$ \,\,\,\,\,\,\,n-dimensional real space\\
$\otimes:$\,\,\,\,\,\,\,\,\,\,\,Tensor product\\
$g_{ij}:$ \,\,\,\,\,\,\,Lorentzian metric\\
$g:$ \,\,\,\,\,\,\,\,\,\,\,Determinant of $g_{ij}$\\
$\tilde{\Gamma}^{\sigma}_{\,\,ij}:$ \,\,\,\,General affine connection\\
$\Gamma^{\sigma}_{\,\,ij}:$\,\,\,\,\,\,\,Levi-Civita connection\\
$\nabla_{i}$: \,\,\,\,\,\,\,\,Covariant derivative w.r.t. Levi-Civita connection\\
$\square$\,:\,\,\,\,\,\,\,\,\,\, D'Alembert operator\\
$R^{\sigma}_{\,\,\lambda ij}:$ \,\,Riemann tensor \\
$R_{ij}:$\,\,\,\,\,\,\,\,Ricci tensor \\
$R:$ \,\,\,\,\, \,\,\,Ricci scalar \\
$z:$\,\,\,\,\, \,\,\,\,\,\,\,Redshift\\
$q:$\,\,\,\,\, \,\,\,\,\,\,\,Deceleration parameter\\
$T:$\,\,\,\,\, \,\,\,\,\,\,Torsion\\
$Q:$\,\,\,\,\, \,\,\,\,\,\,Non-metricity\\
$\Lambda:$\,\,\,\,\, \,\,\,\,\,\,Cosmological constant\\
$T_{ij}:$ \,\,\,\,\,\,\,\,Stress-energy tensor\\
$G:$\,\,\,\,\, \,\,\, \,Newton's gravitational constant\\
$\Omega:$\,\,\,\,\, \,\,\,\,\,\,Density parameter\\
$\chi^{2}:$\,\,\,\,\, \,\,\,\,\,Chi-Square\\
GR:\,\,\,\,\, \,\,\,\,General relativity\\
$\Lambda$CDM:\,\,$\Lambda$ Cold dark matter\\
ECs:\,\,\,\,\,\,\,\,\,\,Energy conditions\\
EoS:\,\,\,\,\,\,\,\,\,\,Equation of state\\
SNeIa:\,\,\,\,\,Type Ia supernovae\\
CMB: \,\,\,\,\,Cosmic microwave background\\
OHD:\,\,\,\,\, Observational Hubble data\\
BAO: \,\, \,\,Baryon acoustic oscillations\\
MCMC:\,\,Markov chain Monte Carlo\\
LSS:\,\,\,\,\, \,\,\,\,\,Large scale structure\\
SPH:\,\,\,\,\, \,\,\,\,Smoothed particle hydrodynamics\\
HMF:\,\,\,\,\, \,\,Halo mass function\\
RSD:\,\,\,\,\, \,\,\,\,Redshift space distortion\\
DE:\,\,\,\,\,\,\,\,\,\,\,\,\,\,Dark energy\\
DM:\,\,\,\,\,\,\,\,\,\,\,\,\,Dark matter\\
MG:\,\,\,\,\,\,\,\,\,\,\,\,Modified gravity\\
TEGR:\,\,\,\,\,Teleparallel equivalent to GR\\
STEGR:\,\,\,Symmetric teleparallel equivalent to GR\\
CAMB:\,\,\,\,\,Code for anisotropies in the microwave background\\
}

\addtocontents{toc}{\vspace{2em}}

%
%


\clearpage 





\mainmatter 

\pagestyle{fancy} 


 \chapter{Introduction and Theoretical Background} 
\label{Chapter1}

\lhead{Chapter 1. \emph{Introduction and Theoretical Background}} 


\section{General Introduction and Motivation}

Cosmology is the large-scale scientific study of the Universe, its elements, and its past and future. In the past two decades, cosmology has advanced significantly as a science, with unexpectedly fast-paced information regarding the creation, structure, and evolution of the Universe. Einstein's general theory of relativity (GR) is a well-known classic field theory of gravity. It incorporates and goes beyond Newton's approach, which applies only to particles travelling in a weak gravitational field with modest velocities (slower than the speed of light). GR is essential for discussing cosmological scenarios or those that require strong gravity, such as compact stellar objects which includes neutron stars, black holes and white dwarfs \cite{Lobato/2022,Moffat/2021,Vilhena/2023,Rahaman/2015,Sokoliuk/2022}. Black holes, where the gravity is so intense that even light cannot escape from them, were predicted to exist within Einstein's theory. In addition, over the past century, GR has continued to achieve significant successes, including the prediction of the existence of Gravitational Waves (GW), which was confirmed by the detection of GW170817 by the Laser Interferometer Gravitational-Wave Observatory (LIGO) and the prediction of the existence of Black Holes (BH), which was observed by the Event Horizon Telescope (EHT) \cite{Abbott/2016,Akiyama/2019}.  Moreover, GR, a theory of dynamical spacetime, naturally provides a framework for an expanding Universe. The first general relativistic cosmological model, which corresponds to a static, homogeneous, and isotropic Universe with spherical geometry, was put forth by Einstein in 1917 \cite{einstein/1917}. But, the gravitational force of matter caused this model to be unstable. Later, Einstein modified his equation of general relativity by adding a new component known as the cosmological constant denoted by $\Lambda$. This new term is used to describe a type of anti-gravity effect because it opposes the gravitational attraction of matter. Despite including the cosmological constant, it turns out that the Einstein static Universe is still sensitive to small perturbations. In the early 1920s, Russian Mathematician Alexander Friedmann discovered that Einstein gravitational field equations contained non-static solutions that might describe an expanding Universe whose size varies with time. The explanations provided by Friedmann demonstrated that our Universe began around $13$ billion years ago in a single event \cite{friedman/1922}. Therefore, according to Friedmann's solutions, the Universe, all matter, space and time itself appeared all at once in a single instant. British scientist Fred Hoyle called this theory the ``Big Bang". Under this term, it came to represent the accepted cosmological paradigm, according to which the Universe was created at a single point at an extremely high density and temperature. After learning about this study, Einstein promptly abandoned the cosmological constant as the greatest error of his life.   

After that Edwin Hubble, in 1929 \cite{Hubble/1929} observed that all galaxies recede from us by measuring the distance between them. Additionally, he introduced the Hubble constant $H_{0}$ in units of $km/s/Mpc$, along with the expansion law \cite{Hubble/1931} $v=H_{0}\,d$, where $v$ is the speed of the galaxy receding away from Earth and $d$ is the distance between the observed galaxy and Earth. It indicates that greater distances translate into greater receding velocities. As a result, the expansion of the Universe appeared to be practically established. The emergence of the Friedmann-Lema\^{i}tre-Robertson-Walker (FLRW) metric by Friedmann, Lema\^{i}tre, Robertson, and Walker independently in the 1920s could be considered as the beginning of this field \cite{friedman/1922,lemaitre/1931}. Soon after, observations revealed that the Universe was expanding, a phenomenon that the FLRW solution could easily account for. The existence of the Cosmic Microwave Background Radiation (CMBR) is another prediction made using the FLRW metric. That is, the CMBR has a thermal black-body spectrum at a uniform temperature of $T \approx 2.7\,K$ at the present epoch, corresponding to microwave part of the electromagnetic spectrum which affects the formation of galaxies \cite{Peebles/1965}. Arno Penzias and Robert Wilson observed this spectrum in 1965 \cite{penzias/1965}. This finding has been recognized as a support to the cosmological model, which asserts that the Universe has expanded since the Big Bang and is homogeneous and isotropic at large scales. Hence, the following years have witnessed remarkable developments in the theoretical and observational sectors. 

According to observational studies of type Ia supernovae (SNeIa) conducted in 1998, the Universe is currently expanding more rapidly than previously thought. Two groups, Riess and Schmidt \cite{riess/1998} (the High-Z Supernova Search Team), and Perlmutter \cite{perlmutter/1999} (Supernova Cosmology Project) acquired the lead in this study and found that the galaxy and its clusters are moving apart from one other in an accelerated rate. Various observations, including data from the Planck satellite \cite{planck/2015}, led to the astounding discovery that the observable Universe contains only 4\% to 5\% of conventional matter composed of baryons and electrons. The remaining balance of the Universe comprises of two fundamentally unidentified components: dark matter (25\%) and dark energy (70\%). The $\Lambda$CDM ($\Lambda$ cold dark matter) paradigm, which assumes that cold dark matter is the predominant component in a Universe whose late-time dynamics are determined by Einstein's cosmological constant, was developed due to these astounding results in modern cosmology. Hence, the accelerated expansion of the Universe has been the most perplexing phenomenon in cosmology for the previous two decades.

But, our current cosmological framework presents a somewhat hazy image. It suffers with severe theoretical issues including dark matter, dark energy, and the Hubble tension. Let us take a quick look at these three issues one by one. Based on the observations, most of the mass is found in the center of galaxy. According to the rotational curves of galaxies, which represent the velocity versus the radial distance from the galaxy's center, the velocity of stars or gas reaches a maximum and remains constant regardless of the radial distance from the galaxy's center. By Kepler's law, one would expect the velocity to begin to fall with distance after reaching a maximum. It has been suggested that the presence of dark matter, a distinct component of matter from baryonic matter, which does not interact via electromagnetism, is required to explain this phenomenon. Dark matter has never been directly detected before. One can check \cite{Carr/1994,Bertone/2018} for further information. High-precision observational data has provided great substantial confirmation that the Universe is going through an accelerated expansion phase, which is pertinent to the issue of dark energy. Invoking the $\Lambda$CDM paradigm is the simplest way to explain cosmic acceleration in recent times. However, suppose that the vacuum energy of the gravitational field is made up of the cosmological constant, then, the theory predicts a discrepancy between the theoretical and observed values of the cosmological constant. The expected theoretical value exceeds the observed value by nearly 60 orders of magnitude. The theoretical value derived from quantum-mechanical processes utilizing the standard model is $10^{-60}\, M_{Pl}^{4}$, where $M_{Pl}$ is the Planck mass, whereas the observed value is $10^{-120}\, M_{Pl}^{4}$. This results in a cosmological constant problem. Also, the Hubble tension \cite{Valentino/2021} is an additional issue that can be addressed in two ways: by directly measuring the speed of the nearest galaxies as they recede or by extrapolating it from inhomogeneity of the CMB. However, the reason why these two do not produce the same Hubble parameter values is unknown. 

The preceding discussion has been independent of any mathematical equations. We shall now establish connections between the mathematical formulation and the underlying physical concepts of some theories. One can now mention the main fundamental ideas on which GR is based on. It is important to keep in mind that, any gravitational field can be understood mathematically and is closely related to a substantial change in the spacetime metric $g_{\mu \nu}$. Geometrically, the metric tensor represents the distance between two adjacent points in the spacetime continuum. In GR, the gravitational field is defined by the quantities that characterize the inherent geometrical characteristics and structure of spacetime. It has been acknowledged that GR has issues at both small and large scales, despite its excellent accord with most observational data. The main problems with the $\Lambda$CDM model in GR to create a cosmological model are described  in the further sections. This paves the way for physics beyond General Relativity, which may address the issues mentioned earlier through modified gravity. Specifically, the new theory must behave similarly to GR on the solar system's scale. Therefore, we have additional ways to modify GR. One might change the matter content by adding new types of matter to the right side of the Einstein equations, or one could change the geometry through modifications to the left side of the Einstein equations.

The most attractive aspect of GR is its geometric interpretation. After understanding the geometrization of gravity, theories employing geometric concepts are of particular interest. Since it can be derived using the metric $g_{\mu \nu}$ and its derivatives, the connection in Einstein's theory of gravity is, in a sense, secondary to the metric. One is then confronted with the well-known Riemannian geometry, in which the structure of the manifold is entirely determined by its metric. In this instance, the connection, which is the Levi-Civita connection, is symmetric and metric compatible ($\nabla_\alpha g_{\mu \nu}=0$). The most basic modification can be achieved by considering a generalized action by substituting an arbitrary function for the gravitational Lagrangian $R$ of the Einstein-Hilbert (EH) action $S=\int \left(\frac{1}{2k}R+L_{m}\right)\sqrt{-g} d^{4}x$, where $k=\frac{8\pi G}{c^{4}}$, $G$ is the gravitational constant and $c$ is the speed of light in vacuum. This led to the development of $f(R)$ theories with some significant works studied in \cite{Nojiri/2006,Sotiriou/2006,Amendola/2007,Sotiriou/2010,Felice/2010}. An alternative approach to $f(R)$ gravity, as discussed in \cite{Harko/2011}, introduces further modifications to the Einstein-Hilbert (EH) action. This is achieved by including a coupling term between matter and gravity, represented by an arbitrary function $f(R,\mathcal{T})$, where $\mathcal{T}$ denotes the trace of the energy-momentum tensor (EMT), a tensorial quantity characterizes the energy and flux of matter.  
It is worth mentioning that the covariant divergence of the energy-momentum tensor in $f(R,\mathcal{T})$ gravity is not conserved, resulting in the non-geodesic motion of massive test particles. The coupling effects between matter and geometry results in the particles experiencing an extra acceleration. Works in \cite{Alvarenga/2013,Shabani/2014,Zare/2016,Moraes/2017,Fisher/2019,Goncalves/2022} have more findings on $f(R,\mathcal{T})$ gravity.  

Another way to extend the geometry would be to relax the Riemannian constraints. There are, however, other gravity theories in which the metric is introduced as a fundamental field variable with a different independent connection where the curvature plays a secondary role. This connection is intended to have a vanishing curvature, in contrast to the Levi-Civita connection, but one that permits non-vanishing torsion, non-metricity, or both. H. Weyl \cite{Weyl/1918} made an innovative approach with the primary goal of geometrically uniting gravity and electromagnetism. Weyl's theory employs a mechanism with two connections: one transmits information about a vector's length, while the other regulates the vector's direction during parallel transport. The non-zero covariant divergence of the metric tensor, a trait that lead to a new geometrical quantity called non-metricity, is the most remarkable aspect of the theory. Meanwhile, the idea of torsion was introduced, which led to another significant advancement in differential geometry, i.e., the Einstein-Cartan theory \cite{Hehl/1976}. Weitzenb\"{o}ck provided a different exquisite geometrical formalism known as Weitzenb\"{o}ck spaces \cite{Weitzenbck/1923}, where the manifold possesses the properties $R^{\mu}_{\,\,\nu \lambda\sigma}=0$, $\mathcal{T}^{\mu}_{\,\,\nu \lambda} \neq 0$, and $\nabla_{\mu}g_{\nu\lambda}=0$. One can notice that, the Weitzenb\"{o}ck space reduces to a Euclidean manifold when $\mathcal{T}^{\mu}_{\,\,\nu \lambda} = 0$. In a Weitzenb\"{o}ck manifold, the Riemann curvature tensor vanishes identically, resulting in distant parallelism, also known as teleparallelism or absolute parallelism. 

The fundamental idea behind the teleparallel formulation of gravity is to replace the spacetime metric, or primary geometrical variable $g_{\mu \nu}$ with a collection of tetrad vectors $e^{i}_{\nu}$. Then, gravitational processes can be described entirely using the torsion tensor produced by the tetrad fields, with the curvature being substituted by torsion. As a result, this leads to the so-called teleparallel equivalent of general relativity (TEGR) \cite{Hayashi/1979}, and its extension is referred to as the $f(T)$ gravity theory \cite{Cai/2016,Krssak/2019}, where $T$ is the torsion scalar. Torsion precisely balances curvature, which has the critical consequence that spacetime becomes a flat manifold, and this is the fundamental characteristic of teleparallel theories. The fact that second-order differential equations are used to describe the gravitational field is another significant characteristic of $f(T)$ gravity theories. More detailed analysis of $f(T)$ theory can be seen in \cite{Jimenez/2018,Capo/2017,Ren/2022,Bahamonde/2023}. Let us now talk about the third equivalent formulation of GR.

One can note that the non-metricity $Q$ disappears in both the curvature based and teleparallel formulations. Geometrically $Q$ depicts the change in a vector's length in parallel transport. Now, a non-vanishing non-metricity $Q$ is considered the fundamental geometrical variable accountable for all varieties of gravitational interactions in a third equivalent formalism of GR, named symmetric teleparallel equivalent to gravity relativity (STEGR) \cite{Nester/1999}. Non-metric gravity is another name for these non-metricity based gravitational theories. While having a Lagrangian represented by the non-metricity scalar $Q$, equivalent to GR description of gravity, a generalization of the Lagrangian with a generic function of the non-metricity $f(Q)$ gives a different description of the grvitational interaction. Various physical and geometrical characteristics of symmetric teleparallel gravity and $f(Q)$ gravity have been studied in different research studies, and interest in this kind of theoretical approach to gravity is increasing rapidly \cite{Jimenez/2018,Harko/2018,Soudi/2019,Lu/2019,Jimenez/2020,Hohmann/2019,Ambrosio/2022,Khyllep/2023,Wang/2022,Hu/2022,Capo/2022a}. Furthermore, the authors in \cite{Harko/2018} established a non-minimal connection between the non-metricity $Q$ and the matter Lagrangian $\mathcal{L}_{m}$ as the foundation for an extension of the $f(Q)$ gravity. The non-minimal interaction between the geometry and matter sectors predictably results in the non-conservation of the energy-momentum tensor and the development of an additional force in the geodesic equation of motion. Another extension of the theory, known as the $f(Q,\mathcal{T})$ gravity, was proposed by Xu. et al. in \cite{Xu/2019}, where the gravity Lagrangian is essentially an arbitrary function of $Q$ and the trace of the energy-momentum tensor $\mathcal{T}$. The authors explored the dynamical evolution for several specific coupling functions.

The current dissertation intends to show several ways to investigate late time cosmology or to look at probable places for future investigations in order to shed more light on the dark sector of the Universe. To explore the accelerated phase of the Universe, we shall mainly concentrate on modified theories of gravity. This chapter includes an overview of the background dynamics, the cosmic observations, the standard model of cosmology, and its extensions.   

\section{Foundations}

This section discusses the fundamental differential geometry concepts that allow for constructing both Riemannian and non-Riemannian spacetime. We will present the basic concepts and principles of GR\footnote{Refer to: Introduction to General Relativity and Cosmology by Christian G. B\"{o}hmer}. This section's primary objective is to introduce differential manifolds, vectors, tensors, and other mathematical concepts used to describe the properties of physical objects \cite{Weinberg/1972,Boh/2016}. 

\subsection{Manifolds}
A manifold is one of the most fundamental concepts in both Mathematics and Physics. Mathematicians developed the theory of analysis in $\mathbb{R}^{n}$ over many years. However, we naturally associate specific other spaces with curved spaces or topological complexity. To address the concept, there is a manifold which refers to a space that may be curved and have a complex topology but which, in certain localized sections, resembles $\mathbb{R}^{n}$. 

We first require the concept of topology and topological space to introduce a manifold. 

\begin{definition}
Let $\mathrm{M}$ be any set, and $\tau=\{\mathrm{U}_{i}, i \in I\}$ denote a collection of subsets of $\mathrm{M}$. The pair $\left(\mathrm{M},\tau\right)$ is a \textbf{topological space} if $\tau$ satisfies  the following conditions:
\begin{enumerate}[label=(\roman*)]
\item $\emptyset$, $\mathrm{M}$ $\in$ $\tau$,
\item If $J$ is any subcollection of $I$, the family $\{\mathrm{U}_{j},j \in J \}$ satisfies $\bigcup\limits_{j \in J} \mathrm{U}_{j} \in \tau$,
\item If $K$ is any finite subcollection of $I$, the family $\{\mathrm{U}_{k},k \in K \}$ satisfies $\bigcap\limits_{k \in K} \mathrm{U}_{k} \in \tau$.
\end{enumerate}
\end{definition}
Here, $\mathrm{U}_{i}$ are known as open sets, $\tau$ is referred to as a topology on $\mathrm{M}$, and $\mathrm{M}$ is a topological space. In fact, a topology on a set provided by a metric function is known as metric\footnote{A metric $d: X \times X \rightarrow \mathbb{R}$ is a function that is symmetric, non-degenerate and obeys a triangle inequality.} topology. Additionally, if the inverse image of an open set is also an open set, then a map between two topological spaces is continuous. Similar to the notion of continuity in topology, the theory of manifolds has smoothness. A function is referred to $\mathrm{C}^{r}$, if its $r^{th}$ derivative exists and is continuous. Thus, $\mathrm{C}^{\infty}$ functions are referred to as smooth. 

\begin{definition}
For any $p$, $q$ $\in$ $\mathrm{M}$, with $p\neq q$, there exists $\mathrm{U}_{1}$, $\mathrm{U}_{2}$ $\in$ $\tau$ such that $p \in \mathrm{U}_{1}$,\, $q \in \mathrm{U}_{2}$ and $\mathrm{U}_{1} \cap \mathrm{U}_{2}= \emptyset$. Topological spaces satisfying such conditions are called \textbf{Hausdorff space}.  
\end{definition}

\begin{definition}
A \textbf{homeomorphism} between topological spaces is a continuous bijective map whose inverse map is also continuous.

\end{definition}

\begin{definition}
An n-dimensional \textbf{differentiable manifold} is a Hausdorff topological space $\mathrm{M}$ that satisfies the following conditions:
\begin{enumerate}[label=(\roman*)]
\item $\mathrm{M}$ is locally homeomorphic to $\mathbb{R}^{n}$. This means that for each $p \in \mathrm{M}$, there is an open set $\mathrm{U}$ such that $p \in \mathrm{U}$ and a homeomorphism $\phi : \mathrm{U} \rightarrow \mathbb{O}$ with $\mathbb{O}$ an open subset of $\mathbb{R}^{n}$.
\item Consider two open sets $\mathrm{U}_{\alpha}$ and $\mathrm{U}_{\beta}$ such that $\mathrm{U}_{\alpha} \cap \mathrm{U}_{\beta} \neq \emptyset $. The corresponding maps  $\phi_{\alpha} : \mathrm{U}_{\alpha} \rightarrow \mathbb{O}_{\alpha}$ and $\phi_{\beta} : \mathrm{U}_{\beta} \rightarrow \mathbb{O}_{\beta}$ are compatible, that means the map $\phi_{\beta} \circ \phi_{\alpha}^{-1}: \phi_{\alpha}\left(\mathrm{U}_{\alpha} \cap \mathrm{U}_{\beta}\right) \rightarrow \phi_{\beta}\left(\mathrm{U}_{\alpha} \cap \mathrm{U}_{\beta}\right)$ is smooth (infinitely differentiable), as is its inverse.
\end{enumerate}
\end{definition}
The maps $\phi_{\alpha}$ are called charts and the collection of charts is called an atlas. The maps $\phi_{\beta} \circ \phi_{\alpha}^{-1}$ take us between different coordinate systems\footnote{The coordinate associated to $p \in \mathrm{U}_{\alpha}$ is $\phi_{\alpha}(p)= \left( x^{1}(p),x^{2}(p),.....,x^{n}(p) \right)$} and are called transition functions. A chart $(\mathrm{U},\phi)$ on $\mathrm{M}$ is said to be compatible with a $C^{\infty}$-atlas $\mathbb{A}$ on $\mathrm{M}$ if $\mathbb{A} \cup {(\mathrm{U},\phi)}$ is a $C^{\infty}$-atlas. So, the two compatible atlases define the same differentiable structure on the manifold.

\subsection{Vectors and Tensors}

In this subsection, we introduce the concepts of vectors and tensors. The most crucial aspect to highlight is that each vector occupies a specific location in spacetime. One can commonly think of vectors as entities that extend between two distinct points in space and even as free vectors that can be effortlessly displaced between various points. The concept above needs more practical utility beyond the confines of flat spaces. The capacity to define preferred curves between points or effectively move vectors across a manifold is lost upon the introduction of curvature. Consequently, the concept of a tangent vector is the foundation of calculus on manifolds. Hereafter, we define the following terms.  

\begin{definition}
Let $\mathrm{F}$ be a collection of $C^{\infty}$ functions from $\mathrm{M}$ to $\mathbb{R}$. We define a \textbf{tangent vector} $\mathrm{V}$ at a point $p \in \mathrm{M}$ to be a map $\mathrm{V}: \mathrm{F} \rightarrow \mathbb{R}$ which is linear and obeys the Leibnitz rule 
\begin{enumerate}[label=(\roman*)]
\item $V(af+bg) =a\, V(f) + b\, V(g)$, \, $f,g \in \mathrm{F}$, and $a,b \in \mathbb{R}$,
\item $V(fg)= f(p)\, V(g) + g(p)\, V(f)$.
\end{enumerate}
\end{definition}
The object $\left. \partial_{\mu} \right|_{p} = \left. \frac{\partial}{\partial x^{\mu}}\right|_p$ acts on functions and obeys all requirements of the tangent vector. For a function $f: \mathrm{M} \rightarrow \mathbb{R}$ and a chart $\phi = \left(x^{1},.....,x^{n}\right)$ in a neighbourhood of \textit{p}, we define a map $f \circ \phi^{-1} : \mathrm{U} \rightarrow \mathbb{R}$ with $\mathrm{U} \subset \mathbb{R}^{n}$. So, we differentiate functions on $\mathrm{M}$ as 
\begin{equation*}
\left. \frac{\partial f}{\partial x^{\mu}}\right|_p = \left.\frac{\partial(f\circ \phi^{-1})}{\partial x^{\mu}}\right|_{\phi(p)}.
\end{equation*}
This depends on the choice of chart $\phi$ and coordinates $x^{\mu}$. However, we need a coordinate independent definition of differentiation. 

\begin{theorem*}
The set of all tangent vectors at point \textit{p} forms an n-dimensional vector space, which is known as \textbf{tangent space} $T_{p}(\mathrm{M})$. The tangent vector $\left. \partial_{\mu} \right|_{p}$ provides a basis for $T_{p}(\mathrm{M})$. So, we can write any tangent vector as 
\begin{equation*}
V_{p}= V^{\mu} \left. \frac{\partial}{\partial x^{\mu}} \right|_{p},
\end{equation*}
with $V^{\mu} = V_{p}(x^{\mu})$. 
\end{theorem*}
The crucial notion is that a certain tangent vector $V_{p}$ exists regardless of the coordinate system used. However, the chosen basis $\{ \partial_{\mu} |_p\}$ depends on our choice of coordinates: charts $\phi$ and coordinates $x^{\mu}$.

Suppose that we chose a different chart $\phi'$ with the coordinates $x'^{\mu}$ near \textit{p}. The tangent vector $V_p$ is expressed as 
\begin{equation*}
V_{p}= \left. V^{\mu}\frac{\partial}{\partial x^{\mu}} \right|_{p} = V'^{\mu} \left. \frac{\partial}{\partial x'^{\mu}} \right|_{p}.
\end{equation*}
Here, the components of the vector vary, while the vector itself remains the same. It should be noted that by applying the chain rule and the tangent vector $V_{p}$ acting on a function $f$, one can express
\begin{equation*}
V_{p}(f)= \left. V^{\mu}\frac{\partial\,f}{\partial x^{\mu}} \right|_{p} = V^{\mu} \left. \frac{\partial x'^{\nu}}{\partial x^{\mu}} \right|_{\phi(p)}  \left. \frac{\partial\, f}{\partial x'^{\nu}} \right|_{p}.
\end{equation*}
 This can be viewed as 
 \begin{equation*}
\left. \frac{\partial}{\partial x^{\mu}} \right|_p = \left. \frac{\partial x'^{\nu}}{\partial x^{\mu}}\right|_{\phi(p)} . \left. \frac{\partial}{\partial x'^{\nu}} \right|_{p},
 \end{equation*}
\begin{equation}
\label{vec}
V'^{\mu} = V^{\mu} \frac{\partial x'^{\nu}}{\partial x^{\mu}}.
\end{equation}
This type of component transformation is referred to as contravariant.

A \textbf{vector field} $V$ is defined to be a smooth assignment of a tangent vector $V_{p}$ to each point $p\in \mathrm{M}$.

\begin{definition}
The dual space to the tangent space $T_{p}(\mathrm{M})$ is called a \textbf{cotangent space} $T_{p}^{*}(\mathrm{M})$. It is the space of all linear maps from $T_{p}(\mathrm{M})$ to $\mathbb{R}$.
\end{definition}

The gradients of the coordinate function $x^{\mu}$ offer a natural basis for the cotangent space, just as the partial derivative along coordinate axes does for the tangent space such that
\begin{equation*}
dx^{\mu} (\partial_{\nu}) = \frac{\partial x^{\mu}}{\partial x'^{\nu}} = \delta^{\mu}_{\,\nu}.
\end{equation*}
which further results in 
\begin{equation}
\label{dual}
W'_{\mu}= \frac{\partial x^{\mu}}{\partial x'^{\mu}} W_{\mu}.
\end{equation}
This type of component transformation is referred to as covariant.
Now, we are ready to define a tensor. 
\begin{definition}
A $\left(m,n\right)$ tensor is a multilinear map from a collection of $m$ dual vectors and $n$ vectors to $\mathbb{R}$. Hence, we write 
\begin{equation*}  
A^{\mu_{1}.....\mu_{m}}_{~~~~~~~~~~ \nu_{1}.....\nu_{n}} = A(dx^{\mu_1}, ......, dx^{\mu_m},\partial_{\nu_1},.....,\partial_{\nu_n}).
\end{equation*}
\end{definition}
Using the transformation laws of vectors and dual vectors in equations  \eqref{vec} and \eqref{dual}, the general coordinate transformation for tensor is 
\begin{equation}
A'^{\mu_{1}'.....\mu_{m}'}_{~~~~~~~~~~ \nu_{1}'.....\nu_{n}'} = \frac{\partial x'^{\mu_{1}'}}{\partial x^{\mu_1}}......\frac{\partial x'^{\mu_{m}'}}{\partial x^{\mu_m}} \frac{\partial x^{\nu_{1}}}{\partial x'^{\nu_{1}'}} ........\frac{\partial x^{\nu_{n}}}{\partial x'^{\nu_{n}'}} A^{\mu_{1}.....\mu_{m}}_{~~~~~~~~~~ \nu_{1}.....\nu_{n}}.
\end{equation}
Now, we define an important aspect to define the distance between two points in the space.

\subsection{The Metric}
The metric structure enables the definition of distance between any two spacetime points. A metric is meant to inform us the infinitesimal squared distance associated with an infinitesimal displacement and is instead an inner product on each vector space $T_{p}(\mathrm{M})$.
\begin{definition}
The metric is a tensor of type $(0,2)$ which is 
\begin{enumerate}[label=(\roman*)]
\item Symmetric: $g\left(V_{1},V_{2}\right) =  g\left(V_{2},V_{1}\right)$,\,\, $V_{1}$, $V_{2}$ $\in$ $T_{p}(\mathrm{M})$,
\item Non-degenerate: For any $p \in \mathrm{M}$, $g(V,V_{1})=0$ for all $V \in T_{p}(\mathrm{M})$, then $V_{1}=0$.
\end{enumerate}
\end{definition}
We may expand $g$ in terms of its components $g_{\mu \nu}$ in a coordinate basis as 
\begin{equation*}
g = g_{\mu \nu}\,\, dx^{\mu} \otimes dx^{\nu}.
\end{equation*}
This further can be written as 
\begin{equation}
ds^{2} = g_{\mu \nu} \,\, dx^{\mu} dx^{\nu},
\end{equation}
with $g = det(g_{\mu \nu})$. 
Furthermore, putting $g_{\mu \nu}$ into its canonical form yields a meaningful characterization\footnote{A square matrix can be reduced to a diagonal matrix.} of the metric as $g_{\mu \nu}=diag\left(-1,-1,.....,-1,+1,+1,...+1,0,......0\right)$. The signature of the metric is the number of positive and negative eigenvalues. If all signs are positive, we refer to the metric as Euclidean or Riemannian. If there is a negative sign, the metric is called Lorentzian or pseudo-Riemannian. Otherwise, it is referred to as indefinite.

\section{General Relativity}
The initial stage in the development of the theory of gravitation involves the establishment of a mathematical framework that enables the formulation of the laws of physics in a covariant fashion. The formulation of the laws of physics can be achieved by utilizing vectors and tensors, which are mathematical entities possessing transformation qualities that embody the concept of covariance in physical laws. Therefore, we commence our discussion by examining the fundamental differential operations that may be derived from vectors and tensors. One can review some concepts in \cite{Carroll/2004}.

\subsection{Some Important Concepts}

\begin{definition}[\textbf{Christoffel Symbol}]
A special connection that we might draw from the metric is represented by an object known as the Christoffel symbol and is provided by 
\begin{equation}
\Gamma^{\alpha}_{\,\,\mu \nu} = \frac{1}{2} g^{\alpha \lambda}\left(\partial_{\mu} g_{\nu \lambda}+ \partial_{\nu} g_{\lambda \mu}- \partial_{\lambda} g_{\mu \nu} \right).
\end{equation}
\end{definition}
Although the three index symbol or Christoffel symbol resembles a tensor, it is not a tensor. It is symmetric in its lower pair of indices by definition. Consequently, the dyad produces $\frac{n(n+1)}{2}$ components in $n$ dimensions. No other symmetries exist, so the third index can take any value, resulting in $\frac{n^{2}(n+1)}{2}$ independent components in $n$ dimensions. The main purpose of this connection is to generate covariant derivatives, a generalization of partial derivatives.

\begin{definition}[\textbf{Covariant Derivative}]
An operator $\nabla$ known as covariant derivative produces a tensor when applied to a tensor in a way entirely independent of coordinates. It is defined as follows 
\begin{equation}
\nabla_{\mu} A^{\nu}= \partial_{\mu} A^{\nu} + \Gamma^{\nu}_{\,\,\mu \lambda} A^{\lambda}.
\end{equation}
Furthermore, the general expression for the covariant derivative is 
\begin{multline}
\nabla_{\lambda} A^{\mu_{1} \mu_{2}....\mu_{n}}_{~~~~~~~~~\nu_{1} \nu_{2}....\nu_{m}} = \partial_{\lambda}A^{\mu_{1} \mu_{2}....\mu_{n}}_{~~~~~~~~~\nu_{1} \nu_{2}....\nu_{m}} + \Gamma^{\mu_{1}}_{\,\,\lambda \sigma}  A^{\sigma \mu_{2}....\mu_{n}}_{~~~~~~~~~\nu_{1} \nu_{2}....\nu_{m}}  + \Gamma^{\mu_{2}}_{\,\,\lambda \sigma}  A^{\mu_{1} \sigma....\mu_{n}}_{~~~~~~~~~\nu_{1} \nu_{2}....\nu_{m}}+....\\
 - \Gamma^{\sigma}_{\,\,\lambda \nu_{1}}  A^{\mu_{1} \mu_{2}....\mu_{n}}_{~~~~~~~~~\sigma \nu_{2}....\nu_{m}} - \Gamma^{\sigma}_{\,\,\lambda \nu_{2}}  A^{\mu_{1} \mu_{2}....\mu_{n}}_{~~~~~~~~~\nu_{1}\sigma....\nu_{m}}- .....
\end{multline}
\end{definition}

\begin{definition}[\textbf{Parallel Transport}]
Parallel transport describes the idea of transporting a vector along a path while keeping its orientation constant. Let $x^{\mu}(\lambda)$ be a  path with tangent vector $\frac{dx^{\mu}}{d\lambda}$ and assume $A^{\mu}$ be a vector. Then, we say $A^{\mu}$ is parallelly transported along a path if 
\begin{equation}
\frac{dx^{\mu}}{d\lambda}. \nabla_{\mu} A^{\nu} = 0.
\end{equation}
\end{definition}
Following the discussion of the definition of parallel transport, we will move on to geodesics. A geodesic is a mathematical construct that generalizes the motion of a straight line across Euclidean space to motion through curved space. If we have two points connected by a curve/path, we may use the arc length to determine the distance between these places along that curve. The curves that minimize the distance between our points are of particular interest to us. Let us summarize this as a definition.

\begin{definition}[\textbf{Geodesics}]
A curve/path $x^{\mu}(\lambda)$ is called a geodesic if it satisfies the equation  
\begin{eqnarray} 
\frac{d^{2} x^{\mu}}{d \lambda^{2}} + \Gamma^{\mu}_{\,\, \delta \sigma} \frac{dx^{\delta}}{d \lambda} \frac{dx^{\sigma}}{d \lambda}&=&0,
\end{eqnarray}
where $\Gamma^{\mu}_{\,\, \sigma \nu}$ is the Christoffel symbol.
\end{definition}

\begin{definition}[\textbf{Riemann Tensor}]
The Riemann tensor or  curvature tensor is a tensor of type $(1,3)$ defined as 
\begin{equation}
R^{\mu}_{\,\,\nu \sigma\lambda} = \partial_{\sigma}\Gamma^{\mu}_{\,\lambda \nu} - \partial_{\lambda}\Gamma^{\mu}_{\, \sigma \nu} + \Gamma^{\mu}_{\,\sigma \delta} \Gamma^{\delta}_{\,\lambda \nu} -  \Gamma^{\mu}_{\,\lambda \delta} \Gamma^{\delta}_{\,\sigma \nu}.
\end{equation}
\end{definition}
It has some important relations as follows.
\begin{enumerate}
\item $R^{\mu}_{\,\, \nu \sigma \lambda} = - R^{\mu}_{\,\, \nu \lambda \sigma}$, \hspace{0.2in} (Antisymmetric)
\item $R^{\mu}_{\,\, \nu \sigma \lambda} + R^{\mu}_{\,\, \sigma \lambda \nu} + R^{\mu}_{\,\, \lambda \nu \sigma} =0 $, \hspace{0.2in} (Cyclic)
\item $R^{\mu}_{\,\, \nu \sigma \lambda ; \delta}+ R^{\mu}_{\,\, \nu  \lambda \delta ; \sigma}+ R^{\mu}_{\,\, \nu \delta \sigma ; \lambda}=0$. \hspace{0.2in} (Bianchi Identity)
\end{enumerate}

\begin{definition}[\textbf{Ricci Tensor}]
The Ricci tensor is a type $(0,2)$ tensor, which is a contracted Riemann tensor. That is 
\begin{equation}
R_{\mu \nu}= R^{\sigma}_{\,\, \mu \sigma \nu}.
\end{equation}
The Ricci tensor associated with the Christoffel connection is symmetric, which is $R_{\mu \nu} = R_{\nu \mu}$.
\end{definition}

\begin{definition}[\textbf{Ricci Scalar}]
The trace of the Ricci tensor is known as the Ricci Scalar. 
\begin{equation}
R = R^{\mu}_{\,\,\mu}= g^{\mu \nu} R_{\mu\nu}.
\end{equation}
\end{definition}

\begin{definition}[\textbf{Einstein Tensor}]
The tensor $G_{\mu \nu}= R_{\mu \nu} -\frac{1}{2} g_{\mu \nu} R$ is the Einstein tensor, which is symmetric and has a  great importance in GR. Furthermore, $\nabla^{\mu} G_{\mu \nu}=0$. It is a fundamental tensor quantity that describes the physical and geometrical properties of the gravitational field.
\end{definition}
So far, we have covered some fundamental mathematical notions and tools that will be required in the mathematical formulation of the General Theory of Relativity. We shall now focus on the aspect of obtaining the Einstein field equations.

\subsection{Einstein Field Equations}
Einstein field equations are a set of differential equations used to characterize the gravitational field within the framework of general relativity. These equations are defined in Riemannian geometry, and they establish a profound relationship between the geometric properties of spacetime and its matter content, providing a complete description of the geometric properties of spacetime and the dynamics of the particles. We seek to find an equation that replaces the Poisson equation for the Newtonian potential. This can be done by the application of the principle of least action to the gravitational field. 

The total action for gravity and any other physical fields present in the system is given by 
\begin{equation}
\label{Action}
S= \frac{1}{2k}\int R \sqrt{-g}\, d^{4}x + \int L_{m} \sqrt{-g}\, d^{4}x,
\end{equation}
where $\sqrt{-g}\, d^{4}x$ is a 4-dimensional volume on the manifold. $g$ is the determinant of the metric tensor, $k = \frac{8 \pi G}{c^{4}}$, $G$ is the Newton's constant and $L_{m}$ is the Lagrangian density of matter fields.

The variation of the above action \eqref{Action} with respect to the metric tensor yields 
\begin{equation}
\label{EF}
G_{\mu \nu}= R_{\mu \nu} -\frac{1}{2} R\, g_{\mu \nu} = \frac{8 \pi G}{c^{4}} \mathcal{T}_{\mu \nu},
\end{equation}
which are known as the Einstein field equations. We also note that the 
\begin{equation}
\label{T}
\mathcal{T}_{\mu \nu}= -\frac{2}{\sqrt{-g}} \frac{\delta \left(\sqrt{-g} L_{m} \right)}{\delta g^{\mu \nu}},
\end{equation}
is named the energy-momentum tensor (which will be discussed in the following sections), quantifying the matter and energy content of spacetime.

Taking the covariant divergence of the equation \eqref{EF}, we can easily find 
\begin{equation}
\nabla^{\mu} G_{\mu \nu} \equiv 0 \equiv \nabla^{\mu} \mathcal{T}_{\mu \nu}.
\end{equation}
This pertains to the important law of conservation of the matter-energy momentum tensor $\mathcal{T}_{\mu \nu}$.

Finally, we may construct general relativistic theoretical models of the Universe using various cosmological assumptions and parameters discussed in the following section.

\section{The Standard Cosmological Model} \label{1.4}
This section describes the standard cosmological model. We discuss essential concepts such as the FLRW metric, the Hubble parameter, and redshift, as well as the successes and flaws of the $\Lambda$CDM model. A more comprehensive introduction to cosmology can be found in Weinberg \cite{Weinberg/1972}.
\subsection{The FLRW Metric} \label{1.4.1}
The \textbf{cosmological principle} states that the Universe is homogeneous and isotropic at all times on large enough scales. In other words, homogeneity suggests that the Universe looks the same to any observer at any place, but isotropy implies no preferred directions from a given position. 
On the assumption that the Universe is homogeneous, it is possible to transform two particles separated by a distance $\vec{r}$ into a coordinate system known as comoving coordinates by using the formula $\vec{r} = a(t)\,\vec{x}$, where $\vec{x}$ is the comoving distance or separation between the same points in the comoving frame. The scale factor $a(t)$ of the Universe governs how the physical separations grow through time. Friedmann, Lema\^{i}tre, Robertson, and Walker (FLRW) developed the following form for the line element in spherical coordinates, considering the geometrical properties of homogeneity and isotropy following the cosmological principle and the fact that the Universe is expanding 
\begin{equation}
\label{metric}
ds^{2}= -c^{2}dt^{2} + a^{2}(t) \left[\frac{dr^{2}}{1-k r^{2}} + r^{2} d\theta^{2} + r^{2} sin^{2}\theta d\phi^{2}\right],
\end{equation}
where $\left(r,\theta,\phi\right)$ are the comoving spatial coordinates in spherical coordinate system, $k$ is the spatial curvature of the Universe. For the sake of simplicity, we will choose units when $c=1$\footnote{Empirically, we know $c=3 \times 10^{8} m/s$, thus we are working in units where 1 second equals $3 \times 10^{8}m$}. Furthermore, $k=-1$, $1$, and $0$ correspond to an open (Hyperbolic), closed (Spherical), and flat Universe (Euclidean), respectively. When examining the geometrical properties of several coordinate systems, it can be helpful to use a new radial-type coordinate $\xi$ defined as 
\begin{eqnarray}
\xi &=& \int \frac{dr}{\sqrt{1-kr^{2}}}
 = \begin{cases}
        arc\, sin\,r, & k=+1\\
        r, & k=0 \\ 
        arc\, sinh\,r, & k =-1
    \end{cases} .
\end{eqnarray}
Hence, the FLRW metric in a new coordinate system becomes 
\begin{equation}
\label{M}
ds^{2}= -dt^{2} + a^{2}(t)\left[ d\xi^{2} + S_{k}^{2}(\xi) \left(d\theta^{2} + sin^{2}\theta d\phi^{2}\right) \right],
\end{equation}
where 
\begin{equation*}
S_{k}(\xi)= \begin{cases}
        sin\,\xi, & k=+1\\
        \xi, & k=0 \\ 
        sinh\,\xi, & k =-1
    \end{cases} .
\end{equation*}
Before determining the Friedmann equations and their solutions, we will introduce additional parameters that play a role in characterizing the expansion of the Universe.

\subsection{The Hubble Parameter} \label{1.4.2}
One can define the rate of change of the distance from the relation $\vec{r} = a(t)\,\vec{x}$ as 
\begin{equation}
\dot{\vec{r}} = \dot{a(t)} \vec{x} = \dot{a(t)} \frac{\vec{r}}{a(t)},
\end{equation}
which gives 
\begin{equation}
\dot{\vec{r}}= \frac{\dot{a(t)}}{a(t)} \vec{r} = H(t)\,\vec{r},
\end{equation}
where dot denotes the time derivative, and $H$ is the Hubble parameter, which provides the rate of change of expansion of the Universe as defined by  
\begin{equation}
\label{Hub}
H(t) = \frac{\dot{a(t)}}{a(t)}.
\end{equation}
From equation \eqref{Hub}, one can obtain the expression of the Hubble law at the present time:  $\vec{v} = H_{0}\,\vec{r}$, where $H_{0}$ represents the Hubble constant at present, and $\vec{v}$ is the velocity at which the objects are moving away from the observer. Edwin Hubble saw it as the first observable foundation for the expansion of the Universe in 1929. According to this principle, a galaxy's recession rate is proportional to its distance from the observer.
\subsection{Redshift} \label{1.4.3}

The fact that practically everything in the Universe appears to be moving away from us and that this apparent movement increases with increasing distance is a significant piece of observational evidence in cosmology. When a photon travels through the Universe, it is inextricably linked to the expansion of space, which causes it to expand or, more precisely, increases its wavelength. Therefore, traveling photons lose energy in the Universe. If these photons were in the visible spectrum, they would begin with a blue hue and a high energy level before turning to the red spectrum as their energy decreases. This is where the term redshift originates.

Since all galaxies are moving away from us, the scientific terminology is redshift $z$, which is defined by 
\begin{equation}
z = \frac{\lambda_{ob} - \lambda_{em}}{\lambda_{em}},
\end{equation}
where $\lambda_{ob}$ and $\lambda_{em}$ are the wavelengths of light at the points of observation and emission, respectively. The Doppler law can now be used to establish that the difference in wavelength between emission and reception
\begin{equation}
\frac{\lambda_{ob} - \lambda_{em}}{\lambda_{em}} = \frac{dv}{c},
\end{equation}
where $\lambda_{ob} > \lambda_{em}$. The time between emission and reception is given by the light travel time $dt = \frac{dr}{c}$, which results 
\begin{equation}
\frac{\lambda_{ob} - \lambda_{em}}{\lambda_{em}} = \frac{\dot{a}}{a} \frac{dr}{c} = \frac{\dot{a}}{a} dt = \frac{da}{a}.
\end{equation}
We obtain that $\lambda \propto a$, where $\lambda$ is the instantaneous wavelength measured at any given time. Hence, one gets the redshift in terms of scale factor as 
\begin{equation}
1+z = \frac{\lambda_{ob}}{\lambda_{em}} = \frac{a(t_{ob})}{a(t_{em})}.
\end{equation}
Next, we turn our attention to the specific energy tensor and different types of matter sources described by it in the following section. 

\subsection{The Stress Energy-Momentum Tensor} \label{1.4.4}

In this section, we go over the properties of the stress-energy-momentum tensor. For modelling the Universe, we assume that the Universe is homogeneous and isotropic and also that the Universe is filled with the perfect fluid\footnote{When a matter component is considered continuous, it cannot support shear stress. That is, we can neglect viscosity.} distribution of matter. The equation for the stress-energy-momentum tensor is 
\begin{equation}
\label{Stress}
\mathcal{T}_{\mu \nu} = \left(\rho + p \right) u_{\mu} u_{\nu} + p g_{\mu \nu},
\end{equation}
where $u_{\mu} = (1,0,0,0)$ is the fluid 4-velocity vector. In addition to stresses it contains information about the energy density and energy fluxes of the matter.
The contraction of $\mathcal{T}_{\mu \nu}$ results in $\mathcal{T}^{\mu}_{\,\nu} = diag(-\rho,p,p,p)$, and the trace $\mathcal{T} = \mathcal{T}^{\mu}_{\mu} = -\rho+3p$, where $\rho$ is the energy density and $p$ is the pressure in all directions of the spacetime.

Further, we can recall that the stress-energy-momentum tensor satisfies the conservation equation $\nabla^{\mu} \mathcal{T}_{\mu \nu} = 0$, which yields 
\begin{equation}
\label{cont}
\dot{\rho} + 3\left(\rho + p\right) \frac{\dot{a}}{a} = 0.
\end{equation}
Often the perfect fluid relevant to cosmology obeys the simple equation of state 
\begin{equation}
p = \omega \rho,
\end{equation}
where $\omega$ is a constant and independent of time. Using equation \eqref{cont}, one gets the relation 
\begin{equation}
\rho \propto a^{-3(1+\omega)}.
\end{equation}
One has $p=0$, that is $\omega =0$ for the non-relativistic matter. The Universe is considered matter-dominated if matter accounts for most of the energy density and is known as dust. The energy density in this case is characterized as 
\begin{equation}
\rho_{m}  = \rho_{m_{0}} a^{-3}(t).
\end{equation}
The Universe in which the majority of the energy density takes the form of radiation, that is $p_{r} = \frac{1}{3} \rho_{r}$ is known as a radiation-dominated Universe. The energy density, in this case, falls off as 
\begin{equation}
\rho_{r}  = \rho_{r_{0}} a^{-4}(t).
\end{equation}
A vacuum-dominated Universe has an equation of state $p_{\Lambda} = - \rho_{\Lambda}$, i.e. $\omega=-1$. The energy density here is constant and defined by
\begin{equation}
\rho_{\Lambda} = \rho_{\Lambda_{0}}.
\end{equation}
Here, $\rho_{m_{0}}$, $\rho_{r_{0}}$, and $\rho_{\Lambda_{0}}$ represents the respective densities at $t=t_{0}$.
Although we have discussed the stress-energy momentum tensor for various matter components, reviewing some general considerations, such as what types of stresses are acceptable at varying levels, is helpful. Consequently, energy conditions are useful for this purpose.

\subsection{Energy Conditions} \label{1.4.5}
Energy conditions (ECs) serve three purposes and impose coordinate invariant restrictions on the stress-energy-momentum tensor of the matter.
First, as Einstein's equation involves no other properties of matter besides its stress tensor, ECs allow us to analyze the behavior of gravitating systems without specifying the behavior of matter in detail. The crucial step that allowed Penrose and Hawking to prove their singularity theorems \cite{Penrose/1965,Hawking/1966} was bypassing a complicated and comprehensive analysis using this method. The second purpose of ECs is to convey a concept of `normal matter' that should apply to various matter kinds. The third objective of the ECs is conceptual simplicity. For example, the positivity of the energy density may be related to the system's stability, at least in the naive sense that systems are stable in classical mechanics when the energy is bounded from below. ECs have great adequacy in classical GR, which considers the singularity problems of spacetime and explains the behavior of spacelike, timelike, or lightlike geodesics \cite{Visser/2000}. These conditions can be derived from the well-known Raychaudhuri equations of the form \cite{Poisson/2004,Hawking/1973}.
\begin{eqnarray}
\label{2.19}
\frac{d\theta}{d\tau} &=& -\frac{1}{3}\theta^2-\sigma_{\mu\nu}\sigma^{\mu\nu}+\omega_{\mu\nu}\omega^{\mu\nu}-R_{\mu\nu}u^{\mu}u^{\nu}\,, \\
\label{2.20}
\frac{d\theta}{d\tau} &=& -\frac{1}{2}\theta^2-\sigma_{\mu\nu}\sigma^{\mu\nu}+\omega_{\mu\nu}\omega^{\mu\nu}-R_{\mu\nu}n^{\mu}n^{\nu}\,,
\end{eqnarray}
where $\theta$, $\sigma^{\mu\nu}$ and $\omega_{\mu\nu}$ are the expansion factor, shear and rotation associated with the geodesic congruence defined by the vector field $u^{\mu}$ and the null vector $n^{\mu}$. 

As a result, there are different ECs, each with distinct advantages and disadvantages in terms of the range of validity, significance and interpretations. We construct scalars from $\mathcal{T}_{\mu \nu}$, which is typically accomplished using arbitrary timelike vectors $t^{\mu}$ or null vectors $n^{\mu}$ \cite{Kontou/2020,Capozziello/2014}.

The weak energy condition (WEC) is probably the most intuitive of the energy conditions. It claims that 
\begin{equation}
\mathcal{T}_{\mu \nu} t^{\mu} t^{\nu} \geq 0,
\end{equation}
for any timelike vector $t^{\mu}$. For a perfect fluid, the WEC suggests that $\rho+ p \geq 0$ and also, $\rho \geq 0$. 
A null energy condition (NEC) is a variation of the WEC, with the timelike vector replaced by a null vector $n^{\mu}$ such that
\begin{equation}
\mathcal{T}_{\mu \nu} n^{\mu} n^{\nu} \geq 0.
\end{equation}
In the perfect fluid description, the NEC asserts $\rho+p \geq 0$. The NEC is crucial because it determines whether the Universe will experience inflation and whether it will evolve into a singularity or bounce solution. The NEC is weaker than the WEC. 

The WEC can be generalized to dominant energy condition (DEC). It indicates that DEC and WEC are equivalent, with the additional requirement that $\mathcal{T}_{\mu \nu} t^{\mu}$ must be either timelike or null. It gives us the perfect fluid condition as $\rho \geq |p|$. 

The strong energy condition (SEC) imposes a bound: 
\begin{equation}
\left(\mathcal{T}_{\mu \nu} - \frac{1}{2}g_{\mu \nu} \mathcal{T}\right) t^{\mu} t^{\nu} \geq 0,
\end{equation}
for every timelike vector $t^{\mu}$. The SEC transforms into $\rho + p \geq 0$ and $\rho + 3p \geq 0$ for a perfect fluid. According to Einstein equation, SEC is strictly geometric, so $R_{\mu \nu} t^{\mu} t^{\nu} \geq 0$. This condition is widely employed and is one of the most important conditions of the Hawking and Penrose singularity theorems.

After determining the fundamental principles underlying the model, many quantities, which are cosmological parameters, may remain undetermined. It is common practice to specify cosmological models with a handful of parameters, which one then attempts to observe in order to determine which version of the model best describes our Universe. In the following section, we will discuss the cosmological parameters that are commonly considered. 

\subsection{Cosmological Parameters}
This section discusses the relevant cosmological parameters for observational cosmology. Consider the Taylor expansion of the scale factor about the present time $t_{0}$. The general form of $a(t)$ is 
\begin{equation}
a(t) = a(t_{0}) + \dot{a(t_{0})} (t-t_{0}) + \frac{1}{2} \ddot{a(t_{0})} (t-t_{0})^{2} + \frac{1}{3!} \dddot{a(t_{0})} (t-t_{0})^{3} + \frac{1}{4!} \ddddot{a(t_{0})} (t-t_{0})^{4}+......
\end{equation}
Dividing by $a(t_{0})$, one gets
\begin{equation}
\frac{a(t)}{a(t_{0})} = 1 + H_{0}(t-t_{0}) - \frac{q_{0}}{2}H_{0}^{2} (t-t_{0})^{2} + \frac{1}{3!} j_{0} H_{0}^{3}(t-t_{0})^{3} + \frac{1}{4!} s_{0} H_{0}^{4} (t-t_{0})^{4}+......
\end{equation}
where $t_{0}$ is the present time. Here, the coefficients are named as Hubble, deceleration, jerk, snap, lerk parameters, respectively. That is 
\begin{eqnarray}
H = \frac{\dot{a}}{a}, \quad q = -\frac{\ddot{a}}{a H^{2}}, \\
j = \frac{\dddot{a}}{a H^{3}}, \quad  s = \frac{1}{a H^{4}} \frac{d^{4}a}{dt^{4}}.
\end{eqnarray}
One can predict whether or not the expansion of the Universe is accelerating or decelerating. So, $q<0$ represents acceleration, while $q>0$ represents deceleration. In addition, the change in the sign of the jerk parameter $j$ in an expanding model denotes an increase or decrease in acceleration.

Now, it is time to talk about the simplest model of the Universe, as we shall see.
\subsection{Friedmann Equations for the Standard Model}

The $\Lambda$CDM model was developed to model universal dynamics such as an accelerated expansion. It consists of a Universe evolving under GR and adding dark energy with a negative equation of state parameter, represented by a cosmological constant. In GR, an accelerated expansion via a cosmological constant is possible, but its origin remains unexplained. This subject will be addressed in subsequent sections. Unless otherwise specified, we will use units such as $c, \hbar =1 $ for the remainder of this work, where $c$ is the speed of light and $\hbar$ is the reduced Planck constant.

In 1915, Einstein's theory of GR was derived from the Einstein-Hilbert action given by
\begin{equation}
S = \int d^{4}x \, \sqrt{-g} \left[\frac{1}{2\kappa}(R-2\Lambda) + L_{m}\right],
\end{equation}
where $\Lambda$ is the cosmological constant and $\kappa = 8\pi G$. The Einstein field equations can be derived by vanishing the variation of the action with respect to metric tensor, that is 
\begin{equation}
\label{LCDM}
G_{\mu \nu} + \Lambda g_{\mu \nu} = \kappa\, \mathcal{T}_{\mu \nu}.
\end{equation}
Furthermore, we examine the cosmological equations of the $\Lambda$CDM model, known as the Friedmann equations. The equations are derived using  equation \eqref{LCDM}, the FLRW metric \eqref{metric}, and the stress-energy momentum \eqref{Stress}, we have
\begin{eqnarray}
\label{F1}
H^{2} &=& \frac{8 \pi G}{3} \rho - \frac{k}{a^{2}} + \frac{\Lambda}{3},\\
\label{F2}
2 \dot{H} + 3 H^{2} &=& -8 \pi G p - \frac{k}{a^{2}} + \Lambda.
\end{eqnarray}
Combining equations \eqref{F1} and \eqref{F2} gives
\begin{equation}
\frac{\ddot{a}}{a} = -\frac{4 \pi G}{3} \left(\rho+3p\right) + \frac{\Lambda}{3},
\end{equation}
which is known as the acceleration equation. If $\ddot{a}>0$, we call the Universe is accelerating, while the Universe is decelerating for $\ddot{a}<0$.

There is a specific density that must exist for a particular value of $H$ in order to make the geometry of the Universe flat, that is $k=0$. This is named as the critical density $\rho_{c}$, which is determined as 
 \begin{equation}
 \rho_{c}= \frac{3H^{2}}{8 \pi G}.
 \end{equation}
Understand that the critical density does not necessarily correspond to the actual density of the Universe since the Universe need not be flat. Therefore, rather than quoting the density of the Universe directly, referring to its value relative to the critical density is frequently more helpful. This quantity is referred to as the density parameter $\Omega$, defined as 
\begin{equation}
\Omega_{i}(t) = \frac{\rho_{i}(t)}{\rho(t)} = \frac{8 \pi G}{3} \frac{\rho_{i}(t)}{H^{2}(t)},
\end{equation}
where $i$ denotes the sum over matter, radiation, spatial curvature and vacuum. One can rewrite the Friedmann equation \eqref{F1} as
\begin{equation}
1 = \Omega_{m}(t) + \Omega_{r}(t) + \Omega_{k}(t) + \Omega_{\Lambda}(t) .
\end{equation}
where $\Omega_{k}(t)$ is the density parameter for curvature defined as 
$\Omega_{k}(t) = -\frac{k}{H^{2}a^{2}}$ and $\Omega_{\Lambda}(t) = \frac{8 \pi G}{3 H^{2}} \rho_{\Lambda}$, where $\rho_{\Lambda} = \frac{\Lambda}{8 \pi G}$. However, if $k=0$, then $\Omega_{m}+\Omega_{r}+\Omega_{\Lambda} =1$, i.e. $\Omega_{total}=1$ at all times. The quantities $\Omega_{i}$ are time-dependent and we can write the values at present time $\Omega_{i,0}$, yielding
\begin{equation}
\label{SH}
H^{2} = H_{0} \left[\Omega_{m,0}\, a^{-3} + \Omega_{r,0}\, a^{-4} + \Omega_{k,0}\, a^{-2} + \Omega_{\Lambda,0}\right].
\end{equation} 
The values for the parameters involved in the above equation \eqref{SH} have been determined by the Planck Collaborations in 2018 \cite{Planck/2018}. We shall now elaborate more on the success and issues of the standard cosmological model in the next section.
\subsection{Successes of $\Lambda$CDM and its Issues}

The standard cosmological model has been extraordinarily successful in characterizing the Universe and is primarily consistent with observations. It can describe the evolution, formation of structures that can only be explained in terms of gravitation within an inflationary, dark energy, dark matter scenario, and the abundance of light elements that can only be explained in primordial nucleosynthesis. Nonetheless, the $\Lambda$CDM approach faces certain shortcomings:

There is a discrepancy between the theoretical and observed values of the cosmological constant. The expected theoretical value exceeds the observed value by nearly 60 orders of magnitude. The theoretical value derived from quantum-mechanical processes utilizing the standard model is $10^{-60}\, M_{Pl}^{4}$, where $M_{Pl}$ is the Planck mass, whereas the observed value is $10^{-120}\, M_{Pl}^{4}$. This discrepancy is referred to as the \textbf{cosmological constant problem} \cite{Joyce/2015}. 

\textbf{The Horizon problem} \cite{Tsujikawa/2013} is a cosmological fine-tuning issue within the Big Bang theory of the Universe, supported by the $\Lambda$CDM model, and is also referred to as the homogeneity problem. It arises from the difficulty of explaining the apparent homogeneity of spatial regions that are casually unconnected in the absence of a process that uniformly sets the initial conditions across all regions. The most widely acknowledged explanation is exponential growth in the early Universe or cosmic inflation.

Another is the \textbf{cosmological coincidence} issue \cite{Velten/2014}, which relates to the fact that the $\Lambda$CDM model predicts that we are in a transitional period between the matter-dominated era and the late-time acceleration era. Observational evidence indicates that the current values of cosmological constant matter densities are comparable in magnitude.

According to current knowledge, \textbf{dark matter (DM)} \cite{Astesiano/2021,Katsuragawa/2017,Zaregonbadi/2016} is non-baryonic and does not interact with other matter components other than gravitational interactions. Without assuming the existence of DM with these characteristics, we cannot explain, for example, the rotational trajectories of galaxies and the formation and distribution of structures in the Universe. Although $\Lambda$CDM assumes the existence of dark matter, neither ground-based nor space-based experiments have detected the signature of DM particles.

There have been disagreements between high and low redshift measurements, such as those based on local measurements and those based on CMB measurements, with the latter assuming $\Lambda$CDM, regarding the estimated value of present Hubble constant $H_{0}$. This tension is known as \textbf{$H_{0}$ tension} \cite{Valentino/2021a} and corresponds to roughly $4.4\sigma$ tension. The Planck data, measurements of weak lensing, and redshift surveys produce a second significant source of tension known as \textbf{$\sigma_{8}$ tension} \cite{Valentino/2021b}. This corresponds to the matter density ($\Omega_{m}$) and the amplitude or growth rate of structure $(\sigma_{8},f\sigma_{8})$. Based on the $\Lambda$CDM, the Planck collaboration estimates $S_{8}= \sigma_{8}\sqrt{\Omega_{m}/0.3} = 0.834 \pm 0.016$, whereas KIDS-450 collaboration estimates $S_{8}= 0.745 \pm 0.039$ \cite{Joudaki/2017}. This results in approximately $2\sigma$ of tension.

These constraints have necessitated the search for an alternative explanation. In the following sections, we update various theories extending beyond $\Lambda$CDM.

\section{Beyond $\Lambda$CDM}
In light of the $\Lambda$CDM issues enumerated in the previous section, one must seek out novel approaches that extend beyond $\Lambda$CDM and, more generally, GR. Possible alternatives to GR will be discussed here. As a result of the flaws in the cosmological constant, new theories to explain cosmic acceleration have been proposed. These can be grouped into two distinct classifications, the first being dark energy (DE) models and the second being modified gravity (MG) models. A DE model is created by adding a DE fluid component parametrized by a static or dynamic equation of state. In MG models, the modification occurs within the gravitational sector, resulting in different field equations from GR. For DE models, modifications are made on the matter side of Einstein equations, ignoring $\Lambda$ and typically adding additional fluid to the set of Einstein equations.
\begin{itemize}
\item It is hypothesized that the dark energy EoS has changed during the evolution of the Universe. As a result, several dynamical dark energy (DDE) models with a time-varying EoS parameter have been proposed. Common DDE models with respect to redshift $z$ include the Chevallier-Polarski-Linder (CPL) model \cite{Linder/2003}, the Jassal-Bagla-Padmanbhan (JBP) model \cite{Jassal/2010}, the Barboza-Alcaniz paramterization \cite{Barboza/2008}, the Wetterich paramterization \cite{Wetterich/2004}, etc.
\item The most widely recognized theory for DDE is the quintessence scalar field model. We consider a scalar field minimally coupled to the matter field, where the scalar field is time-dependent and associated with the potential $V(\phi)$. This $\phi$ exerts a negative pressure, gradually decreasing $V(\phi)$. Quintessence models are categorized into three types based on the nature of potential $V(\phi)$. When $V(\phi) << \dot{\phi}^{2}$, $\omega \approx 1$, which is equivalent to the stiff matter and does not contribute to dark energy. When $V(\phi) >> \dot{\phi}^{2}$, $\omega \approx -1$, corresponds to the cosmological constant. For $ -1<\omega < 1$, we get $\rho \propto a^{-m}$, which gives the accelerated expansion for $0\leq m <2$. The following references help readers familiarise themselves with the wide variety of studies on late-time cosmic acceleration using various quintessence potentials \cite{Gupta/2015,Chiba/2013,Roy/2014,Pantazis/2016}.
\item Caldwell \cite{Caldwell/2002} proposed the phantom field model of DE to explain the late-time cosmic acceleration. In this scenario, the kinetic term exhibits a negative signature. As a result of its negative kinetic energy, a phantom field accelerated expansion of the Universe to an infinite size in a finite amount of time. Big Rip depicts a future in which both the actual volume and expansion rate are infinite. The scalar field models in which the evolution of the EoS parameter resembles that of the phantom field are known as quintom models \cite{Cai/2008,Feng/2005}.
\item The K-essence scalar field model, also known as the K-inflation \cite{Picon/1999,Picon/2001,Chiba/2000}, describes an inflationary model of the early Universe. In contrast to the quintessence models, in which the potential energy term causes an accelerated expansion, the K-essence scalar field models have a dominant kinetic contribution to the energy density, causing the late-time acceleration of the Universe.
\item Another class of theories that modify Einstein's GR provides an alternative explanation for the occurrence of late-time cosmic acceleration. Without relying on a DE component, a wide variety of modified gravity theories are now available in the literature that gives rise to the rapid expansion of the Universe \cite{Tsujikawa/2010,Clifton/2012}. Among these extended theories are $f(R)$ gravity \cite{Carroll/2005,Nojiri/2007,Motohashi/2019,Olmo/2019,Odint/2019,Odint/2020,Oikonomou/2021}, scalar-tensor theories \cite{Sanders/1997,Das/2008,Riazuelo/2002,Boisseau/2000,Crisostomi/2018,Elizalde/2004}, $f(T)$ gravity \cite{Capo/2015,Cai/2018,Nunes/2018,Anagnos/2019,Bahamonde/2020,Bahamonde/2023}, $f(Q)$  gravity \cite{Lazkoz/2019,Barros/2020,Anag/2021,Frusciante/2021,Silva/2022,Hu/2022}, and a few other extended theories \cite{Harko/2011,Myrzakulov/2012,Harko/2010a,Xu/2019,Xu/2020}. Let us discuss some popular theories of modified gravity in the next section.
\end{itemize}

\section{Modified Theories of Gravity} \label{1.6}
Numerous modified theories of gravity extending beyond the standard GR model have been proposed to build a more fundamental framework for explaining dark matter and dark energy and resolving the current observational and theoretical contradictions. MG provides a suitable unification of primordial inflation and cosmic acceleration. The alternative approach to the theory of GR could prove extremely useful for unifying gravity with the theory of quantum mechanics and other fundamental natural interactions.

The Einstein-Hilbert action can be naturally generalized within the framework of Riemannian geometry by substituting the Ricci scalar $R$ with any arbitrary function $f(R)$. This results in the $f(R)$ modified theory of gravity.
There are two approaches to $f(R)$ gravity: the metric formulation, in which the metric is viewed as the only dynamical variable, and the Palatini formulation, in which the connection is considered a fundamental variable together with the metric tensor. Detailed discussions of $f(R)$ gravity can be found in \cite{Sotiriou/2006,Capo/2008a,Motohashi/2019,Stachowski/2017}. The most glaring disadvantage of the $f(R)$ gravity theory is that the scalar field in the Palatini formulation is not dynamic. This implies that no additional degrees of freedom can be introduced, resulting in the existence of physically impossible infinite tidal forces.

\subsection{$f(R)$ Gravity and Its Extensions}

One can generalize the Einstein-Hilbert action by substituting the Ricci scalar $R$ with the generic function $f(R)$ as 
\begin{equation}
\label{R}
S = \frac{1}{2\kappa}\int f(R) \, \sqrt{-g} d^{4}x  + \int  L_{m}\,  \, \sqrt{-g} d^{4}x.
\end{equation}
It is also immediate that GR is recovered at $f(R)=R$.  To obtain the field equations of $f(R)$ gravity, we take into account the variation of action \eqref{R} with respect to the metric tensor, which yields 
\begin{equation}
f'(R) R_{\mu \nu} - \frac{f(R)}{2} g_{\mu \nu} - \left(\nabla_{\mu} \nabla_{\nu} - g_{\mu \nu} \square  \right)f'(R) = 8 \pi G\, \mathcal{T}_{\mu \nu},
\end{equation}
where $\square = \nabla_{\mu}\nabla^{\mu}$ is the D'Alembert operator, $f'(R) =\frac{df(R)}{dR}$, and $\mathcal{T}_{\mu \nu}$ is the stress-energy-momentum tensor defined in \eqref{T}. 
Models of dark energy based on $f(R)$ theories have been intensively studied as the simplest modified gravity scenario to account for the late time acceleration. The model with $f(R) = R + \beta R^{2}$ ($\beta>0$) can cause the Universe to expand at a faster rate due to the presence of $R^{2}$ term. This was the first inflation model proposed by Starobinsky in 1980 \cite{Staro/1980}. Another model with $f(R)= R - \frac{\beta}{R^{n}}$, ($\beta>0, n>0$) was proposed for dark energy in the metric formulation \cite{Nojiri/2003a,Carroll/2004}. However, it was demonstrated that this model suffers from matter instability \cite{Dolgov/2003,Faraoni/2006a} and problems satisfying local gravity constraints \cite{Faraoni/2006b,Erick/2006}. It lacks a conventional matter-dominated era due to a strong coupling between dark energy and dark matter. These results demonstrate how challenging it is to construct reliable models of dark energy.

Henceforth, the $f(R)$ function should meet the following requirements \cite{Sotiriou/2010,Felice/2010}
\begin{itemize}
\item To avoid ghost states, $f(R)>0$, for $R \geq R_{0}$, where $R_{0}$ is the present value of Ricci scalar.
\item To avoid the existence of a scalar degree of freedom with negative mass, i.e. tachyons, we should have $f_{RR}>0$, for $R \geq R_{0}$.
\item $f(R)\rightarrow R-2\Lambda$, for $R \geq R_{0}$. This condition is needed for the presence of a matter-dominated era and for agreement with the local gravity constraints.
\item The condition for stability and late de-sitter limit of the Universe is given by $0< \frac{R f_{RR}}{f_{R}} <1$. 
\end{itemize}
An intriguing extension of gravity is incorporating a non-minimal connection of geometry and matter into the action via the arbitrary function of scalar curvature and Lagrangian density of matter, i.e. $f(R,L_{m})$ gravity or $f(R,\mathcal{T})$ gravity, where $\mathcal{T}$ is the trace of the energy-momentum tensor. Another exciting feature of this theory is that the field equations of $f(R,\mathcal{T})$ gravity reduce to those of $f(R)$ gravity when the energy-momentum tensor is assumed to be traceless, or $\mathcal{T}=0$. 

The action in $f(R,\mathcal{T})$ gravity is of the form \cite{Harko/2011}
\begin{equation}
\label{RT}
S = \frac{1}{2\kappa} \int f(R,\mathcal{T})\, d^{4}x \, \sqrt{-g}  + \int  L_{m}\, d^{4}x \, \sqrt{-g}.
\end{equation}
By varying the action \eqref{RT} with respect to metric tensor, the gravitational field equations of $f(R,\mathcal{T})$ gravity is obtained as 
\begin{equation}
\label{FRT}
f_{R}(R,\mathcal{T}) R_{\mu \nu} -\frac{1}{2} g_{\mu \nu} f(R,\mathcal{T}) + \left(g_{\mu \nu} \square - \nabla_{\mu}\nabla_{\nu} \right) f_{R}(R,\mathcal{T}) = 8\pi G\, \mathcal{T}_{\mu \nu} -f_{\mathcal{T}}(R,\mathcal{T}) \mathcal{T}_{\mu \nu} - f_{\mathcal{T}}(R,T) \Theta_{\mu \nu},
\end{equation}
where $\Theta_{\mu \nu}= g^{\sigma \lambda} \frac{\delta \mathcal{T}_{\sigma \lambda}}{\delta g^{\mu \nu}}$, $f_{R}(R,\mathcal{T}) = \frac{df(R,\mathcal{T})}{dR}$, and $f_{\mathcal{T}}(R,\mathcal{T}) = \frac{df(R,\mathcal{T})}{d \mathcal{T}}$. 
In addition, $f(R,\mathcal{T})$ cosmology has been the subject of substantial research such as the dust fluid is shown to reproduce $\Lambda$CDM, phantom, non-phantom era, Chaplygin gas and scalar field reconstruction, energy conditions, and so on \cite{Jamil/2012,Alvarenga/2013,Fisher/2019}.

One can also write the covariant derivative of equation \eqref{RT} as 
\begin{equation}
\nabla^{\mu} \mathcal{T}_{\mu \nu} = \frac{f_{\mathcal{T}}(R,\mathcal{T})}{8\pi G - f_{\mathcal{T}(R,T)}} \left[(\mathcal{T}_{\mu \nu} + \Theta_{\mu \nu}) \nabla^{\mu} ln\,f_{\mathcal{T}}(R,\mathcal{T}) + \nabla^{\mu} \Theta_{\mu \nu} - \frac{1}{2} g_{\mu \nu} \nabla^{\mu} \mathcal{T} \right].
\end{equation} 
This demonstrates that the energy-momentum tensor is not conserved in the $f(R,\mathcal{T})$ theory. Due to matter-energy coupling, a non-zero value of the covariant divergence of the stress-energy-momentum tensor reflects an additional acceleration that causes massive test particles to follow a non-geodesic motion \cite{Harko/2014b}.
\subsection{Geometrical Representation}

\subsubsection{Geometrical Meaning of Curvature}
We introduce the curvature tensor by considering the parallel displacement of a vector along two distinct paths. Let us consider a parallelogram $ABCD$ with infinitesimally small adjacent sides $AB = dx^{\mu}$ and $AD = \delta x^{\mu}$. Consider a contravariant vector $\left. V^{\mu}\right|_{A}$ defined at point $A$ which is parallelly transported as: first, displace vector $V^{\mu}|_{A}$ parallelly from point $A$ to $B$ as $V^{\mu}|_{B}$. Then, proceed to displace vector $V^{\mu}|_{B}$ parallelly from point $B$ to $C$, resulting in the vector $V^{\mu}|_{CB}$. Secondly, displace vector $V^{\mu}|_{A}$ parallelly from point $A$ to $D$ as $V^{\mu}|_{D}$. Then, proceed to displace vector $V^{\mu}|_{D}$ parallelly from point $D$ to $C$, resulting in the vector $V^{\mu}|_{CD}$. 
Now, we can write the vector $V^{\mu}|_{B}$ obtained by parallel displacement of $V^{\mu}|_{A}$ 
\begin{equation}
\label{C1}
V^{\mu}|_{B} = V^{\mu}|_{A} + dV^{\mu}|_{A}, \quad   \text{where} \quad dV^{\mu}|_{A} = - \Gamma^{\mu}_{\nu \lambda}|_{A}\, V^{\nu}|_{A}\, dx^{\lambda}.
\end{equation} 
Again, the same vector is parallelly displaced from $B$ to $C$ as 
\begin{equation}
\label{C2}
V^{\mu}|_{CB} = V^{\mu}|_{B} + \delta V^{\mu}|_{B}= V^{\mu}|_{B}- \Gamma^{\mu}_{hk}|_{B}\, V^{h}|_{B}\, \delta x^{k}.
\end{equation} 
The connection or the Christoffel symbol depends on the metric tensor, which is a function of coordinates. Hence, for a small displacement, we get 
\begin{equation}
\label{C3}
\Gamma^{\mu}_{hk}|_{B} = \Gamma^{\mu}_{hk}|_{A} + \Gamma^{\mu}_{hk}|_{A,m}\, dx^{m}, \quad \text{where} \quad  \Gamma^{\mu}_{hk}|_{A,m} = \left. \frac{\partial}{\partial x^{m}}\Gamma^{\mu}_{hk}\right|_{A}.
\end{equation}
Substituting equations  \eqref{C3} and \eqref{C1} in \eqref{C2}, we have the following 
\begin{eqnarray}
\nonumber
V^{\mu}|_{CB}&=& V^{\mu}|_{A} - \Gamma^{\mu}_{\nu \lambda}|_{A}\, V^{\nu}|_{A} \,dx^{\lambda} - \left[ \Gamma^{\mu}_{hk}|_{A} + \Gamma^{\mu}_{hk}|_{A,m}\, dx^{m}\right]\left[V^{h}|_{A}-\Gamma^{h}_{\nu \lambda}|_{A}\, V^{\nu}|_{A} \, dx^{\lambda} \right] \delta x^{k},\\ 
 &=& V^{\mu} - \Gamma^{\mu}_{\nu \lambda}\, V^{\nu}\,dx^{\lambda} - \Gamma^{\mu}_{hk}\, V^{h}\, \delta x^{k}  + \Gamma^{\mu}_{hk}\, \Gamma^{h}_{\nu \lambda} V^{\nu} dx^{\lambda} \delta x^{k} - \Gamma^{\mu}_{hk,m} V^{h} dx^{m} \delta x^{k} \label{C4}.
\end{eqnarray}
Here, we ignore the higher-order derivatives. Similarly, one can obtain $V^{\mu}|_{CD}$ by interchanging $dx^{\mu}$ by $\delta x^{\mu}$, that is 
\begin{equation}
\label{C5}
V^{\mu}|_{CD} = V^{\mu} - \Gamma^{\mu}_{\nu \lambda}\, V^{\nu}\,\delta x^{\lambda} - \Gamma^{\mu}_{hk}\, V^{h}\, dx^{k}  + \Gamma^{\mu}_{hk}\, \Gamma^{h}_{\nu \lambda} V^{\nu} \delta x^{\lambda} dx^{k} - \Gamma^{\mu}_{hk,m} V^{h} \delta x^{m} dx^{k}.
\end{equation}
We now subtract equation \eqref{C4} from \eqref{C5} to obtain
\begin{equation}
\label{C6}
V^{\mu}|_{CD} - V^{\mu}|_{CB} = \Gamma^{\mu}_{hk,m} V^{h} dx^{m} \delta x^{k} - \Gamma^{\mu}_{hk,m} V^{h} \delta x^{m} dx^{k} + \Gamma^{\mu}_{hk}\, \Gamma^{h}_{\nu \lambda} V^{\nu} \delta x^{\lambda} dx^{k} - \Gamma^{\mu}_{hk}\, \Gamma^{h}_{\nu \lambda} V^{\nu} dx^{\lambda} \delta x^{k}. 
\end{equation} 
Next, replace $h$ by $\nu$ and $m$ by $\lambda$ in the first two terms of the equation \eqref{C6} on the right-hand side and interchange $\lambda$ by $k$ in the second and third terms. Hence, we obtain
\begin{equation}
V^{\mu}|_{CD} - V^{\mu}|_{CB} = R^{\mu}_{\,\nu \lambda k} V^{\nu} dx^{k} \delta x^{\lambda},
\end{equation}
where 
\begin{equation}
R^{\mu}_{\,\nu \lambda k} = \Gamma^{\mu}_{\nu k,\lambda} - \Gamma^{\mu}_{\nu \lambda,k} + \Gamma^{\mu}_{h \lambda}\, \Gamma^{h}_{\nu k} - \Gamma^{\mu}_{h k}\, \Gamma^{h}_{\nu \lambda}.
\end{equation}
It could be deduced that in a non-Euclidean space, when a tensor is parallelly shifted along a closed curve till returning to the initial point, the resultant vector may not necessarily be the same as the original vector.

\subsubsection{Geometrical Meaning of Torsion}

Let us have a look at a simple example to see how the geometry of torsion works \cite{Sari/2021}. 
Consider the curves $\mathcal{C}:y^{\mu} = y^{\mu}(\lambda)$ and $\tilde{\mathcal{C}}:\tilde{y}^{\mu} = \tilde{y}^{\mu}(\lambda)$. The associated tangent vectors are 
\begin{equation}
u^{\mu} = \frac{dy^{\mu}}{d\lambda}, \quad and \quad \tilde{u}^{\mu} = \frac{d\tilde{y}^{\mu}}{d\lambda}. 
\end{equation}
Assume that $d\tilde{y}^{\mu}$ represent a displacement of $u^{\beta}$ along $\tilde{\mathcal{C}}$ and we get $u'^{\beta}$ as 
\begin{equation}
\label{1.60}
u'^{\beta} = u^{\beta} + (\partial_{\mu} u^{\beta}) d\tilde{y}^{\mu},
\end{equation}
but since $u^{\beta}$ is parallel transported along $\tilde{\mathcal{C}}$, then 
\begin{equation*}
\frac{d\tilde{y}^{\mu}}{d\lambda} \tilde{\nabla}_{\mu} u^{\beta} = 0 = \frac{d\tilde{y}^{\mu}}{d\lambda} \partial_{\mu}u^{\beta} + \tilde{\Gamma}^{\beta}_{\,\,\nu \mu} \frac{d\tilde{y}^{\mu}}{d\lambda} u^{\mu},
\end{equation*}
which means 
\begin{equation}
\partial_{\mu}u^{\beta} d\tilde{y}^{\mu} = - \tilde{\Gamma}^{\beta}_{\,\,\nu \mu} u^{\nu} \tilde{u}^{\mu} d\lambda,
\end{equation}
Henceforth, \eqref{1.60} becomes 
\begin{equation}
u'^{\beta} = u^{\beta} - \tilde{\Gamma}^{\beta}_{\,\,\nu \mu} u^{\nu} \tilde{u}^{\mu} d\lambda. 
\end{equation}
Similarly, for a displacement $dx^{\mu}$ of $\tilde{u}^{\beta}$ along $\mathcal{C}$, we obtain 
\begin{equation}
\tilde{u}'^{\beta} = \tilde{u}^{\beta} - \tilde{\Gamma}^{\beta}_{\,\,\nu \mu} \tilde{u}^{\mu} u^{\nu} d\lambda,
\end{equation}
and 
\begin{equation}
(\tilde{u}^{\beta}+ u'^{\beta})-(u^{\beta}+ \tilde{u}'^{\beta}) = - T^{\beta}_{\,\,\mu \nu} \tilde{u}^{\mu} u^{\nu} d\lambda.  
\end{equation}
It is now apparent that the vectors $(\tilde{u}^{\beta}+ u'^{\beta})$ and $(u^{\beta}+ \tilde{u}'^{\beta})$ are not identical in the presence of torsion, which rules out the existence of the infinitesimal parallelogram. The latter can be expressed as 
\begin{equation}
\tilde{V}^{\beta} = - T^{\beta}_{\,\,\mu \nu} \tilde{u}^{\mu} u^{\nu},
\end{equation}
where $\tilde{V}^{\beta} d\lambda$ is the deviation vector, indicates how much the parallelogram has been broken. Hence, the presence of torsion cracks parallelograms into pentagons.
\subsubsection{Geometrical Meaning of Non-Metricity}

Let us have a look at the geometry of non-metricity. Consider two vectors $u^{\mu}$ and $v^{\mu}$ on a differential manifold with a metric and connection. We define an inner product of these vectors as $u.v = u^{\mu} v^{\nu} g_{\mu \nu}$ and parallel transport both the vectors along a curve $\mathcal{C}: x^{\mu} = x^{\mu}(\lambda)$, we obtain 
\begin{equation}
\tilde{\nabla}_{\lambda}(u.v) =  \frac{dx^{\alpha}}{d\lambda} \left(\tilde{\nabla}_{\alpha} u^{\mu} \right) v_{\mu} + \frac{dx^{\alpha}}{d\lambda} \left(\tilde{\nabla}_{\alpha} v^{\nu} \right) u_{\nu} + \frac{dx^{\alpha}}{d\lambda} \left(\tilde{\nabla}_{\alpha} g_{\mu \nu} \right) u^{\mu} v^{\nu}. 
\end{equation}
The condition of parallel transport of $u^{\mu}$ and $v^{\mu}$ results in 
\begin{equation}
\frac{dx^{\alpha}}{d\lambda} \left(\tilde{\nabla}_{\alpha} u^{\mu} \right) = 0, \quad and \quad \frac{dx^{\alpha}}{d\lambda} \left(\tilde{\nabla}_{\alpha} v^{\mu} \right),
\end{equation}
and hence
\begin{equation}
\tilde{\nabla}_{\lambda}(u.v) = Q_{\alpha \mu \nu} \frac{dx^{\alpha}}{d\lambda}  u^{\mu} v^{\nu}.
\end{equation}
This demonstrates that parallel transporting two vectors around a curve changes the inner product. Equalizing the two vectors requires 
\begin{equation}
\tilde{\nabla}_{\lambda}(|u|^{2}) = Q_{\alpha \mu \nu} \frac{dx^{\alpha}}{d\lambda} u^{\mu} u^{\nu},  
\end{equation}
which shows how the length of the vector's magnitude changes when we parallel transport it along a given curve. Henceforth, the length of a vector changes when being transported in parallel in a space with non-metricity.

\textbf{Note:} The most common combinations of three geometries are defined as 
\begin{itemize}
\item The combination $R^{\mu}_{\,\,\nu \sigma \lambda} \neq 0$,~~ $T^{\beta}_{\,\,\mu \nu} = 0$,~~$Q_{\beta \mu \nu}=0$ with a Levi-Civita connection is known as the GR spacetime.
\item The combination $R^{\mu}_{\,\,\nu \sigma \lambda} \equiv 0$,~~ $T^{\beta}_{\,\,\mu \nu} \neq 0$,~~$Q_{\beta \mu \nu} \neq 0$ gives the Teleparallelism. 
\item The combination $R^{\mu}_{\,\,\nu \sigma \lambda} \equiv 0$,~~ $T^{\beta}_{\,\,\mu \nu} \neq 0$,~~$Q_{\beta \mu \nu} \equiv 0$ gives the Torsional Teleparallelism.
\item The combination $R^{\mu}_{\,\,\nu \sigma \lambda} \equiv 0$,~~ $T^{\beta}_{\,\,\mu \nu} \equiv 0$,~~$Q_{\beta \mu \nu} \neq 0$ shows the non-metricity Teleparallelism.
\item  The combination $R^{\mu}_{\,\,\nu \sigma \lambda} \equiv 0$,~~ $T^{\beta}_{\,\,\mu \nu} \equiv 0$,~~$Q_{\beta \mu \nu} \equiv 0$ results in the Minkowski space. In this case, the connection is fixed upto diffeomorphism and does not carry any gravitational degrees of freedom.
\end{itemize}

The best strategy to further relax the Riemannian constraints is now evident from the above discussion. In other words, a generic affine connection that allows for torsion and non-metricity is required to reach the realm of non-Riemannian geometry. The generic affine connection encompasses torsion and non-metricity degrees of freedom with ease. 
The affine connection can be written as \cite{Jimenez/2018a}
\begin{equation}
\tilde{\Gamma}^{\sigma}_{\,\,\mu \nu} = \Gamma^{\sigma}_{\,\,\mu \nu} + \frac{1}{2} g^{\lambda \sigma} \left(-Q_{\mu \nu \lambda} -Q_{\nu \lambda \mu} + Q_{\lambda \mu \nu}\right) + \frac{1}{2} g^{\lambda \sigma} \left(T_{\nu \lambda \mu} + T_{\mu \lambda \nu}-T_{\lambda \mu \nu}  \right),
\end{equation}
where $\Gamma^{\sigma}_{\,\,\mu \nu} = \frac{1}{2}g^{\sigma \lambda}\left( g_{\lambda \nu,\mu} + g_{\mu \lambda,\nu} - g_{\mu \nu,\lambda} \right)$ represents the usual Levi-Civita connection and the terms associated with the previous combination are known as Disformation and Contortion tensors given by 
\begin{eqnarray}
\label{L}
L^{\lambda}_{\,\,\mu \nu} &=& \frac{1}{2} g^{\lambda \sigma} \left(-Q_{\mu \nu \lambda} -Q_{\nu \lambda \mu} + Q_{\lambda \mu \nu}\right),\\
\label{K}
K^{\lambda}_{\,\,\mu \nu} &=& \frac{1}{2} g^{\lambda \sigma} \left(T_{\nu \lambda \mu} + T_{\mu \lambda \nu}-T_{\lambda \mu \nu}  \right), 
\end{eqnarray}
with $T^{\sigma}_{\,\, \mu \nu} = \tilde{\Gamma}^{\sigma}_{\,\,\nu \mu}-\tilde{\Gamma}^{\sigma}_{\,\,\mu \nu}$ and $Q_{\sigma \mu \nu}= \nabla_{\sigma}g_{\mu \nu}$.
Let us discuss more about torsion teleparallel and non-metricity teleparallel theories.
\subsection{Teleparallel Equivalent to GR and its Extensions}

GR expresses gravitation via the metric tensor and the torsion-free Levi-Civita connection. Curvature is thus displayed through the connection rather than the metric itself. Teleparallel gravity (TG) replaces the standard gravity connection with the Weitzenb\"{o}ck connection, which is curvature-less and satisfies the metricity connection ($\nabla_{\alpha} g_{\mu \nu} =0)$ \cite{Hayashi/1979}. In TG, the metric is a derived quantity that emerges from the tetrad\footnote{At each point of the manifold, the vierbein fields/tetrad form an orthonormal basis for the tangent space, which is presented by the line element of the four-dimensional Minkowski spacetime, i.e. $e_{\mu}e_{\nu} = \eta_{\mu \nu} = diag(-1,+1,+1,+1)$. The tetrads can be used to transform between inertial and non-inertial indices as $g_{\mu \nu} = e^{a}_{\,\,\mu} e^{b}_{\,\,\nu} \eta_{ab}$ with $e^{a}_{\,\,\mu} e_{b}^{\,\,\mu} = \delta^{a}_{b}$}. 
The curvature-less Weitzen\"{o}ck connection in TG is defined by 
\begin{equation}
\tilde{\Gamma}^{\lambda}_{\,\,\mu\nu} =  e^{\lambda}_{\,\,a} \partial_{\nu} e^{a}_{\,\,\mu} = -e^{a}_{\,\,\nu} \partial_{\nu} e^{\lambda}_{\,\,a}.
\end{equation}
The action for TEGR \cite{Sotiriou/2011,Li/2011} is written as 
\begin{equation}
S= -\frac{1}{2\kappa} \int T \, \sqrt{-g}\, d^{4}x + \int L_{m} \, \sqrt{-g}\, d^{4}x,
 \end{equation}
where $T$ is the torsion scalar defined as $T = S_{a}^{\,\,\,\, \mu \nu} T^{a}_{\,\,\,\,\mu \nu}$. For this purpose, we introduce the torsion and contortion tensor as follows
\begin{eqnarray}
\label{T1}
T^{\lambda}_{\,\,\mu \nu} &=& \tilde{\Gamma}^{\lambda}_{\,\,\nu \mu} - \tilde{\Gamma}^{\lambda}_{\,\,\mu \nu} =  e^{\lambda}_{\,\,a} \left( \partial_{\mu} e^{a}_{\,\,\nu} - \partial_{\nu} e^{a}_{\,\,\mu} \right),\\
\label{T2}
K^{\lambda}_{\,\,\mu \nu} &=& \tilde{\Gamma}^{\lambda}_{\,\,\nu \mu} - \Gamma^{\lambda}_{\,\,\mu \nu} = \frac{1}{2} \left({T_{\mu}^{\,\,\lambda}}_{\nu} + {T_{\nu}^{\,\,\lambda}}_{\mu}-T^{\lambda}_{\,\,\mu\nu}   \right) ,
\end{eqnarray}
where $\Gamma^{\lambda}_{\,\,\mu \nu}$ is the Levi-Civita connection.
Subsequently, the superpotential tensor is obtained using equations \eqref{T1} and \eqref{T2}, that is 
\begin{equation}
S_{a}^{\,\, \mu \nu}  = K^{\mu \nu} _{\,\,\,\, a} -e_{a}^{\,\,\nu} T^{\lambda \mu} _{\,\,\,\, \lambda} + e_{a}^{\,\, \mu} T^{\lambda \nu}_{\,\, \,\, \lambda}.
\end{equation} 
Now, we can extend TEGR by taking an arbitrary function of torsion scalar $T$ defined by $f(T)$
\begin{equation}
\label{ATG}
S= \frac{1}{2\kappa} \int f(T) \, \sqrt{-g}\, d^{4}x + \int L_{m} \, \sqrt{-g}\, d^{4}x.
\end{equation}
Varying the action \eqref{ATG} with respect to tetrad, one gets the following field equations for $f(T)$ gravity
\begin{equation}
e^{-1} \partial_{\mu}\left(e\, e^{\lambda}_{\,\,a} S_{\lambda}^{\,\,\,\, \mu \nu} \right) f_{T} + e^{\lambda}_{\,\,a} S_{\lambda}^{\,\,\,\, \mu \nu} \partial_{\mu}(T) f_{TT} - f_{T} e^{\lambda}_{\,\,a} T^{\lambda}_{\,\, \mu \rho} S_{\lambda}^{\,\,\,\, \rho \mu}  + \frac{1}{4} e^{\nu}_{\,\, a} f(T) =  4\pi G \,e^{\lambda}_{\,\,a} \Theta_{\lambda}^{\,\, \nu},
\end{equation}
where $f_{T} = \frac{df}{dT}$, $f_{TT} = \frac{d^{2}f}{dT^{2}}$, and $\Theta_{\lambda}^{\,\,\nu}$ denotes the energy-momentum tensor of the matter sector. 

Since the torsion scalar $T$ is only dependent on the first-order derivatives of the tetrads, this theory is of the second order. One can  review $f(T)$ gravity in \cite{Cai/2016,Nesseris/2013,Basilakos/2019,Nunes/2016a}. 

In addition to an arbitrary function of the torsion scalar, $f(T)$ gravity can be extended by permitting an arbitrary function of both torsion $T$ and energy-momentum tensor $\mathcal{T}$, such as $f(T,\mathcal{T})$ gravity. The action for $f(T,\mathcal{T})$ gravity is given by  \cite{Harko/2014a}
\begin{equation}
S = \frac{1}{2 \kappa} \int \left[T+f(T,\mathcal{T})\right] \sqrt{-g}\, d^{4}x + \int L_{m} \sqrt{-g}\, d^{4}x.
\end{equation}
Varying the action with respect to tetrad yields the field equations given by 
\begin{multline}
(1+f_{T})\left[ e^{-1} \partial_{\mu} \left(e\,e^{\lambda}_{a} S_{\lambda}^{\,\,\,\, \alpha \mu}\right) - e^{\lambda}_{\,\,a} T^{\mu}_{\,\,\nu \lambda} S_{\mu}^{\,\,\,\, \nu \alpha} \right] + \left( f_{TT} \partial_{\mu}T + f_{T\mathcal{T}} \partial_{\mu}\mathcal{T} \right)e^{\lambda}_{\,\,a} S_{\lambda}^{\,\,\,\, \alpha \mu} + e^{\alpha}_{\,\,a} \left(\frac{f+T}{4} \right)\\
 - f_{\mathcal{T}} \left( \frac{e^{\lambda}_{\,\,a} \overset{\mathit{em}}{\mathcal{T}}_{\lambda}^{\,\, \alpha} + p e^{\alpha}_{\,\,a}}{2} \right) = 4\pi G\, e^{\lambda}_{\,\,a} \overset{\mathit{em}}{\mathcal{T}}_{\lambda}^{\,\, \alpha},
\end{multline}
where $f_{T} = \frac{\partial f}{\partial T}$ and $f_{T\mathcal{T}} = \frac{\partial^{2}f}{\partial T \, \partial \mathcal{T}}$. 
\subsection{Symmetric Teleparallel Equivalent to GR and its Extensions} \label{1.6.6}

This section will examine the theory of symmetric teleparallel equivalent to general relativity (STEGR). GR is a metric theory in which the covariant derivative of the metric is zero. In the meantime, non-metricity theories are derived from the non-metricity tensor, which is defined below
\begin{equation}
\label{NM}
Q_{\alpha \mu \nu} = \nabla_{\alpha}g_{\mu \nu}.
\end{equation}
If $Q_{\alpha \mu \nu}(\Gamma,g)=0$, we get a metric-compatible geometry. In non-metric theories, non-metricity measures the amount by which the length of vectors varies when they are parallel transported.

Assuming the relationship in the equation \eqref{NM} gives rise to non-metric theories such as symmetric teleparallelism in which the non-metricity manifests as a flat torsion-free geometry. As the equation \eqref{L} demonstrates, we can derive the disformation tensor from the non-metricity tensor, which measures the Levi-Civita connection's deviation from the symmetric part of the entire connection.

It is useful to introduce the non-metricity conjugate, which is defined as 
\begin{equation}
P^{\alpha}_{\,\,\mu \nu} = -\frac{1}{2}L^{\alpha}_{\,\, \mu \nu} + \frac{1}{4} Q^{\alpha} - \tilde{Q}^{\alpha}_{\,\,\mu \nu} g_{\mu \nu} -\frac{1}{4} \delta^{\alpha}_{(\mu}Q_{\nu)},
\end{equation}
where $Q_{\alpha} = g^{\mu \nu} Q_{\alpha \mu \nu}$ and $\tilde{Q}_{\alpha} = g^{\mu \nu} Q_{\mu \alpha \nu}$ are the two independent traces. Henceforth, one can define the non-metricity scalar as $Q = -Q_{\alpha \mu \nu} P^{\alpha \mu \nu}$. 

The action for STEGR is 
\begin{equation}
S = -\frac{1}{2\kappa} \int Q \sqrt{-g}\, d^{4}x + \int L_{m} \sqrt{-g}\, d^{4}x.
\end{equation}
It is important to note that the difference between the invariant $Q$ and the Ricci scalar is a boundary term. The theory described by $Q$, in which there is no boundary term, is a type of special STG  analogous to an enhanced version of GR. The connection between tangent space and spacetime is entirely trivial and represents a much more straightforward geometric interpretation of gravity; the origins of tangent space and spacetime coincide. This theory is known as coincident GR, i.e., STEGR \cite{Jimenez/2018}. 

We can also extend STEGR by considering an action defined by an arbitrary function $f(Q)$ as
\begin{equation}
\label{fq}
S = -\frac{1}{2\kappa} \int f(Q) \sqrt{-g}\, d^{4}x + \int L_{m} \sqrt{-g}\, d^{4}x.
\end{equation}
Motivating a specific choice of non-metricity scalar and the preceding action is the replication of GR for the choice $f(Q) = Q$.
The variational principle with respect to $g_{\mu \nu}$ yields the field equations
\begin{equation}
\label{fqf}
\frac{2}{\sqrt{-g}} \nabla_{\alpha}\left(\sqrt{-g}f_{Q} P^{\alpha}_{\,\,\mu \nu}\right) + \frac{1}{2}g_{\mu \nu} f + f_{Q}\left(P_{\mu \alpha \beta} Q_{\nu}^{\,\, \alpha \beta} - 2 Q_{\alpha \beta \mu} P^{\alpha \beta}_{\,\,\,\nu}\right) = 8 \pi G\, \mathcal{T}_{\mu \nu},
\end{equation}
where $f_{Q} = \frac{df}{dQ}$. 

Various geometrical and physical aspects of STEGR and $f(Q)$ gravity have been investigated in a number of studies \cite{Lazkoz/2019,Frusciante/2021,Barros/2020}. 

Another specific modified gravity that yields a general class of non-linear gravity model having the action as  \cite{Xu/2019}
\begin{equation}
S = \frac{1}{2\kappa} \int f(Q,\mathcal{T}) \sqrt{-g}\, d^{4}x + \int L_{m} \sqrt{-g}\, d^{4}x,
\end{equation}
where $f(Q,\mathcal{T})$ is a general function of $Q$ and the trace of energy-momentum tensor $\mathcal{T}$. In the presence of geometry-matter coupling, the general field equation describing gravitational phenomena is obtained by varying the action with respect to the metric tensor
\begin{equation}
\label{QTF}
-\frac{2}{\sqrt{-g}} \nabla_{\alpha}\left(\sqrt{-g}f_{Q} P^{\alpha}_{\,\,\mu \nu}\right) - \frac{1}{2}g_{\mu \nu} f + f_{\mathcal{T}} \left( \mathcal{T}_{\mu \nu} + \Theta_{\mu \nu} \right)-f_{Q}\left(P_{\mu \alpha \beta} Q_{\nu}^{\,\, \alpha \beta} - 2 Q_{\alpha \beta \mu} P^{\alpha \beta}_{\,\,\,\,\nu}\right) = 8 \pi G\, \mathcal{T}_{\mu \nu}.
\end{equation}
Broadly speaking, there are two distinct phases. Firstly, it is essential to establish a comprehensive understanding of the model. Ideally, this model should exhibit a high degree of simplicity. Furthermore, once the model has been determined, we utilize our data to quantify the values of the parameters. The forthcoming section provides an overview of diverse observational methodologies.
\section{Cosmological Observations}
This section will introduce fundamental statistical concepts. In cosmological reconstruction, the best-fit parameter values can be derived through either statistical methods: minimizing the function through optimization or maximizing the likelihood function via marginalization. An alternative is to conduct a Markov Chain Monte Carlo (MCMC) analysis, which uses a marginalization of the posterior probability distribution over the parameter space to derive constraints on the parameter values. 
\subsection{Basics of Statistics}
\subsubsection{$\chi^{2}$ Minimization}
Based on observational data, the $\chi^{2}$ statistic has been extensively used to estimate the parameters of theoretical models. If we assume an observational data $(z_{i},f_{i})$ with a standard deviation $\sigma_{i}$, an experimentally measured value ${f_{i}}_{obs}$, and a theoretically anticipated value ${f_{i}}_{th}$, then the $\chi^{2}$ is defined as
\begin{equation}
\chi^{2}= \sum_{i} \frac{\left[{f_{i}}_{obs} - {f_{i}}_{th}(\{\theta\})  \right]^{2}}{\sigma_{i}^{2}},
\end{equation}
where $\{\theta\}$ denotes the set of model parameters.

If the $f_{i}$'s are related to each other, then the $\chi^{2}$ function can be defined as 
\begin{equation}
\chi ^{2} = \sum_{i,j} \left[{f_{i}}_{obs} - {f_{i}}_{th}(\{\theta\}) \right]^{Tr} (Cov^{-1})_{ij} \left[{f_{j}}_{obs} - {f_{j}}_{th}(\{\theta\}) \right],
\end{equation}
where $Tr$ represents the transpose of the matrix, $(Cov^{-1})_{ij}$ is a covariance matrix needed to characterize the errors of the data.

To acquire the parameters of the best-fitting model, we must minimize the $\chi^{2}$ function. We define another quantity, which is the reduced $\chi^{2}_{red}$ given by the expression
\begin{equation}
\chi^{2}_{red} = \frac{\chi^{2}}{d},
\end{equation}
where $d$ signifies the degrees of freedom. The necessary condition that needs to be checked for presenting any over-fitting is $\chi^{2}_{red} < 1$. Moreover, one can determine $\Delta \chi^{2} = \chi^{2}- \chi^{2}_{min}$ for $1\sigma$ (68\%), $2\sigma$ (95\%), and $3\sigma$ (99\%) confidence level ranges of a specific model.
\subsubsection{Maximum Likelihood Analysis}

According to Baye's theorem, the posterior probability distribution of parameters $\{\theta\}$ is obtained as 
\begin{equation}
P\left( \{\theta\}/\mathcal{D}, I\right) = \frac{P \left( \{\theta\}/I \right) P \left( \mathcal{D}/\{\theta\},I \right) }{P\left(\mathcal{D}/I \right)},
\end{equation}
where $I$ represents the prior information and $\mathcal{D}$ is the observational data. $P \left( \{\theta\}/I \right)$ gives the prior probabilities and $P \left( \mathcal{D}/\{\theta\},I \right)$  is the probability of obtaining $\mathcal{D}$ if $\{\theta\}$ is given referenced to $I$, also known as the likelihood $\mathcal{L}$. Further, $P\left(\mathcal{D}/I \right)$ is the global likelihood which serves as a normalization factor
\begin{equation}
P\left(\mathcal{D}/I \right) = \int_{\theta_{1}} .....\int_{\theta_{n}}  P \left( \{\theta\}/I \right)\,P\left( \mathcal{D}/\{\theta\},I \right)\, d\theta_{1}.... d\theta_{n},
\end{equation}
such that 
\begin{equation}
\int_{\theta_{1}} .....\int_{\theta_{n}}  P \left( \{\theta\}/\mathcal{D},I \right)\, d\theta_{1}.... d\theta_{n}=1.
\end{equation}

The likelihood $\mathcal{L}$ is related to $\chi^{2}$ function defined as 
\begin{equation}
\mathcal{L}\left(\{\theta\} \right) = exp\left(-\frac{\chi^{2}}{2} \right).
\end{equation}
Thus, one can note that the minimized value of $\chi^{2}$ corresponds to the maximized likelihood function $\mathcal{L}$. Using the MCMC approach, we sample the posterior probability distribution throughout the parameter space to determine the best-fit parameter values and associated error uncertainties. Using the Metropolis-Hastings (MH) algorithm provides a more effective means of exploring parameter space through random walks from one set of parameter values to the next. The MH rule assesses whether or not a given set of random walks should be accepted by comparing the likelihood of the new and old sets of parameter values. When a fit to the data is optimal, the algorithm is more inclined towards the regions with the highest likelihood. The shape of $\mathcal{L}$ around the maximum is then explored as it wanders around that area of the parameter space. This investigation maps out the posterior probability of each parameter value by maximizing the marginal likelihood function, also known as the integrated likelihood, which results in the marginalized constraints in the parameter space. As a result, the MCMC analysis is comparable to the minimization, where the parameters are marginalized rather than optimized.

\subsection{Observational data} \label{sec1.7.2}
Cosmological observations have been crucial in understanding the history of the expansion of the Universe.  We shall briefly review several cosmological findings in the search for cosmic evolution.
\subsubsection{Hubble Measurements}

The H(z) dataset provides strong evidence for the fine structure of the expansion history of the Universe. Direct measurements of H(z) at different redshifts, based on the ages of the most massive and passively evolving galaxies, generate another standard probe in cosmology. Notably, H(z) measurements are derived from two distinct techniques: galaxy differential age (also known as cosmic chronometer) and radial baryon acoustic oscillations (BAO) size methodologies \cite{Ratra/2018}. 
The authors of \cite{Moresco/2012,Moresco/2016} present 13 H(z) values derived from the BC03 and MaStro SPS models \cite{Bruzual/2003}, which we will refer to as the CCB compilation and the CCM compilation, respectively. The authors of \cite{Maraston/2011,Zhang/2014} provide only 5 H(z) values obtained using the BC03 model; these values have been added to the CCB compilation. In the case of \cite{Moresco/2015}, the combined MaStro/BC03 values for 2 H(z) measurements are available. In \cite{Stern/2010}, an alternative SPS, distinct from the MaStro and BC03 models, is presumed to consist of 11 H(z) values and is subsequently referred to as the CCH compilation and the other 26 points assessed using BAO \cite{Sharov/2018}.
\subsubsection{Type Ia Supernovae}

When a white dwarf star explodes, it causes a tremendous explosion in a large-scale structure known as a Type Ia Supernovae (SNeIa). An explosion occurs when a white dwarf star approaches the Chandrasekhar mass limit after gaining mass from a companion star. Therefore, SNeIa can be used as a standard candle to measure the luminosity distance. 
In 1998, Riess et al. \cite{riess/1998} discovered the accelerated expansion of the Universe using 16 distant and 34 nearby SNeIa from the Hubble telescope observations. In 1999, Perlmutter et al. \cite{perlmutter/1999} confirmed the cosmic acceleration by analyzing 18 nearby supernovae (SNe) from the Calan-Tololo sample and 42-high-redshift SNe. Many research groups have focused on this field, such as the Sloan Digital Sky Survey (SDSS) SNe Survey \cite{Holtzman/2008}, the Lick Observatory Supernova Search (LOSS) \cite{Leaman/2011}, the Carnegie Supernova Project (CSP) \cite{Folatelli/2010}, the Nearby Supernova Factory (NSF) \cite{Copin/2006}, the Supernova Legacy Survey (SNLS) \cite{Astier/2006}, and the Higher-Z Team \cite{Riess/2004}, etc. Moreover, recently the Union 2.1 SNeIa dataset, consisting of 580 SNeIa was released \cite{Visser/2004}. The Pantheon compilation is one of the most up-to-date compilations of data on type Ia supernovae (SNeIa), which contains 1048 points in the redshift range $0.01< z< 2.26$ \cite{Scolnic/2018}. 
\subsubsection{Baryon Acoustic Oscillations}

Baryon Acoustic Oscillations refer to the aggregation or overdensity of baryonic matter at specific length scales caused by acoustic waves propagating in the early universe \cite{Peebles/1970}. Similar to SNeIa, BAO provides a standard candle for length sales in cosmology, which enables us to investigate the expansion history of the Universe. On the matter power spectrum, BAO leaves a distinctive imprint. Consequently, astronomical surveys of galaxy clusters allow for their measurement at low redshifts ($z < 1$) \cite{Alam/2004}. In addition, BAO scales can also be measured through reionization emission, which provides extensive information regarding the early Universe at high redshifts ($1.5 \leq z \leq 20$) \cite{Johnson/2004}. The Hubble parameter and angular diameter distance can be calculated using the apparent magnitude of the BAO as determined by astronomical observations.  Numerous investigations have been conducted for BAO measurements, including the Two-degree-Field Galaxy Redshift Survey (2dFGRS) \cite{Colless/2003}, the Sloan Digital Sky Survey (SDSS) \cite{York/2000}, etc. SDSS is the most successful survey for BAO observation, releasing its eighth data continuously in 2011 \cite{link}.

The following chapters employ the modified gravity theories we have been discussing to some specific challenges with the above observational data.




\chapter{Effective Equation of State in Modified Gravity and Observational Constraints} 

\label{Chapter2} 

\lhead{Chapter 2. \emph{Effective Equation of State in Modified
Gravity and Observational Constraints}} 


\blfootnote{*The work in this chapter is covered by the following publication:\\
\textit{Effective equation of state in modified
gravity and observational constraints}, Classical and Quantum Gravity, \textbf{37}, 205022 (2020).}

The current chapter presents the effective equation of state (EoS) in modified $f(R,\mathcal{T})$ gravity. The detailed study of the work is outlined as follows:
\begin{itemize}
\item We consider an effective equation of state with bulk viscosity to study the cosmological evolution of the Universe. 
\item The present study aims to derive the Hubble parameter and deceleration parameter in terms of redshift $z$ to explain the late time accelerating phase. In order to distinguish the present model from other dark energy models, we also present the statefinder and $Om(z)$ diagnostics analyses. 
\item We eventually provide constraints on the cosmological parameters of our model using 580 points of Type Ia Supernovae (Union 2.1) and the updated version of 57 Hubble datasets.
\end{itemize}

\section{Introduction} 

\textbf{Viscosity}: The $\Lambda$CDM model has received significant attention since it can explain most observations despite its inability to describe the accurate physics of DM and DE. Although the $\Lambda$CDM model agrees with a wide range of experimental data, there exists a substantial phenomenological gap. The two primary issues are the coincidence problem and fine-tuning problem. These issues have prompted the development of numerous dark energy models, including quintessence, perfect fluid models, and scalar fields. In addition, many authors have asserted that cosmic viscosity regulates late time accelerated expansion. The viscosity theories in cosmology are significant in relation to the early Universe when the temperature was approximately $10^4$K (during the neutrino split). In the cosmic fluid, there are two different viscosity coefficients: bulk viscosity and shear viscosity. Due to the accepted spatial isotropy of the Universe, such as the Robertson-Walker metric descriptions, we omit shear viscosity. By contemplating a bulk viscous fluid, the issue of identifying a viable mechanism for the origin of bulk viscosity in an expanding Universe is addressed. Physically, bulk viscosity is viewed as an internal friction resulting from various cooling rates in an expanding matter. By converting the kinetic energy of the particles into heat, its dissipation reduces the effective pressure in an expanding fluid \cite{Okumura/2003}. The existence of viscosity parameters in a fluid is fundamentally attributable to the thermodynamic irreversibility of the motion. If the deviation from reversibility is small, it can be assumed that the momentum shift between various parts of the fluid is linearly dependent on the velocity derivatives. This condition corresponds to the paradigm of constant viscosity. When the viscosity is proportional to the Hubble parameter, on the other hand, the momentum shift entails second-order quantities in the deviation from reversibility, resulting in more intriguing physical outcomes. Consequently, the correct selection of their coefficients may result in crossing the phantom divide line \cite{Brevik/2017}. Other studies indicate that the bulk viscosity is sufficient to drive the cosmic fluid from the quintessence to the phantom region. Sharif and Yousaf \cite{Yousaf/2014} have also investigated stability regions for a non-static restricted class of axially symmetric geometries. Their work includes shearing viscous fluid that collapses non-adiabatically. 

\textbf{Viscosity in modified gravity:} If we consider the problem of cosmic adaptation, i.e., the average stage of low redshift, one can justify the accelerating expansion of the Universe by modifying Einstein equations geometrically. Bulk viscosity can also produce acceleration without a scalar field or cosmological constant if connected to inflation. The bulk viscosity contributes to the pressure term and exerts an additional pressure that accelerates the expansion of the Universe \cite{Odintsov/2020}. Most arguments for standard gravity presume that the fluid of the Universe is perfect and non-viscous. From the standpoint of hydrodynamics, the two viscosity coefficients discussed above enter into play, which indicate a first-order deviation from thermal equilibrium. Thus, this supports Eckart's approach \cite{Eckart/1940} from 1940 due to its non-causal behavior. Therefore, taking second-order deviations from a thermal equilibrium leads to a causal theory respecting special relativity. Now, it is also essential to consider more realistic models, which process due to complex viscosity, for example, research in \cite{Singh/2014} studied the role of bulk viscosity in the evolution of the Universe using a modified $f(R,\mathcal{T})$ gravity model. In addition, many authors have investigated the concept of bulk viscous fluid to explain the accelerating expansion of the Universe \cite{Pady/1987,Cheng/1991,Samanta/2017}. Davood \cite{Davood/2019} also investigated the role of bulk viscosity in $f(T)$ gravity. Eckart originally proposed the form of cosmic pressure $p=(\gamma -1)\rho-3\zeta H$, where $\gamma$ parametrizes the EoS parameter \cite{Ren/2006}. However, Eckart's theory undergoes some anomalies. One of those is the instability of the equilibrium states \cite{Hiscock/1985}. Also, dissipative perturbations propagate at infinite velocities \cite{Israel/1976}.
In 1979, Israel and Stewart \cite{Stewart/1979} developed a more general theory that was causal and stable. The first-order limit of Stewart's viewpoint can also be used to develop the Eckart theory \cite{Titus/2015}.   Fisher and Carlson \cite{Fisher/2019} examined the form $f(R,\mathcal{T})= f_1(R) + f_2(\mathcal{T})$, in which they state that $f(R,\mathcal{T})$ yields a new physics and limits could be placed on the cross-terms by comparing them to observations. Harko and Moreas \cite{Harko/2020a} revised their work to investigate observational constraints on the function $f_2(\mathcal{T})$. The impact of viscosity on the finite-time future singularities in $f(T)$ gravity has been studied by Setare and Houndjo \cite{Setare/2013}. Sharif and Rani \cite{Rani/2013} have also studied dark energy viscosity in $f(T)$ gravity. Viscosity in $f(R)$ gravity is discussed by Brevik \cite{Iver/2012}.

The plan of this chapter is to study the FLRW model with bulk viscosity effects in the modified $f(R,\mathcal{T})$ gravity theory with a general effective equation of state $p=(\gamma-1)\rho +p_{0}+\omega_{H}H+\omega _{H2}H^{2}+\omega _{dH}\dot{H}$. This chapter is divided into different sections. In section \ref{IIa}, we introduce the field equations with bulk viscosity. Then, we describe the general solution and the behavior of various cosmological parameters in section \ref{IIIa}. In section \ref{IVa}, we conduct numerous tests to validate the model, which includes the energy conditions, statefinder, and $Om(z)$ diagnostics. To study all the cosmological parameters, we obtain the best-fit values of model parameters using the observational datasets H(z), SNeIa, and BAO in section \ref{Va}. The final section \ref{VIa} presents the conclusion.

\section{Field Equations}\label{IIa}

Let us consider a FLRW metric in the flat space geometry ($k=0$) by
\begin{equation}
ds^{2}=dt^{2}-a^{2}(t)[dr^{2}+r^{2}d\theta ^{2}+r^{2}\sin ^{2}\theta d\phi
^{2}],  \label{m1}
\end{equation}
with the four-velocity $u^{\mu}= \left(1,0,0,0\right)$ and the cosmic fluid along with a bulk viscosity $\zeta$. So, one can rewrite the energy-momentum tensor for a viscous fluid as follows
\begin{equation} 
\label{2.1}
\mathcal{T}_{\mu \nu}= \rho\, u_{\mu}u_{\nu}- \overline{p} \, h_{\mu \nu},
\end{equation}
where $h_{\mu \nu}= g_{\mu \nu}+ u_{\mu}u_{\nu}$ and the effective pressure is $\overline{p}= p-3\zeta\,H$. If we choose the Lagrangian density as $L_{m}= -\overline{p}$, then the tensor $\Theta_{\mu \nu}$ can be written as
\begin{equation} 
 \label{2.2}
\Theta_{\mu \nu}= -2\,\mathcal{T}_{\mu \nu}-\overline{p}\, g_{\mu \nu}.
\end{equation}
Now, remembering the field equation \eqref{FRT} of $f(R,\mathcal{T})$, we shall find the field equation for the bulk viscous fluid using equations \eqref{2.1}, \eqref{2.2} and the functional form $f(R,\mathcal{T})=R+2f(\mathcal{T})$ (assuming $\kappa=1$) as
\begin{equation}
\label{2.3}
R_{\mu \nu}-\frac{1}{2}R\,g_{\mu \nu}= \mathcal{T}_{\mu \nu}+ 2\,f^{\prime
}(\mathcal{T})\,\mathcal{T}_{\mu\nu}+(2\, \overline{p}\,f^{\prime }(\mathcal{T})+f(\mathcal{T})) g_{\mu\nu}.
\end{equation}
For the particular choice of the function $f(\mathcal{T})=\eta \mathcal{T}$, $\eta$ as a constant and $R=6(2H^{2}+\dot{H})$, one can obtain the following Friedmann equations
\begin{equation}
3H^{2}=\rho +2\eta\, (\rho +\overline{p})+\eta \,\mathcal{T},  \label{2.4}
\end{equation}
and 
\begin{equation}
2\dot{H}+3H^{2}=-\overline{p}+\eta \,\mathcal{T},  \label{2.5}
\end{equation}
where $\mathcal{T}=\rho -3\,\overline{p}$. It can be noted that the field equations reduce to general relativity for $\eta = 0$ ($\eta$ is the coupling constant for modified gravity). Finally, we combine equations \eqref{2.4} and \eqref{2.5} to get a single equation as follows
\begin{equation}
2\dot{H}+(1+2\eta )(p+\rho )-3(1+2\eta)\zeta H=0.  \label{2.6}
\end{equation}

\section{General Solution}\label{IIIa}

The equations \eqref{2.4} and \eqref{2.5} contain four unknown parameters, namely $\rho$, $\overline{p}$, $\zeta$ \& $H$. To obtain an exact solution from the equations, we need two more physically viable equations. As stated in the introduction, we will consider the subsequent EoS (as given in the explicit form in \cite{Ren/2006})
\begin{equation}
\label{2.7}
p =(\gamma -1)\rho +p_{0}+\omega _{H}H+\omega _{H2}H^{2}+\omega _{dH}\dot{H},
\end{equation}
where $p_{0},\omega _{H},\omega _{H2},\omega _{dH}$ are free parameters. In this chapter, we assume that the Universe is filled with a single fluid described by the above defined EoS. Next, we shall make use of the following form of the time-dependent bulk viscosity
\begin{equation}
\label{2.8}
\zeta =\zeta _{0}+\zeta _{1}\frac{\dot{a}}{a}+\zeta _{2}\frac{\ddot{a}}{\dot{a}},  
\end{equation}
to check that this form is effectively equivalent to the form derived by equation \eqref{2.7}, where $\zeta _{0},\zeta _{1},\zeta _{2}$ are constants. The justification behind this is as follows
\begin{align*}
\overline{p}= p-3\,\zeta H &= p-3 \left(\zeta_{0}+\zeta_{1} \frac{\dot{a}}{a}+\zeta_{2} \frac{\ddot{a}}{\dot{a}}\right) H, \\
&= p-3 \,\zeta_{0} H-3\, \zeta_{1} H^{2}-3\, \zeta_{2}(\dot{H}+H^{2}),
\end{align*}
which gives 
\begin{equation}  \label{2.9}
\overline{p}= p-3\, \zeta_{0} H-3(\zeta_{1}+\zeta_{2}) H^{2}-3 \,\zeta_{2} \dot{H}.
\end{equation}
Hence, one can write the corresponding coefficients for the simplicity by
\begin{align*}
& \omega_{H} =-3\, \zeta_{0}, \\
& \omega_{H2} =-3 (\zeta_{1} + \zeta_{2}), \\
& \omega_{dH}=-3\, \zeta_{2}.
\end{align*}
According to the work done in the literature, it is important to mention that the expression for bulk viscosity $\zeta= \zeta_{0} + \zeta_{1}\,H$ indicates that the expansion of scale factor deviates more rapidly from the expansion rate of a perfect fluid for larger bulk viscosity coefficient. Furthermore, for $\zeta=0$ and $\zeta = \zeta_{1}\,H$, the scale factor exhibits a power-law dependence on cosmic time, and the values of $q$ remain constant \cite{Singh/2014}.

The motivation behind investigating the general form of bulk viscosity emerges from the understanding in fluid mechanics that the transport and viscosity phenomenon are intricately connected to the concept of velocity. This velocity, in turn, is associated with the Hubble parameter and acceleration. Given the lack of precise knowledge regarding the specific nature of viscosity, we opt to examine a parameterized bulk viscosity that can be expressed as a linear combination of three distinct factors. The initial term in the equation represents a constant value denoted by $\zeta _{0}$. The second term is directly proportional to the Hubble parameter, which characterizes the relationship between bulk viscosity and velocity. Lastly, the third term is proportional to $\frac{\ddot{a}}{\dot{a}}$, representing the impact of acceleration on bulk viscosity. 

Combining equations \eqref{2.4}, \eqref{2.7}, and \eqref{2.8}, we obtain the explicit form of energy density
\begin{equation}
\rho =\frac{\eta \,p_{0}+2\,\eta\, \omega _{H}H+(2\,\eta\,
\omega_{H2}+3)H^{2}+2\,\eta\, \omega _{dH}\dot{H}}{1+4\,\eta -\eta\, \gamma }.
\label{2.10}
\end{equation}
Subsequently, the bulk viscous pressure using equation \eqref{2.5} is obtained as 
\begin{equation}
\label{2.11}
\overline{p}=\frac{\eta^{2}\,p_{0}+2\,\eta^{2}\omega _{H}H+(2\,\eta^{2}\omega _{H2}-9\,\eta -3+3\,\eta\, \gamma )H^{2}+(2\,\eta^{2}\omega_{dH}-2(1+4\,\eta -\eta\, \gamma ))\dot{H}}{(1+4\,\eta -\eta\, \gamma)(1+3\,\eta)}.  
\end{equation}
Finally, we get the following differential equation using equations \eqref{2.6}, \eqref{2.7}, \eqref{2.11}. That is 
\begin{multline}
\label{2.12} 
\left[ 2+\frac{2(1+2\eta)\omega _{dH}(1+4\eta)}{1+4\eta -\eta\gamma }\right] \dot{H}+\left[ \frac{2(1+2\eta )\omega _{H}(1+4\eta)}{1+4\eta -\eta \gamma }\right] H+\left[ \frac{(1+2\eta)(2\omega_{H2}(1+4\eta)+3\gamma )}{1+4\eta -\eta\gamma }\right] H^{2}+
\\
\left[ \frac{(1+2\eta)p_{0}(1+4\eta)}{1+4\eta -\eta \gamma }\right]
=0.
\end{multline}
We notice that the above equation is highly non-linear, so without loss of generality, we assume $p_0=0$. As a result, the equation becomes simple and the time evolution of the Hubble parameter $H$ can be expressed as 
\begin{equation}
 \label{2.13}
H=\frac{k_{1}}{k_{1}e^{k_{1}t}-k_{2}}, 
\end{equation}
where $k_{1}=\dfrac{\frac{2(1+2\eta)\omega _{H}(1+4\eta)}{
1+4\eta-\eta\gamma }}{2+\frac{2(1+2\eta)\omega _{dH}(1+4\eta)}{1+4\eta-\eta \gamma }}$, $k_{2}=\dfrac{\frac{(1+2\eta )(2\omega
_{H2}(1+4\eta)+3\gamma )}{1+4\eta -\eta\gamma }}{2+\frac{2(1+2\eta
)\omega_{dH}(1+4\eta)}{1+4\eta -\eta \gamma }}$. Using the definition $H=\frac{\dot{a}}{a}$, the scale factor is obtained as  
\begin{equation}  \label{2.14}
a=k_{4} k_{1}^{-\frac{1}{k_{2}}}(k_{1}-k_{2} e^{-k_{1}t})^{\frac{1}{k_{2}}},
\end{equation}
where $k_{4}$ is the constant of integration.

\begin{figure}[]
\centering
\includegraphics[width=9 cm]{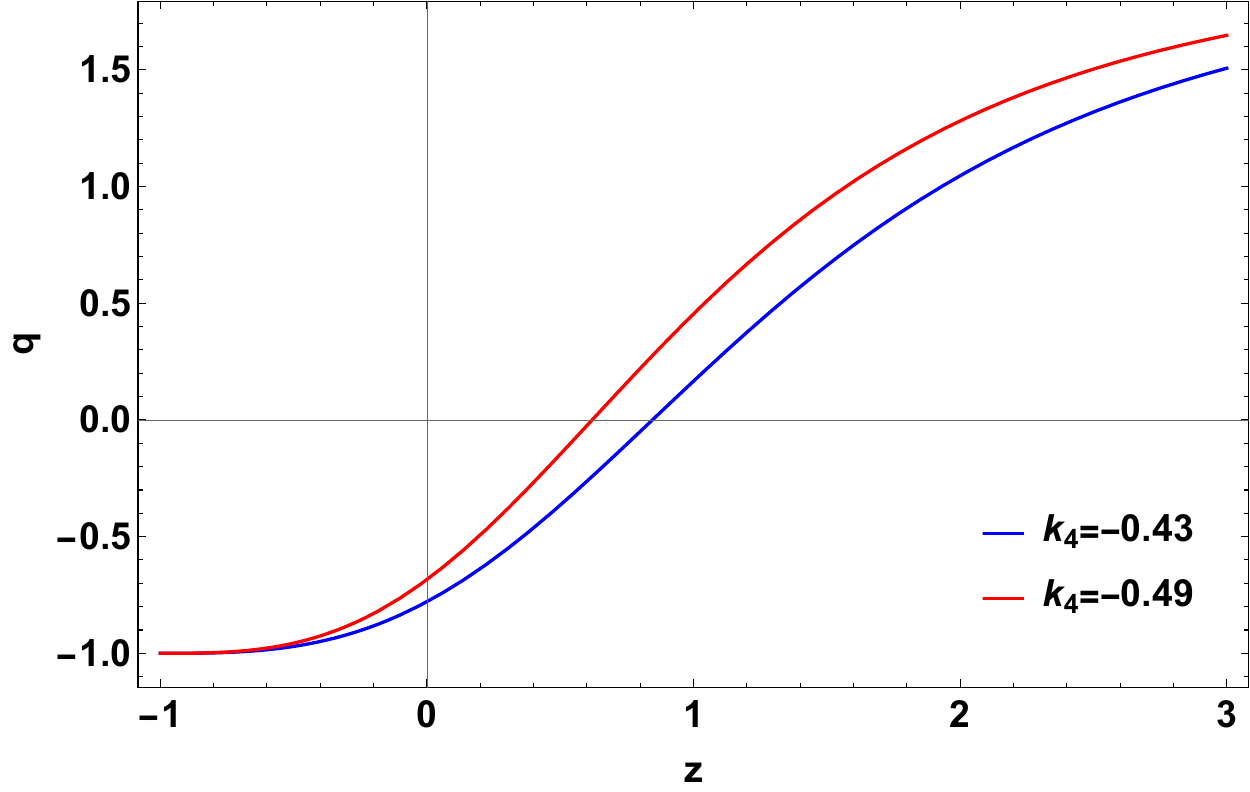}
\caption{Evolution of the deceleration parameter versus redshift $z$ for $k_{2}=3$ and $k_{4}=-0.43, -0.49$.}
\label{f21}
\end{figure}

Next, we make substitutions from the above equation of scale factor to obtain the deceleration parameter ($q=-\frac{a\ddot{a}}{\dot{a}^{2}}$) given by 
\begin{equation}
q=-1+k_{1}e^{k_{1}t}.  \label{2.15}
\end{equation}
The solution set for the formulated system is now readily accessible. The present work aims to examine the dynamics of different cosmological parameters to understand the evolutionary trajectory of the Universe better. Our objective is to investigate the numerous phases of the Universe by imposing constraints on the model parameters, focusing on the transition from a decelerated phase to an accelerated phase. It is understood that a positive value of the parameter $q$ is associated with the decelerating phase of the Universe, whereas a negative value signifies its accelerating behaviour. As a result, we shall present all cosmological parameters by utilizing the redshift $z$ as a reference, as per the established relation, that is 
\begin{equation}
a(t)=\frac{1}{1+z}, \label{2.16}
\end{equation}
with $a_{0}= a(z=0)=1$. The Hubble parameter and deceleration parameter are two observable parameters that can be rewritten in terms of redshift as 
\begin{equation}
H(z)= H_{0} \frac{\left[(k_{4}+k_{4}z)^{k_{2}}-1\right]}{k_{4}^{k_{2}}-1},  \label{2.17}
\end{equation}
and 
\begin{equation}
q(z)=-1+k_{2}\dfrac{(k_{4}+k_{4}z)^{k_{2}}}{(k_{4}+k_{4}z)^{k_{2}}-1}.
\label{2.18}
\end{equation}
According to a recent study by  \cite{Mamon/2017}, it has been shown that the present deceleration parameter, denoted by $q_{0}$ is estimated to be $q_{0}= -0.51_{-0.01}^{+0.09}$. Moreover, the transition redshift from deceleration to acceleration is obtained as $z_{t}= 0.65_{-0.17}^{+0.19}$ \cite{Garza/2019}. The transition of the Universe from a decelerated phase to an accelerated phase at $z_{t}\approx 0.7$ has been depicted in several studies \cite{Knop/2003,Ishida/2008,Cunha/2009,Rani/2015}.
Henceforth, in the current model, we have selected specific values of the free parameters ($k_{1}$, $k_{2}$, $k_{4}$) in order to ensure that our $q_{0}$ and $z_{t}$ values are consistent with values reported in the existing literature. In this discourse, we shall examine a particular model as an illustration and study the cosmic history of the Universe by employing a set of numerical choices of the values of these model parameters. To be specific, we assumed two parameters $\gamma$ and $\eta$ according to the works done previously and found the other parameters using the equations of $k_{1}$, $k_{2}$ and $k_{3}$. Nevertheless, we have  taken the values of $k_{2}$ and $k_{4}$ that observational datasets have constrained in the further section. Figure \ref{f21} illustrates the evolution of $q(z)$ with a suitable choice of the model parameters. It is observed that the deceleration parameter $q$ undergoes a transition from positive to negative at $z_{t}=0.84$ and $z_{t}=0.62$. The corresponding values of $q_{0}$, i.e. at present, are found to be with $q_{0}=-0.78$ and $q_{0}=-0.68$ for two different values of $k_{4}=-0.43$ and $k_{4}=-0.49$, respectively. This observation suggests that the Universe exhibits a transition from early deceleration to the current acceleration in the framework of this model.

\begin{figure}[H]
\begin{center}
$
\begin{array}{c@{\hspace{.1in}}c}
\includegraphics[width=3.1 in, height=2.4 in]{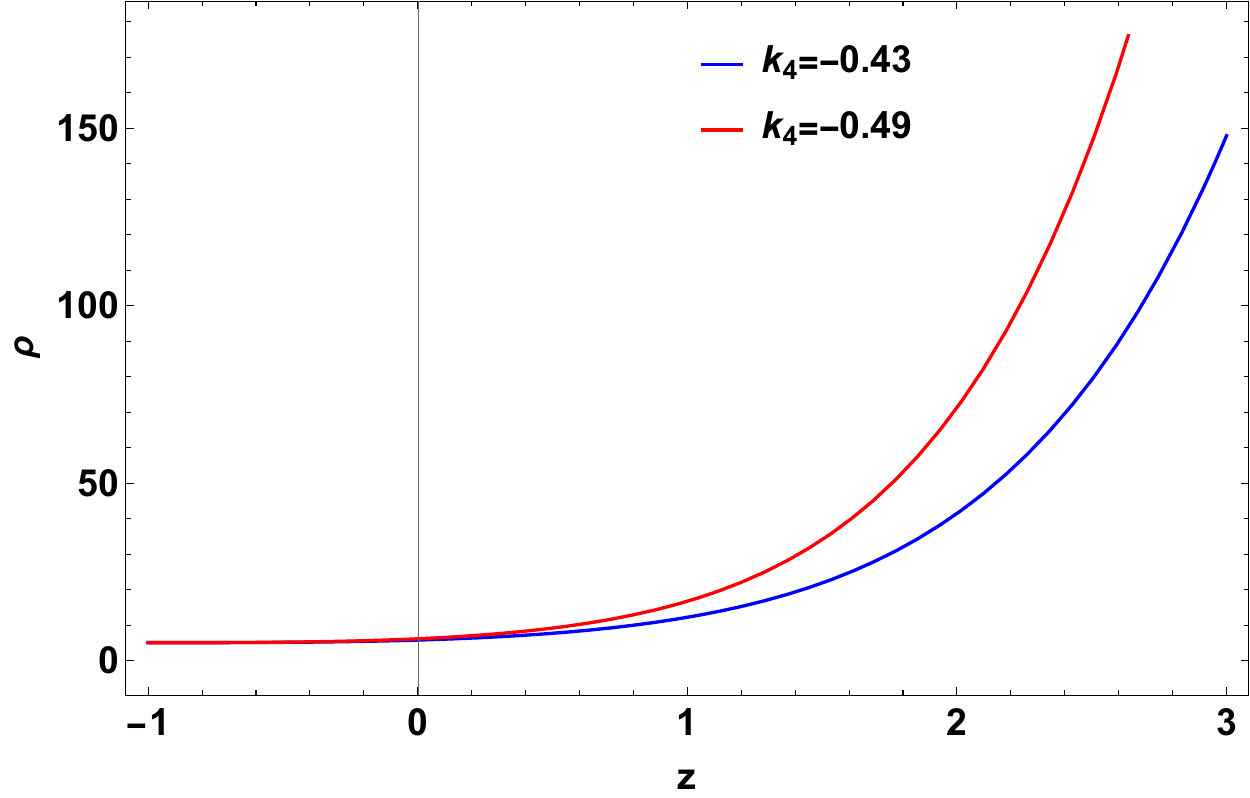} & 
\includegraphics[width=3.1 in, height=2.4 in]{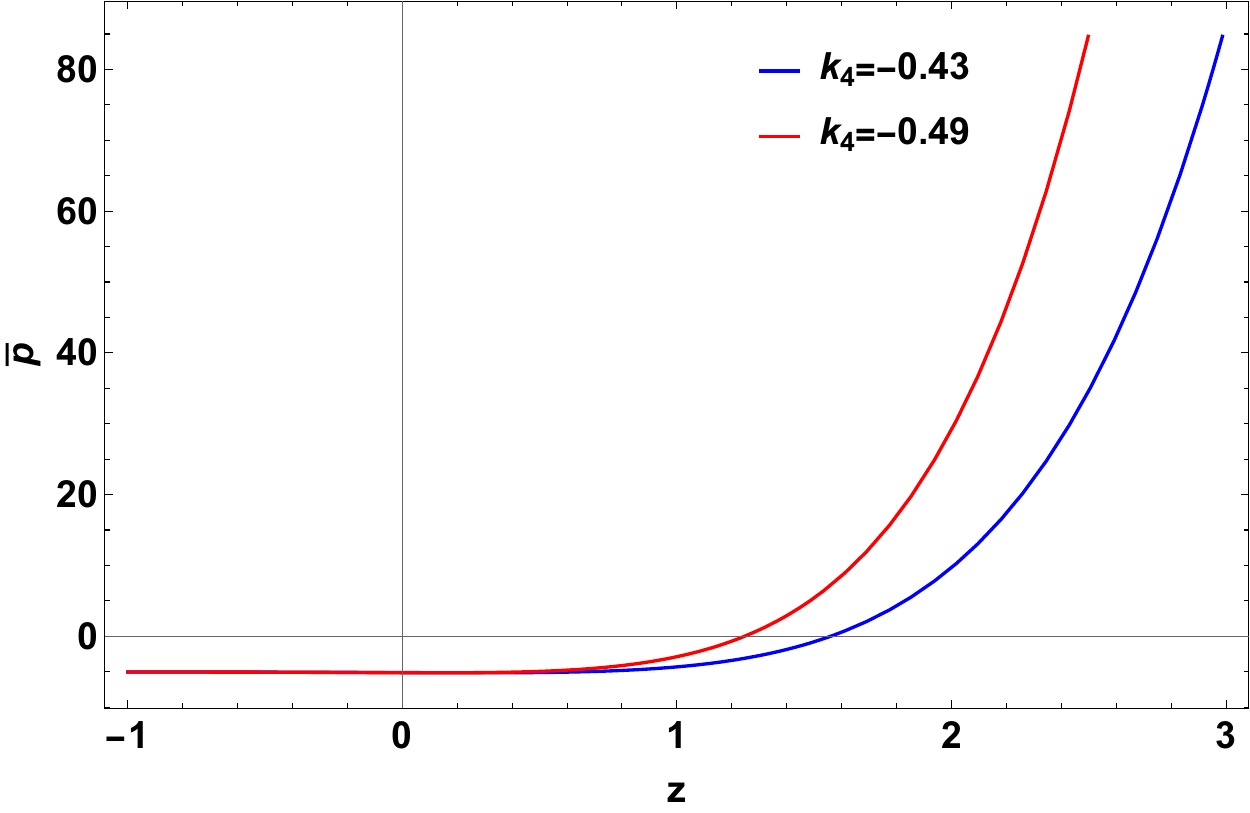}\\
\end{array}
$
\caption{Evolution of the density and pressure versus redshift $z$ for $\protect\eta=-0.1$, $\protect\gamma=1.01$, $\protect\omega_{H}=4.1$, $\protect\omega_{H2}=1.57$, $\protect\omega_{dH}=-0.1$ and $k_{4}=-0.43, -0.49$.}
\label{f22}
\end{center}
\end{figure}

The behavior of $\rho$ and $\bar{p}$ from equations \eqref{2.10} and \eqref{2.11} with respect to redshift $z$ is also depicted in figure \ref{f22}. The observed behavior suggests that the energy density increases as a function of the redshift parameter $z$, whereas the effective pressure now demonstrates a negative behavior, implying the Universe is undergoing an accelerated expansion. 

\begin{figure}[H]
\centering
\includegraphics[width=9 cm]{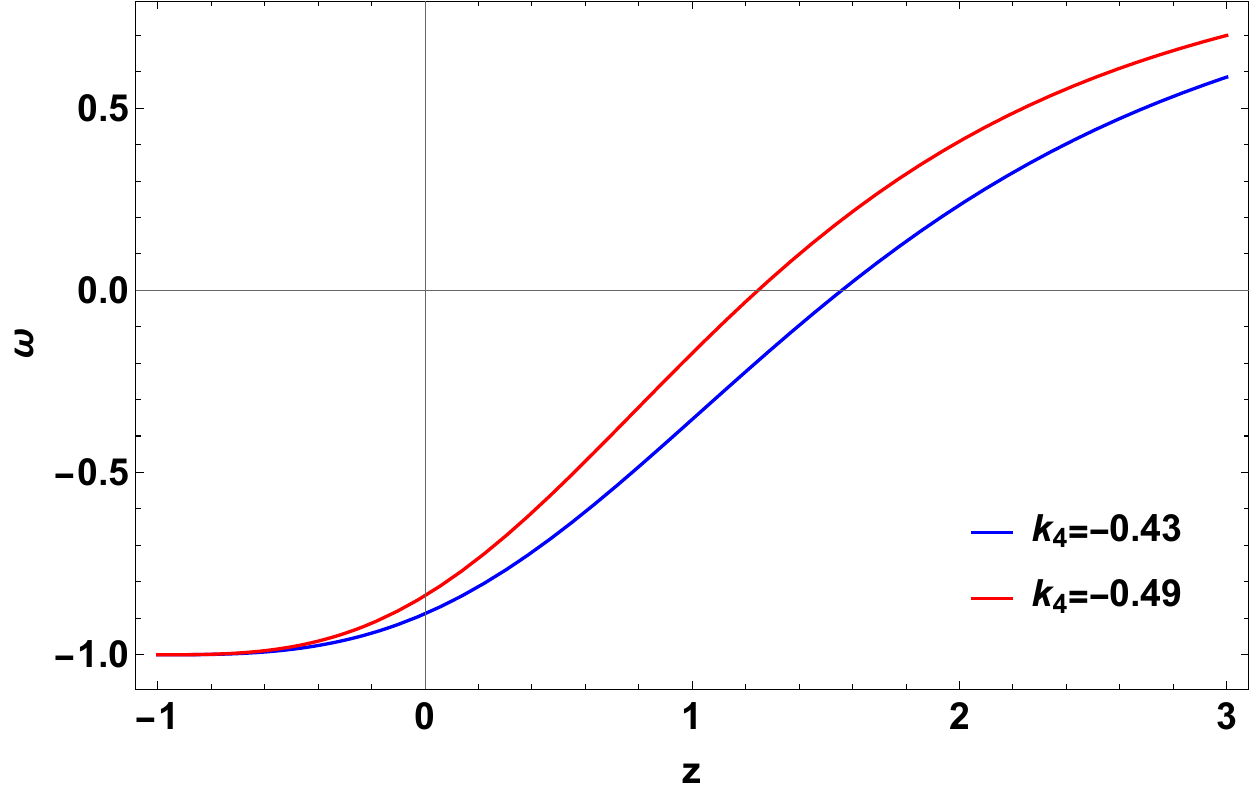}
\caption{Evolution of the equation of state versus redshift $z$ for $\protect\eta =-0.1$, $\protect\gamma =1.01$, $\protect\omega _{H}=4.1$, $\protect\omega _{H2}=1.57$, $\protect\omega_{dH}=-0.1$, $k_{4}=-0.43, -0.49$.}
\label{f24}
\end{figure}

The EoS parameter is employed to categorize the different phases of the Universe, distinguishing between the decelerating and accelerating eras. It assigns numerous epochs into several classifications as outlined: when $\omega = 1$, it signifies the presence of stiff fluid. If $\omega = 1/3$, the model depicts a radiation-dominated phase while $\omega =0$ represents a matter-dominated phase. In the present accelerated stage of evolution, $-1 < \omega < 0$ indicates the quintessence phase, $\omega = -1$ represents the cosmological constant, i.e., $\Lambda$CDM model, and $\omega < -1$ yields the phantom era. In figure \ref{f24}, the plot displays the EoS parameter versus redshift $z$, utilising the identical model parameter values as previously described. 

The graph in figure \ref{f24} shows that as $z\rightarrow -1$, $\omega \rightarrow -1$ in the future. This also illustrates a transition from positive to negative in due course of evolution. It signifies the earlier decelerating phase of the Universe with positive pressure, which is favourable to the structure formation. Subsequently, it indicates the present accelerating phase of the evolution characterized by negative pressure. It is clear that the present model does not cross the phantom divide line, thereby ensuring that the model is free from Big Rip singularity. The present values of the EoS parameter is obtained as $\omega _{0}=-0.88$ for $k_{4}=-0.43$ and $\omega _{0}=-0.84$ for $k_{4}= -0.49$ together with the stated values of other model parameters. In the following section, we will discuss cosmological diagnostics and observational datasets.

\begin{figure}[H]
\begin{center}
$
\begin{array}{c@{\hspace{.1in}}c}
\includegraphics[width=3.1 in, height=2.4 in]{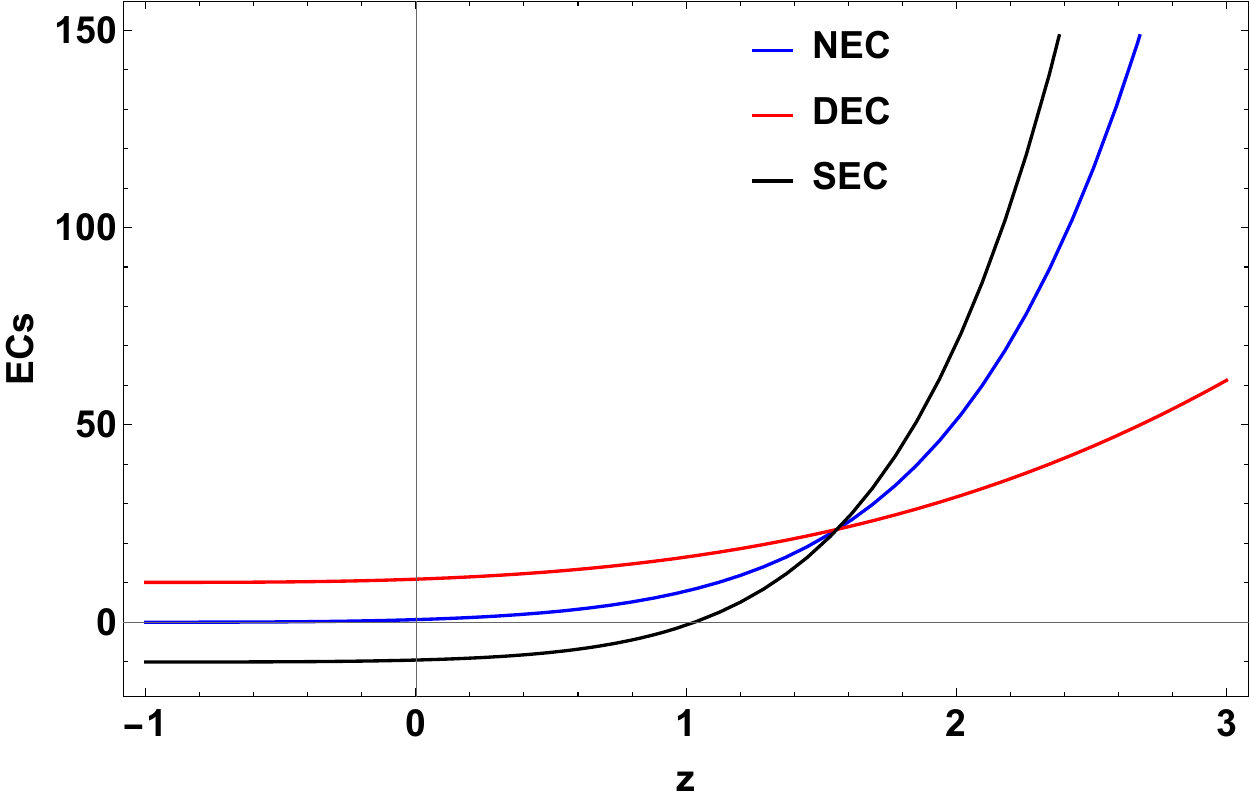} & 
\includegraphics[width=3.1 in, height=2.4 in]{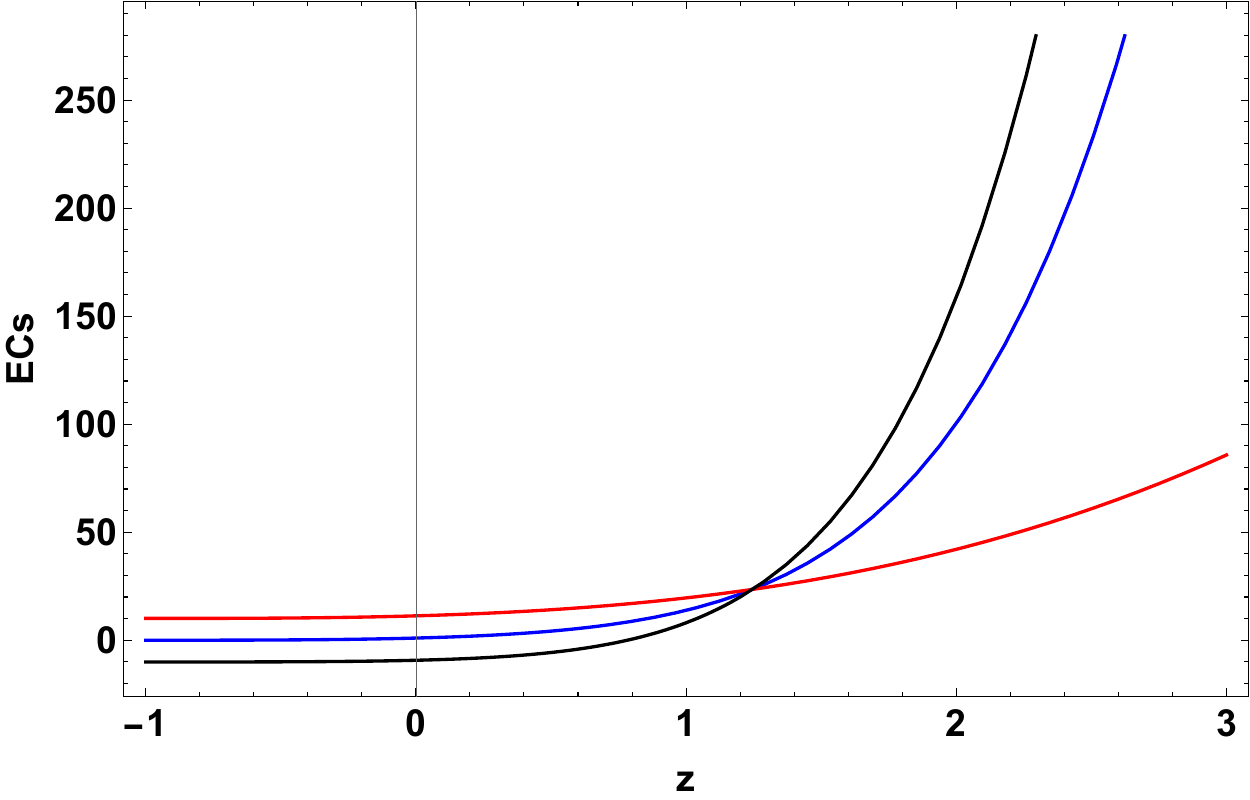}\\
\end{array}
$
\caption{Evolution of energy conditions versus redshift $z$ for $\protect\eta =-0.1$, $\protect\gamma =1.01$, $\protect\omega _{H}=4.1$, $\protect\omega _{H2}=1.57$, $\protect\omega_{dH}=-0.1$, $k_{4}=-0.43, -0.49$, respectively.}
\label{f25}
\end{center}
\end{figure}

Next, we present various energy conditions as mentioned in section \ref{1.4.5}. The violation of NEC implies that none of the mentioned ECs are validated. The SEC is currently a subject of much discussion for the current accelerated expansion of the Universe \cite{Barcelo/2002, moraes/2017}. SEC must be violated in cosmological scenarios both during the inflationary expansion and in the current epoch. The graph of these energy conditions is depicted in figure \ref{f25}.
The analysis focused on the observation that the NEC and DEC satisfy their conditions, whereas the violation of the SEC immediately leads to the rapid expansion of the Universe.

\section{Tests for validation of the Model}\label{IVa}

There are some theoretical and observational tests to check the validity of any cosmological model. In this discussion, we will examine some cosmological tests that might be employed to authenticate the validity of our derived model.

\subsection{Statefinder Diagnostic} \label{sec2.4.2}
\begin{figure}[]
\begin{center}
$
\begin{array}{c@{\hspace{.1in}}c}
\includegraphics[width=3.0 in, height=2.5 in]{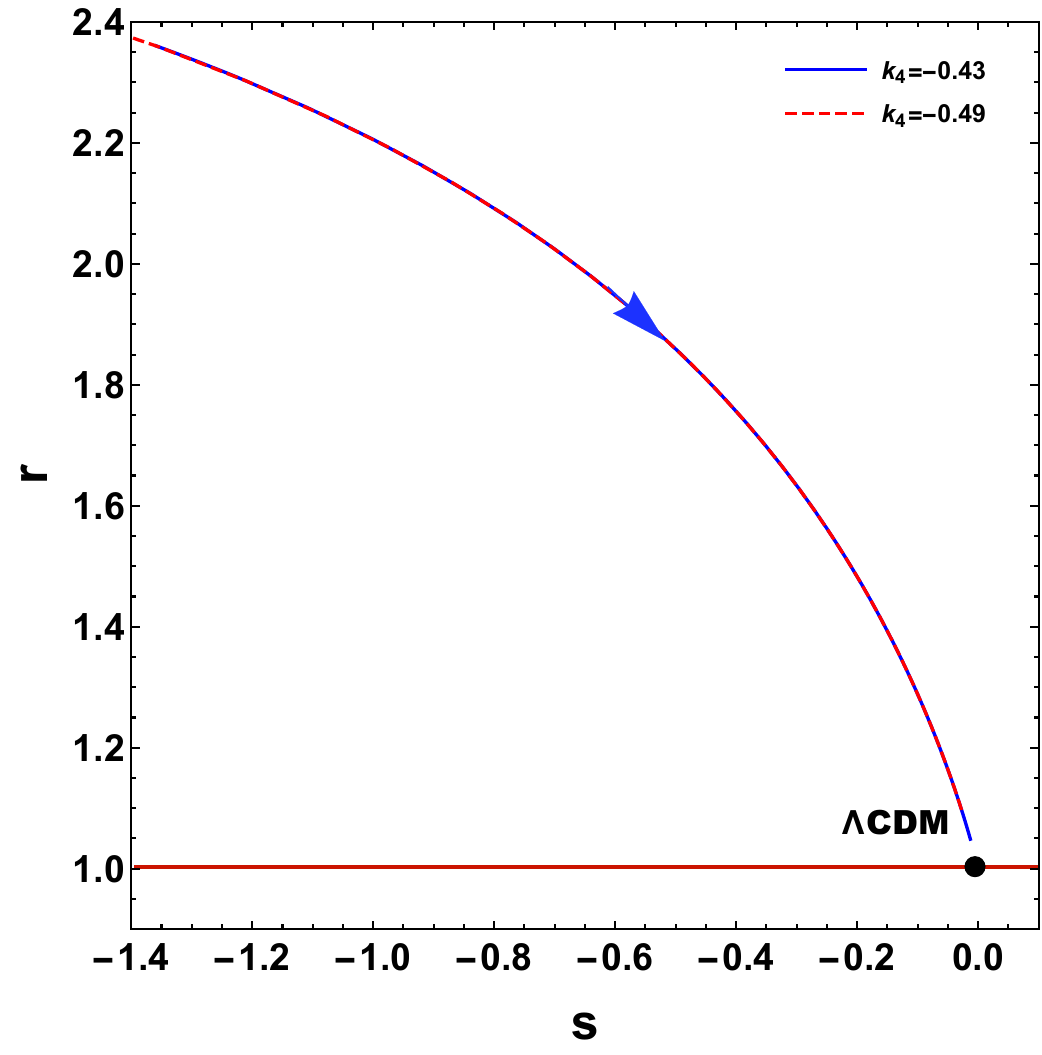} & 
\includegraphics[width=3.0 in, height=2.5 in]{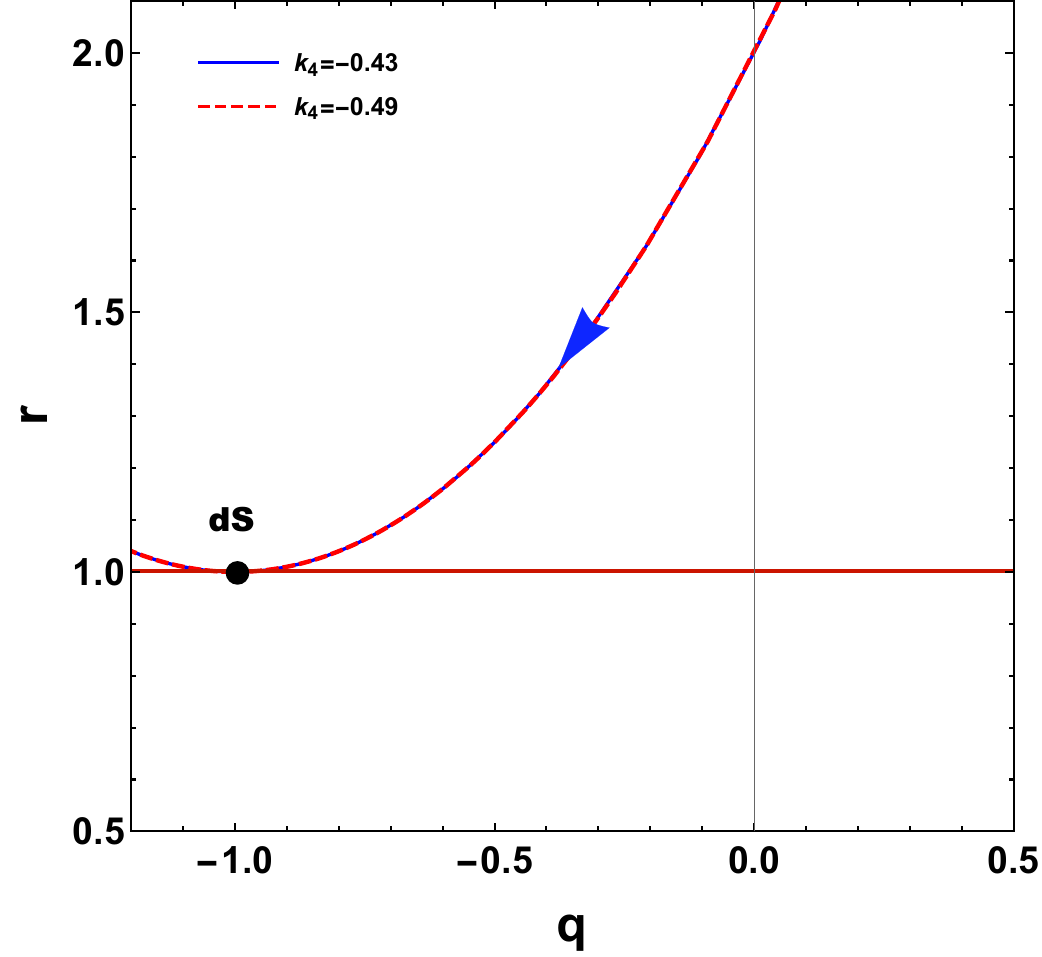}\\
\end{array}
$
\caption{Evolution of $r-s$ and $r-q$ planes for $\protect\eta =-0.1$, $\protect\gamma =1.01$, $\protect\omega _{H}=4.1$, $\protect\omega _{H2}=1.57$, $\protect\omega_{dH}=-0.1$, $k_{4}=-0.43, -0.49$.}
\label{f28}

\end{center}
\end{figure}

In order to distinguish our model from a variety of existing dark energy models, it is possible to introduce a set of parameters known as statefinder pairs \cite{Sahni/2003,Alam/2003,Pasqua/2017}
\begin{eqnarray}
    r &=& \frac{\dddot{a}}{aH^3},\\
    s &=&\frac{r-1}{3(q-1/2)}, \quad q \neq \frac{1}{2}.
\end{eqnarray}
For the sake of simplicity, we could redefine the statefinder pair $\{r,s\}$ fully in terms of the deceleration parameter
\begin{eqnarray}
    r(z) &=& q(z)(1+2q(z))+q'(z)(1+z),\\
    s(z) &=&\frac{r(z)-1}{3(q(z)-1/2)},  \quad q(z) \neq \frac{1}{2}.
\end{eqnarray}
We will proceed with the construction of the phase plane denoted as $r-s$ and $r-q$, where distinct points on the plane represent different states of the Universe, adhering to the following conditions:
\begin{itemize}
    \item $\Lambda$CDM corresponds to $(s=0,r=1)$,
    \item Chaplygin gas (CG) corresponds to $(s<0,r>1)$,
    \item Standard cold dark matter (SCDM) corresponds to $(r=1,q=0.5)$,
    \item Quintessence corresponds to $(s>0,r<1)$.
\end{itemize}
The analysis focuses on investigating the departure of any dark energy model from the specified coordinates in the $r- s$ and $r-q$ planes. The plot below depicts statefinder pairs of the model we have developed. So the present model resembles the $\Lambda$CDM model in the future. The $r-s$ plane trajectories of the model are depicted in figure \ref{f28}. The trajectories in the $r-s$ plane are confined to the region $r > 1,s < 0$, akin to the generalized Chaplygin gas model of dark energy as described in \cite{Ya/2007}. It is noted that the model will reach $\Lambda$CDM and may lie in quintessence at late times. The present position of the bulk viscous model in the $r-s$ plane corresponds to $\{1.04,-0.013\}$ and $\{1.09,-0.02\}$ for $k_{4}=-0.43$ and $k_{4}=-0.49$. The present model can also be differentiated from the holographic dark energy model with event horizon as the infrared cutoff, in which the $r-s$ evolution starts from a region $r \sim 1$, $s \sim 2/3$ and terminates at the $\Lambda$CDM point \cite{Liu/2008}. This observation suggests that the current model exhibits notable differences when compared to the $\Lambda$CDM model.
The central solid line within the $r-q$ plane highlights the trajectory of the $\Lambda$CDM Universe, effectively bisecting the plane. The upper region is representative of the Chaplygin gas model, while the lower area corresponds to the Quintessence model. In a similar fashion, the plot on the right in figure \ref{f28} illustrates the convergence of our model towards the de-Sitter point ($q=-1,r=1$) while also exhibiting deviations from the SCDM model.

\subsection{$Om(z)$ Diagnostic}

In the following section, we will analyze the $Om$ diagnostic denoted as $Om(z)$. This diagnostic tool distinguishes the standard $\Lambda$CDM model from other dark energy models such as quintessence and phantom. The study of the $Om(z)$ diagnostic mainly relies on the utilization of first-order derivatives, as it encompasses the Hubble parameter. In reference with Sahni et al. \cite{Sahni/2003,Alam/2003,Pasqua/2017}, $Om(z)$ for a flat Universe is defined as 
\begin{equation}
Om(z)=\frac{\left( \frac{H(z)}{H_{0}}\right) ^{2}-1}{(1+z)^{3}-1}.
\end{equation}
According to the definition, we obtain $Om(z)$ for our model as 
\begin{equation}
Om(z)=\frac{\frac{\left( (k_{4}z+k_{4})^{k_{2}}-1\right)^{2}}{\left(
k_{4}^{k_{2}}-1\right)^{2}}-1}{(z+1)^{3}-1}.
\end{equation}
Thus, we have different values of $Om(z)$ for the $\Lambda$CDM,
phantom and quintessence cosmological models. The behavior of dark energy may be categorized into three types: quintessence, phantom, and $\Lambda$CDM. The quintessence type, characterized by $\omega>-1$, is associated with negative curvature. The phantom type, described by $\omega<-1$, corresponds to positive curvature, and lastly, the $Om(z)$=$\Lambda$CDM is associated with zero curvature. 

\begin{figure}[H]
\centering
\includegraphics[width=8.5cm]{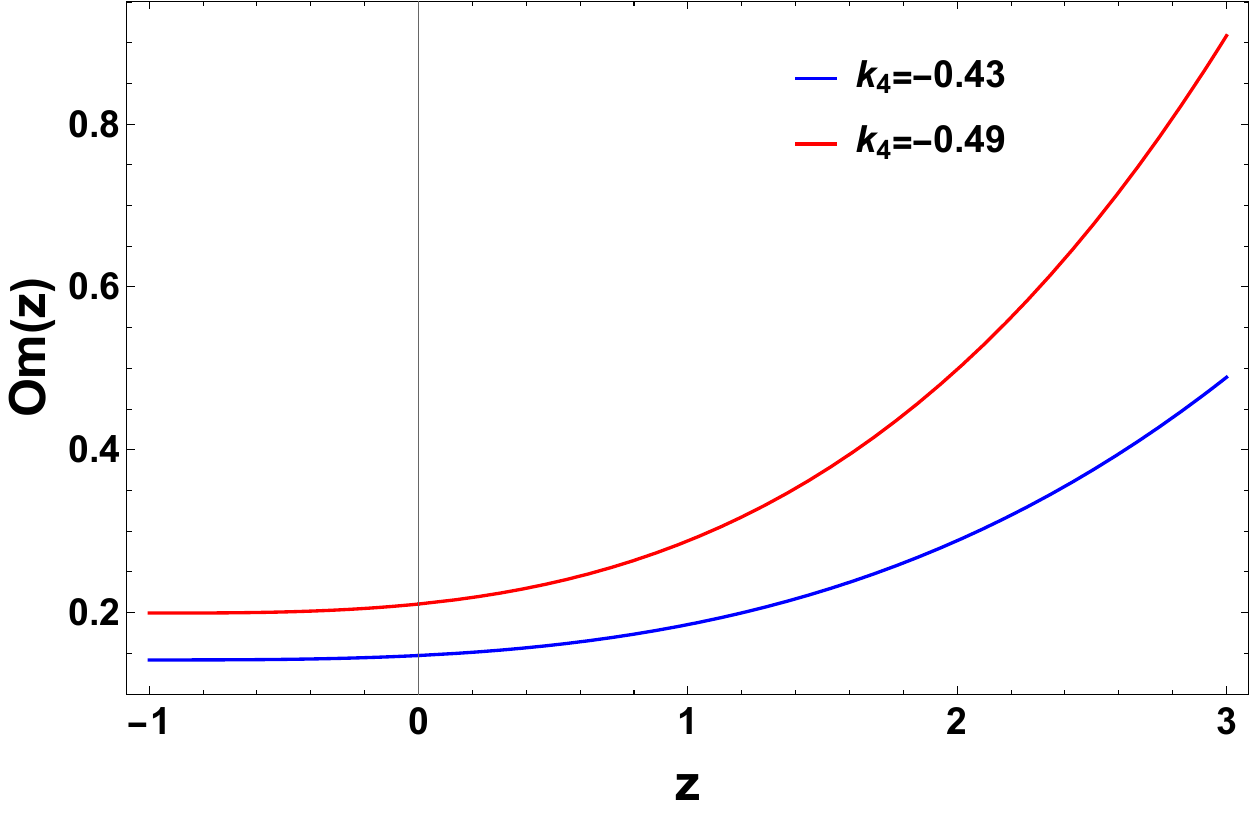}
\caption{Evolution of $Om(z)$ versus redshift $z$ for $\protect\eta =-0.1$, $\protect\gamma =1.01$, $\protect\omega _{H}=4.1$, $\protect\omega _{H2}=1.57$, $\protect\omega_{dH}=-0.1$, $k_{4}=-0.43, -0.49$.}
\label{f29}
\end{figure}
The authors in \cite{Zhao/2018} have presented a parametrization of $Om(z)$ that allows the examination of the compatibility or discrepancy between the $\Lambda$CDM and observations by using the latest and enhanced observations of H(z) and SNeIa.
The behavior can be easily seen in figure \ref{f29}, illustrating that the growth of $Om(z)$ at later stages tends to support the viability of decaying dark energy models (quintessence dark energy at late time) as discussed in \cite{Shafieloo/2009}.

\section{Observational Constraints}\label{Va}
A variety of observational datasets are currently available, as discussed in the section \ref{sec1.7.2}. To assess the feasibility of our acquired model, we shall evaluate its performance using these datasets. Here, we consider $57$ points of H(z) data, of which $31$ points are derived from the differential age method and $26$ points are from BAO and other methods \cite{Sharov/2018} (given in the appendix). Secondly, we consider $580$ points of type Ia Supernovae corresponding to Union $2.1$ compilation dataset \cite{Union2.1 DATA, Ritika/2018} (\href{https://supernova.lbl.gov/Union/}{https://supernova.lbl.gov/Union/}) to achieve our objective of determining the best-fit values of the model parameters and comparing them to the $\Lambda$CDM model.
\subsection{Fitting the Model with H(z) \& SNeIa Datasets} \label{2.5.1}
\begin{figure}[H]
\centering
\includegraphics[width=7.5cm]{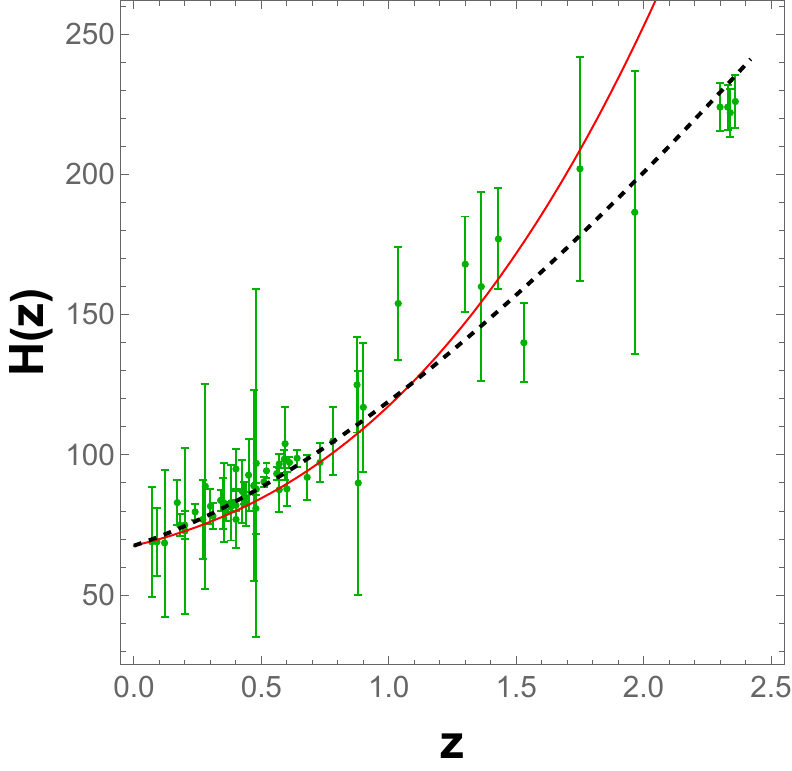}
\caption{The plot shows $57$ points of the H(z) dataset (green dots) with corresponding error bars along with the present model (solid red line). The $\Lambda$CDM model is also shown in the black dashed line.}
\label{f210}
\end{figure}

First, let us initiate a discussion regarding the Hubble data. The Hubble rate is usually defined by the following formula
\begin{equation}
  H(z)=\frac{-1}{1+z}\frac{dz}{dt}.
\end{equation}
The ratio $dz/dt$ can be obtained by calculating the ratio $\Delta z/\Delta t$, where $\Delta z$ is the redshift separation in the galaxy sample. This value can be estimated with a high value of precision and accuracy using spectroscopic techniques. However, the determination of the value of $\Delta t$ is much more challenging and requires some standard clocks. For that purpose, it is possible to utilize massive, passively evolving, and old stellar populations that are present across a wide range of redshifts and therefore could be considered as cosmic chronometers \cite{Ratra/2018,Moresco/2015}.
Now, we shall define the $\chi^{2}$ function for the $H(z)$ dataset by 
\begin{equation}
\label{Chih}
\chi_{H}^{2}=\sum_{i=1}^{N_{H}}\dfrac{\left[H_{obs}(\theta,z_{i})-H_{th}(z_{i})\right]^{2}}{\sigma(z_{i})^{2}},
\end{equation}
where $H_{obs}$ and $H_{th}$ are the observed and theoretical values of $H$, $\theta$ is the parameter space, $\sigma (z_{i})$ is the standard error in the measured value of $H$, and $N_{H}$ is the number of data points. The plot displayed in figure \ref{f210} exhibits a significant relationship between the $H(z)$ dataset and the model parameter values, indicating a favourable match. This fit is then compared to the $\Lambda$CDM model. For our computation, we choose a value of $H_{0}=67.8$ $km/s/Mpc$ as given by Planck2015 \cite{planck/2015}.

\begin{figure}[H]
\centering
\includegraphics[width=7.5cm]{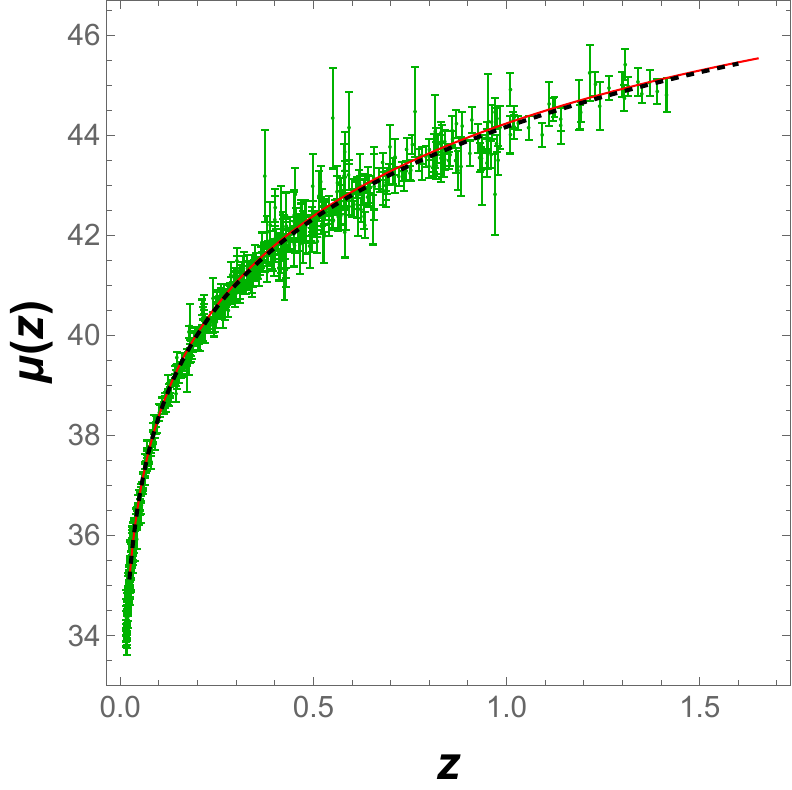}
\caption{The plot shows $580$ points of the SNeIa dataset (green dots) with corresponding error bars along with the present model (solid red line). The $\Lambda$CDM model is also shown in the black dashed line.}
\label{f211}
\end{figure}

Next we define the corresponding chi-square function for SNeIa as
\begin{equation}
\label{Chisn}
    \chi^2_{SN}= \sum^{580}_{i=1} \left(\frac{\Delta \mu_i}{\sigma_{\mu(z_i)}}\right)^{2},
\end{equation}
where $\Delta\mu_i=\mu_{th}(\theta,z_{i})-\mu_{obs}(z_{i})$, and $\mu_{obs}$, $\mu_{th}$, $\sigma_{\mu(z_{i})}$, denote the observed and theoretical distance moduli of the model, and the standard error in the measurement of $\mu(z)$, respectively. 
And the distance moduli is expressed as follows
\begin{equation}
    \mu^{th}=5\log_{10}D_L(z)+\mu_0,\quad \mu_0 = 5 \log_{10} \frac{H_0^{-1}}{\mathrm{Mpc}}+25,
\end{equation}
\begin{equation}
    D_L(z)=\frac{c(1+z)}{H_0}S_K\bigg(H_0\int^z_0\frac{d\overline{z}}{H(\overline{z})}\bigg).
\end{equation}
Here, the function $S_k(x)$ is given by
\begin{equation}
S_k(x)=    \begin{cases}
      \sinh(x\sqrt{\Omega_k})/\Omega_k,\quad \Omega_k >0\\
      x,\quad\quad\quad\quad\quad\quad\quad\;\;\; \Omega_k=0\\
      \sin (x \sqrt{|\Omega_k|})/|\Omega_k|,\quad \Omega_k<0
    \end{cases}\,.
\end{equation}
It is widely acknowledged that the spatial curvature of our Universe is flat, resulting in a value of $\Omega_K=0$. The $\chi_{SN}^{2}$ function and the distance $D_{L}(z)$ are calculated in order to quantify the differences between the SNeIa observational data and predictions made by our model. The plot presented in figure \ref{f211} shows a nice fit to the SNeIa dataset with suitable model parameter values compared to the $\Lambda$CDM model.
\subsection{Estimation of Model Parameters with H(z), SNeIa \& BAO Datasets} \label{2.5.2}
Here, in this subsection, we shall find the constraints with the above discussed datasets i.e. H(z) and SNeIa together with one more external dataset, the BAO dataset for our analysis. From very large scales, BAO measures the structures in the Universe. In this chapter, we have considered a sample of BAO distance measurements from surveys of SDSS(R) \cite{padn/2012}, 6dF Galaxy survey \cite{6df/2011}, BOSS CMASS \cite{boss/2014} and WiggleZ \cite{wig/2012}. So, the distance redshift ratio $d_{z}$ is expressed as $d_{z}=\frac{r_{s}(z_{\ast })}{D_{v}(z)}$, where $r_{s}(z_{\ast})$ is the comoving sound horizon at the time photons decouple and $z_{\ast}$ is the photon decoupling redshift. In accordance to Planck 2015 results \cite{planck/2015}, the value of $z_{\ast }$ is determined to be $1090$. We have taken $r_{s}(z_{\ast })$ as considered in \cite{waga}. Also, dilation scale $D_{v}(z)$ is  given by $D_{v}(z)=\big(\frac{d_{B}^{2}(z)z}{H(z)}\big)^{\frac{1}{3}}$, where $d_{A}(z)$ is the angular diameter distance. The chi-square value for BAO measurements is written as \cite{gio/2012} 
\begin{equation}
\chi_{BAO}^{2}=B^{T}C^{-1}B.
\end{equation}
The matrix $B$ in the chi-square formula of BAO datasets is obtained by $d_{B}(z_{\ast })/D_{V}(z_{BAO})$ 
\begin{equation*}
B=\left( 
\begin{array}{c}
\frac{d_{B}(z_{\star })}{D_{V}(0.106)}-30.95 \\ 
\frac{d_{B}(z_{\star })}{D_{V}(0.2)}-17.55 \\ 
\frac{d_{B}(z_{\star })}{D_{V}(0.35)}-10.11 \\ 
\frac{d_{B}(z_{\star })}{D_{V}(0.44)}-8.44 \\ 
\frac{d_{B}(z_{\star })}{D_{V}(0.6)}-6.69 \\ 
\frac{d_{B}(z_{\star })}{D_{V}(0.73)}-5.45%
\end{array}%
\right) \,,
\end{equation*}
and the inverse covariance matrix $C^{-1}$ is defined in \cite{gio/2012}
\begin{equation*}
C^{-1}=\left( 
\begin{array}{cccccc}
0.48435 & -0.101383 & -0.164945 & -0.0305703 & -0.097874 & -0.106738 \\ 
-0.101383 & 3.2882 & -2.45497 & -0.0787898 & -0.252254 & -0.2751 \\ 
-0.164945 & -2.454987 & 9.55916 & -0.128187 & -0.410404 & -0.447574 \\ 
-0.0305703 & -0.0787898 & -0.128187 & 2.78728 & -2.75632 & 1.16437 \\ 
-0.097874 & -0.252254 & -0.410404 & -2.75632 & 14.9245 & -7.32441 \\ 
-0.106738 & -0.2751 & -0.447574 & 1.16437 & -7.32441 & 14.5022%
\end{array}
\right) \,.
\end{equation*} 

\begin{figure}[H]
\centering
\includegraphics[width=8cm]{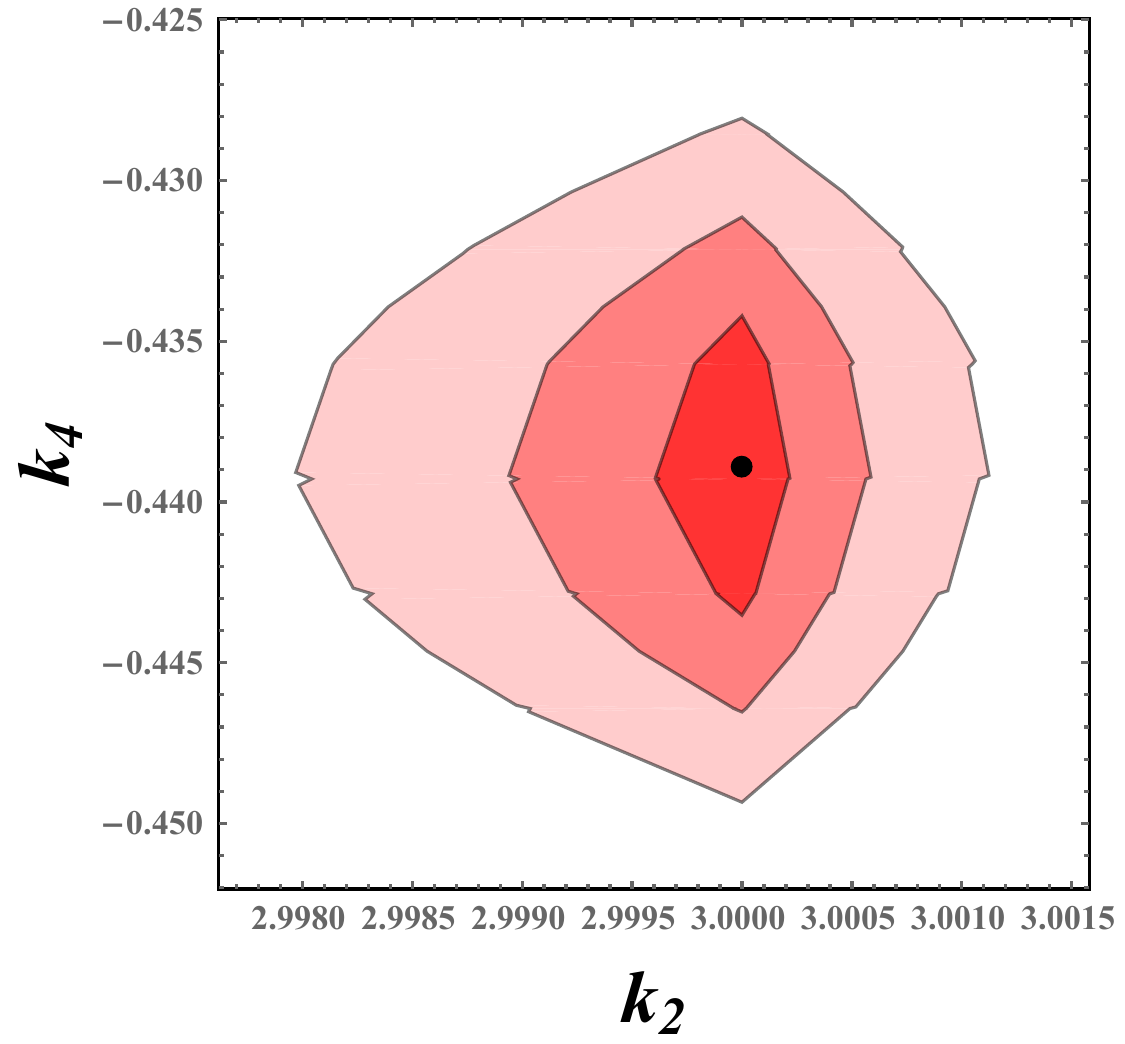}
\caption{The plot shows the contour plot for the model parameters $k_{2}$
and $k_{4}$ for independent H(z) dataset at $1\sigma $, $2\sigma $ and $3\sigma $ level in $k_{2}$-$k_{4}$ plane.}
\label{f212}
\end{figure}

\begin{figure}[H]
\centering
\includegraphics[width=8cm]{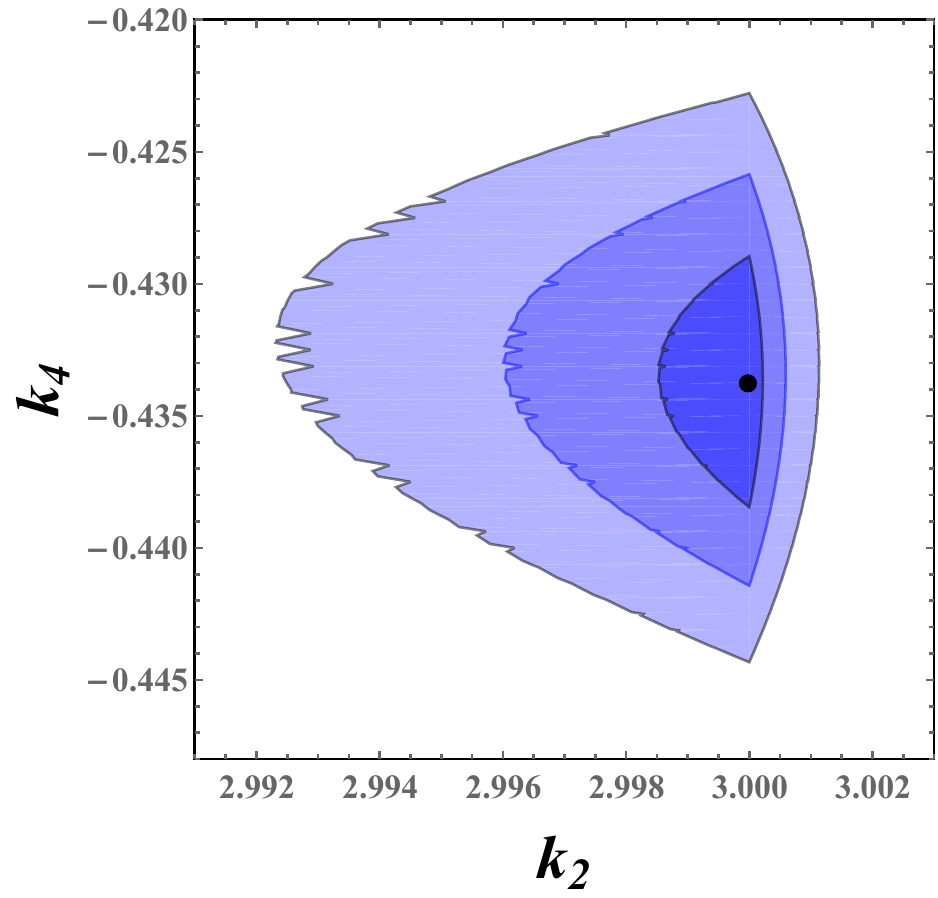}
\caption{The plot shows the contour plot for the model parameters $k_{2}$
and $k_{4}$ for combined H(z)+SNeIa+BAO dataset at $1\sigma$, $2\sigma$ and $3\sigma $ level in $k_{2}$-$k_{4}$ plane.}
\label{f213}
\end{figure}

Using the three samples of datasets, we have determined the likelihood contours for the model parameters $k_{2}$ and $k_{4}$ at $1\sigma $, $2\sigma$ and $3\sigma $ confidence levels. These contours are displayed in figures \ref{f212} and \ref{f213}, representing the $k_{2}$-$k_{4}$ plane. We have found constraints with the independent H(z) dataset and combined H(z)+SNeIa+BAO dataset. The best estimated values of the model parameters $k_{2}$ and $k_{4}$ are found to be $k_{2}=3$, $k_{4}=-0.4389$ and $k_{2}=3$, $k_{4}=-0.43374$, respectively for independent H(z) datasets and joint H(z)+SNeIa+BAO datasets. The values obtained for both the H(z) as well as joint data are almost identical. So, we have considered the second value of $k_4$ in the neighbourhood, that is $k_4=-0.49$.

\section{Conclusions}\label{VIa}

The present chapter analyzed a cosmological model that describes the phenomenon of cosmic acceleration through a viscous fluid. The dynamical equation of the Hubble parameter can be entirely integrated using the effective viscosity equation of state (EoS). This integration leads to an exact solution for Einstein's field equation in the FLRW background for modified $f(R,\mathcal{T})$ gravity. Without introducing a cosmological constant or dark energy, the effective EoS (pressure with added bulk viscosity) characterizes the late-time acceleration of the Universe.

In section \ref{IIIa}, the deceleration parameter demonstrated a transition from early deceleration to present acceleration at $z_t = 0.84$ and $z_t = 0.62$ for different values of $k_4=-0.43$ and $k_4=-0.49$, respectively. Additionally, the corresponding values of the current deceleration parameter are $q_0 = -0.78$ and $q_0 = -0.62$, which is depicted in figure \ref{f21}. Figure \ref{f24} illustrated the evolution of effective equation of state $\omega$ specifically $\omega_0 = -0.88$ for $k_4=-0.43$ and $\omega_0 = -0.84$ for $k_4=-0.49$.  The cosmological model under consideration continues within the quintessence phase, characterized by a value of the equation of state parameter $\omega$ that does not cross the phantom line $\omega = -1$. Furthermore, in the future, the model approaches a value of $\omega$ equal to $-1$, resulting in the emergence of the Einstein-de-Sitter space.

It is also important to note that the EoS parameter $\omega$ undergoes a transition from a positive regime in the past to a negative regime in the present, implying that the incorporation of a bulk viscous pressure term is crucial for achieving a decelerated expansion in the past (suitable for structure formation) and an accelerated expansion in the present. This is depicted in figure \ref{f24}, demonstrating that standard cosmology confirms the existence of the latter regime when $\omega < -\frac{1}{3}$. 
In section \ref{IVa}, we have discussed some physical properties of the model, the evolution of cosmological parameters, and energy conditions. It has been noticed that NEC and DEC do not violate their conditions, whereas SEC is violated, resulting in a repulsive force that causes the Universe to accelerate. The violation of SEC depicted in figure \ref{f25} demonstrates the viability of our model, as described in \cite{Barcelo/2002}. In addition, statefinder parameter analysis and $Om(z)$ diagnostic have been performed and compared to the $\Lambda$CDM model. It has been observed that the model will converge to the $\Lambda$CDM model in the distant future but exhibits deviations from it in the present. We have conducted model fitting utilizing the updated $57$ points of Hubble, $580$ points of Union $2.1$ compilation supernovae, and the BAO datasets in section \ref{Va}.

ECs have provided us with unique insights into the spacetime evolution processes underlying the spacetime structure. The NEC and DEC are satisfied in the current paradigm, but the SEC is violated due to cosmic acceleration. As it is known, the formation of wormholes necessitates the explicit violation of the null energy condition, which compels us to investigate further the confrontation of dark energy with an unusual spacetime structure. With this effective viscosity EoS in non-minimally coupled gravity, additional research can be conducted. However, the model addressed in the present chapter may also offer an alternative approach that could provide some insights into the physical interpretation of the theories with curvature-matter coupling and for observationally testing alternative gravity theories.

In the following two chapters, we will assess the validity of $f(Q,\mathcal{T})$ gravity in a cosmological context by utilizing advanced and latest observational datasets.



\chapter{$f(Q,\mathcal{T})$ Gravity Models with Observational Constraints} 

\label{Chapter3} 

\lhead{Chapter 3. \emph{$f(Q,\mathcal{T})$ Gravity Models with Observational Constraints}} 

\blfootnote{*The work in this chapter is covered by the following publication:\\
\textit{$f(Q,\mathcal{T})$ gravity models with observational constraints}, Physics of the Dark Universe, \textbf{30}, 100664 (2020).}

This chapter presents the cosmology of late time in $f(Q,\mathcal{T})$ gravity, where dark energy is purely geometric. The detailed study of the work is outlined as:
\begin{itemize}
\item We begin by utilizing a well-motivated $f(Q,\mathcal{T})$ gravity model with $f(Q,\mathcal{T})= \xi \,Q^{n}+b\,\mathcal{T}$ where $\xi$, $n$ and $b$ are model parameters.
\item In addition, we assume that the Universe is dominated by pressure-less matter, resulting in a power-law scale factor.
\item We use the 580 points from the Union 2.1 compilation supernovae dataset and the revised 57 points from the Hubble dataset to constrain the model parameters to test the model's cosmological viability. 
\item We investigate the nature of geometrical dark energy modeled by the parametrization of $f(Q,\mathcal{T})=\xi\,Q^{n}+b\,\mathcal{T}$ using the statefinder diagnostic in $r-s$ and $r-q$ planes, as well as the $Om(z)$ diagnostic analysis.
\end{itemize}

\section{Introduction}\label{Ic}
Extended theories of gravity such as $f(R)$ gravity, $f(G)$ gravity, $f(R,\mathcal{T})$ gravity, etc. are widely employed in modern cosmology (For a recent review on modified gravity see \cite{review}). Also see \cite{Sahoo/2020} for some interesting cosmological applications of
modified gravity to address the late-time acceleration and other
shortcomings of $\Lambda$CDM cosmologies. $f(Q,\mathcal{T})$ gravity is a recently proposed extended theory of gravity in which the Lagrangian density of the gravitational field is a function of the non-metricity $Q$ and the trace of energy-momentum tensor $\mathcal{T}$ \cite{Xu/2019}. A very recent study by Yixin et al. \cite{Xu/2020} demonstrates the non-minimal coupling between $Q$ and $\mathcal{T}$ taking into account a variety of cosmological models with specific functional forms of $f(Q,\mathcal{T})$. Different forms of $f(Q,\mathcal{T})$ can result in obtaining a large variety of cosmological evolution, including the decelerating and accelerating expansion. Also, according to the study in \cite{Sne/2020}, $f(Q,\mathcal{T})$ gravity yields an excellent theoretical estimate of baryon-to-entropy ratios and therefore, could solve the puzzle of over-abundance of matter over anti-matter. With that in mind, we intend to investigate the cosmological viability of $f(Q,\mathcal{T})$ gravity in sufficing the conundrum of late-time acceleration without incorporating dark energy.

The present chapter is organized as follows: In section \ref{IIc}, a concise overview of $f(Q,\mathcal{T})$ is presented. In the very next section \ref{IIIc}, we discuss the cosmological model in $f(Q,\mathcal{T})$ framework and derive some necessary cosmological parameters. In accordance, the non-parametric method is sometimes more beneficial, as the evolution of our Universe can be found directly from the observational data. Thus, in section \ref{IVc}, we constrain the model parameters using H(z) and SNeIa datasets. We present geometrical diagnostics that allow for distinguishing between various dark energy models and $\Lambda$CDM in section \ref{Vc}. Further, the behavior of energy density is demonstrated in section \ref{VIc}. In section \ref{VIIc}, we distinguish between the non-linear model under consideration and a specific linear form of $f(Q,\mathcal{T})$. Lastly, in section \ref{VIIIc}, we explain our findings and draw our conclusions.

\section{Overview of $f(Q,\mathcal{T})$ Gravity}\label{IIc}
In order to obtain the generalized Friedmann equations, one considers a flat FLRW metric \eqref{metric} (for $k=0$) and the equation \eqref{QTF}, which leads to the following equations
\begin{eqnarray}  
\label{4.1}
8\pi \rho &=& \frac{f}{2}-6FH^{2}-\frac{2\widetilde{G}}{1+\widetilde{G}}(\dot{F}H+F\dot{H}),\\
\label{4.2}
8\pi p &=& -\frac{f}{2}+6FH^{2}+2(\dot{F}H+F\dot{H}),
\end{eqnarray}
where dot represents a derivative with respect to time and $F= f_{Q}$ and $8 \pi \widetilde{G}=f_{\mathcal{T}}$ are the derivatives of $f$ with respect to $Q$ and $\mathcal{T}$, respectively. The non-metricity scalar using the FLRW metric is obtained as $Q=6H^{2}$. In light of the previous work done by \cite{Xu/2019}, we use the assumption $G=1$ in the action \eqref{fq} to maintain consistency. 

We can then combine equations \eqref{4.1} \& \eqref{4.2} and arrive at the following equation 
\begin{equation}  \label{4.3}
\dot{H}+\dfrac{\dot{F}}{F} H= \dfrac{4\pi}{F}(1+\widetilde{G})(\rho+p).
\end{equation}
In the present discussion, our attention shall be directed in obtaining expressions for different cosmological parameters. Thus, we begin our analysis by considering the energy density. For generality, we assume that the cosmological matter satisfies an equation of state of the form $p=(\gamma -1)\rho$, where $\gamma$ is a constant, and $0\leq \gamma \leq 2$. By solving equations \eqref{4.1} and \eqref{4.3}, we obtain the energy density $\rho$ 
\begin{equation}  \label{4.4}
\rho= \dfrac{f-12F H^{2}}{16\pi(1+\gamma \widetilde{G})}.
\end{equation}

\section{Cosmological Model with $f(Q,\mathcal{T}) = \xi \,Q^{n}+b\,\mathcal{T}$}\label{IIIc}

Given that the functional form is arbitrary, we focus on the power-law model of $f(Q,\mathcal{T})$ \cite{Xu/2019} represented by 
\begin{equation}  \label{4.5}
f(Q,T)=\xi\,Q^{n}+b\,\mathcal{T},
\end{equation}
where $\xi$, $b$ and $n$ are model parameters. We choose this form as it considers a $Q^{n}$ term along with the term $T$ representing
deviations from $\Lambda$CDM and $f(Q)$. Also note that $F= f_{Q}= \xi\,n\, Q^{n-1}$ and $8 \pi\,\widetilde{G}=f_{\mathcal{T}}=b$.
We determine the solution for zero pressure (dust matter), for which $\gamma =1$ and the equation \eqref{4.4} reduces to 
\begin{equation}  \label{4.6}
\rho =\dfrac{\xi6^{n}H^{2n}(1-2n)}{16\pi +3b}.
\end{equation}
The dynamical equation \eqref{4.3} that characterises the dynamics of the
model is now presented 
\begin{equation}
\dot{H}+\dfrac{3(8\pi +b)}{n(16\pi +3b)}H^{2}=0,  \label{4.7}
\end{equation}
which integrates easily to yield the time evolution of the Hubble parameter $H(t)$ 
\begin{equation}  \label{4.8}
H(t)=\frac{1}{At+c_{1}}\text{, where }A=\dfrac{3(8\pi +b)}{n(16\pi +3b)},
\end{equation}
and $c_{1}$ is the constant of integration. From equation \eqref{4.8}, we
obtain the explicit form of scale factor as a simple power law type solution given by
\begin{equation}  \label{4.9}
a(t)=c_{2}(At+c_{1})^{\frac{1}{A}},
\end{equation}
where $c_{2}$ is another constant of integration. As we are dealing with
zero pressure and attempting to explain the present cosmic acceleration of the Universe, we shall express all the above cosmological parameters in terms of redshift $z$ defined by $z=\frac{a_{0}}{a}-1$, where $a_{0}$ is the present value (at time $t=t_{0}$) of the scale factor. In addition, we will consider the normalized value $a_{0}=c_{2}(At_{0}+c_{1})^{\frac{1}{A}}=1$ for which the $t$-$z$ relationship is established as
\begin{equation}  \label{4.10}
t(z)=-\frac{c_{1}}{A}+\frac{1}{A}\left[ c_{2}(1+z)\right]^{-A}.
\end{equation}
Henceforth, using equation \eqref{4.10}, we obtain the Hubble parameter in terms of $z$. That is 
\begin{equation}  \label{4.11}
H(z)=H_{0}(1+z)^{A}=H_{0}(1+z)^{\frac{3(8\pi +b)}{n(16\pi +3b)}},
\end{equation}
containing only two model parameters $n$ and $b$. Following the expression of the Hubble parameter, our focus is to determine the deceleration parameter $q=-1-\frac{\dot{H}}{H^{2}}$, which turns out to be
\begin{equation}
q(t)=-1+\dfrac{3(8\pi +b)}{n(16\pi +3b)}, \label{4.12}
\end{equation}
One can observe that the expression gives us a constant value for specific  $n$ and $b$ as expected due to the power-law type expansion of the model. 
The values of the model parameters $n$ and $b$ in the $H(z)$ function are currently unknown. To determine these values, we aim to impose constraints by analyzing the observational data discussed in the subsequent section.
\section{Parameters of the Model \& Observational Constraints}\label{IVc}

\begin{figure}[H]
\centering
\includegraphics[width=8 cm]{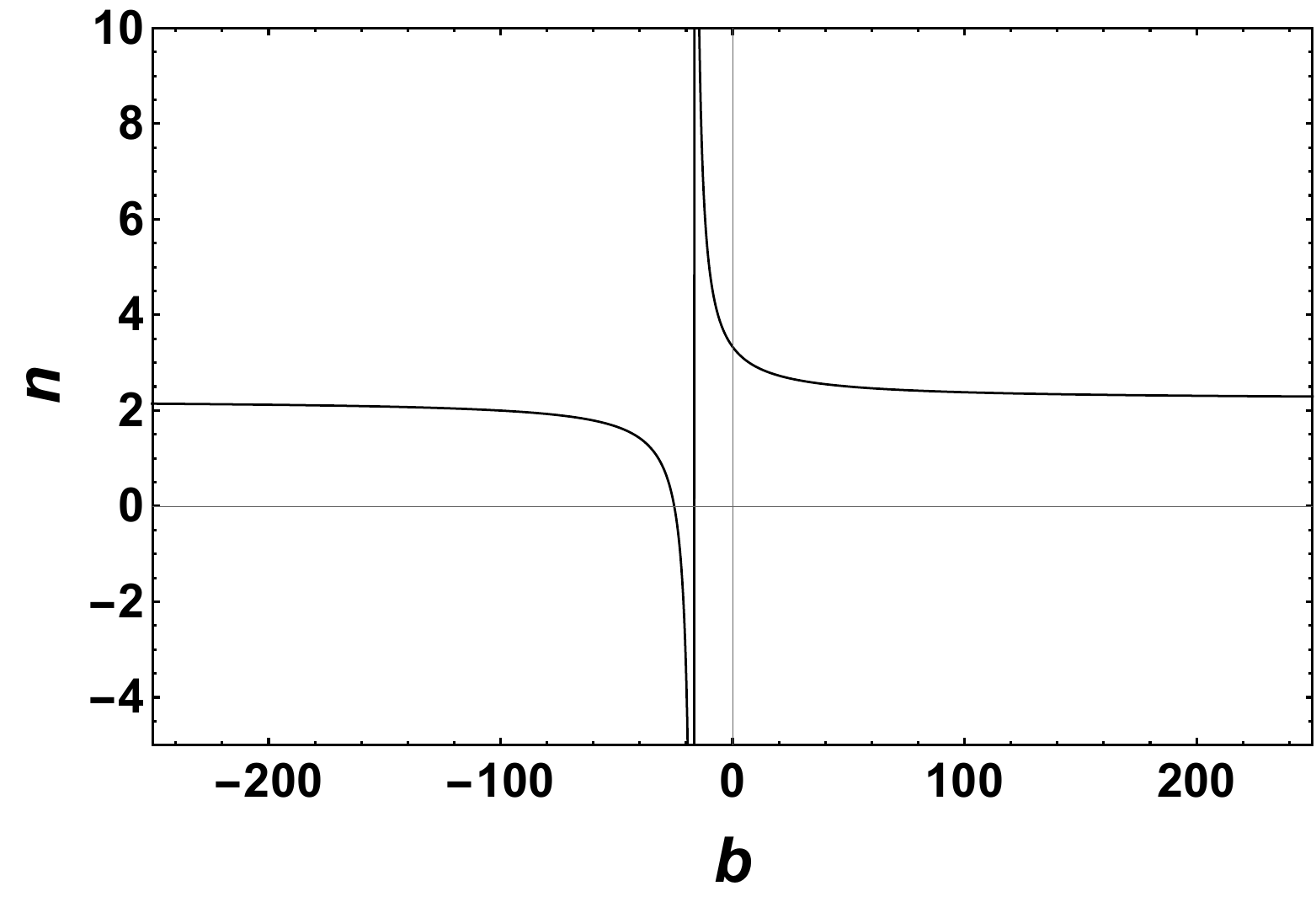}
\caption{The variation of model parameters $n$ and $b$ in the shown ranges to have a $q_{0}$ value consistent with observations.}
\label{f41}
\end{figure}

The expressions in equations \eqref{4.11} and \eqref{4.12} illustrate the inclusion of two model parameters that govern the dynamics of the model. The model parameter $n$ is more significant than $b$, as determined by the expression  $\dfrac{24\pi +3b}{16\pi +3b}$, which contains the homogeneous term $3b$ in both numerator and denominator. The choice of model parameters $n$ and $b$ must be such that the deceleration parameter attains a negative value at present and is consistent with the observed value $q_{0}\simeq -0.54$ \cite{Almada/2020,Garza/2019,Akarsu/2019}. This requires $n=\frac{25.142+b}{7.542+0.45b}$. In order for a model to be consistent with the observations and the $q_{0}$ value of around $-0.54$, the model parameters $n$ and $b$ must satisfy the relation
\begin{equation}  \label{4.13}
n=\frac{25.142+b}{7.542+0.45b}.
\end{equation}
The graph in figure \ref{f41} illustrates the choice of these two model parameters. Based on the information in figure \ref{f41}, we can make a rough estimate for the range of the model parameters. The best estimate could be $n\in (1,3)$ and $b\in (0,2)$. We have examined two observational datasets to achieve more precise estimates for the model parameters $n$ and $b$. The first dataset consists of $57$ points of Hubble data, while the second dataset comprises $580$ points of Union 2.1 compilation supernovae dataset, as discussed in section \ref{Va}.

\begin{figure}[]
\begin{center}
$
\begin{array}{c@{\hspace{.1in}}c}
\includegraphics[width=3.0 in, height=2.5 in]{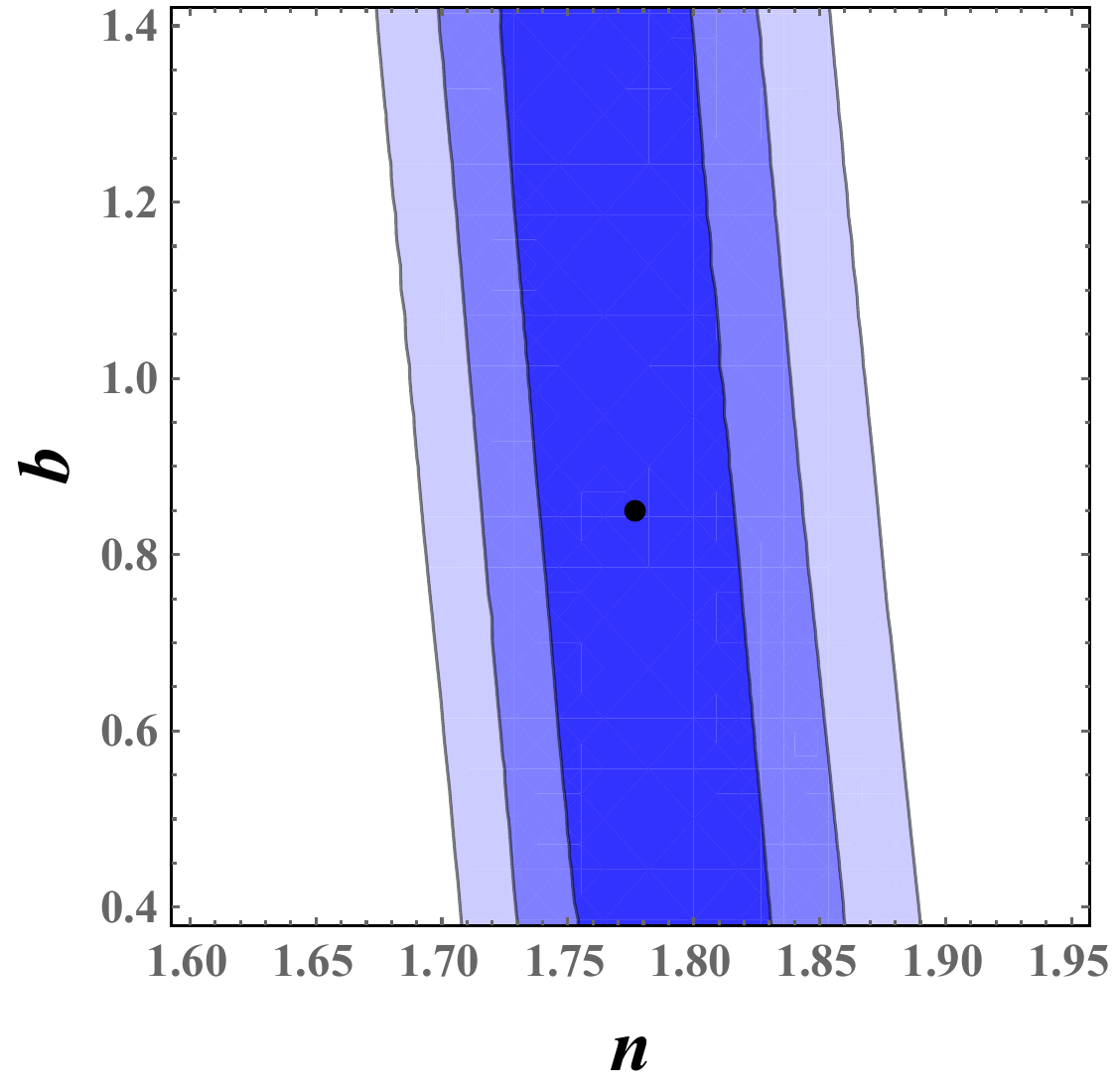} & 
\includegraphics[width=3.0 in, height=2.5 in]{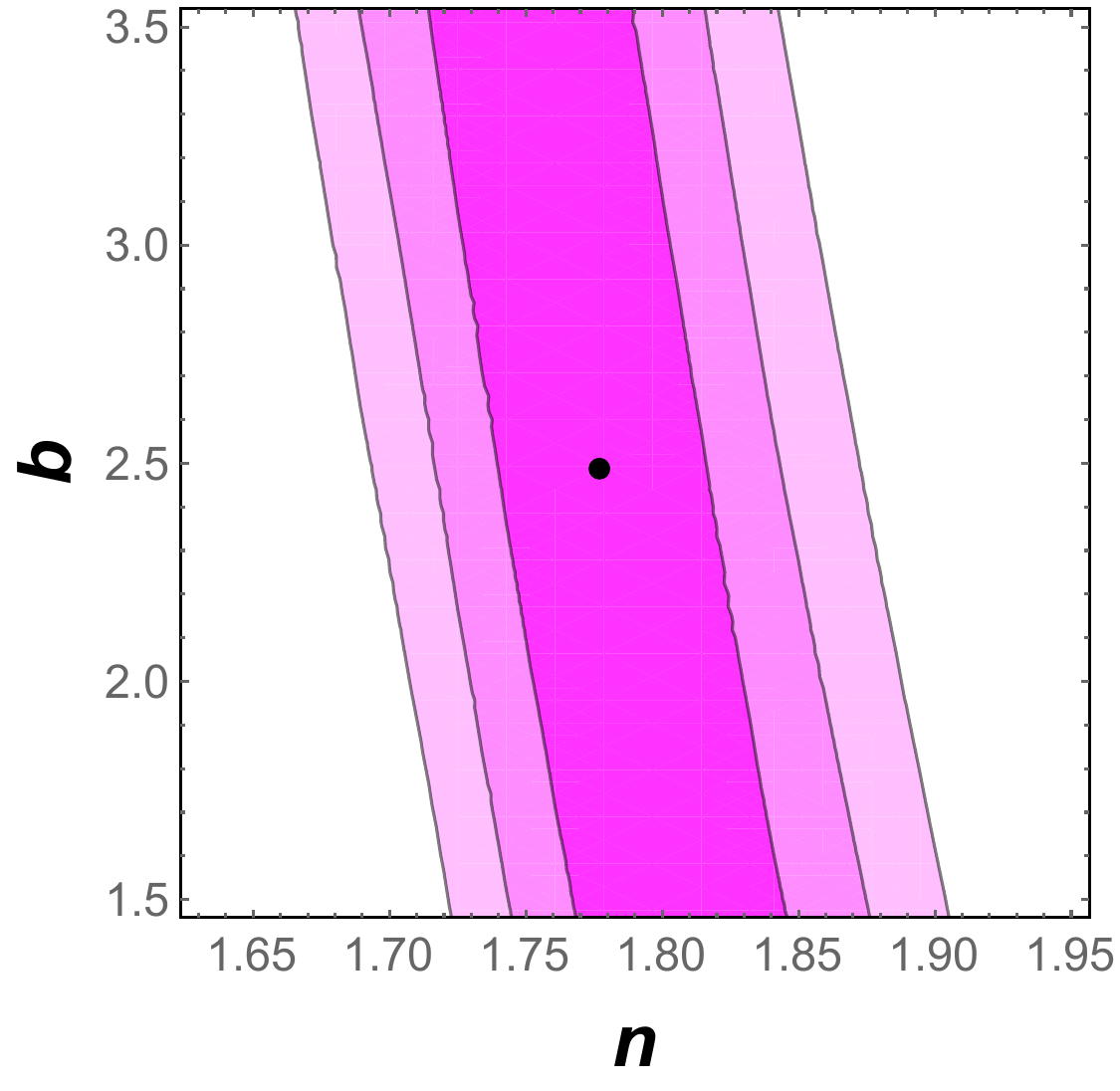}\\
\end{array}
$
\caption{The contour plots using $57$ points of H(z) dataset and $580$ points of SNeIa dataset displaying the likelihood values of $n$ and $H_{0}$ at $1\sigma $, $2\sigma $ and $3\sigma $ level in the $n$-$H_{0}$ plane along with the constrained values illustrated by black dots.}
\label{f43}
\end{center}
\end{figure}

Since, the values of the model parameter $b$ ranges from $-\infty $ to $
+\infty $, the $b$-axis in the above contour plots is unbounded but the
model parameter $n$ is in its fixed range $n\in (1,4)$ in both the contour plots. We have obtained the best fitting pair $(n,b)$ of model parameter values in figure \ref{f43}.

\begin{figure}[H]
\begin{center}
$
\begin{array}{c@{\hspace{.1in}}c}
\includegraphics[width=3.0 in, height=2.5 in]{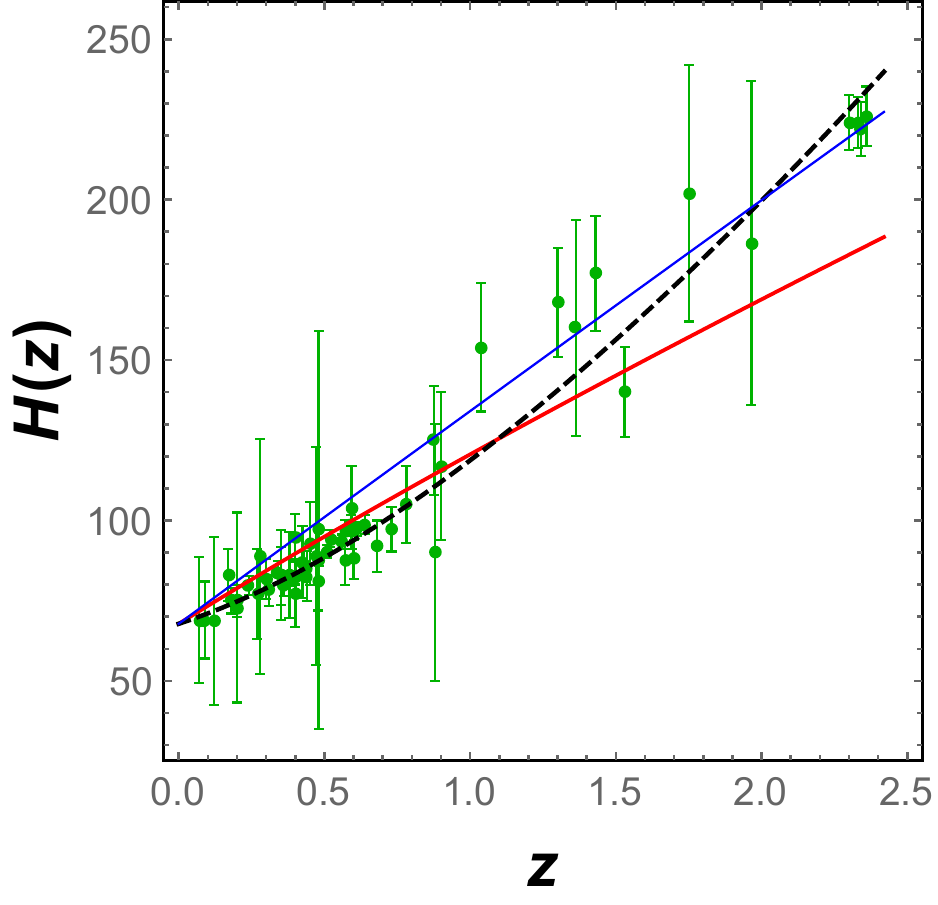} & 
\includegraphics[width=3.0 in, height=2.5 in]{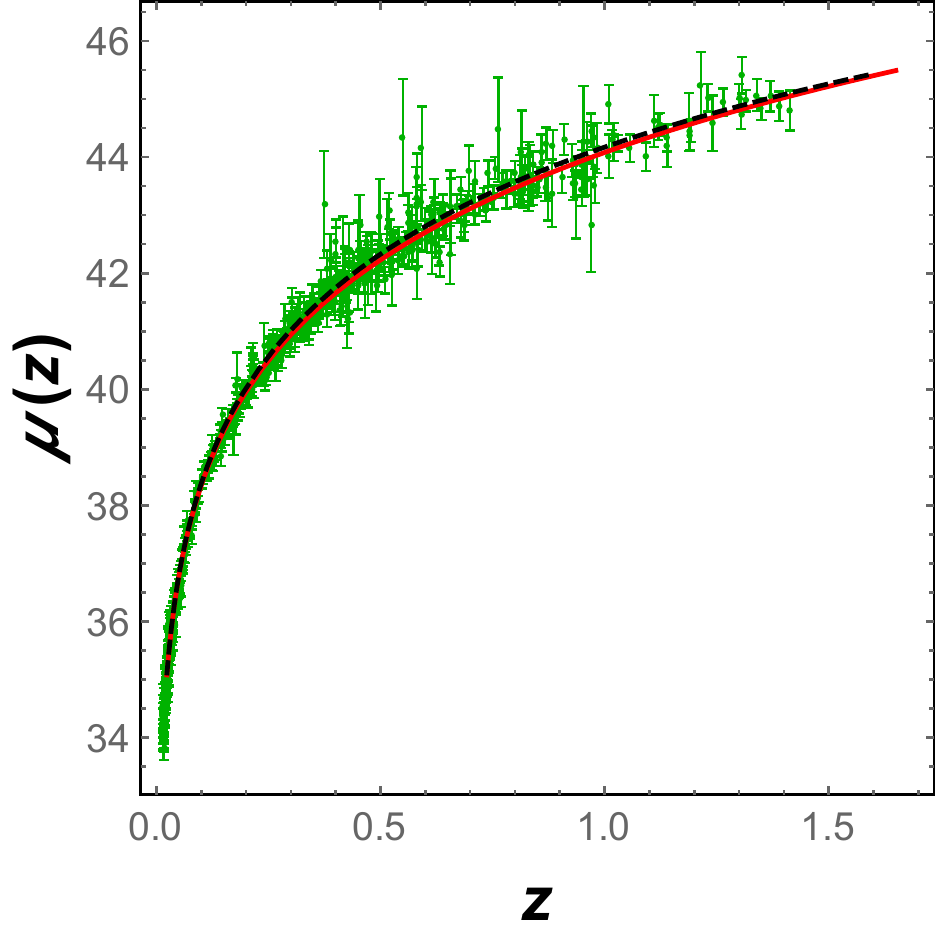}\\
\end{array}
$
\caption{The error bar plots of $57$ points of H(z) dataset, $580$ points of SNeIa dataset with the $\Lambda$CDM model.}
\label{f42}
\end{center}
\end{figure}

The left panel in figure \ref{f42} shows the error bar plot of $57$ points of the Hubble dataset together with the presented model shown in solid red line with $n=1.7763$ and $b=0.8491$ compared with the $\Lambda$CDM model shown in black dashed line showing a poor fit at higher redshift but better at lower redshift. The blue line depicted in the figure serves as a fiducial model, intended solely for comparison with the values of $n=1.5$ and $b=0.8491$, which fall outside the contour. The right panel shows the error bar plot of $580$ points of Union 2.1 compilation supernovae dataset together with the present model shown in solid red line with $n=1.7769$ and $b=2.4889$ compared with the $\Lambda $CDM model shown in black dashed line.

Moreover, the left panel in figure \ref{f43} depicts the contour plot due to the $57$ points of the Hubble dataset displaying likelihood values of the model parameters $n$ and $b $ in $n$-$b$ plane at $1\protect\sigma $, $2\protect\sigma $ \& $3\protect\sigma $ level. The black dot represents the constrained values of model parameters as $n=1.7763$ and $b=0.8491$. The right panel shows the contour plot due to the $580$ points of Union 2.1 compilation supernovae dataset displaying likelihood values of the model parameters $n$ and $b$ in $n$-$b$ plane at $1\sigma $, $2\protect\sigma $ \& $3\protect\sigma $ level. The black dot shows the constrained values of model parameters found as $n=1.7769$ and $b=2.4889$.

\section{Statefinder and $Om(z)$ Diagnostic Analysis}\label{Vc}
\begin{figure}[]
\begin{center}
$
\begin{array}{c@{\hspace{.1in}}c}
\includegraphics[width=3.0 in, height=2.5 in]{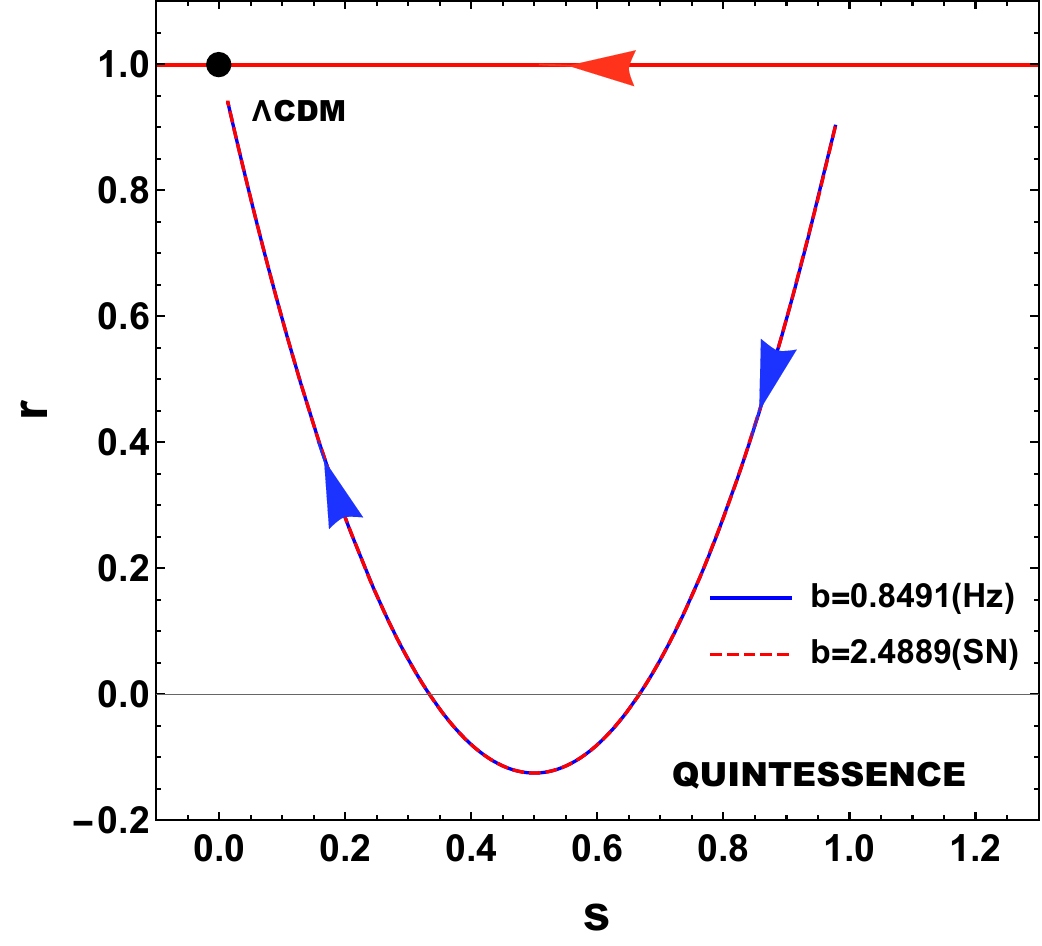} & 
\includegraphics[width=3.0 in, height=2.5 in]{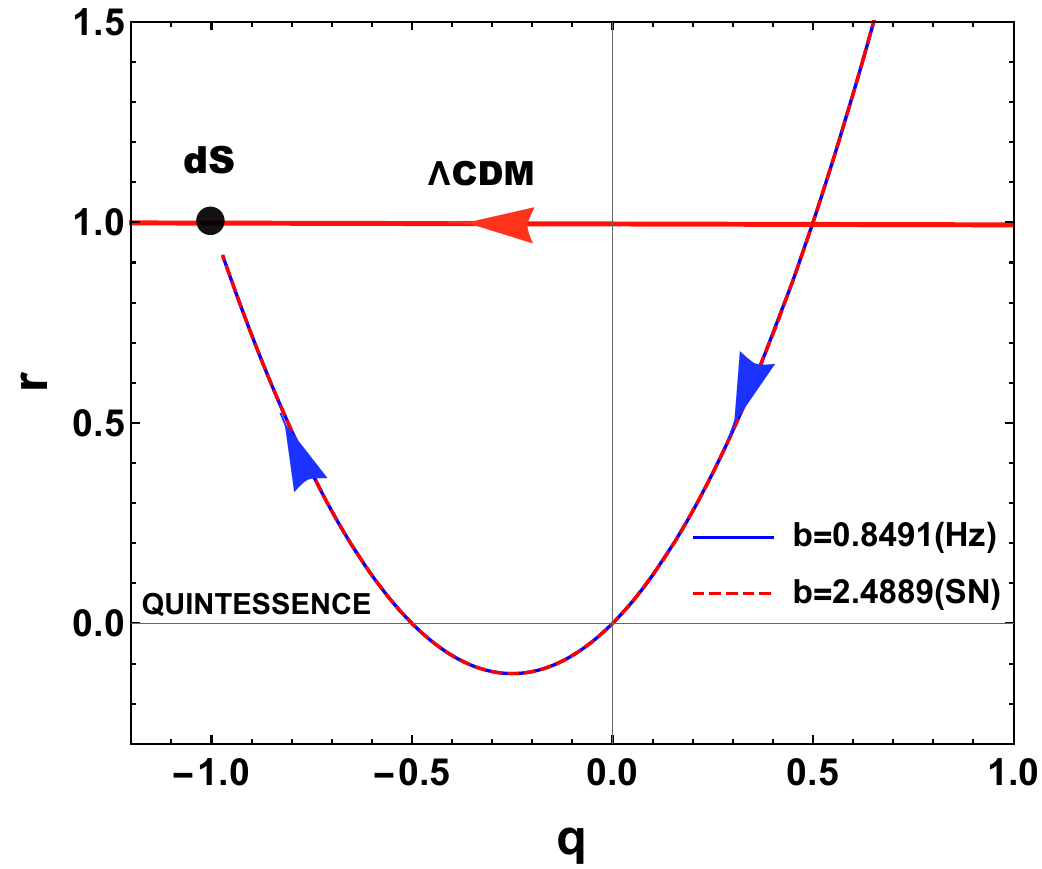}\\
\end{array}
$
\caption{The left panel shows the $r-s$ plane for our model with $b=0.8491$ and $b=2.4889$ and varying $n$. The right panel shows the $r-q$ plane for $b=0.8491$ and $b=2.4889$ and varying $n$.}
\label{f44}
\end{center}
\end{figure}

\begin{figure}[H]
\centering
\includegraphics[width=9 cm]{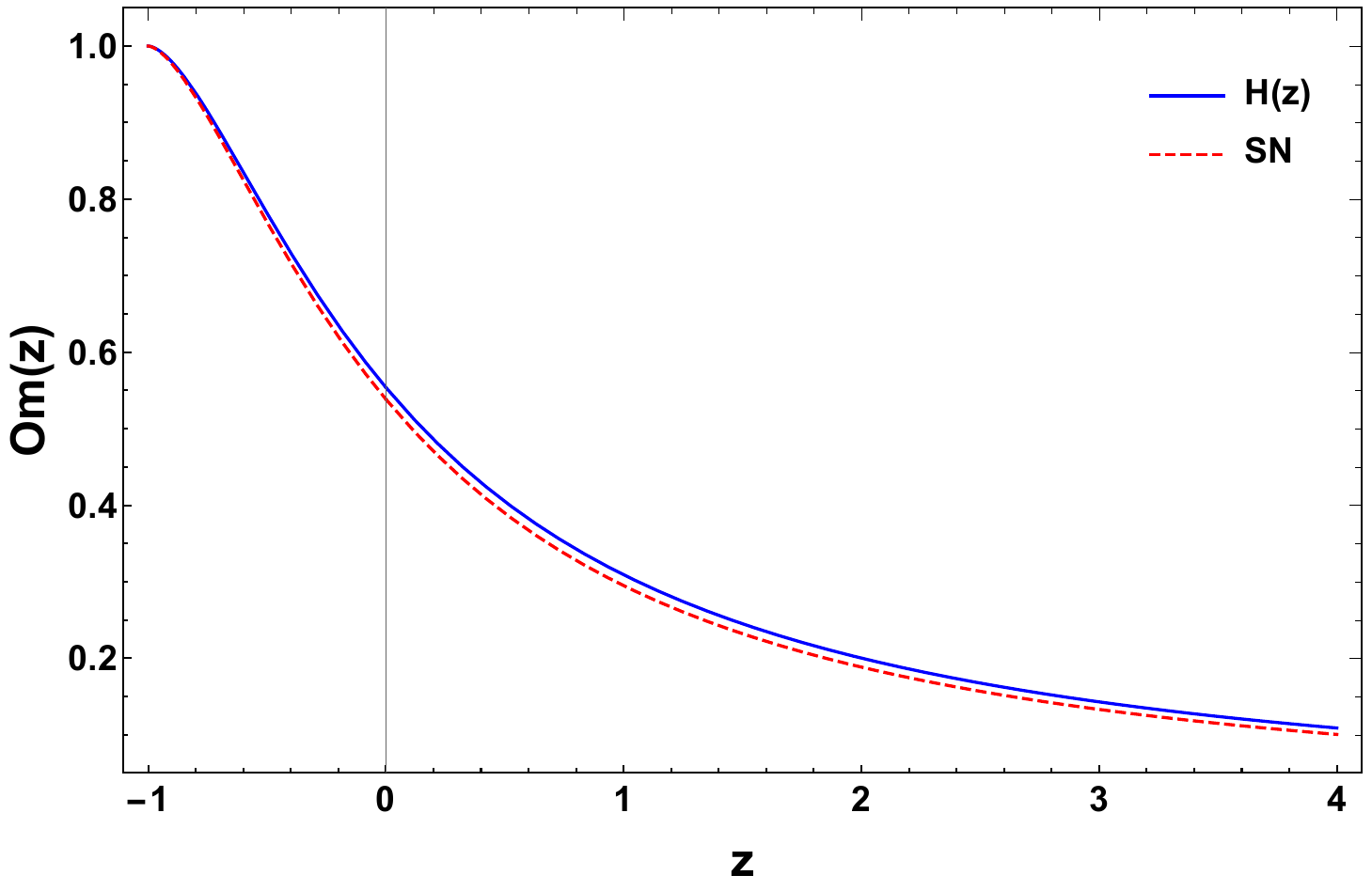}
\caption{Evolution of $Om(z)$ for $n=1.7763$, $b=0.8491$ and $n=1.7769$, $b=2.4889$.}
\label{f45}
\end{figure}

This section discusses the diagnostic tools used to distinguish between various dark energy models, as discussed in section \ref{sec2.4.2}. 
The plot of $r-s$ and $r-q$ planes is depicted in figure \ref
{f44}. Different dark energy models can be represented by different trajectories in the $r-s$ plane. Figure \ref{f44} reveals that the point $(s,r)=(0,1)$ represents the $\Lambda$CDM, whereas $(q,r)=(-1,1)$ corresponds to the de Sitter (dS) point. That is, our constructed model will finally approach the $\Lambda$CDM passing from the quintessence phase. The red line divides the plane into two parts, depicting the Quintessence phase as a lower half. The statefinder plots have been done for the values of $n$ and $b$ constrained by the H(z) and SNeIa datasets. The phenomenon described in this study is also evident in previous research conducted by \cite{Rani/2015,Suresh/2012}. We note that for the H(z) dataset, the $r$ and $s$ parameters at the present epoch are $r_{0}=-0.112$ and $s_{0}=0.554$, while for the SNeIa dataset, we have $r_{0}=-0.118$ and $s_{0}=0.538$. However, according to reference \cite{Albert}, these parameters have the potential to be inferred by forthcoming discoveries, thereby significantly helping in the explanation of the nature of dark energy.
Additionally, we have generated a plot of the function $Om(z)$, as shown in figure \ref{f45}. This plot visually represents a quintessence type at present and $\Lambda$CDM at late times, based on the constrained values of the model parameters derived from the H(z) and SNeIa datasets.

\section{Evolution of $\rho(z)$} \label{VIc}
\begin{figure}[]
\begin{center}
$
\begin{array}{c@{\hspace{.1in}}c}
\includegraphics[width=3.0 in, height=2.5 in]{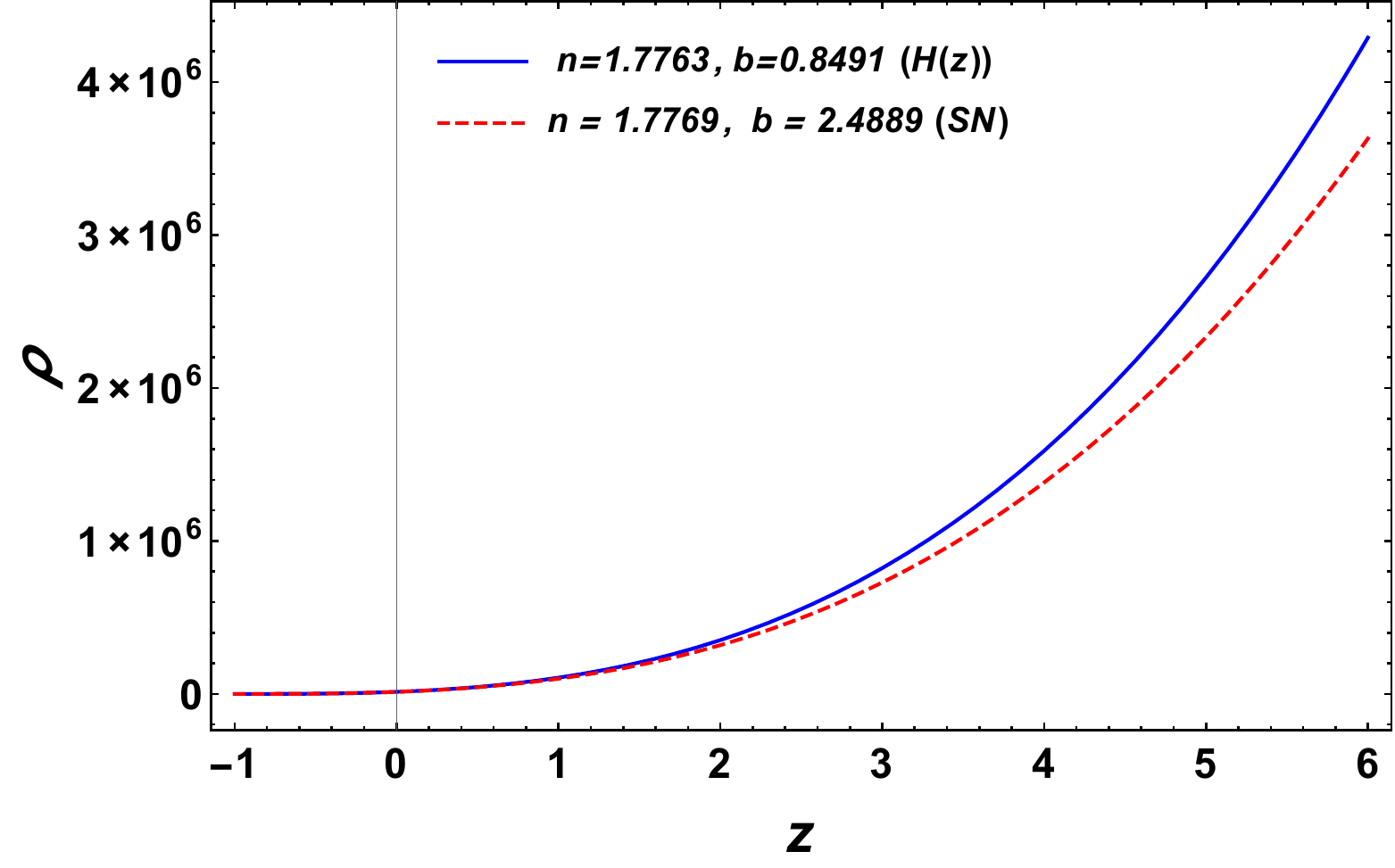} & 
\includegraphics[width=3.0 in, height=2.5 in]{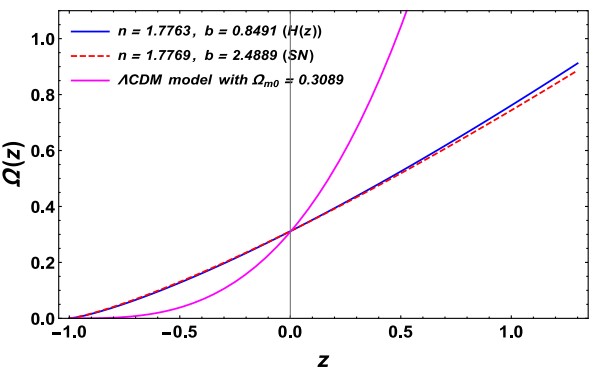}\\
\end{array}
$
\caption{The evolution of energy density (left panel) and the
density parameter (right panel) for the obtained model versus redshift $z$
with $\xi=-0.003$.}
\label{f46}
\end{center}
\end{figure}

The energy density also plays a significant role in cosmological theories. The energy density expression in equation \eqref{4.6} can be formulated in terms of the redshift $z$
\begin{equation}
\rho (z)=\dfrac{6^{n}\xi(1-2n)}{16\pi +3b}H_{0}^{2n}(1+z)^{2\left( \frac{24\pi+3b}{16\pi +3b}\right) }.  \label{4.14}
\end{equation}
We define the density parameter, $\Omega =\frac{8\pi \rho }{3H^{2}}$ using the equation \eqref{4.14} as
\begin{equation}
\Omega (z)=\frac{\xi\,\pi\, 2^{n+3} 3^{n-1} (1-2 n)\, (z+1)^{-\frac{6 (b+8 \pi )}{(3 b+16 \pi ) n}} \left(H_{0} (z+1)^{\frac{3 b+24 \pi }{3 b n+16 \pi  n}}\right)^{2 n}}{(3 b+16 \pi ) H_{0}^2},  \label{4.15}
\end{equation}
with $\Omega (0)=\frac{8\pi \,6^{n}\xi(1-2n)}{3(16\pi +3b)}H_{0}^{2n-2}$. Since the expressions in equations \eqref{4.14} and \eqref{4.15} contain a term $(1-2n)$ that will be negative for the discussed range of $n$ values, we must adopt the adjustable free parameter $\xi$ so that $\rho$ assumes a positive value. In figure \ref{f46} (left panel), we depict the evolution of the energy density with respect to redshift $z$ for the constrained numerical values of the model parameters $n$ and $b$. Also, we have shown the evolution of the density parameter (right panel) for our model together with the evolution of density parameter of matter density $\Omega_{m}=0.3089(1+z)^{3}$ as in the $\Lambda$CDM model $H(z) = \sqrt{\Omega_{m}(1+z)^{2}+\Omega_{\Lambda}}$ for comparison (where $\Omega_{m0} = 0.3089$ and $\Omega_{\Lambda} = 0.6911$ as suggested by Planck2015 results.

Now, let us analyse the simplest instance of the functional form, specifically the linear functional form of $f(Q,\mathcal{T})$ in the next section.

\section{Linear Form of $f(Q,\mathcal{T}) = \xi\,Q + b\,\mathcal{T}$} \label{VIIc}
We can see that for $n=1$, the considered functional form of $f(Q,\mathcal{T})=\xi\,Q^{n}+b\,\mathcal{T}$ reduces to the linear form $f(Q,\mathcal{T})=\xi\,Q+b\,\mathcal{T}$, which is similar to the linear functional form considered in \cite{Xu/2019}.
One can obtain the expression for the energy density from equation \eqref{4.6}
\begin{equation}
\rho =\frac{-6\xi H^{2}}{16\pi +3b},  \label{4.16}
\end{equation}
and the dynamical equation in $H$ reads as
\begin{equation} \label{4.17}
\dot{H}+\frac{3(8\pi +b)}{16\pi +3b}H^{2}=0\text{,}
\end{equation}
yielding a Hubble parameter in the form 
\begin{equation}
H=\frac{1}{Bt+k_{1}}.  \label{4.18}
\end{equation}
Here, $B=\frac{24\pi +3b}{16\pi +3b}$ and $k_{1}$ is the constant of
integration. Using the equation \eqref{4.18}, we can also easily find the scale factor given by
\begin{equation}
a(t)=k_{2}(Bt+k_{1})^{\frac{1}{B}},  \label{4.19}
\end{equation}
where $k_{2}$ is the constant of integration. As discussed above, we
must express all the above cosmological parameters in terms of redshift $z$, for which the $t-z$ relationship is obtained as
\begin{equation} \label{4.20}
t(z)=-\frac{k_{1}}{B}+\frac{1}{B}\left[ k_{2}(1+z)\right] ^{-B}.
\end{equation}
Henceforth, the Hubble parameter in terms of $z$ can be written as follows
\begin{equation}
H(z)=H_{0}(1+z)^{B}=H_{0}(1+z)^{\left( \frac{24\pi +3b}{16\pi +3b}\right) }\text{,}  \label{4.21}
\end{equation}
which contains only one model parameter $b$. Moreover, $q(t)=-1+\frac{24\pi +3b}{16\pi+3b}$ is also constant with only one model parameter $b$. Since, the term $\frac{24\pi +3b}{16\pi +3b}$ assumes a value $\simeq \frac{3}{2}$ for any value of $b\in (-\infty ,\infty )$, we have $q(t)=0.5$, showing a constant deceleration. The model reduces to the standard lore with $H(z)=H_{0}(1+z)^{\frac{3}{2}}$ and $a(t)\varpropto \left( \frac{3}{2}t+k_{1}\right) ^{\frac{2}{3}}$. It is noted that the energy density reduces to $\rho (z)=\frac{-6\xi}{16\pi +3b}H_{0}^{2}(1+z)^{3}$.

\section{Conclusions}\label{VIIIc}

This chapter addressed the late time cosmology using a well-motivated $f(Q,\mathcal{T})$ gravity model using the functional form $f(Q,\mathcal{T})=\xi\,Q^{n}+b\,\mathcal{T}$, where $\xi$, $n$ and $b$ are model parameters proposed in \cite{Xu/2019}. By constraining the free parameters with observational datasets: 57 points of the Hubble dataset and 580 points of the Union 2.1 compilation supernovae dataset, we found that the deceleration parameter at $z=0$ to be negative, that is $q_{0}=-0.17$ and $q_{0}=-0.19$, respectively, which is consistent with the present scenario of an accelerating Universe. Previous works in power law cosmology also reported similar constraints (see for example \cite{Rani/2015,Suresh/2012}). The solution and data analysis have already been discussed for the model considered in \cite{Xu/2019} with the $f(Q,\mathcal{T})$ function $f(Q,\mathcal{T})=-\sigma Q-\delta \mathcal{T}^{2}$ ($\sigma$ and $\delta$ are model parameters). Another model is also considered in the same paper with the $f(Q,\mathcal{T})$ function $f(Q,\mathcal{T})=\alpha \,Q+\beta \mathcal{T}$ ($\alpha $ and $\beta $ are model parameters) which is similar to the linear case with $n=1$ in our considered $f(Q,\mathcal{T})=\xi\,Q^{n}+b\,\mathcal{T}$ form, that is, $\xi \,Q+b\,\mathcal{T}$. In the linear case, the solution mimicked the power law expansion model with $a(t)\varpropto (Bt+c_{1})^{\frac{1}{B}}$, where $B=\frac{24\pi +3b}{16\pi +3b}$. This model relies on a single parameter $b$. It is observed that the parameter $b$ contribute significantly less in the evolution as the term $B=\frac{24\pi +3b}{16\pi +3b}\approx \frac{3}{2}$ for $b\in (-\infty ,\infty )$. Consequently, the model behaved similarly to the standard lore of $a(t)\sim t^{\frac{2}{3}}$ with a constant deceleration $q=0.5$.

We have thoroughly investigated the nature of dark energy modelled by the
parametrization $f(Q,\mathcal{T})=\xi\,Q^{n}+b\,\mathcal{T}$ using the statefinder diagnostic as $r-s$ and $r-q$ planes and the $Om(z)$ diagnostic analysis for our model. For the values of the model parameters $n$ and $b$ obtained by $57$ points of $H(z)$ dataset, the present statefinder parameter values are obtained as $r_{0}=-0.112$ and $s_{0}=0.554$, whereas the Union 2.1 compilation dataset resulted $r_{0}=-0.118$ and $s_{0}=0.538$.

The model investigated here is good at explaining the current observations but may not adequately explain the early evolution (as it is inconsistent with the constraints derived from the BAO dataset, which are not considered here). One important thing to notice is that the model parameter $b$, which is the coefficient of the trace $\mathcal{T}$ in the $f(Q,\mathcal{T})= \xi\,Q^n+b\,\mathcal{T}$ form considered, contributes very little to the evolution, indicating that the linear trace $\mathcal{T}$ does not affect the evolution. The behavior of energy density and the density parameter with respect to redshift $z$ for the constrained values of the model parameters $n$ and $b$ are depicted in figure \ref{f46} with $\Omega_{m0} = 0.3089$. Some more functional form of $f(Q,\mathcal{T})$ could be explored in the same way and is deferred to our future works.
 
One can also consider an effective equation of state which could distinguish and discriminate between the standard model and various theories of gravity. Henceforth, in the upcoming chapter, we will investigate $f(Q,\mathcal{T})$ gravity using an effective equation of state.


\chapter{Constraining Effective Equation of State in $f(Q,\mathcal{T})$ Gravity} 

\label{Chapter4} 

\lhead{Chapter 4. \emph{Constraining Effective Equation of State in $f(Q,\mathcal{T})$ Gravity}} 

\blfootnote{*The work in this chapter is covered by the following publication:\\
\textit{Constraining Effective Equation of State in $f(Q,\mathcal{T})$ Gravity}, European Physical Journal C, \textbf{81}, 555 (2021).}

The present chapter discusses a parameterized effective equation of state with two parameters in $f(Q,\mathcal{T})$ gravity. The detailed study of the work follows as
\begin{itemize}
\item We use the recently proposed $f(Q,\mathcal{T})$ gravity to investigate the accelerated expansion of the Universe, where $Q$ is the non-metricity and $\mathcal{T}$ is the trace of the energy-momentum tensor.
\item The investigation uses a parameterized effective equation of state with two parameters $m$ and $n$. 
\item We consider the linear form $f(Q,\mathcal{T})= Q+b\,\mathcal{T}$, where $b$ is a constant. By confining the model with the recently published 1048 Pantheon sample, we could determine the parameters $b$, $m$, and $n$ with the best fit.
\end{itemize}

\section{Introduction}\label{Id}

Several studies have investigated cosmological models and the phenomenon of rapid cosmological expansion within the framework of $f(Q,\mathcal{T})$ gravity, employing various functional forms. Yang et al. \cite{Yang/2021} developed the geodesic deviation and Raychaudhuri equations in the $f(Q,\mathcal{T})$ gravity based on the observation that the curvature-matter coupling considerably modifies the nature of tidal forces and the equation of motion in the Newtonian limit. Hence, it is possible to investigate the feasibility of the recently proposed $f(Q,\mathcal{T})$ gravity within various cosmological scenarios. Thus, utilizing the post-Newtonian limit seems suitable for comparing theoretical predictions with observations made within the solar system. Therefore, a rationale exists for examining multiple aspects of the $f(Q,\mathcal{T})$ theory from a theoretical, observational and cosmological perspective. To review $f(Q,\mathcal{T})$ gravity in different aspects, one can check references \cite{Sahoo/2020,Najera/2022}.

It is commonly known that the EoS parameter defines the relationship between pressure and energy density. The EoS parameter is used to classify various phases of the Universe, specifically distinguishing between decelerated and accelerated expansion. Understanding the late time acceleration in $f(Q,\mathcal{T})$ gravity is the focus of our work, and we do so by reconstructing a parameterized equation of state parameter. The present value of the effective equation of state parameter is established by model parameters that are constrained by observational evidence. CMB \cite{Eisenstein/2005}, SNeIa \cite{riess/1998,perlmutter/1999}, BAO \cite{Spergel/2007} and other observational datasets are currently available for various measurements and are providing robust evidence for the accelerating Universe. Consequently, we will use Pantheon datasets \cite{Scolnic/2018} to constrain the model parameters. The recently proposed supernovae Pantheon sample contains 1048 points covering the redshift range $0.01<z<2.26$. We use the MCMC ensemble sampler given by the emcee library.\\
The chapter has been divided into various sections. In section \ref{IId}, we engage in a comprehensive examination of the cosmological model employing the parameterized equation of state. We derive an expression for the Hubble parameter using the Friedmann equations of $f(Q,\mathcal{T})$. The brief discussion on observational data used to constrain the model parameters is presented in section \ref{IIId}. Section \ref{IVd} includes the behavior of cosmological parameters such as the deceleration parameter and EoS parameter. The last section \ref{Vd} encompasses the concluding remarks. 

\section{Cosmological Model and Equation of State}\label{IId}

In a flat FLRW metric, we have already obtained the generalized Friedmann equations in \eqref{4.1} and \eqref{4.2}.
Since, $f(Q,\mathcal{T})$ is an arbitrary function of $Q$ and $\mathcal{T}$, we consider the simplest functional form $f(Q,\mathcal{T})= Q+b\,\mathcal{T}$, where $b$ is a constant. In order to elaborate upon our prior research on the linear trace term, we assumed a specific functional form. One can note that $F= f_{Q}= 1$ and $8\pi \widetilde{G}= b$.
Further solving equations \eqref{4.1} and \eqref{4.2} for $p$ and $\rho$ allow us to determine the equation of state parameter $\omega= \frac{p}{\rho}$ given by
\begin{equation}
\label{5.3}
\omega= \frac{3 H^{2}(8\pi + b)+ \dot{H}(16 \pi+ 3b)}{ b \dot{H} - 3 H^{2}(8\pi+ b)}.
\end{equation}
To establish a comprehensive understanding, it is imperative to express all the cosmological parameters in terms of redshift $z$. Consequently, the relation for $t$ and $z$ using $\frac{a_{0}}{a}= 1+z$ can be derived as 
\begin{equation}
\label{5.4}
\frac{d}{dt}= \frac{dz}{dt}\frac{d}{dz}= -(1+z) H(z)\frac{d}{dz}.
\end{equation}
Normalizing the present value of scale factor to be $a_{0}= a(0)=1$. The Hubble parameter can be written in the form
\begin{equation}
\label{5.5}
\dot{H}= -(1+z) H(z)\frac{dH}{dz}.
\end{equation}
We need one more alternate equation to solve equation \eqref{5.3} for $H$. So, we assume a well-motivated parametric form of the equation of state parameter as a function of redshift $z$ \cite{Ankan/2016} which is defined by
\begin{equation}
\label{5.6}
\omega= -\frac{1}{1+ m(1+z)^{n}},
\end{equation}
where $m$ and $n$ are model parameters. Mukherjee describes the behavior of the considered equation of state parameter in \cite{Ankan/2016}. At the epoch of recent acceleration, it has a negative value of less than $-\frac{1}{3}$. For positive values of the model parameters $m$ and $n$, the value of $\omega$ tends to zero at a high redshift $z$ and depends on the model parameter at $z = 0$. The effective equation of state assumed for the current reconstruction conveniently accommodates these two phases of evolution. A positive model parameter frequently establishes a lower limit on the value of $\omega$ and maintains it in the non-phantom regime.
Using equations \eqref{5.3}, \eqref{5.5} and \eqref{5.6}, we obtain the Hubble parameter $H$ in terms of redshift $z$ as
\begin{equation}\label{5.7}
H(z)= H_{0} \left( \frac{b+(16\pi+ 3b)(1+m(1+z)^{n})}{b+(16\pi+3b)(1+m)}\right) ^{l},
\end{equation}
where $l= \frac{3(8\pi+b)}{n(16 \pi+3 b)}$, $H_{0}$ is the Hubble value at $z=0$.

In the following section, we attempt to obtain the best possible values for the model parameters associated with the equation of the Hubble parameter.

\section{Observational Constraints}\label{IIId}

This section describes the observational dataset used to constrain the model parameters $b$, $m$, and $n$ after obtaining the solutions for our model. We use the MCMC sampling technique to explore the parameter space and mainly employ Python's emcee \cite{Mackey/2013} library. To estimate the parameters, one need not compute the evidence, which is a normalizing constant. Instead, the prior and likelihood are sufficient to determine the posterior distributions of the parameters.\\
We use the recent Pantheon dataset for our work. One can find the dataset in the link\footnote[8]{\href{https://cdsarc.cds.unistra.fr/viz-bin/cat?J/ApJ/859/101}{https://cdsarc.cds.unistra.fr/viz-bin/cat?J/ApJ/859/101}}; it consists of 1048 Supernova Type Ia experiment results discovered by the Pan-STARRS1(PS1) Medium Deep Survey, the Low-z, SDSS, SNLS, and HST surveys \cite{Scolnic/2018,Chang/2019}, in the redshift range $z \in (0.01, 2.26)$ described in section \ref{3.3.2}.

\begin{figure}[H]
\centering
\includegraphics[scale=0.7]{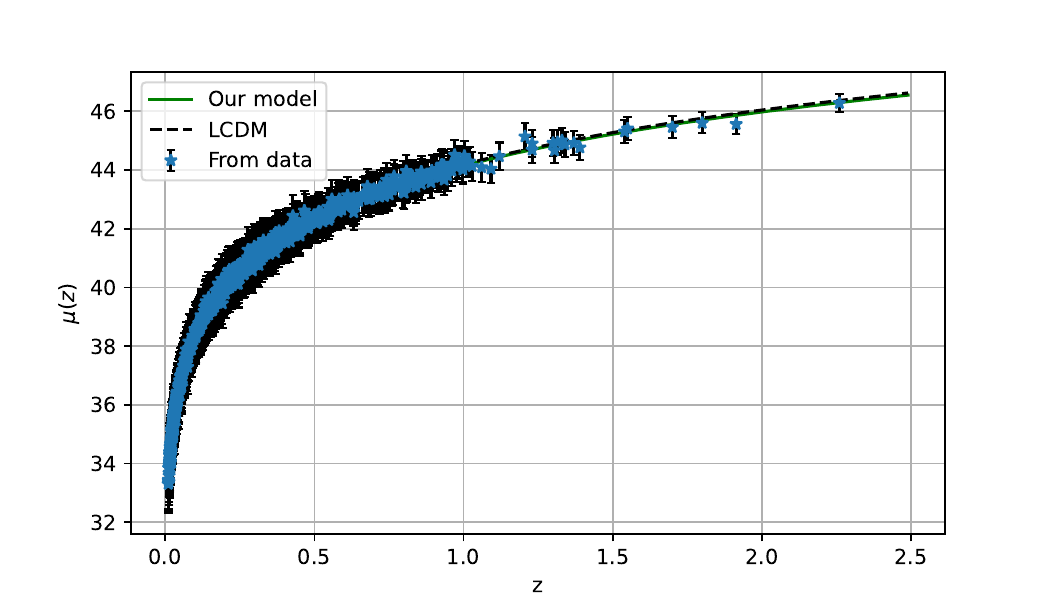}
\caption{The evolution of distance modulus $\mu(z)$ versus redshift $z$ for our model shown in red line and $\Lambda$CDM in black dotted line which shows nice fit to the 1048 points of the Pantheon dataset with its error bars.}
 \label{f51}
\end{figure}

As the functional form $f = Q + b\mathcal{T}$ contains $b$ as a model parameter and $m$, $n$ are the parameters in the parametric functional form of the equation of state $\omega$. The bounds for the parameters from our analysis are obtained as $b = 0.2^{+2.7}_{-2.9} \ , m = 0.47^{+0.27}_{-0.21},\ n = 3.2^{+1.8}_{-2.0}$. The comparison between our model and the widely accepted $\Lambda$CDM model is depicted in figure \ref{f51}. For the plot, we assume $\Omega_{m0}= 0.3$, $\Omega_{\Lambda 0} = 0.7$, and $H_0= 69\,km/s/Mpc$ \cite{Planck/2018}. The figure also includes the Pantheon experimental results, 1048 data points along with their error, and allows for a clear comparison between the two models.

\begin{figure}[]
\centering
\includegraphics[scale=0.9]{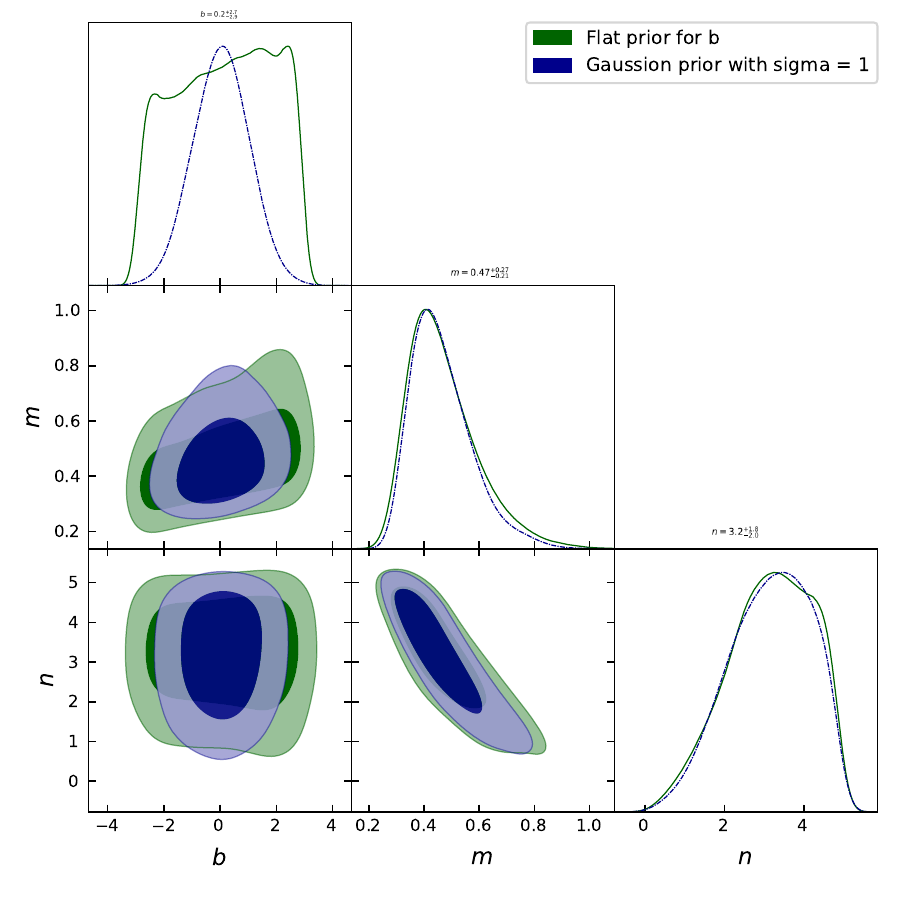}
\caption{The plot compares the two numerical analyses with different prior distributions for the parameter $b$. For either case, we considered a uniform distribution for the other two parameters, $m$ and $n$. The counter represent $1\sigma$ and $2\sigma $ confidence intervals.} \label{f52}
\end{figure}

Initially, we perform the analysis considering a flat prior for all the parameters. However, we notice the marginalized distribution for parameter $b$ to be roughly uniform in the range. We then motivate our work to study the result in the neighbourhood of $b=0$. This approach intends to find any deviation from GR, which accounts for a local minimum for the function in equation \eqref{5.7}. We also perform the numerical analysis with a Gaussian prior for the parameter $b$ with $\sigma=1.0$ as dispersion. The results are presented in figure \ref{f52}. There is no significant difference in the marginalized distributions of the remaining parameters, $m$ and $n$. 

\begin{figure}[]
\centering
\includegraphics[scale=0.9]{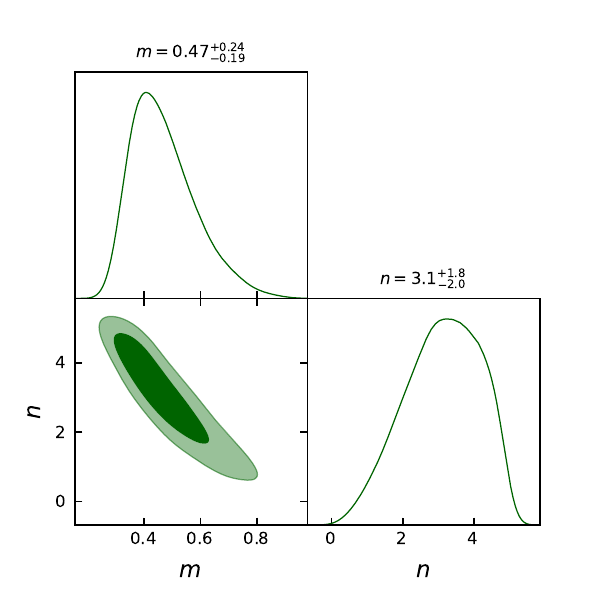}
\caption{The plot showing the best fit values of the model parameters $m$ and $n$ obtained with 1048 points of Pantheon dataset at $1\sigma$ and $ 2\sigma $ confidence level with $b=0$.} \label{f53}
\end{figure}

If we consider the case of $b=0$, the model reduces to $f(Q,\mathcal{T})= f_{\Lambda}(Q)=Q$, i.e. it has a direct link to $\Lambda$CDM model. Therefore, the equation of Hubble parameter $H$ reduces to 
\begin{equation}
\label{5.8}
H(z)=H_{0}\left(\frac{m (z+1)^n+1}{m+1}\right)^{\frac{3}{2 n}},
\end{equation}
where $m$ and $n$ are model parameters. One interesting point regarding this expression of the Hubble parameter is that for $n = 3$, this becomes exactly like the $\Lambda$CDM model as stated in \cite{Ankan/2016}.
The constraints for $m$ and $n$ using the Pantheon SNeIa dataset are shown in  figure \ref{f53}.

\section{Cosmological Parameters} \label{IVd}

In this section, we examine the behavior of the deceleration and equation of state parameters. The equation of deceleration parameter $q = -1-\frac{\dot{H}}{H^{2}}$ , according to our model, reads
\begin{equation} \label{5.9}
q = -1-\frac{3 (b +8 \pi ) m (-z-1) (z+1)^{n-1}}{b +(3 b +16 \pi ) \left(m (z+1)^n+1\right)}.
\end{equation}

\begin{figure}[H]
\centering
\includegraphics[scale=0.4]{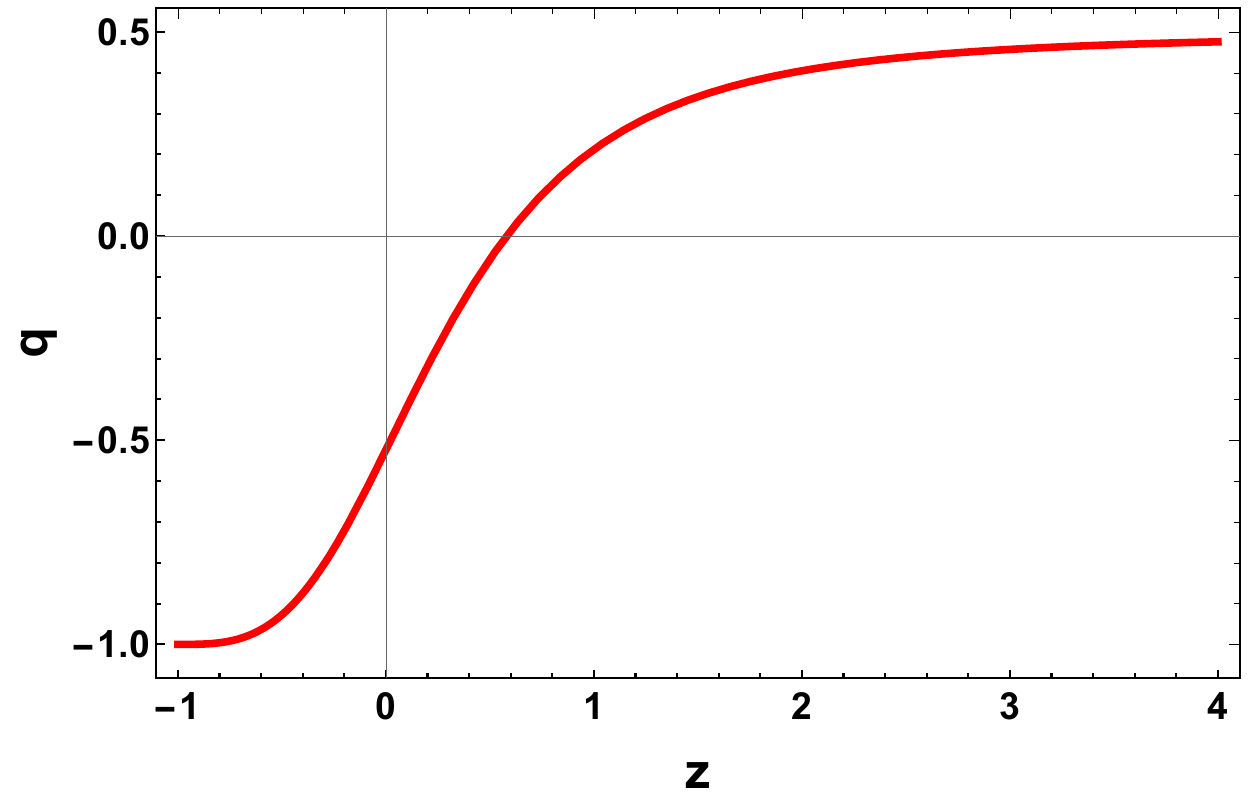}
\caption{Evolution of the deceleration parameter for the best fit values $b=0.2$, $m=0.47$ and $n=3.2$ from the analysis of SNeIa Pantheon samples.} \label{f54}
\end{figure}


Figure \ref{f54} depicts the behavior of $q(z)$ based on the estimated values of model parameters $b$, $m$, and $n$ obtained from the Pantheon sample. We can observe a well-behaved transition from deceleration to acceleration phase at redshift $z_{t}$. The value of the transition redshift has been determined to have a value of $z_{t}= 0.58\pm 0.30$ \cite{Farooq/2013}. The findings align with several works in literature \cite{Cunha/2009, JV/2008}. Also, we can note that the value of $q_{0}$ is obtained as $-0.52$ \cite{Christine/2014}, which is negative at present, indicating that the Universe is undergoing acceleration.

\begin{figure}[H]
\centering
\includegraphics[scale=0.4]{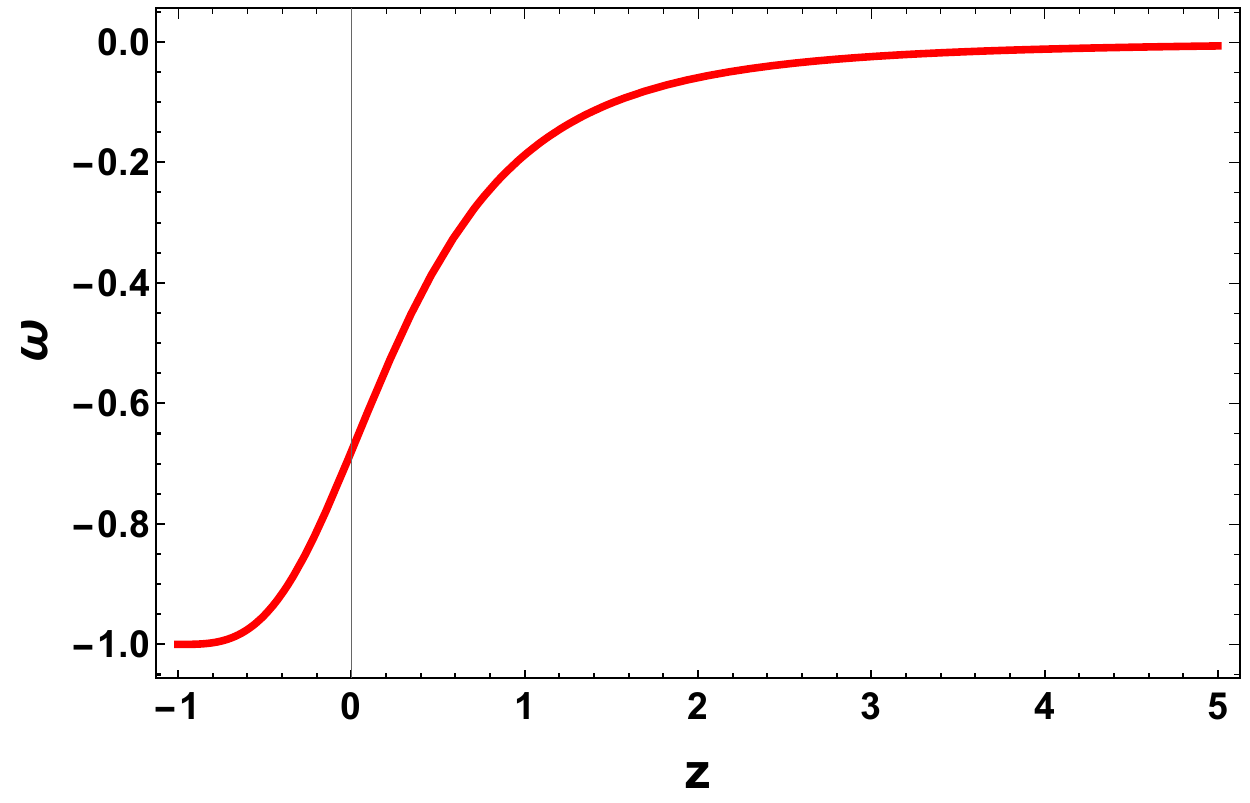}
\caption{Evolution of the equation of state parameter for the best fit values $b=0.2$, $m=0.47$ and $n=3.2$ from the analysis of SNeIa Pantheon samples.} \label{f55}
\end{figure}

In addition, we have analyzed the effective EoS parameter, as shown in figure \ref{f55}. The value of the EoS parameter at $z=0$ is obtained as $\omega_{0}= -0.68^{+0.10}_{-0.11}$ \cite{Christine/2014}, which indicates an accelerating phase, exhibiting a quintessence type behavior.

\section{Conclusions}\label{Vd}
As new theories of gravity emerge in the literature, it is vital to put them to the test to determine if they are viable in describing the dark sector of the Universe. To begin, we considered the functional form $f(Q,\mathcal{T}) = Q+b\,\mathcal{T}$, where $b$ is a free parameter. 

We used a well-motivated parametric form of the equation of state parameter as a function of redshift $z$ to solve the field equations for $H$. At the epoch of recent acceleration, the value of the EoS parameter is seen to be negative, specifically less than $-\frac{1}{3}$. At a high redshift $z$, the value of $\omega$ tends to zero for positive values of the model parameters $m$ and $n$, and its value is influenced by the model parameter at $z = 0$. The Pantheon study, a recently proposed observational dataset, was used to constrain the parameter space. The parameters from our study have bounds of $b = 0.2^{+2.7}_{-2.9}$, $m = 0.47^{+0.27}_{-0.21}$, and $n = 3.2^{+1.8}_{-2.0}$.
The error bar plot depicting the 1048 points from the Pantheon dataset, together with our derived model and the $\Lambda$CDM model assuming $\Omega_{m0}= 0.3$, $\Omega_{\Lambda0} = 0.7$ and $H_{0}= 69\,km/s/Mpc$ demonstrates a good match to the observational results as seen in section \ref{IIId}.

Finally, we observed the behavior of the deceleration parameter and equation of state parameter in section \ref{IVd}. We see a well-behaved transition from deceleration to acceleration phase at redshift $z_{t}$. The value of the transition redshift is obtained as $z_{t}= 0.58 \pm 0.30$, while the present deceleration parameter is estimated to be $q_{0}= -0.52$. Furthermore, it is worth noticing that the value of the EoS parameter at $z=0$ exhibits a value of $\omega_{0}= -0.68^{+0.10}_{-0.11}$, which provides strong evidence in favour of accelerating phase. The current analysis provides a justification and motivation for investigating these extensions, potentially explaining the deviation from the $\Lambda$CDM. In this case, the model does not deviate much from the $\Lambda$CDM.

In addition to considering the cosmic evidence, it is imperative to ascertain the stability of the newly proposed $f(Q,\mathcal{T})$ theory. This is necessary to facilitate the analysis of the perturbation level differences between this theory and the $\Lambda$CDM framework. The investigation of $f(Q,\mathcal{T})$ can be further explored in future research endeavors. The study mentioned previously examines modified gravity models as a potential alternative to dark energy models. Our subsequent work aims to explore an alternative perspective on the enigma of dark matter, as given by modified theories. In this study, we investigate the impact of modified gravity on the dynamics of large-scale structures.

\chapter{On the Impact of $f(Q)$ Gravity on the Large Scale Structure} 

\label{Chapter5} 

\lhead{Chapter 5. \emph{On the Impact of $f(Q)$ Gravity on the Large Scale Structure}} 

\blfootnote{*The work in this chapter is covered by the following publication:\\
\textit{On the impact of $f(Q)$ gravity on the large scale structure}, Monthly Notices of the Royal Astronomical Society, \textbf{522}, 252-267 (2023).}


The current chapter presents the exponential $f(Q)$ gravity to probe structure formation and viable cosmology. A detailed study of the work follows:
\begin{itemize}
\item We consider the exponential $f(Q)$ form namely $f(Q) = Q + \alpha\,Q_{0}\left(1-e^{-\beta \sqrt{Q/Q_{0}}}\right)$ to probe the structure formation with box sizes $L_{\mathrm{box}}=10/100$ Mpc$/h$ and middle resolution $N_p^{1/3}=512$. 
\item The present study aims to reproduce viable cosmology within the 
aforementioned modified gravity theory using MCMC sampling on H(z)/BAO/Pantheon datasets and constrain a  parameter space.
\item While carrying out N-body+SPH simulations, we derive CDM+baryons overdensity/temperature/mean molecular weight fields, matter power spectrum (both 2/3D, with/without redshift space distortions), bispectrum, two-point correlation function and halo mass function.
\end{itemize}


\section{Introduction} \label{sec:3.1}

It is well-known that GR is a quite successful theory on various cosmological scales, and it can describe the recent accelerated expansion of the Universe by introducing the so-called cosmological constant (or $\Lambda$ term) in the Einstein-Hilbert action integral. Since the gravitational Lagrangian is not practically restricted by only the linear Ricci scalar term, one can introduce additional terms to emulate effective dark energy and reproduce different Universe evolutionary phases, such as cosmological inflation or late time accelerated expansion.

There are many ways to modify GR, for example, by introducing some matter fields (canonical scalar field, vector and gauge boson fields, Dirac spinors, etc.). Another way is to present an entirely different notion of the Lorentzian 4-manifold curvature by adjusting the metric-affine connection \cite{Felice/2010,capozziello2011extended}. For example, one could use the so-called torsion or non-metricity, which are constructed based on Weitzenb\"ock and metric incompatible affine connections, respectively. Consequently, GR has two analogs, namely, TEGR and STEGR as mentioned in section \ref{1.6}. In the current work, we will focus on the arbitrary parameterization of an extended STEGR ($f(Q)$ gravitation).

A key aspect of $f(Q)$ theory is the usage of a flat connection pertaining to the existence of affine coordinates in which all of its components vanish, converting covariant derivatives into partial derivatives \cite{hohmann2021general,dimakis2022flrw}. So, it is possible to distinguish gravity from inertial effects in the $f(Q)$ theory. For many modified gravity theories, the development of the $f(Q)$ theory provides a fresh starting point. It offers a straightforward formulation in which self-accelerating solutions spontaneously appear in both the early and late Universe. Compared to other geometric extensions of GR, both $f(T)$ and $f(Q)$ theories have a substantial benefit in that the background field equations are always of second order.

Many studies have been incorporated on $f(Q)$ gravity and is a very promising theory, that can reproduce the behaviour of both early and late Universe, satisfy constraints from Cosmic Microwave Background (CMB), SNeIa, BAO, H(z) datasets and primordial scalar index $n_s$, standard sirens from LIGO/VIRGO/ET \cite{DAgostino:2022tdk,Ferreira:2022jcd}. For instance, exponential gravitation was constrained in the study by \cite{ANAGNOSTOPOULOS2021136634,atayde2021can}, and the authors found out that such a theory can challenge concordance $\Lambda$CDM theory. Additionally, observational constraints on the $f(Q)$ gravity have been established for a number of parameterizations of the $f(Q)$ function using various observational probes \cite{Lazkoz/2019,Jimenez/2018}. Aside from these findings, $f(Q)$ gravity has been the focus of several investigations in various studies \cite{Hu/2022,alb/2022,wang/2022,esp/2022}. In the current chapter, we are going to investigate exponential $f(Q)$ gravity in terms of observational constraints using MCMC methodologies and high-resolution N-body simulations, which will be discussed in the following subsections.

\subsection{N-body Simulation as a Probe of Modified Gravity}

To probe the validity of a particular modified theory of gravitation, one needs to incorporate various cosmological observables, ranging from cosmic expansion rate to clustering and structure formation history. The latter could be most effectively studied with the use of the so-called N-body simulations, which are well-known to be the best theoretical probe of the large scale structure of the Universe, that provide information on the matter power spectrum/bispectrum, $N$-point correlation functions and halo mass function, void size function, etc. Over the last few years, such an approach has attracted some interest in modified gravity (see the work in \cite{Hassani/2020}). Authors of the paper \cite{Wilson/2023} developed a pipeline to differentiate modified gravity theories from the $\Lambda$CDM model and constrain those theories properly using voids in the N-body simulations, that are known to be less affected by the non-linear and baryonic physics in relation to the dark matter halos. Besides, in addition to voids, intrinsic shape alignments of massive halos and galaxy/halo angular clustering could be used to discriminate modified gravity theories from $\Lambda$CDM in the presence of massive neutrino (see \cite{Lee/2023} and \cite{Drozda/2022} respectively). The aforementioned Halo Mass Function (further-HMF) was examined for $f(Q)$ and Dvali-Gabadadze-Porrati (DGP) gravities in \cite{Gupta/2022}. Widely used code \texttt{MG-Gadget}, introduced and developed in \cite{Puchwein/2013} were employed to study $f(R)$ Hu-Sawicki theory \cite{Arnold/2016,Giocoli/2018}, and conformally coupled gravity \cite{Ruan/2022}. In turn, we are going to use \texttt{ME-GADGET} code (for documentation, check \cite{Zhang/2018}) to study $f(Q)$ gravity behaviour. Such code was applied to the case of $f(T)$ theory \cite{Huang/2022}, interacting dark energy \cite{Zhang/2018} and cubic vector gallileon \cite{Chen/2023}.

This chapter is organized as follows: In section \ref{sec:3.1}, we briefly introduce modified theories of gravity and N-body simulations. Consequently, in section \ref{sec3.2}, we present the foundations of symmetric teleparallel gravity and we adopt the FLRW isotropic line element to derive field equations for our exponential choice of $f(Q)$ function. In section \ref{sec:3.3}, we introduce each observational dataset of our consideration and perform MCMC analysis. In section \ref{sec3.4}, we analyze the provided constraints deriving theoretical predictions for the deceleration parameter, statefinder pair and $Om(z)$. In the following section \ref{sec3.5}, we set up the \texttt{ME-GADGET} suite and study the N-body output for a small simulation box size. We therefore compare the results with the ones above obtained for large $L_{\mathrm{box}}$ in section \ref{sec3.6}. Finally, in the last section \ref{sec3.7}, we present the concluding remarks on the key topics of our study.

\section{Modified Symmetric Teleparallel Gravitation}\label{sec3.2}

As we already mentioned in section \ref{1.6.6}, both Ricci scalar and torsion terms vanish, and therefore we are left with only non-metricity. Therefore, to proceed with the STEGR case, one could derive the non-metricity scalar from the non-metricity tensor and its independent traces in section \ref{1.6.6} defined by 
\begin{equation}
    Q=-g^{\mu\nu}(L^\alpha_{\,\,\,\beta\nu}L^\beta_{\,\,\,\mu\alpha}-L^\beta_{\,\,\,\alpha\beta}L^\alpha_{\,\,\,\mu\nu})=-P^{\alpha \beta \gamma}Q_{\alpha \beta \gamma}.
\end{equation}
The condition of symmetric teleparallelism makes the generic affine connection to be inertial. The most general connection is 
\begin{equation}
 \tilde{\Gamma}^{\alpha}_{\,\,\, \mu\nu}= \frac{\partial x^{\alpha} }{\partial \xi^{\sigma}} \frac{\partial^2 \xi^{\sigma} }{\partial x^{\mu} \partial x^{\nu}}, 
\end{equation}
where $\xi^{\sigma}$ is an arbitrary function of spacetime position. We can always choose a coordinate $x^{\alpha}=\xi^{\sigma}$ by utilizing a general coordinate transformation, where the general affine connection $\tilde{\Gamma}^{\alpha}_{\,\,\, \mu\nu}=0$. We call this coordinate the coincident gauge \cite{Jimenez/2018}. Thus, in the coincident gauge, we will have $Q_{\alpha \mu \nu}= \partial_{\alpha}g_{\mu \nu}$, i.e. all the covariant derivatives are identical to ordinary derivatives.

Now we are going to present the formalism of modified symmetric teleparallel cosmology. Einstein-Hilbert action of the $f(Q)$ theory of gravity is therefore could be written as mentioned in equation \eqref{fq}.

The reason for the above action and specific selection of the non-metricity scalar is that GR is recreated, up to a boundary term for the choice $f=Q$, i.e., for this choice, we recover the allegedly ``symmetric teleparallel equivalent of GR". By varying the action \eqref{fq} with respect to the metric tensor (using least action principle $\delta \mathcal{S}=0$), we could obtain the corresponding field equations in equation \eqref{fqf}.
The connection equation of motion can be computed by noticing that the variation of the connection with respect to $\xi^{\alpha}$ is equivalent to performing a diffeomorphism so that $\partial_{\xi} \Gamma^{\alpha}_{\,\, \mu\nu}=-\mathcal{L}_{\xi}  \Gamma^{\alpha}_{\,\, \mu\nu} = -\nabla_{\mu} \nabla_{\nu} \, \xi^{\alpha}$ \cite{Jimenez/2020}. Besides this, in the absence of hypermomentum, one can take the variation of equation \eqref{fq} with respect to the connection 
\begin{equation}
\label{con}
    \nabla_{\mu}\nabla_{\nu}\left( \sqrt{-g} f_{Q} P^{\mu \nu}_{\,\,\, \,\,\alpha} \right)=0.
\end{equation} 
For the metric and connection equations, one can notice that $\nabla_{\mu}T^{\mu}_{\,\,\,\,\nu}=0$, where $\nabla_{\mu}$ is the metric-covariant derivative.

Since, we have already defined all of the necessary quantities, we could proceed further and set up the background spacetime.

\subsection{FLRW Cosmology}
In order to study the evolution of our Universe, it will be useful to assume that background spacetime is isotropic and homogeneous, namely FLRW spacetime\footnote[9]{Notice that we have used the Diff gauge freedom to fix the coincident gauge and, therefore, setting the lapse ($N(t)$ in metric) to 1 is not, in principle, a permitted choice. It happens however that the $f(Q)$ theories retain a time-reparametrization invariance that allows to get rid of the lapse \cite{Jimenez/2018}. }.
Consequently, with the assumption of FLRW spacetime, the non-metricity scalar is written as follows
\begin{equation}
    Q=6H^2.
\end{equation}
Finally, one could evaluate the FLRW field equations of the $f(Q)$ theory:
\begin{align}
&3H^2 = \kappa \left(\rho_\text{m}+\rho_{\text{eff}}\right)\, , \\
&3H^2 + 2\dot{H}= -\kappa \left(p_\text{m}+p_{\text{eff}}\right)\, . 
\end{align}
Here, $\rho_\text{m}$ and $p_\text{m}$ are matter energy density and isotropic pressure, respectively. Moreover, in the equation above $\rho_{\text{eff}}$ and $p_{\text{eff}}$ are effective energy density and pressure that define the contribution of $f(Q)$ gravity to the field equations. For modified STEGR, field equations with exact forms of effective quantities read
\begin{equation}
    3H^2 = \frac{ \kappa}{2f_Q}\bigg(\rho_m+\frac{f}{2}\bigg),
    \label{eq:17}
\end{equation}
\begin{equation}
    \left(12 H^{2} f_{QQ} + f_{Q}\right) \dot{H} = -\frac{ \kappa}{2} \left( \rho_{m} + p_{m} \right).
    \label{eq:18}
\end{equation}
where $f_Q=\frac{\partial f(Q)}{\partial Q},\quad f_{QQ}=\frac{\partial^2 f(Q)}{\partial Q^2}$.

The energy-momentum tensor of the cosmological fluid which is given in equation \eqref{Stress}
leads to conservation equation as $\dot{\rho}+ 3 H \left(\rho+p\right)=0$.
In symmetric teleparallel gravity and its extensions, the conservation law $T^{\mu}_{\,\,\,\nu;\mu}=0$ holds for the matter energy-momentum tensor. The $T^{\mu}_{\,\,\,\nu;\mu}=0$ holds through \eqref{con} for the connection as well \cite{Jimenez/2018a,dimakis2022flrw}. 

\subsection{Exponential $f(Q)$ Gravity}
This work particularly aims at investigating the $f(Q)$ gravity model, namely modified exponential $f(Q)$ gravity (which is built from the linear and exponential terms, respectively). In $f(Q)$ theory, numerous cosmic possibilities have been examined using various exponential models, notably inflationary cosmology, BBN constraints, and dynamic system analysis  \cite{Harko/2018,Anagnos/2023,Khyllep/2023}. For that kind of gravity, $f(Q)$ function reads (we adapt the work of \cite{Linder/2010} for modified STEGR)
\begin{equation}
    f(Q) = Q+\alpha Q_0(1-e^{-\beta\sqrt{Q/Q_0}}),
\end{equation}
where $\alpha$, $\beta$ are free parameters corresponding to additional degrees of freedom and $Q_{0}=6H_{0}^{2}$ is the present value of $Q$. We can reduce the number of degrees of freedom by matching first Friedmann equation \eqref{eq:17} at the present time (i.e. assuming $z=0$)
\begin{equation}
\alpha = -\frac{e^\beta(-1+\Omega_{m0})}{-1+e^\beta-\beta}.   
\end{equation}
Thus, the complexity of this form is just one step more than the standard $\Lambda$CDM. The exponential modified gravity could satisfy stability, validity and not cross the phantom divide line \cite{Arora/2022}. In order to solve the field equations and obtain the numerical form of the Hubble parameter, we use
\begin{equation}
    \dot{H}=aH\frac{dH}{da}.
\end{equation}
We will solve the field equation numerically, as already stated, with \texttt{Mathematica} numerical ODE solver \texttt{NDSolve}. Initial conditions at the vanishing redshift for $\dot{H}$ could be therefore set up (as a cosmographical quantity)
\begin{equation}
\dot{H}_0=-H_0^2(1+q_0),
\end{equation}
where $q_0$ is the current deceleration parameter, we fix it to $q_0=-0.55$ \cite{Reid/2019}. Additionally, for MCMC, as truths, we assume that the present value of the Hubble parameter is $H_0=69 \, \mathrm{km/s/Mpc}$ and that matter mass fraction at present is $\Omega_{m0}=0.315\pm 0.007$, following the observational constraints of \textit{Planck2018} \cite{Planck/2018}.

\section{MCMC Constraints}\label{sec:3.3}
In this section, we constrain our $f(Q)$ gravity model via observational datasets. To explore the parameter space, we will be using the MCMC methodology and Python package \texttt{emcee} \citep{Mackey/2013}.

\subsection{Observational Hubble Data (OHD)}

To determine the priors and likelihood functions (which are necessary), we use the H(z) dataset consisting of 31 points from cosmic chronometers described in section \ref{2.5.1}. We introduce the chi-square function as mentioned in equation \eqref{Chih} to constrain our modified gravity model. The likelihood function for MCMC sampling has its usual exponential form $\mathcal{L}=\exp(-\chi^2/2)$.

\subsection{Pantheon SNeIa Sample} \label{3.3.2}
We also use the Pantheon dataset to constrain our modified gravity with dark energy, which consists of 1048 SNeIa (discovered by the Pan-STARRS1 (PS1) Medium Deep Survey, Low $z$, SNLS, SDSS and HST \cite{Scolnic/2018}). In this case, we have used the binned Pantheon sample\footnote[10]{$\href{https://github.com/dscolnic/Pantheon/tree/master/Binned_data}{https://github.com/dscolnic/Pantheon/tree/master/Binned_data}$}. The corresponding chi-square function reads as in equation \eqref{Chisn}. The nuisance parameters in the Tripp formula \cite{tripp/1998} $\mu= m_{B}-M_{B}+\alpha x_{1}-\beta c+ \Delta_{M}+\Delta_{B}$ were retrieved using the novel method known as BEAMS with Bias Correction (BBC) \cite{Kessler/2017}, and the observed distance modulus is now equal to the difference between the corrected apparent magnitude $M_{B}$ and the absolute magnitude $m_{B}$ $\left(\mu= m_{B}-M_{B}\right)$. Additionally, one can define the chi-square function in terms of the covariance matrix as follows \cite{Deng/2018}:
\begin{equation}
\chi^2_{\mathrm{SN}}=\Delta \boldsymbol{\mu}^T \mathbf{C}^{-1} \Delta \boldsymbol{\mu},
\end{equation}
where the covariance matrix consists of statistical and systematic uncertainties respectively \cite{Conley/2011}:
\begin{equation}
\mathbf{C} = \mathbf{D}_{\mathrm{stat}}+\mathbf{C}_{\mathrm{sys}}.
\end{equation}
In the current work, we assume that the diagonal matrix of statistical uncertainties looks like $\mathbf{D}_{\mathrm{stat},ii}=\sigma^2_{\mu(z_i)}$. Besides, systematic uncertainties are derived using the Bias Corrections (BBC) method, introduced and developed in \cite{Scolnic/2018}:
\begin{equation}
\mathbf{C}_{ij,\mathrm{sys}} = \sum^K_{k=1}\bigg(\frac{\partial \mu^{obs}_i}{\partial S_k}\bigg)
\bigg(\frac{\partial \mu^{obs}_j}{\partial S_k}\bigg)\sigma^2_{S_k}.
\end{equation}
Indexes $\{i,j\}$ denote the redshift bins for distance modulus, $S_k$ here denotes the magnitude of systematic error, $\sigma_{S_k}$ is its standard deviation uncertainty.

\subsection{Baryon Acoustic Oscillations}

Consequently, we also use the chi-square function for the BAO dataset as described in section \ref{2.5.2}. In addition, we consider that the photon decoupling epoch arises at the redshift \cite{planck/2015}
\begin{equation}
    z_* = 1048[1+0.00124(\Omega_bh^2)^{-0.738}][1+g_1(\Omega_mh^2)^{g_2}],
\end{equation}
where,
\begin{equation}
    g_1=\frac{0.0783(\Omega_bh^2)^{-0.238}}{1+39.5(\Omega_bh^2)^{-0.763}},
\end{equation}
\begin{equation}
    g_2=\frac{0.560}{1+21.1(\Omega_bh^2)^{-1.81}}.
\end{equation}
This dataset was gathered from the works of \cite{Basilakos/2012, Eisenstein/2005,gio/2012}.

\begin{figure}[H]
    \centering
    \includegraphics[width=\textwidth]{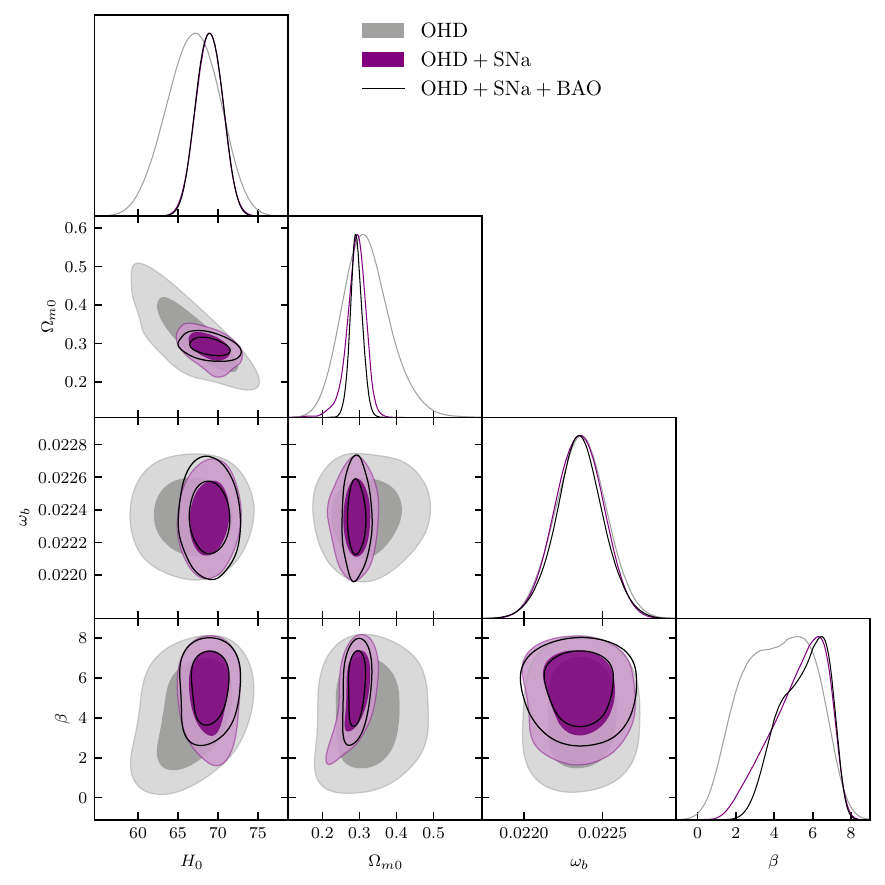}
    \caption{MCMC best fits from H(z) (OHD), Pantheon, and BAO datasets and joint distribution for exponential $f(Q)$ model. Here $\omega_{b} = {\Omega_{b0}}\,h^{2}$.}
    \label{f32}
\end{figure}

We performed the joint analysis for the combined H(z)+SNeIa+BAO by minimizing $\chi^{2}_{H} +\chi^{2}_{SN} +\chi^{2}_{BAO}$. The results are, therefore, numerically derived from MCMC trained on H(z), Pantheon, BAO, and joint datasets. Results are placed in the Table \ref{t1e} for model parameters $H_0$, $\beta$ and $\Omega_{m0}$. Furthermore, the $1\sigma$ and $2\sigma$ likelihood contours for the possible subsets of parameter space  are presented in figure \ref{f32}.

\subsection{Statistical Evaluation}
To evaluate the success of our MCMC analysis, one should perform the statistical evaluation using the so-called Akaike Information Criterion (AIC) and Bayesian Information Criterion (BIC). The first quantity, namely AIC can be expressed as follows \cite{Akaike/1974}:
\begin{equation}
    \mathrm{AIC} = \chi^2_{\mathrm{min}}+2d.
\end{equation}
with $d$ being the number of free parameters in a chosen model. To compare our results with the well-known $\Lambda$CDM model, we are going to use the AIC difference between our modified gravity model and the $\Lambda$CDM $\Delta\mathrm{AIC}=|\mathrm{AIC}_{\Lambda\mathrm{CDM}}-\mathrm{AIC}_{\mathrm{MG}}|
$. In that case, if $\Delta\mathrm{AIC}<2$, there is strong evidence in favor of the modified gravity model, while for $4<\Delta\mathrm{AIC}\leq 7$ there is little evidence in favor of the modified gravity model of our consideration. Finally, for the case with $\Delta \mathrm{AIC}>10$, there is practically no evidence in favor of the modified gravity \cite{Liddle/2007}. In addition, BIC is defined through the relation, written down below:
\begin{equation}
    \mathrm{BIC} =\chi^2_{\mathrm{min}}+d\ln N.
\end{equation}
Here, $N$ is the number of data points used for MCMC. For BIC, if $\Delta \mathrm{BIC}<2$, there is no strong evidence against a chosen model that deviates from $\Lambda$CDM, if $2\leq \Delta \mathrm{BIC}<6$ there is evidence against the modified gravity model and finally for $\Delta\mathrm{BIC}>6$, there is strong evidence against the modified gravity model. We therefore store the $\chi^2_{\mathrm{min}}$/AIC/BIC data for the modified gravity model of our consideration in the Table \ref{t1e}. We see that $\Delta\rm AIC=1.25$ and $\Delta\rm BIC=3.01$ so our model can precisely mimic $\Lambda$CDM one.

\begin{table}[H]
	\renewcommand\arraystretch{1.1}
	\caption{Best-fit values of model parameters and statistical analysis	}
	\begin{center}
	\begin{tabular} { l |c| c |c }
		\hline
		\hline
		Datasets    & $H_{0}$      & $\Omega_{m0}$     & $\beta$     \\ \hline
		Hubble (OHD)  & $66.9\pm3.3$   & $0.320^{+0.055}_{-0.070}$ &$4.3\pm1.9$  \\
		OHD+SNeIa & $68.9\pm1.7$   & $0.290^{+0.028}_{-0.020}$ & $5.3^{+1.8}_{-1.0}$  \\
		OHD+SNeIa+BAO &  $68.9\pm1.6$ & $0.292\pm0.016$   & $5.6\pm 1.25$   \\ \hline
	$\rm Models$ & $\chi_{\rm min}^2$ & $\rm AIC$ & $\rm BIC$ \\
$\Lambda$CDM & 58.700 & 67.248 & 76.127     \\ 
$f(Q)$   & 57.616 &  68.499 &  79.137  \\
	    \hline
		\hline
	\end{tabular}
	\end{center}
	\label{t1e}
\end{table}

\section{Validity of Cosmological Constraints}\label{sec3.4}

\begin{figure}[H]
\begin{center}
$
\begin{array}{c@{\hspace{.1in}}c}
\includegraphics[width=3.1 in, height=2.4 in]{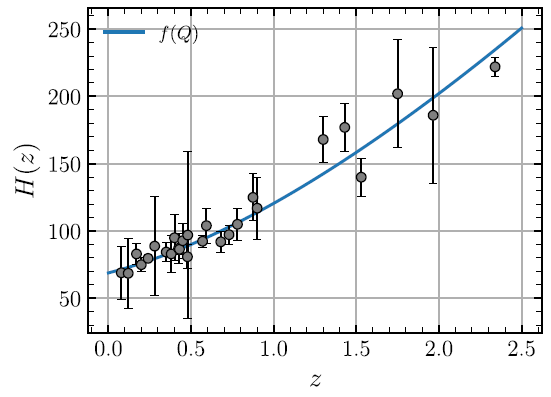} & 
\includegraphics[width=3.1 in, height=2.5 in]{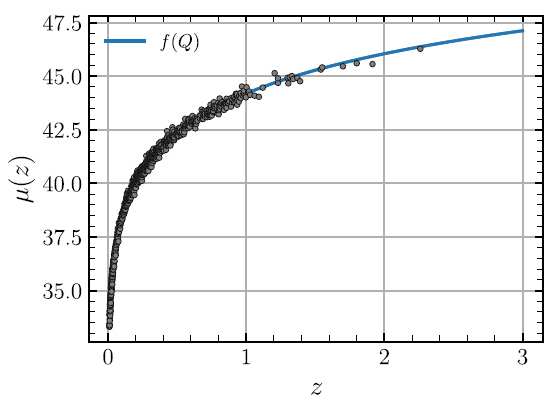}\\
\end{array}
$
\caption{Evolution of the Hubble parameter and distance modulus for exponential $f(Q)$ gravity with the best fit values from MCMC.}
\label{f31}
\end{center}
\end{figure}

\begin{figure}[H]
\begin{center}
$
\begin{array}{c@{\hspace{.1in}}c}
\includegraphics[width=3.1 in, height=2.4 in]{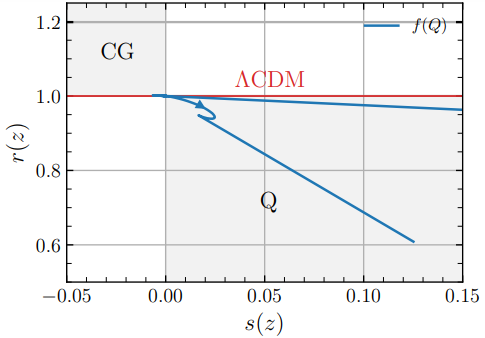} & 
\includegraphics[width=3.1 in, height=2.5 in]{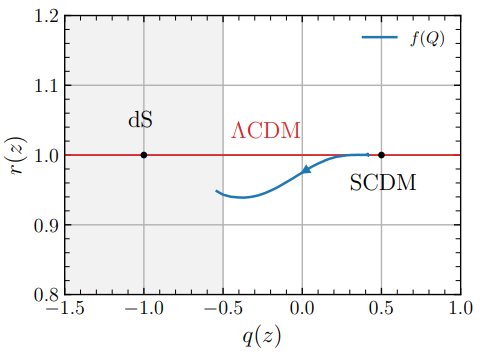}\\
\end{array}
$
\caption{Evolution of statefinder pairs for exponential $f(Q)$ gravity.}
\label{f33}

\end{center}
\end{figure}

\begin{figure}[H]
\centering
\includegraphics[width=9 cm]{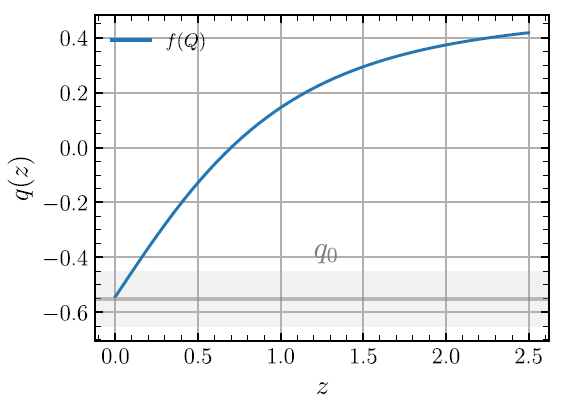}
\caption{Evolution of the deceleration parameter versus redshift $z$ for exponential $f(Q)$ gravity.}
\label{f34}
\end{figure}

Consequently, we plot both statefinder parameter phase portraits, deceleration parameter, and additionally H(z) as probes of model validity in the cosmological sense in figures \ref{f31} and \ref{f33}. Statefinder diagnostics and $q(z)$ are performed only for the joint dataset, since other datasets show similar behavior. Remarkably, a transition from the deceleration to an acceleration phase is seen in figure \ref{f34}. A valid interval for $q_{0}$ is marked as a gray area. As stated already, one may check the Universe evolutionary scenario using statefinder pairs $\{r,s\}$ and $\{r,q\}$. From the $r-s$ plane of our model, one could observe that the early Universe was filled with quintessence, then passed the $\Lambda$CDM phase, and is currently reverting towards the quintessence scenario. On the other hand, in the $r-q$ plane, it is evident that the Universe once passed through the $\Lambda$CDM phase. However, now that our spacetime is generally filled with quintessential fluid, it is expected that the future universe will eventually turn to the de-Sitter state (when the $\Lambda$ term will fully dominate). The point on quintessential fluid also coincides with MCMC observational constraints.

\begin{figure}[H]
\begin{center}
$
\begin{array}{c@{\hspace{.1in}}c}
\includegraphics[width=3.1 in, height=2.4 in]{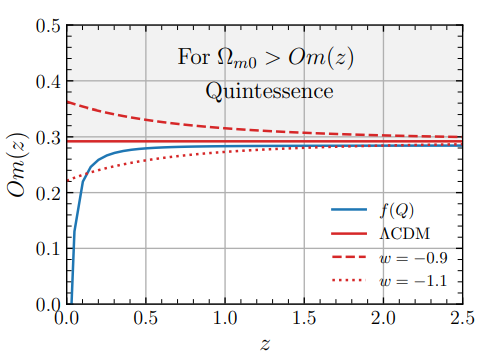} & 
\includegraphics[width=3.1 in, height=2.3 in]{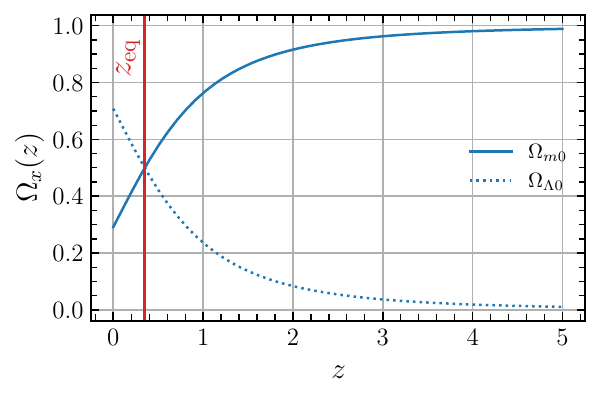}\\
\end{array}
$
\caption{Evolution of $Om(z)$ function and dimensionless mass density for matter and effective dark energy for exponential $f(Q)$ gravity, $\Lambda$CDM, and $\omega$ varying $\Lambda$CDM cosmologies.}
\label{f35}
\end{center}
\end{figure}

Consequently, we place the numerical solution of $Om(z)$ function for $f(Q)$ model in figure \ref{f35}. For the sake of comparison, we as well plot $Om(z)$ solutions within the classical $\Lambda$CDM model and within $\omega$ varying $\Lambda$CDM cosmologies. As one could easily notice, for our $f(Q)$ model, $Om(z)$ shows only $Om(z)<\Omega_{m0}$ behavior in the distant past, which could lead to the presence of phantom fluid (for more information on the subject, see \cite{Mostaghel/2016}).

However, at $z\approx2$, our model transits $\Lambda$CDM and has a constantly growing trend. Therefore, in the near past, the quintessential fluid appears, which converges well with the statefinder diagnostics and MCMC. 
Finally, we also analyse both matter and effective dark energy mass densities for our model to confirm its validity. Corresponding results are plotted in figure \ref{f35}. One can easily notice both $\Omega_{m0}\land\Omega_{\Lambda 0}\in[0,1]$ and their sum always converges to unity, an epoch of the equality appears at redshift $z\approx 0.35$, which is very close to the $\Lambda$CDM estimate.

The numerical fitting findings are utilized as input parameters for simulation purposes, enabling an investigation into the effect of the interaction between the dark sector and the structure formation process. This investigation is conducted through the implementation of N-body simulations. The parameters utilized in our computational analysis are presented in Table \ref{t1e}.

\section{N-body Simulations of LSS with Small $\mathrm{L_{\mathrm{Box}}}$}\label{sec3.5}
As we already remarked, the primary purpose of this chapter is to perform N-body simulations of the comoving box that contains DM+baryonic matter and dark energy in exponential $f(Q)$ gravitation and compare our results with the Large Scale Structure of concordance $\Lambda$CDM cosmology. For that aim, we will use the publicly available code \texttt{ME-GADGET}, a modification of the well-known hydrodynamical N-body code \texttt{GADGET2}. It has been modified for generality to perform simulations for practically any cosmological model. The code above is described in the pioneering works of \cite{Rui/2019,Zhang/2019}, whereas the tests are provided in \cite{Zhang/2018a}. This code as an input needs tables with Hubble flow $H/H_0$ and the deviation of effective gravitational constant from the Newtonian one $G_{\mathrm{eff}}/G_N$ (in some models of modified gravity, namely screened ones, such deviation exists only up to some scale $k_{\mathrm{screen}}$ because of the so-called fifth force). One can find the effective gravitational constant exact form in \cite{Jimenez/2020} as
\begin{equation}
    G_{\mathrm{eff}}=\frac{G_N}{f_Q}.
\end{equation}
The equation above is being numerically solved, assuming appropriate best-fit values for the free parameters of our model. As one could easily notice, at the very early time (high-$z$ epochs), $f(Q)$ gravity has Newtonian-like gravitational constant and then, at approximately $a\approx0.1$, $G_{\mathrm{eff}}$ is being separated from $G_{\mathrm{N}}$ for our model.
Since we have already defined needed inputs for \texttt{ME-GADGET} code, we could proceed further to fine-tuning our simulation setup.

\subsection{Simulation Setup}
One needs to define various parameters to produce the simulations and initial conditions based on the second-order Lagrangian Perturbation Theory (namely, 2LPT). We want to obtain the mid-resolution simulations, and henceforth, particle number is $N=512^3$ and mesh size is respectively $N_{\mathrm{mesh}}=2\times512^3$. The simulation box has periodic vacuum boundary conditions and sides with length $10\,\mathrm{Mpc/h}$. Initial conditions were produced with the \texttt{Simp2LPTic} code (see GitHub repository \href{https://github.com/liambx/Simp2LPTic}{https://github.com/liambx/Simp2LPTic}), and glass files (pre-initial conditions) were generated with the use of \texttt{ccvt-preic} (check \href{https://github.com/liaoshong/ccvt-preic}{https://github.com/liaoshong/ccvt-preic}). The capacity constrained Voronoi tessellation (CCTV) method, which is an alternative method to produce a uniform and isotropic particle distribution to generate pre-initial conditions. More details are explained in the appendix.

\begin{figure}[H]
    \centering
    \includegraphics[width=\textwidth]{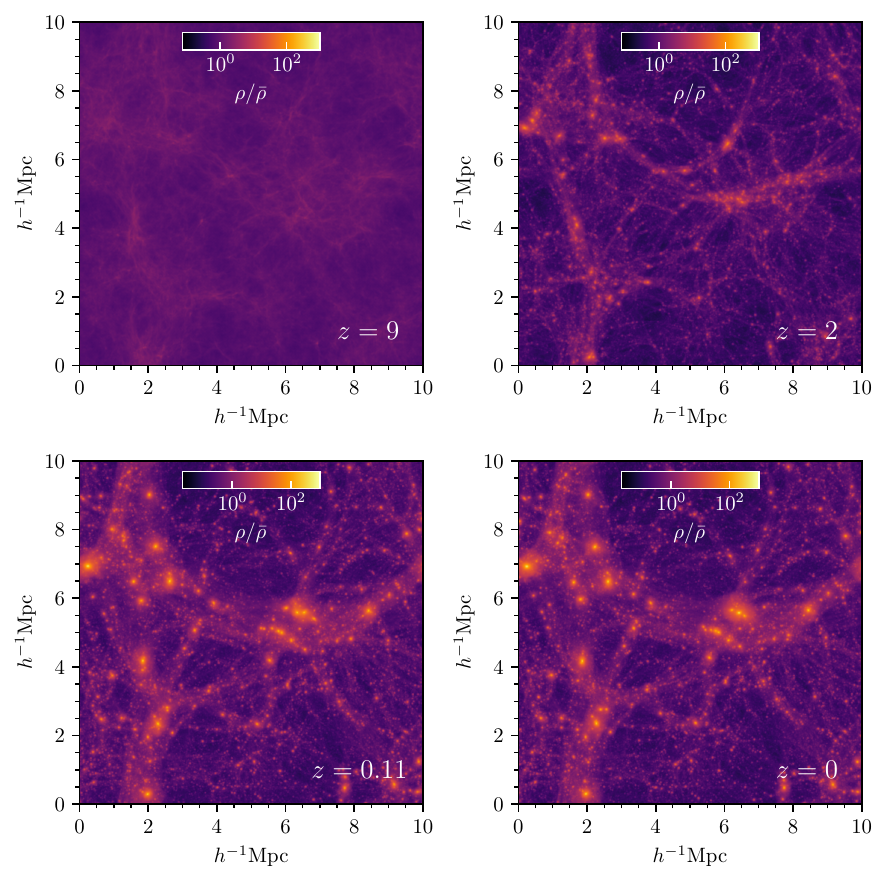}
    \caption{N-body simulations snapshot of overdensity for $f(Q)$ gravity with best fit MCMC values on different redshifts.}
    \label{f36}
\end{figure}

We assumed that the glass tile fraction is unitary, since the final configuration can be scaled to any box size. Moreover, cosmological parameters were borrowed from our MCMC constraints, discussed earlier: $h=H_0/100=0.689\pm 0.016$ (so-called ``little-h"), $\Omega_{m0}=0.292 \pm 0.016$, leading to $\Omega_{\Lambda0}=0.708$, if one will not take into account radiation and massive neutrino species. On the other hand, baryon mass density equals $\Omega_b=0.0493$ (the relation between total matter density and baryon mass density decides how many gas particles are present in the simulation). Moreover, matter power spectrum amplitude at $k=8\mathrm{Mpc/h}$ is assumed to be $\sigma_8=0.811 \pm 0.006$. The initial power spectrum is linear, constructed from the Eisenstein \& Hu transfer function \cite{Eisenstein/1998} (power spectrum were constructed using code \texttt{CAMB}, see \cite{Lewis/2011}). Initial conditions have been generated at the redshift $z=10$, and we use the spectrum index of scalar perturbations as $n_s=0.9649\pm 0.0042$ obtained from Planck \cite{PlanckInf/2020}.

\begin{figure}[H]
    \centering
    \includegraphics[width=\textwidth]{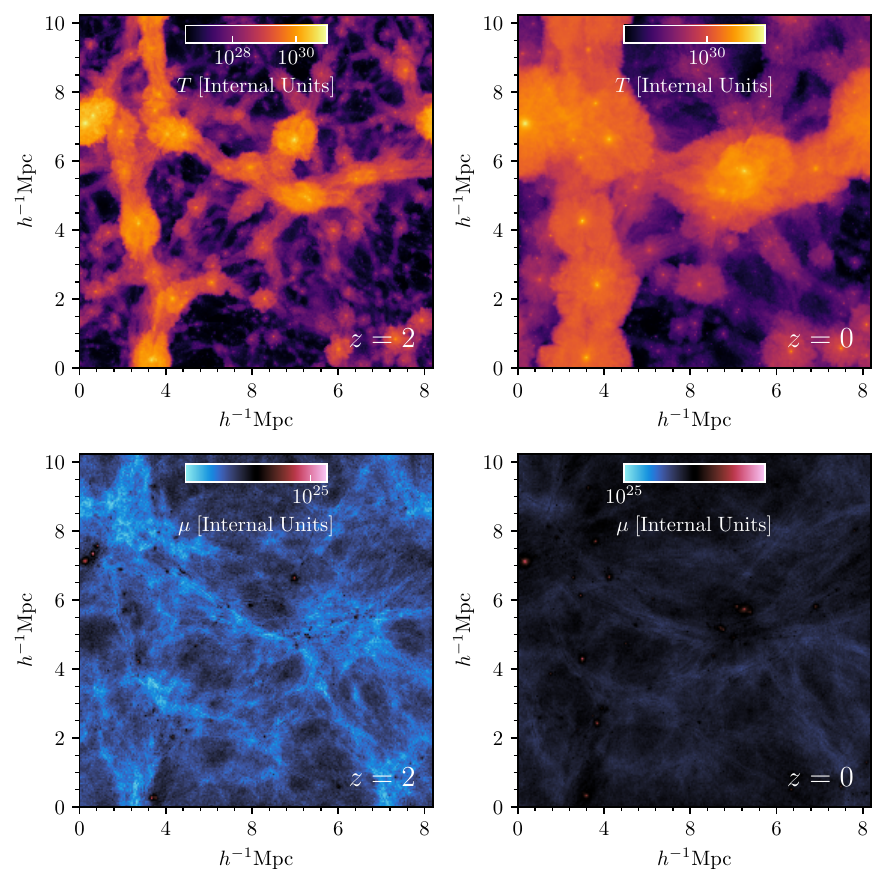}
    \caption{SPH simulation snapshots of $f(Q)$ gravity for gas temperature $T$ and mean molecular weight $\mu$.}
    \label{f37}
\end{figure}

\subsection{Results}
The current subsection will discuss the main results, obtained from the N-body simulations of the Large Scale Structure of the Universe. Firstly, we demonstrate the spatial slices of CDM overdensity $\delta_{\mathrm{CDM}}=\rho_{\mathrm{CDM}}/\overline{\rho}_{\mathrm{CDM}}$ for our $f(Q)$ gravity model with different values of redshift $z$ in the figure \ref{f36}. In addition to the overdensity measurements, we also show the temperature of gas $T$ that arises from Smoothed Particle Hydrodynamics (SPH) and mean molecular weight $\mu=m_{\mathrm{HI}}/\overline{m}$, which defines the relation between mean particle mass and neutral hydrogen particle mass on the figure \ref{f37}, respectively. As one can easily notice, the DM walls are represented by the smaller value of mean molecular weight. Besides, temperature maps show the well-known hot ``bubbles" within the Inter-Galactic Medium (IGM) that are formed due to impinging galactic winds.

Now we investigate the matter power spectrum for our model. For comparison, we are going to use the $\Lambda$CDM cosmology power spectrum, generated with the use of \texttt{CAMB} code\footnote[11]{Documentation for this code is stored in \href{https://camb.readthedocs.io/en/latest/}{camb.readthedocs.io}} \cite{Lewis/2011,Lewis/2002,Lewis/2000,Howlett/2012}. In order to extract $P(k)$ for some value of redshift within our N-body framework, we used Python-based code \texttt{Pylians3}\footnote[12]{For installation procedure and full documentation, refer to the \href{https://pylians3.readthedocs.io/en/master/}{pylians3.readthedocs.io}} \cite{Pylians}. 

\begin{figure}
    \centering
    \includegraphics[width=10cm]{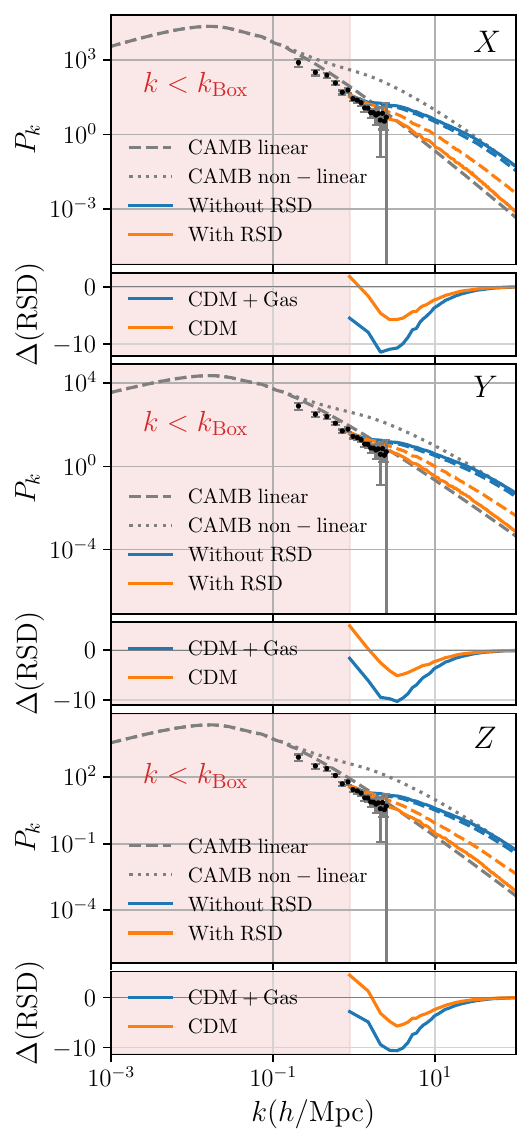}
 \caption{Matter power spectrum with/without RSDs for $f(Q)$ gravity    vs. \texttt{CAMB} linear/non-linear $P(k)$ for $\Lambda$CDM. Dashed N-body $P(k)$ represents the CDM-only power spectrum, while the solid line represent CDM+Gas $P(k)$. Error bars represent Ly $\alpha$ forest observations on high $z$.}
  \label{f38}
\end{figure}

We consequently compare the matter power spectrum on the figure \ref{f38} with/without Redshift-Space Distortions (RSDs) directed along both $X$, $Y$ and $Z$ axes. As we noticed during numerical analysis, up to some $k$ near $k_{\mathrm{Box}}$ limit for our simulation, $P(k)$ spectrum in Fourier space does reconstruct non-linear matter power spectrum, given by \texttt{CAMB}, while Redshift-Space Distorted (RSD) one behaves like the linear matter power spectrum, as expected. 
Also, it is worth noticing that the difference between RSD and regular matter power spectrum is bigger for the CDM+Gas case. Finally, the effect of RSDs in our simulations is almost isotropic, so that $\Delta(\mathrm{RSD})$ differs only by a few percent with the change of RSD direction axis.

\subsection{Halo Mass Function}
\begin{figure}[H]
    \centering
    \includegraphics[width=9 cm]{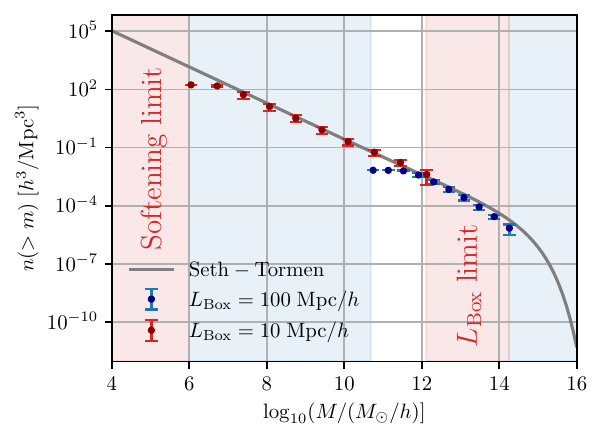}
    \caption{Halo mass function for $f(Q)$ gravitation with $L_\mathrm{Box}=10\mathrm{\;Mpc}/h$ and theoretical prediction for HMF by Seth-Tormen}
    \label{f340}
\end{figure}

Now, we employ the halo mass function (HMF) to demonstrate the relative frequency of dark matter halos across a range of mass values that is, it defines the number of halos at a certain mass. It is a good measure of the structure formation. There is an accepted notion that galaxies are situated within dark matter halos, and it is valuable to analyze halo statistics for establishing connections between simulations and observations. First, we built the halo catalog for all of our snapshots with the use of halo/subhalo structure finder, namely \texttt{ROCKSTAR} (for more information on the subject, refer to the documentation paper \cite{Behroozi/2013}). Consequently, we built the binned halo mass function, which is based on the values of $M_{\mathrm{200c}}$ (the mass of enclosed halo volume with an energy density 200 times bigger than the critical density of the Universe $\rho_{\mathrm{cr}}$). We plot the respective results in figure \ref{f340} with the added Seth-Tormen theoretical prediction for halo mass function, based on \textit{Planck2018} fiducial cosmology and \texttt{CAMB} power spectrum at the $z=0$. Seth-Tormen HMF was computed using the python package \texttt{pynbody} \cite{Potnzen/2013}.

From the figure \ref{f340} shown, one could easily notice that, in general, our prediction for halo mass function from the modified gravity N-body simulation shows values of $n$ that approximately coincide with the ones that are theoretically predicted by the Seth-Tormen HMF.

\section{LSS with Large $L_{\mathrm{Box}}$: Comparison}\label{sec3.6}
\begin{figure}[H]
    \centering
    \includegraphics[width= 15cm]{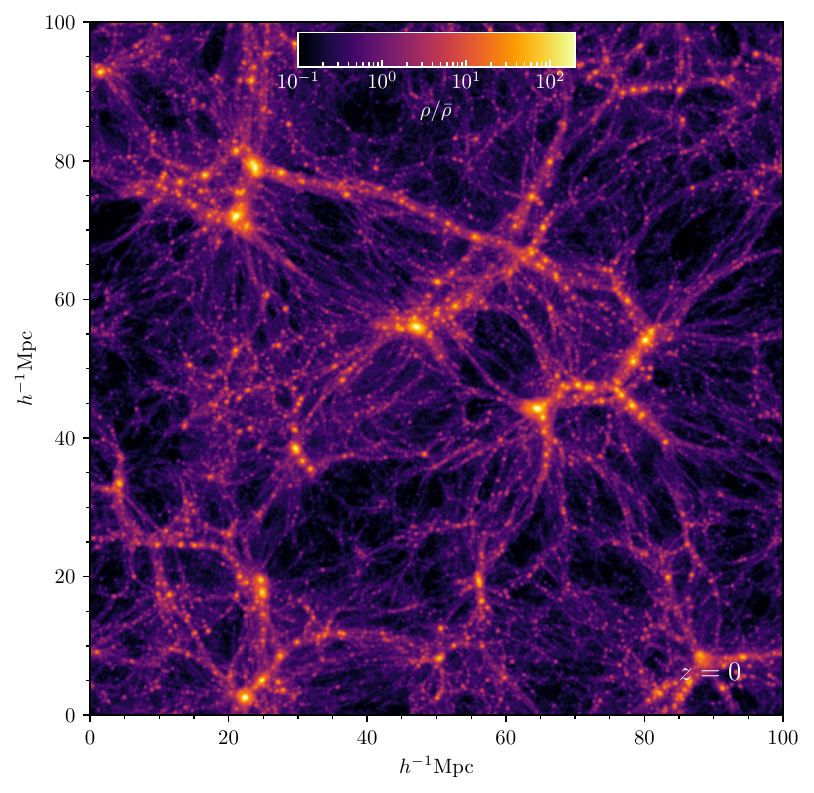}
    \caption{Present day snapshot of CDM over density for $L_{\mathrm{box}}=100h^{-1}$Mpc run.}  
    \label{f341}
\end{figure}

We are now going to perform an analysis of N-body simulation for a bigger simulation box size, namely with $L_{\mathrm{box}}=100h^{-1}$Mpc. In that case, we only differ in force resolution ($\epsilon=3.9$\,kpc), while other cosmological parameters are assumed to be the same. At first, we, as usual, plot the CDM overdensity field for vanishing redshift in figure \ref{f341}.

\begin{figure}
\centering
\includegraphics[width=10cm]{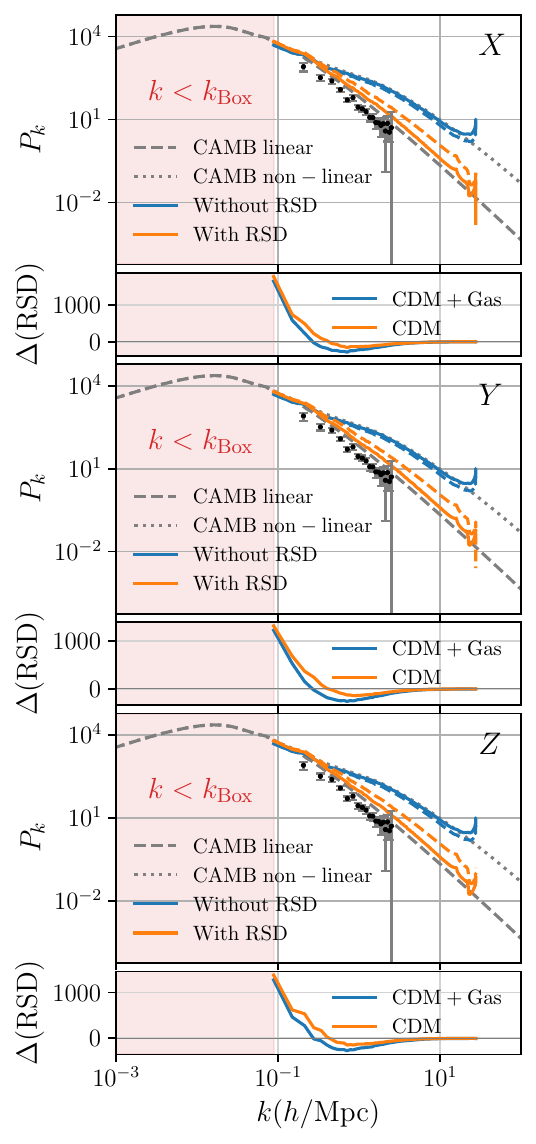}
\caption{Matter power spectrum with/without RSDs for $f(Q)$ gravity vs. \texttt{CAMB} linear/non-linear $P(k)$ for $\Lambda$CDM. Dashed N-body $P(k)$ represents the CDM-only power spectrum, while the solid line represents CDM+Gas $P(k)$. Error bars represent Ly $\alpha$ forest observations on high $z$. For this case, we have assumed a large simulation box size of 100$h^{-1}$Mpc.} 
\label{f342}
\end{figure}

In addition, we also plot the matter power spectrum for CDM, CDM+baryons in figure \ref{f342}. As an obvious consequence of a larger box size, one can notice that maximum wavenumber $k$ grew to $k_{\mathrm{max}}\approx20\;h/$Mpc. Even at such big scales, our matter power spectrum, derived from the corresponding N-body simulation, converges with the theoretical prediction from \texttt{CAMB} code with up to sub-percent accuracy. As we noticed previously for the small simulation box, the axis of redshift-space distortions had a very small impact on the matter power spectrum. This statement also holds for large $L_{\mathrm{box}}$.

In the previous section, we discussed the halo mass function for the case with a smaller simulation box. Now we can discuss the same matter for the larger $L_{\rm Box}$. As it appears, HMF extracted from the simulation replicates the Seth-Tormen HMF almost perfectly up to $M\approx 10^{14}M_{\odot}$ (see figure \ref{f340}). However, at bigger halo masses, our simulation HMF slightly differs from the theoretical prediction, usually observed in N-body simulations.

\subsection{Reduced Bispectrum from 3PCF}
\begin{figure}[H]
\centering
\includegraphics[width=9 cm]{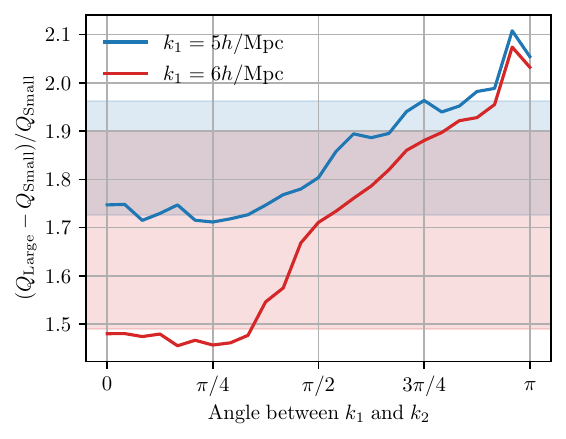}
\caption{Relation of reduced matter power spectrum $Q(k_1,k_2,k_3)$ for both small and large simulation volumes with $1\sigma$ deviation, varying $k_1=2k_2$.}    
\label{f343}
\end{figure}

Finally, we are going to introduce the so-called reduced bispectrum, which is derived from the regular bispectrum and matter power spectrum via the following relation, written below
\begin{equation}
    Q = \frac{B}{P_1P_2+P_2P_3+P_1P_3},
\end{equation}
where $P_i=P_m(k_i)$. We plot the relation between the bispectrum of large and small cosmological volumes in figure \ref{f343}. It is easy to notice that for smaller wavenumber ($k_1=5$), relation between $Q(k_1,k_2,k_3)$ for both the cases has a mean value $\approx1.9$ for all bins of angle $\theta$ (where its maximum value is $\theta=\pi$, which is the angle between two sides of triangle $k_1$ and $k_2$). In contrast, when considering a relatively big $k_1$ (in our case, it is $k_1=6h/\rm Mpc$), the deviation between the reduced bispectrum of a large $L_{\mathrm{box}}$
and that of a small $L_{\mathrm{box}}$ is small. This can be attributed to the fact that the range of wavenumbers is shifted towards higher values.
Conversely, in the first scenario, where $k_1=5h/\rm Mpc$, the wave numbers were at the limit of the box size for the smaller simulation, resulting in distorted outcomes and causing the deviation to grow.

\subsection{2PCF for $f(Q)$ gravitation}
\begin{figure}[]
    \centering
    \includegraphics[width=9 cm]{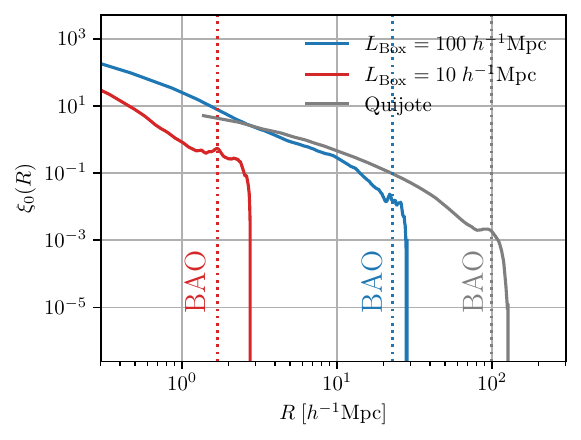}
    \caption{Monopole redshift-space distorted two point correlation function with $L_{\mathrm{Box}}=10/100h^{-1}$Mpc for $f(Q)$ modified gravity. In relation we plot the Quijote simulation correlation function for the Planck $\Lambda$CDM cosmology with a Gpc-wide box. Also, for each case we display the scale, at which the BAO bump occurs.}
\label{f39}
\end{figure}

In addition to the matter power spectrum/reduced bispectrum and halo mass function, we derive the two-point correlation function (further - just 2PCF) in a real space for CDM halos. Generally, 2PCF is defined as follows
\begin{equation}
    \xi(|\mathbf{x}_1-\mathbf{x}_2|) = \langle \delta_m(\mathbf{x}_1)\delta_m(\mathbf{x}_2)\rangle,
\end{equation}
where $\mathbf{x}_i$ is three-dimensional position of an $i$-th CDM halo and $\delta_m$ is CDM overdensity parameter. We show the monopole redshift-space distorted two-point correlation functions for both large and small simulations boxes on the figure \ref{f39}, where we added Quijote simulations \cite{Francisco/2020} 2PCF, that admits \textit{Planck} $\Lambda$CDM cosmology. We additionally marked BAO bumps for each case with color-coded dotted lines for completeness. It is obvious, that in the case of a small simulation box size, the permitted range of $R$ is very small (up to $R\approx 2\times10^0h^{-1}$Mpc) and because of the small box size, the correlation function is undersampled and does not coincide with the Quijote one. On the other hand, for $L_{\rm Box}=100h^{-1}$Mpc simulation, the correlation function corresponds to the Quijote one with sub-percent accuracy for the range $R\in[2\times10^0,10^1]$. Now, we can proceed to the latest topic of our consideration, namely two-dimensional power spectra.

\subsection{2D matter power spectrum}
\begin{figure}[]
    \centering
    \includegraphics[width=\textwidth]{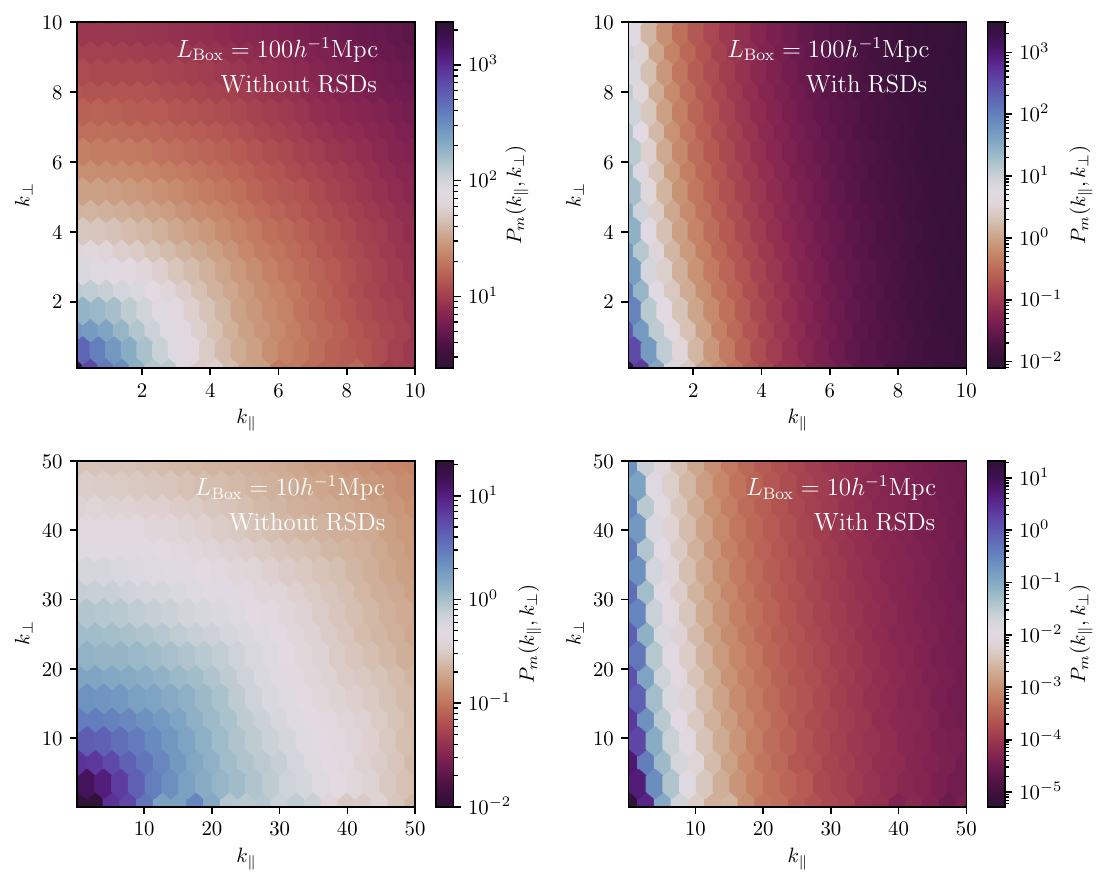}
    \caption{Two-dimensional matter power spectrum for small and large simulations within $f(Q)$ gravitation with/without RSDs.}
    \label{fig:2D}
\end{figure}

We plot the two-dimensional matter power spectrum for small and large boxes in figure \ref{fig:2D} with/without redshift-space distortions. As one can easily notice, on the plots with RSDs, the so-called ``Finger of God" effect is observed\footnote[13]{At smaller scales, there is a significant scattering in the velocity of galaxies. This dispersion, in addition to the cosmological redshift, contributes to a broader distribution of redshift values. Consequently, the elongation of the galaxy distribution along the line of sight occurs, leading to a weakened correlation between galaxies. This phenomenon is commonly referred to as the Finger of God effect.}, which arises because of the large scatter of galaxies' recessional velocities at the small scales. Also, it is worth informing that 2D matter power spectra for both box sizes are similar. Now, since we discussed all of the topics for both simulation volumes within the modified theory of gravitation, we can proceed to the concluding remarks on the key findings within our study.

\section{Conclusions}\label{sec3.7}
One can describe gravity using several geometric bases. The STEGR, which attributes gravity to the non-metricity tensor, has recently drawn much attention. A fascinating method for studying modified gravity is $f(Q)$ gravity, an extension of STGR. This study examined large-scale structure formation observables using N-body simulations of $f(Q)$ gravitation for the first time to assess the theory's validity in the cosmological context. Simulations were run with the use of \texttt{ME-GADGET} code, modification of the widely known \texttt{GADGET-2} code for two simulation boxes, namely $L_{\rm Box} = 10h^{-1}$Mpc and $L_{\rm Box} = 100h^{-1}$Mpc to decide on the optimal box size and compare the results for both simulation volumes.

We first performed Markov Chain Monte Carlo (MCMC) analysis for our exponential $f(Q)$ model to obtain best-fit values of free parameters in section \ref{sec:3.3}. To test the fits provided by MCMC, we obtained theoretical predictions for the dimensionless mass densities $\Omega_{m0}$ and $\Omega_{\Lambda0}$, the Hubble parameter H(z), deceleration parameter $q(z)$ and statefinder pair $\{r,s\}$, $Om(z)$ parameter, placing graphical results on the figures \ref{f33}, \ref{f34}, and \ref{f35}, respectively. As we noticed, the Hubble parameter respected low redshift observations, and the deceleration parameter provided correct values of $q_0$ and transitional redshift within the constrained range. Moreover, statefinder diagnostics predict that the Universe was initially in the Quintessence phase, passing the $\Lambda$CDM state and returned to Quintessence again. Finally, $Om(z)$ demonstrated that at the high-$z$ range, our Universe was filled with phantom fluid, passed $\Lambda$CDM EoS at $z\approx2$ and now again has a phantom equation of state. After theoretical predictions, we started working on the N-body simulations whose primary findings corresponding to the quantities of interest are as follows:
\begin{itemize}
    \item Three-dimensional matter power spectrum monopole $P_k$: This was the first probe of a large-scale structure we studied in the present work. We plotted non-linear matter power spectra (with/without RSDs) for both small and large simulation volumes on the figures \ref{f38} and \ref{f342}, respectively, where we plotted CAMB linear/non-linear $\Lambda$CDM power spectra and observational data from Ly-$\alpha$ forest for the sake of comparison. One can notice that within the permitted range of wavenumber $k$ (limited by mean inter-particle separation and simulation box length), non-linear matter power spectra from small/large N-body simulations coincide with the CAMB one. However, for $L_{\rm Box}=10h^{-1}$Mpc, non-linear $P_k$ coincide with linear CAMB prediction too early because of the small box size.
    \item Halo Mass Function: the second significant quantity that can solely conclude the validity of a simulation. We place Seth-Tormen's theoretical HMF and the ones extracted from our N-body simulations in figure \ref{f340}. As we found, our small box size cannot provide sufficient halo masses and reproduce viable halo mass function at all mass ranges up to the resolution limit, while large simulation follows Seth-Tormen prediction very precisely within the large span of halo masses $\log_{10}M/M_{\odot}\in[10,14]$, but gets slightly smaller than the theoretical prediction for higher masses.
    \item Two-Point Correlation Function monopole $\xi_0(r)$: we also investigated the redshift-space distorted correlation function monopoles in figure \ref{f39}, where Quijote simulations correlation function is plotted to compare our results to fiducial $\Lambda$CDM cosmology. As remarked during numerical analysis, small box simulation fails to predict correct CDM halo correlations. On the other hand, in the range, $R\in[2\times10^0,10^1]h^{-1}$Mpc large box simulation precisely reconstructs Quijote data.
    \item Reduced bispectrum $Q(k_1,k_2,k_3)$: for the reduced bispectra case, we plotted the relation $(Q_{\rm Large}-Q_{\rm Small})/Q_{\rm Small}$ with different $k_1$ values (which acts as a triangle side length) in figure \ref{f343}. We observed that this relation is generally close to $\approx 1.5$ across all bins of the angle between $k_1$ and $k_2$ (namely $\theta$) if one assumes the value of $k_1$ that is not on the resolution limit for both cases (while it is worth to notice that we only adopted the case, where $k_2=2k_1$).
    \item Two-dimensional matter power spectrum $P_m(k_\parallel,k_\perp)$: This is the last quantity extracted from our N-body simulations. We plotted 2D power spectra for both simulations in figure \ref{fig:2D} with/without redshift-space distortions. From the plots, we noticed the so-called ``Finger of God" effect in the RSD case because of the elongated positions of CDM halos.
\end{itemize}
In conclusion, considering all the above points, the small simulation volume experiment failed to recreate the matter power spectrum and correlation function correctly. However, the second one, namely more extensive N-body simulation provided both viable 3D/2D matter power spectrum, correlation function, and halo mass functions and, therefore, we can consider exponential $f(Q)$ model to be a viable substitution of $\Lambda$CDM cosmology, since it not only satisfies many large scale structure constraints mentioned above but also provide correct distance modulus up to high redshift values, where $\Lambda$CDM fails.

In future works, it will be interesting to investigate this modified gravity model using hybrid N-body and SPH simulations that incorporate supernova/AGN feedback, star and galaxy formation, jets, etc. using code \texttt{GIZMO} that allows the use of tabulated Hubble parameter and the effective gravitational constant. It will, however, require a lot more computational resources (on the scale of millions of CPU hours).




\chapter{Conclusions and Future Perspectives} 

\label{Chapter6} 

\lhead{Chapter 6. \emph{Conclusions}} 


Let us take a moment to summarise the findings brought to light by this thesis. The primary objective of this thesis is to examine the development of dark energy models to assess the compatibility of our cosmological theories with observational evidence, intending further to understand the phenomenon of accelerated expansion in the Universe. We need access to a diverse range of independent methods to evaluate and validate our theories and models. 

In chapter \ref{Chapter1}, we discussed the mathematical notations, fundamental elements, applications in cosmology, fundamental theories of gravity, and cosmological observations. Aside from this, it is well-known that the fundamental theory of gravity, like general relativity, fails to address certain issues, such as fine-tuning and the flatness problem. Therefore, its modifications and generalizations are more effective in addressing these concerns, and the chapter concludes with a summary of the modified gravity theories. 

Let us discuss the results obtained in this dissertation. In chapter \ref{Chapter2}, we developed a dark energy model using the effective equation of state in $f(R,\mathcal{T})$ gravity. Since the precise form of viscosity is unknown, we considered a parametrized bulk viscosity consisting of three linear terms. The first term is a constant, while the second and third are proportional to the Hubble parameter and $\frac{\ddot{a}}{\dot{a}}$, respectively. Using the observational datasets: 57 points of Hubble data and 580 points of Union 2.1 supernovae, we examined the model's viability and discussed the evolution of various cosmological parameters. 
We observed a transition in the deceleration parameter from positive to negative at $z= 0.84$ and $z=0.62$ with $q=-0.78$ and $q=-0.68$ for $k_{4}=-0.43$ and $k_{4} = 0.49$, respectively. It is also seen that as $z \rightarrow -1$, $\omega \rightarrow -1$, Universe approaches $\Lambda$CDM in the near future. Additionally, we found that the incorporation of bulk viscosity supported a decelerated expansion in the past and an accelerated expansion in the present. We presented the validation tests for our model using statefinder diagnostics and energy conditions. We discovered that the current paradigm satisfies the NEC and DEC but violates the SEC when describing cosmic acceleration. Moreover, statefinder diagnostics allowed us to observe our model's deviation from the $\Lambda$CDM. According to the $r-s$ plane, the current model lies in the Chaplygin gas region and resembles the $\Lambda$CDM or may be quintessence in the future. This indicates that the current model is distinct from the $\Lambda$CDM at present. 

Consideration of non-minimal coupling of the non-metricity $Q$ and the trace of the energy-momentum tensor $\mathcal{T}$, i.e., $f(Q,\mathcal{T})$ gravity, is another way to go beyond GR. Chapter \ref{Chapter3} examined the $f(Q,\mathcal{T})$ theory, according to which dark energy is solely geometric. However, there is no doubt that the coupling should have a positive value to reproduce the solar system scale. Employing MCMC techniques, we have addressed $f(Q,\mathcal{T})$ gravity using $f(Q,\mathcal{T}) =  \xi \, Q^{n} + b\, \mathcal{T}$ in a matter-dominated Universe. The 57 Hubble data points and 580 points of Union 2.1 supernovae data points were used to reconstruct the dark sector as a function of redshift. The deceleration parameter appears negative, which is consistent with the current scenario of an accelerating Universe. The statefinder parameters exhibit a quintessence phase and converge to $\Lambda$CDM in the future using the constrained values of the model parameters $n$ and $b$ derived from the datasets. However, the linear model $f(Q,\mathcal{T}) =  \xi \,Q + b\, \mathcal{T}$ resembles the power-law expansion model $a(t) \propto \left(B\,t+c_{1}\right)^{1/B}$, where $B=\frac{24 \pi +3b}{16 \pi + 3b}$. One can notice that the model behaves similar to the standard lore of $a(t) \propto t^{2/3}$ with a constant deceleration of $q = 0.5$. Hence, it reveals that the model parameter $b$ contributes less to the evolution, suggesting that the linear trace does not affect the evolution much.

To study more on the contribution of the linear trace, the parameterized equation of state in $f(Q,\mathcal{T})$ gravity is considered in chapter \ref{Chapter4} which represents an additional theoretical scenario beyond GR. The Pantheon sample was used to constrain the parameter space $\{b,m,n\}$, as the functional form $f$ includes $b$ as a model parameter, while $m$ and $n$ are the parameters of the considered parametrized equation of state. The bounds on the parameter space for our analysis are $b=0.2^{+2.7}_{-2.9}$, $m=0.47^{+0.27}_{-0.21}$, and $n=3.2^{+1.8}_{-2.0}$. Further, we focus on the neighborhood of $b=0$, which seeks to identify any deviation from GR. It is seen that the approach does not result in a statistically significant difference between the marginalized distributions of parameters $m$ and $n$. Lastly, we observed the behavior of the deceleration parameter, which demonstrates a transition from deceleration to acceleration with $q_0=-0.52$. In addition, value of the EoS parameter at $z=0$ is $\omega_0=-0.68^{+0.10}_{-0.11}$, which supports a phase of acceleration. Henceforth, to the cosmic evidence, we must ensure that the $f(Q,\mathcal{T})$ theory is stable to perform the perturbation level difference in approach to $\Lambda$CDM. 

In chapter \ref{Chapter5}, we focused on the extension of the newly proposed geometrical framework of GR, named as the $f(Q)$ gravity. We proposed and investigated the structure formation and viable cosmology for exponential $f(Q)$ gravity. However, we provided cosmological constraints for the model using Hubble data, a reduced Pantheon sample, and the BAO dataset. In order to compare our model with $\Lambda$CDM model results, we conducted AIC and BIC analysis. We obtained that there is a strong evidence in favor of our $f(Q)$ gravity model. We observed that the deceleration parameter provided a consistent value of $q_{0}$ and a  transition within the allowed range. We discovered that the statefinder diagnostics initially predicted quintessence behavior, then the $\Lambda$CDM state, and then quintessence behavior again.

After theoretical predictions, N-body simulations were conducted. We investigated the 3D monopole power-spectrum using small/large N-body simulations, coincides with CAMB one within the permitted range of wavenumber, whereas for $L_{box}= 10\,h^{-1}Mpc$, non-linear $P_{k}$ coincides with linear CAMB prediction too early due to the small box size.
Then, we moved on to the second significant number, the Halo mass function, that alone can determine the viability of simulations. We found that the small box size could not provide enough halo masses to produce a valid halo mass function at all mass ranges up to the resolution limit. In contrast, the large simulation follows the Seth-Tormen prediction precisely within the large span of halo masses $\log_{10}M/M_{\odot}\in[10,14]$ but gets slightly smaller than the theoretical prediction at higher masses. In addition, we examined the redshift-space distorted correlation function monopoles, where the correlation function from the Quijote simulations is plotted in order to compare our results to fiducial cosmology. Observations indicate that small box simulation fails to predict CDM halo correlations accurately. In contrast, $R\in[2\times10^0,10^1]h^{-1}$Mpc large box simulation precisely reconstructs Quijote data in the range. Moreover, 2D power spectra were depicted for both simulations with and without redshift-space distortions. We observed the so-called ``Finger of God" effect in the RSD case due to the elongated CDM halo positions.

This chapter concludes that the small simulation volume experiment did not recreate the matter power spectrum and correlation function. Nevertheless, a realistic 3D/2D matter power spectrum, correlation function, and halo mass function were obtained from the second, more extended N-body simulation. The exponential $f(Q)$ model may therefore be viewed as an alternative to the fiducial $\Lambda$CDM cosmology, as it not only satisfies the above-mentioned large-scale structure constraints but also provides correct distance modulus up to high redshift values, where $\Lambda$CDM fails.

As observed, the theories of gravitational extension and modification, which effectively incorporate advanced observations and attract the cosmological community's interest, do not seem to alter the foundations of GR. Gravity is still classical and is related to the geometry of spacetime. These findings indicate that we are not at an endpoint. In the future, several large-scale surveys, such as the dark energy spectroscopic instrument (DESI), Square kilometer array (SKA), and the large-aperture synoptic survey telescope (LSST), will cover the entire redshift range in which accelerated expansion played a significant role over the next decade. We will use the newly collected observational data to investigate the $H_{0}$ and $\sigma_{8}$ tensions in these modified theories of gravity. In anticipation of having actual data, forecast analysis enhances our understanding of cosmology.







\addtocontents{toc}{\vspace{2em}} 

\backmatter

%
\cleardoublepage
\pagestyle{fancy}
\label{References}
\lhead{\emph{References}}


\begin{thebibliography}{100}


\bibitem{Lobato/2022} R. V. Lobato et al., \textit{Eur. Phys. J. C} \textbf{82}, 540 (2022).

\bibitem{Moffat/2021} J.W. Moffat, \textit{J. Cosmol. Astropart. Phys.} \textbf{02}, 017 (2021). 

\bibitem{Vilhena/2023} S.G. Vilhena et al., \textit{J. Cosmol. Astropart. Phys.} \textbf{04}, 044 (2023). 

\bibitem{Rahaman/2015} F. Rahaman et al.,  \textit{Phys. Lett. B} \textbf{746}, 73-78 (2015). 

\bibitem{Sokoliuk/2022} O. Sokoliuk et al., \textit{Eur. Phys. J. C} \textbf{137}, 1077 (2022). 


\bibitem{Abbott/2016} B. P. Abbott et al., \textit{Phys. Rev. Lett.} \textbf{116}, 061102 (2016). 

\bibitem{Akiyama/2019} K. Akiyama et al., \textit{Astrophys. J. Lett.} \textbf{875}, L6 (2019).  

\bibitem{einstein/1917} A. Einstein, Sitzungsber. Preuss. Akad. Wiss. Berlin (Math.Phys.) \textbf{1917}, 142 (1917).


\bibitem{friedman/1922} A. Friedman, \textit{Zeitschrift für Physik} \textbf{10}, 377 (1922).

\bibitem{Hubble/1929} E. Hubble, \textit{Proc. N. A. S.} \textbf{15}, 168 (1929).

\bibitem{Hubble/1931} E. Hubble, M. L. Humason, \textit{Astrophys. J.} \textbf{74}, 43 (1931). 

\bibitem{lemaitre/1931} A. G. Lemaitre, \textit{Mon. Not. Roy. Astron. Soc.} \textbf{91}, 483 (1931).

\bibitem{penzias/1965} A. A. Penzias, R. W. Wilson, \textit{Astrophys. J.} \textbf{142}, 419 (1965).

\bibitem{riess/1998} A. G. Riess et al., \textit{Astron. J.} \textbf{116}, 1009 (1998).
 
\bibitem{perlmutter/1999} S. Perlmutter et al., \textit{Astrophys. J.} \textbf{517}, 565 (1999).

\bibitem{planck/2015} P. A. R. Ade et al., \textit{Astron. Astrophys.} \textbf{594}, A13 (2015).

\bibitem{Peebles/1965} P. J. E. Peebles, \textit{Astrophys. J.} \textbf{142}, 1317 (1965).

\bibitem{Carr/1994} B. Carr, \textit{Annu. Rev. Astron. Astrophys.} \textbf{32}, 531-590 (1994). 

\bibitem{Bertone/2018} G. Bertone, D. Hooper, \textit{Rev. Mod. Phys.} \textbf{90}, 045002 (2018). 

\bibitem{Valentino/2021} E. Di Valentino et al., \textit{Class. Quantum Grav.} \textbf{38}, 153001 (2021). 

\bibitem{Nojiri/2006} S. Nojiri, S. D. Odintsov, \textit{Phys. Rev. D} \textbf{74}, 086005 (2006).

\bibitem{Amendola/2007} L. Amendola, D. Polarski, S. Tsujikawa, \textit{Phys. Rev. Lett.} \textbf{98}, 131302 (2007).

\bibitem{Sotiriou/2006} T. P. Sotiriou, \textit{Class. Quantum Grav.} \textbf{23}, 5117 (2006).

\bibitem{Sotiriou/2010} T. P. Sotiriou, V. Faraoni, \textit{Rev. Mod. Phys.} \textbf{82}, 451 (2010).

\bibitem{Felice/2010}  A. De Felice, S. Tsujikawa, \textit{Liv. Rev. Relat.} \textbf{13}, 1-161 (2010).
 
\bibitem{Harko/2011} T. Harko et al., \textit{Phys. Rev. D} \textbf{84}, 024020 (2011).

\bibitem{Shabani/2014} H. Shabani, M. Farhoudi, \textit{Phys. Rev. D} \textbf{90}, 044031 (2014).

\bibitem{Zare/2016} R. Zaregonbadi, M. Farhoudi, N. Riazi, \textit{Phys. Rev. D} \textbf{94}, 084052 (2016).

\bibitem{Moraes/2017} P. H. R. S. Moraes, P. K. Sahoo, \textit{Phys. Rev. D} \textbf{96}, 044038 (2017).

\bibitem{Goncalves/2022} T. B. Goncalves, J. L. Rosa, F. S. N. Lobo, \textit{Phys. Rev. D} \textbf{105}, 064019 (2022).

\bibitem{Alvarenga/2013} F. G. Alvarenga et al., \textit{Phys. Rev. D} \textbf{87}, 103526 (2013).

\bibitem{Fisher/2019} S. B. Fisher, E. D. Carlson, \textit{Phys. Rev. D} \textbf{100}, 064059 (2019).


\bibitem{Weyl/1918} H. Weyl, Sitzungsberichter der K\"{o}niglich Preussischen Akademie der Wissenschaften zu Berlin \textbf{465}, (1918).  

\bibitem{Hehl/1976} F. W. Hehl et al., \textit{Rev. Mod. Phys.} \textbf{48}, 393 (1976).

\bibitem{Weitzenbck/1923} R. Weitzenbock, Invariantentheorie (Noordhoff, Groningen, 1923). 

\bibitem{Hayashi/1979} K. Hayashi, T. Shirafuji, \textit{Phys. Rev. D} \textbf{19}, 3524 (1979).

\bibitem{Cai/2016} Yi-Fu Cai et al., \textit{Rep. Prog. Phys.} \textbf{79}, 106901 (2016). 

\bibitem{Krssak/2019} M. Krssak et al., \textit{Class. Quantum Grav.} \textbf{36}, 183001 (2019). 


\bibitem{Capo/2017} S. Capozziello, G. Lambiase, E. N. Saridakis, \textit{Eur. Phys. J. C} \textbf{77}, 576 (2017). 

\bibitem{Ren/2022} X. Ren et al., \textit{Astrophys. J.} \textbf{932}, 131 (2022). 

\bibitem{Bahamonde/2023} S. Bahamonde et al., \textit{Eur. Phys. J. C} \textbf{83}, 193 (2023).

\bibitem{Jimenez/2018} J. B. Jimenez, L. Heisenberg, T. Kovisto, \textit{Phys. Rev. D} \textbf{98}, 044048 (2018).

\bibitem{Nester/1999} J. M. Nester, H-J. Yo, \textit{Chin. J. Phys.} \textbf{37}, 113 (1999).

\bibitem{Harko/2018} T. Harko et al., \textit{Phys. Rev. D} \textbf{98}, 084043 (2018).

\bibitem{Soudi/2019} I. Soudi et al., \textit{Phys. Rev. D} \textbf{100}, 044008 (2019).

\bibitem{Lu/2019} J. Lu, X. Zhao, G. Chee, \textit{Eur. Phys. J. C} \textbf{79}, 530 (2019).

\bibitem{Hohmann/2019} M. Hohmann et al., \textit{Phys. Rev. D} \textbf{99}, 024009 (2019).

\bibitem{Ambrosio/2022} F. D'Ambrosio et al., \textit{Phys. Rev. D} \textbf{105}, 024042 (2022).

\bibitem{Wang/2022} W. Wang, H. Chen, T. Katsuragawa, \textit{Phys. Rev. D} \textbf{105}, 024060 (2022).

\bibitem{Capo/2022a} S. Capozziello, R. D'Agostino, \textit{Phys. Lett. B} \textbf{832}, 137229 (2022). 

\bibitem{Hu/2022} K. Hu, T. Katsuragawa, T. Qiu, \textit{Phys. Rev. D} \textbf{106}, 044025 (2022). 

\bibitem{Khyllep/2023} W. Khyllep et al., \textit{Phys. Rev. D} \textbf{107}, 044022 (2023).

\bibitem{Jimenez/2020} J. B. Jimenez et al., \textit{Phys. Rev. D} \textbf{101}, 103507 (2020).


\bibitem{Xu/2019} Y. Xu et al., \textit{Eur. Phys. J. C} \textbf{79}, 708 (2019). 

\bibitem{Weinberg/1972} S. Weinberg, Gravitation and cosmology: principles and applications of the general theory of relativity, Wiley, New York, (1972).

\bibitem{Boh/2016} Christian G. B\"{o}hmer, Introduction to general relativity and cosmology, World Scientific, London, (2016).

\bibitem{Carroll/2004} S. Carroll, Spacetime and Geometry, An Introduction to GR, Addison Wesley, Boston, (2004).

\bibitem{Penrose/1965} R. Penrose, \textit{Phys. Rev. Lett.} \textbf{14}, 57-9 (1965). 

\bibitem{Hawking/1966} S. Hawking, \textit{Proc. R. Soc. A} \textbf{294}, 571 (1966). 

\bibitem{Visser/2000} M. Visser, C. Barcelo, \textit{COSMO-99} 98 (2000).

\bibitem{Hawking/1973} S. W. Hawking, G. F. R. Ellis,  The Large Scale Structure of Space-Time (Cambridge: Cambridge University Press) (1973).

\bibitem{Poisson/2004}  J. Ehlers, \textit{Int. J. Modern Phys. D}   \textbf{15}, 1573 (2006); S. Nojiri, S. D. Odintsov, \textit{Int. J. Geom. Methods Mod. Phys.} \textbf{4}, 115 (2007). 

\bibitem{Kontou/2020} E. A. Kontou, Ko Sanders, \textit{Class. Quantum Grav.} \textbf{37}, 193001 (2020).

\bibitem{Capozziello/2014} S. Capozziello, F. S. N. Lobo, J. P. Mimoso, \textit{Phys. Lett. B} \textbf{730}, 280-283 (2014).


\bibitem{Planck/2018} N. Aghanim et al., \textit{Astron. Astrophys.} \textbf{641}, A6 (2020).

\bibitem{Joyce/2015} A. Joyce et al., \textit{Phys. Rept.} \textbf{568}, 1-98 (2015).

\bibitem{Tsujikawa/2013} S. Tsujikawa. Introductory review of cosmic inflation. In 2nd Tah Poe School on Cosmology: Modern Cosmology, 4 (2003).

\bibitem{Velten/2014} H. E. S. Velten, R. F. Vom Marttens, W. Zimdahl, \textit{Eur. Phys. J. C} \textbf{74(11)}, 3160 (2014).

\bibitem{Astesiano/2021} D. Astesiano, M. L. Ruggiero, \textit{Phys. Rev. D} \textbf{106}, 044061 (2021).

\bibitem{Katsuragawa/2017} T. Katsuragawa, S. Matsuzaki,  \textit{Phys. Rev. D} \textbf{95}, 044040 (2017).

\bibitem{Zaregonbadi/2016} R. Zaregonbadi, M. Farhoudi, N. Riazi,  \textit{Phys. Rev. D} \textbf{94}, 084052 (2016).


\bibitem{Valentino/2021a} E. Di Valentino et al., \textit{Astropart. Phys.} \textbf{131}, 102605 (2021).

\bibitem{Valentino/2021b} E. Di Valentino et al., \textit{Astropart. Phys.} \textbf{131},102604 (2021).

\bibitem{Joudaki/2017} S. Joudaki et al., \textit{Mon. Not. Roy. Astron. Soc.} \textbf{471}, 1259 (2017).
 
\bibitem{Linder/2003} E. Linder, \textit{Phys. Rev. Lett.} \textbf{90}, 091301 (2003).

\bibitem{Jassal/2010} H. K. Jassal, J. S. Bagla, T. Padmanabhan, \textit{Mon. Not. R. Astron. Soc.} \textbf{405}, 2639 (2010).

\bibitem{Barboza/2008} E. M. Barboza, J. S. Alcaniz, \textit{Phys. Lett. B} \textbf{666}, 415 (2008).

\bibitem{Wetterich/2004} C. Wetterich, \textit{Phys. Lett. B} \textbf{594}, 17 (2004). 

\bibitem{Gupta/2015} G. Gupta, R. Rangarajan, A. A. Sen, \textit{Phys. Rev. D} \textbf{92}, 123003 (2015). 

\bibitem{Chiba/2013} T. Chiba, A. D. Felice, S. Tsujikawa, \textit{Phys. Rev. D} \textbf{87}, 083505 (2013).

\bibitem{Roy/2014} N. Roy, N. Banerjee, \textit{Eur. Phys. J. Plus} \textbf{129}, 162 (2014).

\bibitem{Pantazis/2016} G. Pantazis, S. Nesseris and L. Perivolaropoulos, \textit{Phys. Rev. D} \textbf{93}, 103503 (2016).

\bibitem{Caldwell/2002} R. R. Caldwell, \textit{Phys. Lett. B} \textbf{545}, 23 (2002).

\bibitem{Cai/2008} Y.-F. Cai, J. Wang, \textit{Class. Quant. Grav.} \textbf{25}, 165014 (2008). 

\bibitem{Feng/2005} B. Feng, X. Wang, X. Zhang, \textit{Phys. Lett. B} \textbf{607}, 35 (2005).

\bibitem{Picon/1999} C. Armendariz-Picon, T. Damour, V. Mukhanov, \textit{Phys. Lett. B} \textbf{458}, 209 (1999).

\bibitem{Picon/2001} C. Armendariz-Picon, V. Mukhanov, P. J. Steinhardt, \textit{Phys. Rev. D} \textbf{63}, 103510 (2001).

\bibitem{Chiba/2000} T. Chiba, T. Okabe, M. Yamaguchi, \textit{Phys. Rev. D} \textbf{62}, 023511 (2000).

\bibitem{Tsujikawa/2010} S. Tsujikawa, Modified gravity models of dark energy, Lectures on Cosmology: Accelerated Expansion of the Universe, G. Wolschin ed., pp. 99-145, Springer Berlin Heidelberg (2010).

\bibitem{Clifton/2012} T. Clifton, P.G. Ferreira, A. Padilla, C. Skordis, \textit{Phys. Rep.} \textbf{513}, 1 (2012).

\bibitem{Carroll/2005} S. M. Carroll et al., \textit{Phys. Rev. D} \textbf{71}, 063513 (2005).

\bibitem{Nojiri/2007} S. Nojiri, S. D. Odintsov, \textit{Phys. Lett. B} \textbf{652}, 343 (2007).

\bibitem{Motohashi/2019} H. Motohashi, A. A. Starobinsky, \textit{J. Cosmol. Astropart. Phys.} \textbf{11}, 025 (2019).

\bibitem{Olmo/2019} G. J. Olmo, D. R. Garcia, A. Wojnar, \textit{Phys. Rev. D} \textbf{100}, 044020 (2019).

\bibitem{Odint/2019} S. D. Odintsov, V. K. Oikonomou, \textit{Phys. Rev. D} \textbf{99}, 064049 (2019).

\bibitem{Odint/2020} S. D. Odintsov, V. K. Oikonomou, \textit{Phys. Rev. D} \textbf{101}, 044009 (2020).

\bibitem{Oikonomou/2021} V. K. Oikonomou, \textit{Phys. Rev. D} \textbf{103}, 044036 (2021).

\bibitem{Sanders/1997} R. H. Sanders, \textit{Astrophys. J.} \textbf{480}, 492 (1997).

\bibitem{Das/2008} S. Das, N. Banerjee, \textit{Phys. Rev. D} \textbf{78}, 043512 (2008).

\bibitem{Riazuelo/2002} A. Riazuelo, J. P. Uzan, \textit{Phys. Rev. D} \textbf{66}, 023525 (2002).

\bibitem{Boisseau/2000} B. Boisseau, G. Esposito-Farese, D. Polarski, A. A. Starobinsky, \textit{Phys. Rev. Lett.} \textbf{85}, 2236 (2000).

\bibitem{Crisostomi/2018} M. Crisostomi, K. Koyama, \textit{Phys. Rev. D} \textbf{97}, 084004 (2018). 

\bibitem{Elizalde/2004} E. Elizalde, S. Nojiri, S. D. Odintsov, \textit{Phys. Rev. D} \textbf{70}, 043539 (2004).

\bibitem{Capo/2015} S. Capozziello, O. Luongo, E. N. Saridakis, \textit{Phys. Rev. D} \textbf{91}, 124037 (2015). 

\bibitem{Cai/2018} Yi-Fu Cai et al., \textit{Rep. Prog. Phys.} \textbf{79}, 106901 (2016).

\bibitem{Nunes/2018} R. C. Nunes, \textit{J. Cosmol. Astropart. Phys.} \textbf{05}, 052 (2018). 

\bibitem{Anagnos/2019} F. K. Anagnostopoulos, S. Basilakos, E. N. Saridakis, \textit{Phys. Rev. D} \textbf{100}, 083517 (2019).

\bibitem{Bahamonde/2020} S. Bahamonde, J. L. Said, M. Zubair, \textit{J. Cosmol. Astropart. Phys.} \textbf{10}, 024 (2020). 


\bibitem{Lazkoz/2019} R. Lazkoz et al., \textit{Phys. Rev. D} \textbf{100}, 104027 (2019).

\bibitem{Barros/2020} B. J. Barros et al., \textit{Phys. Dark Univ.} \textbf{30}, 100616 (2020).

\bibitem{Anag/2021} F. K. Anagnostopoulos, S. Basilakos, E. N. Saridakis, \textit{Phys. Lett. B} \textbf{822}, 136634 (2021).


\bibitem{Silva/2022} J. E. G. Silva et al., \textit{Phys. Rev. D} \textbf{106}, 024033 (2022).


\bibitem{Frusciante/2021} N. Frusciante, \textit{Phys. Rev. D} \textbf{103}, 144021 (2021).



\bibitem{Myrzakulov/2012} R. Myrzakulov, \textit{Eur. Phys. J. C} \textbf{72}, 2203 (2012). 

\bibitem{Harko/2010a} T. Harko, F. S. N. Lobo, \textit{Eur. Phys. J. C} \textbf{70}, 373–379 (2010).  


\bibitem{Xu/2020} Y. Xu et al., \textit{Eur. Phys. J. C} \textbf{80}, 449 (2020).  


\bibitem{Capo/2008a} S. Capozziello, C. Corda, M. F. De Laurentis, \textit{Phys. Rev. D} \textbf{669}, 255--259 (2008).


\bibitem{Stachowski/2017} A. Stachowski, M. Szydowski, A. Borowiec, \textit{Eur. Phys. J. C} \textbf{77}, 406 (2017). 


\bibitem{Staro/1980} A. A. Starobinsky, \textit{Phys. Lett. B} \textbf{91}, 99-102 (1980).

\bibitem{Nojiri/2003a} S. Nojiri, S. D. Odintsov, \textit{Phys. Rev. D} \textbf{68}, 123512 (2003). 

\bibitem{Dolgov/2003} A. D. Dolgov, M. Kawasaki, \textit{Phys. Lett. B} \textbf{573}, 1-42 (2003).

\bibitem{Faraoni/2006a} V. Faraoni, \textit{Phys. Rev. D} \textbf{74}, 104017 (2006).

\bibitem{Faraoni/2006b} V. Faraoni, \textit{Phys. Rev. D} \textbf{74}, 023529 (2006).

\bibitem{Erick/2006} A. L. Erickcek, T. L. Smith, M. Kamionkowski, \textit{Phys. Rev. D} \textbf{74}, 121501 (2006).


\bibitem{Jamil/2012} M. Jamil et al., \textit{Eur. Phys. J. C} \textbf{72}, 1999 (2012).


\bibitem{Harko/2014b} T. Harko, \textit{Phys. Rev. D} \textbf{90}, 044067 (2014).

\bibitem{Jimenez/2018a} J. B. Jimenez, L. Heisenberg, T. S. Koivisto, \textit{J. Cosmol. Astropart. Phys.} \textbf{2018}, 039 (2018).

\bibitem{Sari/2021} E. N. Saridakis et al., Modified Gravity and Cosmology: An Update by the CANTATA, Springer, (2021).

\bibitem{Sotiriou/2011} T. P. Sotiriou, B. Li, J. D. Barrow, \textit{Phys. Rev. D} \textbf{83}, 104030 (2011).

\bibitem{Li/2011} B. Li, T. P. Sotiriou, J. D. Barrow, \textit{Phys. Rev. D} \textbf{83}, 064035 (2011).


\bibitem{Nesseris/2013} S. Nesseris et al., \textit{Phys. Rev. D} \textbf{88}, 103010 (2013).

\bibitem{Basilakos/2019} F. K. Anagnostopoulos, S. Basilakos, E. N. Saridakis, \textit{Phys. Rev. D} \textbf{100}, 083517 (2019).

\bibitem{Nunes/2016a} R. C. Nunes, S. Pan, E. N. Saridakis, \textit{J. Cosmol. Astropart. Phys.} \textbf{08}, 011 (2016).

\bibitem{Harko/2014a} T. Harko et al., \textit{J. Cosmol. Astropart. Phys.} \textbf{12}, 021 (2014).



\bibitem{Moresco/2012} M. Moresco et al., \textit{J. Cosmol. Astropart. Phys.} \textbf{08}, 006 (2012).

\bibitem{Moresco/2016} M. Moresco et al., \textit{J. Cosmol. Astropart. Phys.} \textbf{05}, 014 (2016).

\bibitem{Bruzual/2003} G. Bruzual, S. Charlot, \textit{Mon. Not. R. Astron. Soc.} \textbf{344}, 1000 (2003).

\bibitem{Maraston/2011} C. Maraston, G. Strömbäck, \textit{Mon. Not. R. Astron. Soc.} \textbf{ 418}, 2785 (2011).

\bibitem{Zhang/2014} C. Zhang et al., \textit{Res. Astron. Astrophys.} \textbf{14}, 1221 (2014). 

\bibitem{Moresco/2015} M. Moresco, \textit{Mon. Not. R. Astron. Soc.} \textbf{450}, L16 (2015).

\bibitem{Stern/2010} D. Stern et al., \textit{J. Cosmol. Astropart. Phys.} \textbf{02}, 008 (2010).

\bibitem{Sharov/2018} G. S. Sharov, V. O. Vasiliev, \textit{Mathematical Modelling and Geometry} \textbf{6}, 1 (2018).

\bibitem{Ratra/2018} H. Yu, B. Ratra, F-Y Wang, \textit{Astrophys. J.} \textbf{856}, 3 (2018).

\bibitem{Holtzman/2008} J. A. Holtzman et al., \textit{Astron. J.} \textbf{136}, 2306 (2008); R. Kessler et al., ApJS. \textbf{185},  32 (2009).

\bibitem{Leaman/2011} J. Leaman et al., \textit{Mon. Not. R. Astron. Soc.} \textbf{412}, 1419 (2011); W. D. Li et al., \textit{Mon. Not. R. Astron. Soc.} \textbf{412}, 1441 (2011).

\bibitem{Folatelli/2010} G. Folatelli et al., \textit{Astron. J.}  \textbf{139}, 120 (2010); G. Folatelli et al., \textit{Astron. J.}  \textbf{139} , 519 (2010).

\bibitem{Copin/2006} Y. Copin et al., \textit{New Astronomy Rev.} \textbf{50}, 436 (2006); R. A. Scalzo et al., \textit{Astrophys. J.}  \textbf{713}, 1073 (2009).

\bibitem{Astier/2006} P. Astier et al., \textit{Astron. Astrophys.} \textbf{447}, 31 (2006); S. Baumont et al., \textit{Astron. Astrophys.} \textbf{491}, 567 (2008).

\bibitem{Riess/2004} A. G. Riess et al., \textit{Astrophys. J.} \textbf{607}, 665 (2004); A. G. Riess et al., \textit{Astrophys. J.} \textbf{659}, 98 (2007).

\bibitem{Visser/2004} M. Visser, \textit{Class. Quantum Grav.} \textbf{21}, 2603 (2004). 

\bibitem{Scolnic/2018} D. M. Scolnic et al., \textit{Astrophys. J.} \textbf{859}, 101 (2018). 

\bibitem{Peebles/1970} P. J. E. Peebles, J. T. Yu, \textit{Astrophys. J.} \textbf{162}, 815 (1970).

\bibitem{Alam/2004} U. Alam, V. Sahni, A. A. Starobinsky, \textit{J. Cosmol. Astropart. Phys.} \textbf{06}, 008 (2004). 

\bibitem{Johnson/2004} J. Johnson et al., \textit{J. Cosmol. Astropart. Phys.} \textbf{0409}, 007 (2004).

\bibitem{Colless/2003} M. Colless et al., arXiv:astro-ph/0306581, (2003).

\bibitem{York/2000} D. G. York et al., \textit{Astrophys. J.} \textbf{120}, 1579 (2000);  M. Tegmark et al., \textit{Phys. Rev. D} \textbf{74}, 123507 (2006).

\bibitem{link} http://www.sdss3.org/dr8/.

\bibitem{Eckart/1940} C. Eckart, \textit{Phys. Rev.} \textbf{58}, 919 (1940).

\bibitem{Okumura/2003} H. Okumura, F. Yonezawa, Physica A: Statistical Mechanics and its Applications, \textbf{321}, 207-219 (2003).

\bibitem{Brevik/2017} I. Brevik et al., \textit{Int. J. Mod. Phys. D} \textbf{26}, 1730024 (2017).

\bibitem{Yousaf/2014} M. Sharif, Z. Yousaf, \textit{J. Cosmol. Astropart. Phys.} \textbf{06}, 019 (2014).

\bibitem{Odintsov/2020} S. D. Odintsov, D. S.-C. Gomez, G.S. Sharov, \textit{Phys. Rev. D} \textbf{101}, 044010 (2020).

\bibitem{Singh/2014} C. P. Singh, P. Kumar, \textit{Eur. Phys. J. C} \textbf{74}, 3070 (2014).

\bibitem{Misner/1968} W. Misner, \textit{Astrophys. J.} \textbf{151}, 431 (1968).


\bibitem{Pady/1987} T. Padmanabhan, S. Chitre, \textit{Phys. Lett. A} \textbf{120}, 433
(1987).

\bibitem{Cheng/1991} B. Cheng, \textit{Phys. Lett. A} \textbf{160}, 329 (1991).

\bibitem{Samanta/2017} G. C. Samanta, R. Myrzakulov, \textit{Chin. J. Phys.} \textbf{55}, 1044 (2017).

\bibitem{Davood/2019} S. Davood Sadatian, \textit{EPL} \textbf{126}, 30004 (2019).

\bibitem{Ren/2006} J. Ren, X. H. Meng, \textit{Phys. Lett. B} \textbf{633}, 1 (2006).

\bibitem{Hiscock/1985} W. A. Hiscock, L. Lindblom, \textit{Phys. Rev. D} \textbf{31}, 725 (1985).

\bibitem{Israel/1976} W. Israel, \textit{Ann. Phy.} \textbf{100}, 310 (1976).

\bibitem{Stewart/1979} W. Israel, J. M. Stewart, \textit{Ann. Phys.} \textbf{118}, 341 (1979).

\bibitem{Titus/2015} A. Sasidharan, T. K. Mathew, \textit{Eur. Phys. J. C.} \textbf{75}, 348 (2015).

\bibitem{Harko/2020a} T. Harko, P. H. R. S. Moreas, \textit{Phys. Rev. D} \textbf{101}, 108501 (2020).

\bibitem{Setare/2013} M. R. Setare, M. J. S Houndjo, \textit{Can. J. Phys.} \textbf{91}, 260-267 (2013).

\bibitem{Rani/2013} M. Sharif, S. Rani, \textit{Mod. Phys. Lett. A} \textbf{27}, 1350118 (2013).

\bibitem{Iver/2012} I. Brevik, \textit{Entropy} \textbf{14}, 2302-2310 (2012).

\bibitem{Mamon/2017} A. A. Mamon, S. Das, \textit{Eur. Phys. J. C} \textbf{77}, 495 (2017).

\bibitem{Garza/2019} J. R. Garza et al., \textit{Eur. Phys. J. C} \textbf{79}, 890 (2019). 

\bibitem{Knop/2003} R. A. Knop et al., \textit{Astrophys. J.} \textbf{598}, 102 (2003).

\bibitem{Ishida/2008} E. E. O. Ishida et al., \textit{Astropart. Phys.} \textbf{28}, 547 (2008).

\bibitem{Cunha/2009} J. V. Cunha, \textit{Phys. Rev. D} \textbf{79}, 047301 (2009).

\bibitem{Rani/2015} S. Rani et al., \textit{J. Cosmol. Astropart. Phys.} \textbf{1503}, 031 (2015).

\bibitem{Barcelo/2002} C. Barcelo, M. Visser, \textit{Int. J. Mod. Phys. D} \textbf{11}, 1553 (2002).

\bibitem{moraes/2017} P. H. R. S. Moraes, P. K. Sahoo, \textit{Eur. Phys. J. C} \textbf{77}, 480 (2017).

\bibitem{Sahni/2003} V. Sahni et al., \textit{J. Exp. Theor. Phys.} \textbf{77}, 201--206, (2003).

\bibitem{Alam/2003} U. Alam et al., \textit{Mon. Not. R. Astron. Soc.} \textbf{344}, 1057, (2003).

\bibitem{Pasqua/2017} A. Pasqua et al., \textit{J. Cosmol. Astropart. Phys.} \textbf{04}, 015 (2017).


\bibitem{Ya/2007} B. Wu Ya et al., \textit{Gen. Relativ. Gravity} \textbf{39}, 653 (2007).

\bibitem{Liu/2008} D. J. Liu, W.Z. Liu, \textit{Phys. Rev. D} \textbf{77}, 027301 (2008).

\bibitem{Zhao/2018} J-Zhao Qi et al., \textit{Res. Astron. Astrophys.} \textbf{18}, 066 (2018).

\bibitem{Shafieloo/2009} A. Shafieloo, V. Sahni, A. A. Starobinsky, \textit{Phys. Rev. D} \textbf{80}, 101301 (2009).

\bibitem{Union2.1 DATA} N. Suzuki et al., \textit{Astrophys. J.} \textbf{746}, 85 (2012).

\bibitem{Ritika/2018} R. Nagpal et al., \textit{Eur. Phys. J. C} \textbf{78}, 946 (2018).


\bibitem{padn/2012} N. Padmanabhan et al., \textit{Mon. Not. Roy. Astron. Soc.} \textbf{427}, 2132 (2012).

\bibitem{6df/2011} F. Beutler et al., \textit{Mon. Not. Roy. Astron. Soc.} \textbf{416}, 3017 (2011).

\bibitem{boss/2014} BOSS collaboration, L. Anderson et al., \textit{Mon. Not. Roy. Astron. Soc.} \textbf{441}, 24 (2014).

\bibitem{wig/2012} C. Blake et al., \textit{Mon. Not. Roy. Astron. Soc.} \textbf{425}, 405 (2012).

\bibitem{waga} M. Vargas dos Santos, R. R. R. Reis, \textit{J. Cosmol. Astropart. Phys.} \textbf{1602}, 066 (2016).

\bibitem{gio/2012} R. Giostri et al., \textit{J. Cosmol. Astropart. Phys.} \textbf{1203}, 027 (2012).


\bibitem{review} S. Nojiri et al., \textit{Phys. Rept.} \textbf{692}, 1 (2017).

\bibitem{Sahoo/2020} P. K. Sahoo, S. Bhattacharjee, \textit{New Astronomy} \textbf{77}, 101351 (2020); R. Zaregonbadi, et al., \textit{Phys. Rev. D} \textbf{94}, 084052 (2016); G. Sun, Y. C. Huang, \textit{Int. J. Mod. Phys. D} \textbf{25}, 1650038 (2016); P. H. R. S. Moraes et al. \textit{J. Cosm. Astrop. Phys.}, \textbf{06}, 005 (2016); P. H. R. S. Moraes, P.K. Sahoo, \textit{Eur. Phys. J. C} \textbf{79}, 677 (2019); E. Elizalde, M. Khurshudyan, \textit{Phys. Rev. D} \textbf{99}, 024051 (2019).

\bibitem{Sne/2020} S. Bhattacharjee, P. K. Sahoo, \textit{Eur. Phys. J. C} \textbf{80}, 289 (2020).




\bibitem{Almada/2020} A. H. Almada et al. \textit{Phys. Rev. D} \textbf{101}, 063516 (2020).

\bibitem{Akarsu/2019} O. Akarsu et al. \textit{Eur. Phys. J. C}  \textbf{79}, 846 (2019).

\bibitem{Suresh/2012} S. Kumar, \textit{Mon. Not. Roy. Astron. Soc.} \textbf{422}, 2532 (2012).

\bibitem{Albert} J. Albert et al. [SNAP Collaboration]: arXiv:0507458; J.
Albert et al. [SNAP Collaboration]: arXiv:0507459. 




\bibitem{Yang/2021} J. Z. Yang et al., \textit{Eur. Phys. J. C} \textbf{81}, 111 (2021).

\bibitem{Najera/2022} A. Najera, A. Fajardo, \textit{J. Cosm. Astrop. Phys.} \textbf{03}, 020 (2022).

\bibitem{Eisenstein/2005} D. J. Eisenstein et al., \textit{Astrophys. J.} \textbf{633}, 560 (2005).

\bibitem{Spergel/2007} D. N. Spergel et al., \textit{Astrophys. J. Suppl. Ser.} \textbf{170}, 377 (2007).

\bibitem{Chang/2019} Z. Chang et al., \textit{Chin. Phys. C} \textbf{43}, 125102 (2019).

\bibitem{Ankan/2016} A. Mukherjee, \textit{Mon. Not. R. Astron. Soc.} \textbf{460}, 273 (2016). 

\bibitem{Mackey/2013} D. F. Mackey et al., \textit{Publ. Astron. Soc. Pac.} \textbf{125}, 306 (2013). 

\bibitem{Farooq/2013} O. Farooq, B. Ratra, \textit{Astrophys. J.} \textbf{766}, L7 (2013).

\bibitem{JV/2008} J. V. Cunha, J. A. S. Lima, \textit{Mon. Not. R. Astron. Soc.} \textbf{390}, 210 (2008).


\bibitem{Christine/2014} C. Gruber,  O. Luongo, \textit{Phys. Rev. D} \textbf{89}, 103506 (2014).


\bibitem{capozziello2011extended} S. Capozziello, M. De Laurentis, \textit{Phys. Rep.} \textbf{509}, 167-321 (2011).

\bibitem{hohmann2021general} M. Hohmann, \textit{Phys. Rev. D} \textbf{104}, 124077 (2021).

\bibitem{dimakis2022flrw} N. Dimakis et al., \textit{Phys. Rev. D} \textbf{106}, 043509 (2022).

\bibitem{DAgostino:2022tdk} R. D'Agostino, R.C. Nunes, \textit{Phys. Rev. D} \textbf{106}, 124053 (2022).

\bibitem{Ferreira:2022jcd} J. Ferreira et al., \textit{Phys. Rev. D} \textbf{105}, 123531 (2022).

\bibitem{ANAGNOSTOPOULOS2021136634} F. K. Anagnostopoulos, S. Basilakos, E. N. Saridakis, \textit{Phys. Lett. B} \textbf{822}, 136634, (2021).

\bibitem{atayde2021can} L. Atayde, N. Frusciante, \textit{Phys. Rev. D} \textbf{104}, 064052 (2021).


\bibitem{alb/2022} I. S. Albuquerque, N. Frusciante, \textit{Phys. Dark Univ.} \textbf{35}, 100980 (2022).

\bibitem{wang/2022} W. Wang, H. Chen, T. Katsuragawa, \textit{Phys. Rev. D} \textbf{105}, 024060 (2022).

\bibitem{esp/2022} F. Esposito et al., \textit{Phys. Rev. D} \textbf{105}, 084061 (2022).

\bibitem{Hassani/2020} F. Hassani, L. Lombriser, \textit{Mon. Not. R. Astron. Soc.} 497, 1885 (2020).
 
\bibitem{Wilson/2023} C. Wilson, R. Bean, \textit{Phys. Rev. D} \textbf{107}, 124008 (2023).

\bibitem{Lee/2023} J. Lee et al., \textit{Astrophys. J.} \textbf{945}, 15 (2023). 

\bibitem{Drozda/2022} P. Drozda et al., \textit{Phys. Rev. D} \textbf{106}, 043513 (2022).

\bibitem{Gupta/2022} S. Gupta et al., \textit{Phys. Rev. D} \textbf{105}, 043538 (2022).

\bibitem{Puchwein/2013} E. Puchwein, M. Baldi, V. Springel, \textit{Mon. Not. R. Astron. Soc.} \textbf{436}, 348 (2013).

\bibitem{Arnold/2016} C. Arnold, V. Springel, E. Puchwein, \textit{Mon. Not. R. Astron. Soc.} \textbf{462}, 1530 (2016).

\bibitem{Giocoli/2018} C. Giocoli, M. Baldi, L. Moscardini, \textit{Mon. Not. R. Astron. Soc.} \textbf{481}, 2813 (2018).

\bibitem{Ruan/2022} C-Z. Ruan et al., \textit{J. Cosmol. Astropart. Phys.} \textbf{05}, 018 (2022).

\bibitem{Zhang/2018} J. Zhang et al., \textit{Phys. Rev. D} \textbf{98}, 103530 (2018).

\bibitem{Huang/2022} Y. Huang et al., \textit{Phys. Rev. D} \textbf{106}, 064047 (2022).

\bibitem{Chen/2023} S. Chen et al., \textit{Astrophys. J.} \textbf{951}, 64 (2023).



\bibitem{Anagnos/2023} K. F. Anagnostopoulos et al., \textit{Eur. Phys. J. C} \textbf{83}, 58 (2023).

\bibitem{Linder/2010} Eric V. Linder et al., \textit{Phys. Rev. D} \textbf{81}, 127301 (2010).

\bibitem{Arora/2022} S. Arora, P.K. Sahoo, \textit{Ann. Phys.}  \textbf{534}, 2200233 (2022).

\bibitem{Reid/2019} M. J. Reid, D. W. Pesce, A. G. Riess, \textit{Astrophys. J. Lett.} \textbf{886}, L27 (2019).

\bibitem{tripp/1998} R. Tripp, \textit{Astron. Astrophys.} \textbf{331}, 815-820, (1998).

\bibitem{Kessler/2017} R. Kessler, D. Scolnic, \textit{Astrophys. J.} \textbf{836}, 56, (2017).

\bibitem{Deng/2018} H. K. Deng, H. Wei, \textit{Eur. Phys. J. C} \textbf{78}, 755 (2018).

\bibitem{Conley/2011} A. Conley et al., \textit{ApJS} \textbf{192}, 1 (2011).

\bibitem{Basilakos/2012} S. Basilakos, A. Pouri, \textit{Mon. Not. R. Astron. Soc.} \textbf{423}, 3761, (2012).


\bibitem{Akaike/1974} H. Akaike, \textit{IEEE Transactions on Automatic Control} \textbf{19}, 716 (1974).

\bibitem{Liddle/2007} A. R. Liddle, \textit{Mon. Not. R. Astron. Soc.} \textbf{377}, L74, (2007).

\bibitem{Mostaghel/2016} B. Mostaghel, H. Moshafi, S. M. S. Movahed, \textit{Eur. Phys. J. C} \textbf{77}, 541, (2017).

\bibitem{Rui/2019} R. An et al., \textit{Mon. Not. R. Astron. Soc.} \textbf{489}, 297 (2019).

\bibitem{Zhang/2019} J. Zhang et al., \textit{Astrophys. J. Lett.} \textbf{875}, L11 (2019).

\bibitem{Zhang/2018a} J. Zhang et al., \textit{Phys. Rev. D} \textbf{98}, 103530 (2018).


\bibitem{Eisenstein/1998} D. J. Eisenstein, W. Hu, \textit{Astrophys. J.} \textbf{496}, 605-614 (1998).

\bibitem{PlanckInf/2020} Y. Akrami et al., \textit{Astron. Astrophys.} \textbf{641}, A10 (2020).

\bibitem{Lewis/2011} A. Lewis, A. Challinor, CAMB: Code for Anisotropies in the Microwave Background, Astrophysics Source Code Library, record ascl:1102.026, (2011).

\bibitem{Lewis/2002} A. Lewis, S. Bridle, \textit{Phys. Rev. D} \textbf{66}, 103511 (2002).

\bibitem{Lewis/2000} A. Lewis, A. Challinor, A. Lasenby, \textit{Astrophys. J.} \textbf{538}, 473-476 (2000).

\bibitem{Howlett/2012} C. Howlett, A. Lewis, A. Hall, \textit{J. Cosmol. Astropart. Phys.} \textbf{1204}, 027 (2012).

\bibitem{Pylians} F. Villaescusa-Navarro, Pylians: Python libraries for the analysis of numerical simulations, Astrophysics Source Code Library, record ascl:1811.008, (2018).

\bibitem{Behroozi/2013} P. S. Behroozi, R. H. Wechsler, H-Yi Wu, \textit{Astrophys. J.} \textbf{762}, 109 (2013).

\bibitem{Potnzen/2013} A. Pontzen et al., pynbody: N-Body/SPH analysis for python, Astrophysics Source Code Library, record ascl:1305.002, (2013).

\bibitem{Francisco/2020} V. N. Francisco et al., ApJS, \textbf{250}, 2 (2020).




\bibitem{h14} D. Stern et al., \textit{J. Cosmol. Astropart. Phys.} \textbf{02}, 008 (2010).

\bibitem{h7} E. Gaztaaga, A. Cabre, L. Hui, \textit{Mon. Not. Roy. Astron. Soc.} \textbf{399}, 1663 (2009).

\bibitem{h13} J. Simon, L. Verde, R. Jimenez, \textit{Phys. Rev. D} \textbf{71},
123001 (2005).

\bibitem{h10} A. Oka et al., \textit{Mon. Not. Roy. Astron. Soc.} \textbf{439}, 2515 (2014).

\bibitem{h6} Y. Wang et al., \textit{Mon. Not. Roy. Astron. Soc.} \textbf{469}, 3762 (2017).

\bibitem{h16} M. Moresco et al., \textit{ J. Cosmol. Astropart. Phys.} \textbf{08},
006 (2012).

\bibitem{h1} C. H. Chuang, Y. Wang, \textit{Mon. Not. Roy. Astron. Soc.} \textbf{435}, 255 (2013).

\bibitem{h15} C. Zhang et al., \textit{Research in Astron. and Astrop.} \textbf{14}, 1221 (2014).

\bibitem{h11} S. Alam et al., \textit{Mon. Not. Roy. Astron. Soc.} \textbf{470},
2617 (2017).

\bibitem{h8} C. Blake et al., \textit{Mon. Not. Roy. Astron. Soc.} \textbf{425}, 405
(2012).

\bibitem{h18} M. Moresco et al., \textit{J. Cosmol. Astropart. Phys.} \textbf{05},
014 (2016).

\bibitem{h19} A.L. Ratsimbazafy et al., \textit{Mon. Not. Roy. Astron. Soc.} \textbf{467}, 3239 (2017).

\bibitem{h2} C. H. Chuang et al., \textit{Mon. Not. Roy. Astron. Soc.} \textbf{433}, 3559 (2013).

\bibitem{h5} L. Anderson et al., \textit{Mon. Not. Roy. Astron. Soc.} \textbf{441}, 24 (2014).

\bibitem{h9} N. G. Busca et al., \textit{Astron. Astrop.} \textbf{552}, A96 (2013).

\bibitem{h12} J. E. Bautista et al., \textit{Astron. Astrophys.} \textbf{603}, A12
(2017).

\bibitem{h4} T. Delubac et al., \textit{Astron. Astrophys.} \textbf{574}, A59
(2015).

\bibitem{h3} A. Font-Ribera et al., \textit{J. Cosmol. Astropart. Phys.} \textbf{05}, 027 (2014).

\bibitem{h17} M. Moresco, \textit{Mon. Not. Roy. Astron. Soc.: Letters.} \textbf{450}, L16 (2015).

\bibitem{Baldi/2010} M. Baldi et al., \textit{Mon. Not. Roy. Astron. Soc.: Letters.} \textbf{403}, 1684 (2010).

\bibitem{Liao/2018} S. Liao, \textit{Mon. Not. Roy. Astron. Soc.: Letters.} \textbf{481}, 3750 (2018).

\bibitem{Polarski/2016} D. Polarski, A. A. Starobinsky, H. Giacomini, \textit{J. Cosmol. Astropart. Phys.} \textbf{12}, 037 (2016).

\bibitem{Xu/2013} L. Xu, \textit{Phys. Rev. D} \textbf{88}, 084032 (2013).

\bibitem{Kazantzidis/2018} L. Kazantzidis, L. Perivolaropoulos, \textit{Phys. Rev. D} \textbf{97}, 103503 (2018).

\bibitem{Ryden} B. Ryden, Introduction to Cosmology, Cambridge University Press, (2017).









 











\end{thebibliography}
\cleardoublepage
\clearpage
\lhead{\emph{Appendices}}
\chapter{Appendices}


\section*{ME-GADGET}
When considering the simulation technique, it is crucial to acknowledge four significant distinctions from the $\Lambda$CDM simulation \cite{Baldi/2010}. Firstly, the expansion rate of the Universe, denoted as $H(a)$, shall be modified accordingly. Secondly, the interaction of energy exchange between dark matter and dark energy, as well as the mass of the simulated particles that represent the energy density of dark matter, ought to be modified in accordance with the scale factor. Thirdly, in the simulation, the particles experience an increase in drag force due to their interaction with the background dark energy field. Lastly, the additional gravitational force caused by the perturbation of the dark energy field can be treated as an effective gravitational constant. As a result, the DM particles in the simulations will experience an additional force, also called the fifth force. The fifth force is caused by the perturbation of DE. Hence, the fifth force can be understood as a modification to the Poisson equation in harmonic space, as computed by the modified CAMB. These four modifications were carried out in the original N-body simulation code GADGET2, resulting in a modified version referred to as ME-GADGET.

\section*{CCVT Algorithm}
The methodology involves initializing a random distribution of $N_p$ particles within the designated region $R$, followed by an iterative relaxation process to achieve a Centroidal Voronoi Tessellation (CCVT) distribution \cite{Liao/2018}. To satisfy the first constraint of the CCVT distribution, we use $N_s$ spatial points to uniformly sample the region $R$ and assign $c = N_s/N_p$ of them to each particle. Subsequently, a fraction of these points is allocated to each particle. Using this approach, an assignment A is generated where each particle is allocated an equal number of spatial locations, denoted as the capacity $c$ for every particle. In the subsequent relaxing phase, the capacity of each particle remains constant while solely altering the specific assignment. It should be noted that in the first assignment, spatial locations can be arbitrarily allocated to particles, provided that each particle receives an equal capacity. In order to accelerate the process of relaxing, it is advisable to employ an initial assignment that is compact based on the distance between particles in spatial points.

Empirical evidence has demonstrated that the CCVT distribution is associated with the minimal value of the energy function, that is 
\begin{equation}
E = \sum_{i \in N_{s}}|x_{i} - r_{j=A(i)}|^{2},
\end{equation}
where $r_{j}$ and $x_{i}$ are the coordinates of the $j-th$ particle and $i-th$ spatial point respectively. The assignment function $A(i)$ tells us the index of the particle that the $i-th$ spatial point is assigned to.
Hence, the relaxing process can be understood as minimizing the energy function associated with the system comprising particles and spatial points. It should be noted that the energy function computation can readily accommodate the periodic boundary condition. Consequently, the CCVT configuration possesses periodicity and can be seamlessly employed in cosmological simulations. 

\section*{Tools of Large-Scale Structure} 
This section describes the theoretical frameworks that provide the link between the current cosmological model and observed structure in the present day Universe \cite{Ryden}.
Consider some component of the Universe whose energy density $\rho(\vec{r},t)$ is a function of position as well as time. At a given $t$, the spatially averaged energy density is 
\begin{equation}
\overline{\rho}(t) = \frac{1}{V}\int_{V} \rho(\vec{r},t) d^{3}r.
\end{equation}
The volume $V$ over which we are averaging must be large compared to the size of the biggest structure in the Universe. It is useful to define a dimensionless density fluctuations
\begin{equation}
\delta(\vec{r},t)= \frac{\rho(\vec{r},t) - \overline{\rho}(t)}{\overline{\rho}(t)}.
\end{equation}
The value of $\delta$ is thus negative in underdense regions and positive in overdense regions.

An attempt to describe the physics of structure formation is concerned with the statistical rather than the individual properties of the distribution of density perturbations. We can expand the dimensionless density enhancement at some position $\vec{r}$ as a three-dimensional Fourier series defined as 
\begin{equation}
\delta(\vec{r})= \frac{V}{(2\pi)^{3}}\int \delta_{\vec{k}} e^{-i\vec{k}.\vec{r}} d^{3} k,
\end{equation}
where the individual Fourier components $\delta_{\vec{k}}$ are given by 
\begin{equation}
\delta_{\vec{k}} = \frac{1}{V} \int \delta(\vec{r})e^{i\vec{k}.\vec{r}} d^{3} k.
\end{equation}
When conducting the Fourier transform, the function $\delta(\vec{r})$ is decomposed into an infinite number of sine waves, each with comoving wavenumber $\vec{k}$ and comoving wavelength $\lambda = \frac{2\pi}{k}$.  When $\vert \delta_{\vec{k}}\vert \ll 1$, each Fourier component obeys the perturbation equation. In this scenario, the perturbation equation for gravity governed by the function $f(Q)$ is employed.

The power spectrum is defined as the mean square of the Fourier components, that is 
\begin{equation}
P(k) = \langle \vert \delta_{\vec{k}}\vert^{2} \rangle,
\end{equation}
where the average is taken over all possible orientations of the wavenumber $\vec{k}$. 

The most effective approach for identifying low-amplitude overdensities on a spatial scale of 160\,Mpc involves analyzing the correlation function of galaxies. The correlation function $\xi(r)$ is a quantitative measure indicating the expected number of probable galaxies to be observed inside a certain infinitesimal volume $dV$. The expected $dN$ is given by the relation
\begin{equation}
dN = n_{g}\left[1+\xi(r)\right]dV,
\end{equation}
where $n_{g}$ is the average number of galaxies at the present day. So, the correlation function $\xi(r)$ is the Fourier transform of the power spectrum $P(k)$ 
\begin{equation}
\xi(r) = \frac{V}{(2\pi)^{3}} \int P(k) e^{-i\vec{k}.\vec{r}} d^{3} k.
\end{equation}

\section*{Redshift-Space Distortion Effect}

Redshift-space distortions (RSD) are a phenomenon in observational cosmology where the spatial distribution of galaxies appears distorted when their positions are viewed as a function of their redshift in contrast to their distances \cite{Polarski/2016,Xu/2013,Kazantzidis/2018}. The distribution of galaxies in observations is known to deviate from that in real space due to the peculiar velocity of each galaxy, resulting in the redshift space distortion (RSD) phenomenon. Since we cannot calculate
the percentage of the measured velocity that arises from the Hubble flow or from the peculiar velocity, the resulting distance measurement becomes inaccurate by a certain value, denoted as $\Delta D$, that is 
\begin{equation}
D = \frac{v_{Hubble} + v_{pec}}{H_{0}} = D_{real} + \Delta D.
\end{equation}
Therefore, it is necessary to examine the disparity in the power spectrum between actual space and redshift space in both three-dimensional (3D) and two-dimensional (2D) contexts. The Pylians python package is employed to compute the overdensity of dark matter, both with and without redshift-space distortions (RSD). 

\section*{Datasets}

\begin{table}[H]
\begin{center}
\begin{tabular}{|c|c|c|c|}
\hline
$z$ & $H(z)$ & $\sigma _{H}$ & Ref. \\  \hline
 $0.24$ & $79.69$ & $2.99$ & \cite{h7} \\ 
 $0.30$ & $81.7$ & $6.22$ & \cite{h10} \\ 
 $0.31$ & $78.18$ & $4.74$ & \cite{h6} \\ 
 $0.34$ & $83.8$ & $3.66$ & \cite{h7} \\ 
 $0.35$ & $82.7$ & $9.1$ & \cite{h1} \\ 
 $0.36$ & $79.94$ & $3.38$ & \cite{h6}\\ 
 $0.38$ & $81.5$ & $1.9$ & \cite{h11} \\ 
 $0.40$ & $82.04$ & $2.03$ & \cite{h6}\\ 
 $0.43$ & $86.45$ & $3.97$ & \cite{h7} \\ 
 $0.44$ & $82.6$ & $7.8$ & \cite{h8} \\ 
 $0.44$ & $84.81$ & $1.83$ & \cite{h6}\\
 $0.48$ & $87.79$ & $2.03$ & \cite{h6}\\ 
 $0.51$ & $90.4$ & $1.9$ & \cite{h11}\\ 
 $0.52$ & $94.35$ & $2.64$ & \cite{h6} \\ 
 $0.56$ & $93.34$ & $2.3$ & \cite{h6} \\ 
 $0.57$ & $87.6$ & $7.8$ & \cite{h2} \\ 
 $0.57$ & $96.8$ & $3.4$ & \cite{h5}\\ 
 $0.59$ & $98.48$ & $3.18$ & \cite{h6}\\ 
 $0.60$ & $87.9$ & $6.1$ & \cite{h8} \\ 
 $0.61$ & $97.3$ & $2.1$ & \cite{h11} \\ 
 $0.64$ & $98.82$ & $2.98$ & \cite{h6}\\ 
 $0.73$ & $97.3$ & $7.0$ & \cite{h8}\\
 $2.30$ & $224$ & $8.6$ & \cite{h9} \\ 
 $2.33$ & $224$ & $8$ & \cite{h12} \\ 
 $2.34$ & $222$ & $8.5$ & \cite{h4} \\ 
 $2.36$ & $226$ & $9.3$ & \cite{h3} \\ 
\hline
\end{tabular}
\caption*{Details of $H(z)$ dataset: $26$ points of Hubble parameter values $H(z)$ with errors $\sigma _{H}$ from the BAO and other methods}
\end{center}
\end{table}

\begin{table}[H]
\begin{center}
\begin{tabular}{|c|c|c|c|}
\hline
$z$ & $H(z)$ & $\sigma _{H}$ & Ref. \\  \hline
$0.070$ & $69$ & $19.6$ & \cite{h14} \\ 
$0.90$ & $69$ & $12$ & \cite{h13} \\ 
$0.120$ & $68.6$ & $26.2$ & \cite{h14}\\ 
$0.170$ & $83$ & $8$ & \cite{h13} \\ 
$0.1791$ & $75$ & $4$ & \cite{h16} \\ 
$0.1993$ & $75$ & $5$ & \cite{h16}\\ 
$0.200$ & $72.9$ & $29.6$ & \cite{h15} \\ 
$0.270$ & $77$ & $14$ & \cite{h13}\\ 
$0.280$ & $88.8$ & $36.6$ & \cite{h15} \\ 
$0.3519$ & $83$ & $14$ & \cite{h16} \\ 
$0.3802$ & $83$ & $13.5$ & \cite{h18} \\
$0.400$ & $95$ & $17$ & \cite{h13} \\ 
$0.4004$ & $77$ & $10.2$ & \cite{h18} \\ 
$0.4247$ & $87.1$ & $11.2$ & \cite{h18} \\ 
$0.4497$ & $92.8$ & $12.9$ & \cite{h18} \\ 
$0.470$ & $89$ & $34$ & \cite{h19} \\ 
$0.4783$ & $80.9$ & $9$ & \cite{h18}\\ 
$0.480$ & $97$ & $62$ & \cite{h14}\\ 
$0.593$ & $104$ & $13$ & \cite{h16} \\ 
$0.6797$ & $92$ & $8$ & \cite{h16} \\ 
$0.7812$ & $105$ & $12$ & \cite{h16}\\ 
$0.8754$ & $125$ & $17$ & \cite{h16}\\
$0.880$ & $90$ & $40$ & \cite{h14} \\ 
$0.900$ & $117$ & $23$ & \cite{h13} \\ 
$1.037$ & $154$ & $20$ & \cite{h16} \\ 
$1.300$ & $168$ & $17$ & \cite{h13} \\ 
$1.363$ & $160$ & $33.6$ & \cite{h17} \\ 
$1.430$ & $177$ & $18$ & \cite{h13}  \\ 
$1.530$ & $140$ & $14$ & \cite{h13} \\
$1.750$ & $202$ & $40$ & \cite{h13}  \\
$1.965$ & $186.5$ & $50.4$ & \cite{h17} \\ \hline
\end{tabular}
\caption*{Details of $H(z)$ dataset: $31$ points of Hubble parameter values $H(z)$ with errors $\sigma _{H}$ from the differential age method}
\end{center}
\end{table}

\begin{table}[H]
\begin{tabular}{|c|c|c|c|c|c|c|} 
\hline
$z_{BAO}$ & $0.106$ & $0.2$ & 0.35 & $0.44$ & $0.6$ & $0.73$  \\  \hline
$\frac{d_{A}(z_*)}{D_{v}(z)}$ & $30.95 \pm 1.46$ & $17.55 \pm 0.60$ & $10.11 \pm 0.37$ & $8.44 \pm 0.67$ & $6.69 \pm 0.33$ & $5.45 \pm 0.31$ \\ \hline
\end{tabular}
\caption*{Measurements of BAO observable for MCMC sampling procedure}
\end{table}

\cleardoublepage
\pagestyle{fancy}

\label{Publications}
\lhead{\emph{List of Publications}}

\chapter{List of Publications}
\section*{Thesis Publications}
\begin{enumerate}

\item \textbf{Simran Arora}, Xin-he Meng, S.K.J. Pacif, P.K. Sahoo, \textit{Effective equation of state in modified gravity and observational constraints}, \textcolor{blue}{Classical and Quantum Gravity} \textbf{37}, 205022 (2020).

\item Oleksii Sokoliuk, \textbf{Simran Arora}, Subhrat Praharaj, Alexander Baransky, P.K. Sahoo, \textit{On the impact of $f(Q)$ gravity on the large scale structure}, \textcolor{blue}{Monthly Notices of the Royal Astronomical Society} \textbf{522}, 252-267 (2023).

\item \textbf{Simran Arora}, S.K.J Pacif, Snehasish Bhattacharjee, P.K Sahoo, \textit{$f(Q,\mathcal{T})$ gravity models with observational constraints}, \textcolor{blue}{Physics of the Dark Universe} \textbf{30}, 100664 (2020).

\item \textbf{Simran Arora}, A Parida, P.K. Sahoo, \textit{Constraining effective equation of state in $f(Q,\mathcal{T})$ gravity}, \textcolor{blue}{European Physics Journal C} \textbf{81}, 555 (2021).
\end{enumerate}

\section*{Other Publications}
\begin{enumerate}

\item \textbf{Simran Arora}, A. Bhat, P. K. Sahoo, \textit{Squared torsion $f(T,\mathcal{T})$ gravity and its cosmological implications,} \textcolor{blue}{Fortschritte der Physik-Progress of Physics} \textbf{71}, 2200162 (2023).

\item G. N. Gadbail, \textbf{Simran Arora}, P. K. Sahoo, \textit{Reconstruction of $f(Q, \mathcal{T})$ Lagrangian for various cosmological scenario,} \textcolor{blue}{Physics Letters B} \textbf{838}, 137710 (2023).

\item G. N. Gadbail, \textbf{Simran Arora}, P. K. Sahoo, \textit{Cosmology with viscous generalized Chaplygin gas in $f(Q)$ gravity,} \textcolor{blue}{Annals of Physics} \textbf{451}, 169269 (2023).

\item G. N. Gadbail, \textbf{Simran Arora}, P. K. Sahoo, \textit{Dark energy constraint on equation of state parameter in the Weyl type $f(Q,\mathcal{T})$ gravity,} \textcolor{blue}{Annals of Physics} \textbf{451}, 169244 (2023).

\item M. Koussour, \textbf{Simran Arora}, D. J. Gogoi, M. Bennai, P. K. Sahoo, \textit{Constant sound speed and its thermodynamical interpretation in $f(Q)$ gravity,} \textcolor{blue}{Nuclear Physics B} \textbf{990}, 116158 (2023).

\item \textbf{Simran Arora}, P. K. Sahoo, \textit{Crossing Phantom Divide in $f(Q)$ Gravity,} \textcolor{blue}{Annalen der Physik} \textbf{534}, 2200233 (2022).

\item \textbf{Simran Arora}, S. Mandal, S. Chakraborty, G. Leon, P. K. Sahoo, \textit{Can f (R) gravity isotropize a pre-bounce contracting universe?,} \textcolor{blue}{Journal of Cosmology and Astroparticle Physics} \textbf{09}, 042 (2022).

\item \textbf{Simran Arora}, S. K. J. Pacif, A. Parida P. K. Sahoo, \textit{Bulk viscous matter and the cosmic acceleration of the universe in $f(Q,\mathcal{T})$ gravity,} \textcolor{blue}{Journal of High Energy Astrophysics} \textbf{33}, 1-9 (2022).

\item Tee-How Loo, Avik De, \textbf{Simran Arora}, P. K. Sahoo, \textit{Impact of curvature based geometric constraints on $F(R)$ theory,} \textcolor{blue}{European Physical Journal C} \textbf{82}, 705 (2022).

\item Avik De, \textbf{Simran Arora}, U. C. De, P. K. Sahoo, \textit{A complete study of conformally flat pseudo-symmetric spacetimes in the theory of $F(R)$ gravity,} \textcolor{blue}{Results in Physics} \textbf{32}, 105053 (2022).

\item G. N. Gadbail, \textbf{Simran Arora}, P. K. Sahoo, \textit{Generalized Chaplygin gas and accelerating universe in $f(Q,\mathcal{T})$ gravity,} \textcolor{blue}{Physics of the Dark Universe} \textbf{37}, 101074 (2022).

\item S. K. J Pacif, \textbf{Simran Arora}, P. K. Sahoo, \textit{Late-time acceleration with a scalar field source: Observational constraints and statefinder diagnostics,} \textcolor{blue}{Physics of the Dark Universe} \textbf{32}, 100804 (2021).

\item \textbf{Simran Arora}, J. R. L. Santos, P. K. Sahoo,  \textit{Constraining $f(Q,\mathcal{T})$ gravity from energy conditions,} \textcolor{blue}{Physics of the Dark Universe} \textbf{31}, 100804 (2021).

\item Avik De, Tee-How Loo, \textbf{Simran Arora}, P. K. Sahoo, \textit{Energy conditions for a $(WRS)_4$ spacetime in $F(R)$ gravity,} \textcolor{blue}{The European Physical Journal Plus} \textbf{136}, 218 (2021).

\item G. N. Gadbail, \textbf{Simran Arora}, P. K. Sahoo, \textit{Viscous cosmology in the Weyl-type $f(Q,\mathcal{T})$ gravity,} \textcolor{blue}{European Physics Journal C} \textbf{81}, 1088 (2021).

\item G. N. Gadbail, \textbf{Simran Arora}, P. K. Sahoo, \textit{Power law cosmology in Weyl-type $f(Q,\mathcal{T})$ gravity,} \textcolor{blue}{The European Physical Journal Plus} \textbf{136}, 1040 (2021).

\item P. K. Sahoo, Sanjay Mandal, \textbf{Simran Arora}, \textit{Energy conditions in non-minimally coupled $f(R,\mathcal{T})$ gravity,} \textcolor{blue}{Astronomische Nachrichten} \textbf{342}, 89-95 (2021).

\item R. Solanki, \textbf{Simran Arora}, P. K. Sahoo, P. H. R. S. Moraes, \textit{Bulk viscous fluid in symmetric teleparallel cosmology: theory versus experiment,} \textcolor{blue}{Universe} \textbf{9}, 12 (2021).

\item \textbf{Simran Arora}, P. K. Sahoo, \textit{Energy conditions in $f(Q,\mathcal{T})$ gravity,} \textcolor{blue}{Physica Scripta} \textbf{95}, 095003 (2020).

\item \textbf{Simran Arora}, S. Bhattacharjee, P. K. Sahoo,  \textit{Late-time viscous cosmology in $f(R,\mathcal{T})$ gravity,} \textcolor{blue}{New Astronomy} \textbf{82}, 101452 (2020).

\end{enumerate}
\cleardoublepage
\pagestyle{fancy}
\lhead{\emph{Biography}}

\chapter{Biography}

\section*{Brief Biography of the Candidate:}
\textbf{Ms. Simran Arora} received her Bachelor's degree in 2016 and her Master's degree in 2018 in Mathematics from Kirori Mal College, University of Delhi, New Delhi. She received an academic award for obtaining the first rank in her Bachelor's. She has experience as the Assistant Professor at Hindu College of Engineering (Deemed to University), Haryana. She passed the Council of Scientific and Industrial Research (CSIR), the National Eligibility test (NET) for assistant professor and lecturership in June 2019 (All India Rank-14), and the Graduate Aptitude Test in Engineering (GATE) 2018 \& 2019. She received the ``Young Relativist Award" at the International Conference of Differential Geometry and Relativity in collaboration with the Tensor Society, 2022, Department of Mathematics, Kuvempu University, Shivamoga-Karnataka, India. She also participated in the thematic six-week program at Institut Henri Poincare, Paris, France, entitled ``Quantum gravity, Random Geometry and Holography" from January 09 - February 17, 2023. She has published 24 research articles in renowned international journals during her Ph.D. research career. She has presented her research at several National and International conferences.

\section*{Brief Biography of the Supervisor:}

\textbf{Prof. Pradyumn Kumar Sahoo} has more than 20 years of immense research experience in the field of Applied Mathematics, Cosmology, General Theory of Relativity, Modified Theories of Gravity, and Astrophysical Objects. He received his Ph.D. degree from Sambalpur University, Odisha, India in 2004. He joined the Department of Mathematics, BITS Pilani, Hyderabad Campus, as an Assistant Professor in 2009 and is currently a Professor. He also holds the position of Head of the Department since October 2020. He has been awarded with visiting professor fellowship at Transilvania University of Brasov, Romania. He has been placed among the top 2\% of scientists worldwide, according to a survey by researchers from
Stanford University, USA, in Nuclear and Particle Physics published in a reputed international journal. His study has resulted in 200 scientific articles published in reputable journals. He is guiding 17 Ph.D. students as supervisor and co-supervisor (4 completed and 13 Ongoing) and many M.Sc. theses.

Prof. Sahoo is also a COST (CA21136) member: Addressing observational tensions in cosmology with systematics and fundamental physics. As a visiting scientist, he got an opportunity to visit The European Organization for Nuclear Research CERN, Geneva, Switzerland, a well-known research centre for scientific research. He has participated in many
international and national conferences, most of which he has presented his work as an invited speaker. He has various research collaborations at both the national and international levels. He contributed to BITS through five sponsored research projects from University Grants
Commission (UGC 2012-2014),  DAAD-Research Internships in Science and Engineering (RISE) Worldwide (2018, 2019 and 2023), Council of Scientific and Industrial Research (CSIR 2019-2022), National Board for Higher Mathematics (NBHM 2022-2025), Science and Engineering Research Board (SERB), Department of Science and Technology (DST 2023-2026). He is an expert reviewer of Physical Science Projects, SERB, DST, Govt. of India, and University Grants Commission (UGC) research schemes. He serves today's research society as an Editorial board member of various reputed journals.

\end{document}